\documentclass[authoryear,12pt]{elsarticle}      
\usepackage[english]{babel}
\usepackage{amsmath}
\usepackage[toc,page]{appendix}
\usepackage{amssymb}        
\usepackage{amsthm}         
\usepackage{natbib}         
\usepackage{graphicx}       
\usepackage{subfigure}      
\usepackage{color}

\journal{Advances in Applied Mechanics}

\begin{document}

\begin{frontmatter}
\title{Computational Homogenization of Polycrystals}
\author{Javier Segurado$^{1, 2}$}
\author{Ricardo A. Lebensohn$^{3}$}
\author{Javier LLorca$^{1, 2, }$\corref{cor1}}

\address{$^1$ IMDEA Materials Institute, 28906, Getafe, Madrid, Spain. \\  $^2$ Department of Materials Science, Polytechnic University of Madrid, 28040 - Madrid, Spain. \\ $^3$ Los Alamos National Laboratory, Los Alamos, NM 87545, USA.}
\cortext[cor1]{Corresponding author}

\begin{abstract}

This paper reviews the current state-of-the-art in the simulation of the mechanical behavior of polycrystalline materials by means of computational homogenization. The key ingredients of this modelling strategy are presented in detail starting with the parameters needed to describe polycrystalline microstructures and the digital representation of such microstructures in a suitable format to perform computational homogenization. The different crystal plasticity frameworks that can describe the physical mechanisms of deformation in single crystals (dislocation slip and twinning) at the microscopic level are presented next. This is followed by the description of computational homogenization methods based on mean-field approximations by means of the viscoplastic self-consistent approach, or on the full-field simulation of the mechanical response of a representative polycrystalline volume element by means of the finite element method or the fast Fourier transform-based method. Multiscale frameworks based on the combination of mean-field homogenization and the finite element method are presented next to model the plastic deformation of polycrystalline specimens of arbitrary geometry under complex mechanical loading. Examples of application  to predict the strength, fatigue life, damage, and texture evolution under different conditions are presented to illustrate the capabilities of the different models. Finally, current challenges and future research directions in this field are summarized.

\end{abstract}


\end{frontmatter}
\clearpage

\section{Introduction}

Physically-based modelling of the mechanical behavior of materials is an integral part of Materials Science and Engineering, as an indispensable tool to understand the relationship between microstructure and properties. However, the complexity of the deformation mechanisms that need to be represented, in addition to limitations in computational resources, have until recently restricted the goal of these formulations to provide qualitative trends or to complement phenomenological models that need to be adjusted from experimental data obtained under different conditions. Recent advances in simulation techniques, computational power, and multiscale modelling strategies are rapidly overcoming these limitations. These new paradigms are aimed at developing formulations that can quantitatively predict relationships between processing, microstructure and properties, and contribute to the design of materials with a given set of properties {\it in silico}, before they are manufactured in the laboratory. Successful examples of the latter can be found in the realm of metallic alloys \citep{ALW06, O13} or structural composites \citep{LGM11, GVM17}, and the roadmaps for implementation of these strategies in different industrial sectors are clearly delineated \citep{ICME13}. Although the number of success stories is still limited, they show a dramatic reduction in the time necessary to develop new materials with optimized properties for specific applications, as well as a large return-of-investment. Thus, there is a manifest interest in both the scientific community and industry to improve and develop new multiscale modelling strategies.

Within this scenario, homogenization theory stands out as one of the most important tools to relate microstructure with effective properties in heterogeneous materials \citep{NH93, T01}. Homogenization theory assumes that the characteristic length scale of the microscopic domains (the average size of the heterogeneities) is much larger than the molecular dimensions (and, thus, continuum mechanics is applicable), and much smaller than the characteristic length scale of the macroscopic sample. Under these conditions, the macroscopic or effective properties of the material can be determined from geometrical features of the microstructure and properties of the different heterogeneities in the material. Following the pioneer work of \cite{E57}, different homogenization mean-field formulations were developed first for linear materials \citep{K58, HS63, H65, MT73}. These methods were later extended to deal with non-linear mechanical behaviors (nonlinear elasticity, plasticity and viscoplasticity) by means of different linearization approximations \citep{H65b, BZ79, TW88, P91, P96}.

Among these mean-field approaches, the elastic self-consistent (SC) scheme \citep{H54} arises as the most reliable approach to obtain estimations of the elastic response of polycrystals \citep{H65}. This approximation assumes that each domain in a set that statistically represents the heterogeneous material is embedded in a homogenous effective medium whose elastic properties are not known {\it a priori}, but need to be obtained as an average over those heterogeneities. The linear elastic SC formulation was extended to non-linear viscoplastic polycrystals by \cite{MCA87}, taking into account that plastic deformation in each grain is accommodated by dislocation glide, resulting in local mechanical responses determined by the crystallography and lattice orientation of each grain. This methodology provides a robust framework to establish physically-based relationships between texture, microstructure and mechanical properties of plastically deforming polycrystals, accounting for their evolving microstructure and the anisotropy of the single crystal grains. The viscoplastic self-consistent (VPSC) approximation was later improved by \cite{LT93, LTP07}, and the VPSC formulation is nowadays widely used to simulate the mechanical behavior of polycrystalline aggregates.

However, homogenization strategies based on mean-field approximations have two main limitations. The first one is that these approaches are based on a description of the microstructure based on average values of grain sizes, shapes and orientations, and therefore cannot take into account the effects of local heterogeneities such as clustering of grains with different size, shape, or orientation, or to represent a given misorientation distribution. While these local microstructure features has often negligible effects on properties that depend on the average values of the micromechanical fields (like, for instance, the elastic modulus), this is not the case for properties that depend on the extremal values (e.g. damage) \citep{SGL03, SL04, SL05}. The second limitation arises from assuming that the micromechanical fields are constant in each phase/grain. Under these circumstances, it is difficult to simulate phenomena in which these fields are localized in narrow bands in one or several phases/grains, as a result of plastic deformation, damage progression, low strain rate sensitivity (i.e. high non-linearity), strong mechanical contrasts, etc. \citep{GSL04, GL07, TGL08}. More sophisticated non-linear homogenization methods have been developed in recent years, using linearization schemes at phase/grain level that also incorporate information on the second moments of the field fluctuations in each phase/grain, to ameliorate this second limitation, at the cost of more complex and expensive numerical algorithms \citep{BP95, LP04}.

These limitations of mean-field homogenization approaches can be overcome by means of full-field homogenization. Using this strategy, the effective properties of the polycrystal are determined by means of the full-field solution of a boundary value problem of a microstructural Representative Volume Element (RVE) under homogeneous boundary conditions. It has been established that the success and accuracy of computational homogenization approaches for polycrystals rely on the adequacy of three factors: the representation of the microstructure, the constitutive description of the single crystal behavior, and the numerical approach to solve the boundary value problem. From the RVE perspective, the first requirement is that the size of the RVE has to be large enough to provide an accurate statistical representation of the polycrystal's microstructure, as well as to lead to effective properties that are independent of the RVE dimensions. Obviously, the larger the RVE and the more complex the single crystal constitutive behavior, the higher the computational cost and, thus, fast and efficient numerical procedures to solve the boundary value problem are essential.

Full-field homogenization has been traditionally carried out using the finite element method (FEM) to solve the governing partial-differential equations (PDEs) of micromechanics, in combination with the CP formalism developed by \cite{PAN82}. This model takes into account the process of plastic deformation by crystallographic slip and provides an accurate and physically-based representation of this phenomenon at the microscopic level within each grain \citep{B91, MSS99, DRC00, RSZ01, BEK01}. Alternatively, \cite{MS94} proposed a spectral formulation based on the efficient Fast Fourier Transform (FFT) algorithm to perform full-field homogenization of a periodic RVE. The governing PDEs were solved in this case by means of the computation of convolution integrals involving the periodic Green's function associated with the displacement (or velocity) field of an homogeneous linear reference material. This strategy was initially applied to composite materials  \citep{MS98, EM99, MMS00}, in which the source of heterogeneity is related to the spatial distribution of phases with different mechanical properties. It was later adapted to polycrystals, in which the heterogeneity is related to the spatial distribution of crystals with anisotropic mechanical properties \citep{L01, LCB05, LBC08, LMM09}. It should also be noted that grain boundaries (GBs) of very different character are also ubiquitous in polycrystals and have to be modelled appropriately.

As a consequence of the above developments, simulation of the mechanical behavior of polycrystals has significantly improved in the last 3 decades and mature strategies based on either mean-field or full-field homogenization approaches are nowadays available. The state-of-the-art of Computational Homogenization of Polycrystals (CHP) is summarized in this review, which is structured as follows. Section 2 is focused on the description of polycrystalline microstructures, including the experimental techniques available for their determination, and the digital representation of microstructures in a suitable format to perform the computational homogenization. Section 3 presents the Crystal Plasticity (CP) framework used by the different homogenization methods to describe the plastic deformation of each single crystal within the polycrystal. Both phenomenological and physically-based models are described, as both are of interest for different applications. A detailed description of the Green's function-based VPSC homogenization approach is given in Section 4, while full-field homogenization strategies based on the numerical solution of a boundary value problem of an RVE using either the FEM or FFTs are presented in sections 5.1 and 5.2, respectively. While a detailed description of CPFEM can be found elsewhere \citep{Roters20101152, REB10} and, therefore, is omitted in this review, the efficient FFT-based formulation is described in detail. In particular,  the underlying commonalities and differences between both Green's function-based CP formulations (VPSC and FFT-based models) are presented with an unified notation.
Furthermore, section 6 presents the formulation of multiscale frameworks for modeling the plastic deformation of polycrystalline specimens of arbitrary geometry under complex mechanical loading. These frameworks are based on the combination of polycrystal homogenization and FEM, by connecting the single crystal (microscopic) behavior with the specimen (macroscopic) response through an intermediate (mesoscopic) scale. The behavior at the mesoscopic scale is obtained from the homogenized response of a polycrystal, that represents each (polycrystalline) material point in the FE mesh. A few applications of the described models are presented in Section 7, showing the capabilities of the different simulation strategies. These examples illustrate virtual design strategies, in which the mechanical response of  polycrystalline microstructures can be accurately computed and optimized {\it in silico} before the material is actually manufactured. Finally, potential new applications of CHP are summarized in the last section, together with current limitations and topics that should be addressed in the future.

Throughout this paper, vectors are denoted by bold lowercase roman letters $\mathbf{a}$, second-rank tensors by bold capital roman letters or bold Greek letters ($\mathbf{A}$ or $\boldsymbol{\sigma}$) and fourth-rank tensors by $\mathbb{A}$. In addition, bold capital roman ($\mathbf{E}$) and greek letters ($\mathbf{\Omega}$) are used to express volume-average vectors, such as stress, strain or velocities.  A Cartesian coordinate system is used with respect to the orthonormal basis $\left(\mathbf{e}_{1},\mathbf{e}_{2},\mathbf{e}_{3}\right)$. The notations for tensor product, contraction and double contraction products are: $ \mathbf{a}\otimes\mathbf{b}=a_ib_j\mathbf{e}_{i}\otimes\mathbf{e}_{j}; \mathbf{A}\mathbf{B}=A_{ik}B_{kj}\left(\mathbf{e}_{i}\otimes\mathbf{e}_{j}\right) $;  $ \mathbf{a} \cdot \mathbf{b}=a_ib_i$ and 
$\mathbf{A}:\mathbf{B=}A_{ij}B_{ij} $. Finally $\mathbf{1}$ and $\mathbb{I}$ stand for the second- and fourth-rank identity tensors, respectively.

\section{Representation of the microstructure}

The first step to perform computational homogenization is the generation of synthetic RVEs of the microstructure. The definition of an RVE can be approached from two different perspectives. On the one hand, it can be defined as the minimum volume of the microstructure whose properties (obtained using either homogeneous traction or displacement boundary conditions) are equal to the effective properties of the heterogeneous solid. On the other hand, the RVE can be defined as a the minimum volume of the microstructure whose statistical descriptors (that define the features of the heterogeneous microstructure) are equivalent to those found in the heterogeneous solid. Another important concept is that of Statistical Representative Volume Element (SRVE) \citep{JKF04}. The SRVE does not fulfill one or both conditions imposed to the RVE by itself, but it does fulfill them in a statistical sense: the effective properties and statistical descriptors of the microstructure obtained by averaging the results obtained for many SRVE are equivalent to those obtained for an RVE. The use of SRVE is sometimes very useful in computational homogenization because the simulation of very large RVEs is computationally much more expensive than that of many smaller SRVEs.

\subsection{Microstructural parameters}

Polycrystals are aggregates of single crystal grains, which are the building blocks of the RVE. Crystals are modelled as homogeneous polyhedra with well-defined crystallographic orientations separated by grain boundaries. In addition, internal discontinuities in the lattice of a grain due to the presence of twinned regions can also be included, due to the importance of these features to reproduce the mechanical response at the local level (for instance, in the nucleation of fatigue cracks). The main features to generate an RVE of a polycrystalline microstructure are the grain size and shape distribution, together with the Orientation Distribution Function (ODF) that defines the crystallographic texture of the polycrystal. In addition, RVEs can include local microstructure heterogeneities or specific grain misorientation distributions.

Optical microscopy of polished sections of the microstructure is the standard technique to measure the grain size and shape distribution. Chemical etching can be used to reveal the grain boundaries and the statistical parameters that characterize the grain distribution can be automatically obtained using image-analysis software tools. Of course, the 2-D information obtained has to be transformed into 3-D and this operation may or may not be simple, depending on the microstructural features. For instance, there are algorithms available that can infer the 3-D grain size distribution from the data obtained in 2-D for equiaxed grain distributions \citep{HB98}. Similarly, grain shapes can be accurately obtained from 2-D images in different orientations in the case of materials with well-defined anisotropy, as it is the case of rolled and extruded sheets. From the viewpoint of grain orientation, the ODF can be easily obtained from X-ray diffraction using texture goniometers. The information obtained can be directly used to model grain orientation distribution in the RVE.

More sophisticated methods have been developed in recent years to provide a more accurate quantification of the microstructure in polycrystals. Automated serial sectioning by means of a microtome \citep{AV01} or mechanical milling \citep{SMP03, S06} can be used in combination with software tools to build the 3-D microstructure and extract microstructural features, to determine grain size and shape distributions in 3-D. If higher resolution is required, sequential milling can be carried out using a focus ion beam or a femtosecond laser within a scanning electron microscope \citep{UGD06, EHN11}. One additional advantage of this latter approach is that 3-D chemical and crystallographic information can be obtained during the process. In the particular case of polycrystals, the use of Electron Back-Scattered Diffraction (EBSD) provides information about the crystallographic orientation of each grain, the orientation relationship at the grain boundaries as well as the presence of twins within the grains \citep{UGW04, KZR06, FJG13}. While serial sectioning techniques are destructive, non-destructive characterization of the 3-D structure of polycrystals can be obtained by means X-ray tomography. Standard phase-contrast tomography provides information about porosity and second phases. Moreover, information about grain size and crystallographic orientation is presently available by means of diffraction-contrast tomography \citep{LSL08, PLR15}. Thus, detailed quantification of the microstructure of polycrystals is nowadays possible, but the acquisition time and the management and interpretation of the information contained in the massive datasets generated by 3-D characterization techniques may be problematic. Thus, the microstructural information necessary to build the RVE for computational homogenization should be assessed beforehand, to ensure efficient data collection and reconstruction.

It should be finally noted that the relevant microstructural information can be captured from 2-D or 3-D datasets by statistical descriptors, such as the 1-point and 2-point statistics \citep{T01}. 1-point statistics capture the probability associated with finding a specific phase or orientation at a point thrown randomly into the microstructure. They can be used to characterize the volume fraction of different phases or the texture of the polycrystal. 2-point statistics indicate the probability that both ends of a segment of given length thrown randomly into the microstructure lie in the same phase. They can be used to characterize the spatial distribution, size and shape of the different grains in the polycrystal. Once the particular statistical descriptors of a microstructure have been determined, phase-recovery algorithms can be used to generate RVEs, which are statistically equivalent, i.e. they share the same statistical descriptors \citep{FNK08, NTF10,dovskavr2014aperiodic}. The use of 2-point correlation function is particularly interesting for describing composites or two phase-materials. In the case of polycrystals \cite{NTF10} each orientation should be considered as a different phase and the number of 2-point correlation functions to describe the microstructure becomes large. In this case, alternative microstructure descriptors such as the probability distribution of grain sizes, or grain principal axis distribution are commonly used.

Motivated by this observation, we further explore the potential of Wang tiles to represent long-range spatial correlations in disordered microstructures, a problem common to materials science [11], geostatistics [12], or image analysis [13].
In this regard, two closely related applications can be dis- tinguished, namely, the microstructure reconstruction [14?16] based on given spatial statistics and microstructure compres- sion [17?19] aiming at efficient representation of material structures in multiscale computations

\subsection{Digital representation of the microstructure}
\label{microstructure_representation}
\begin{figure}[h]
\includegraphics[scale=0.25]{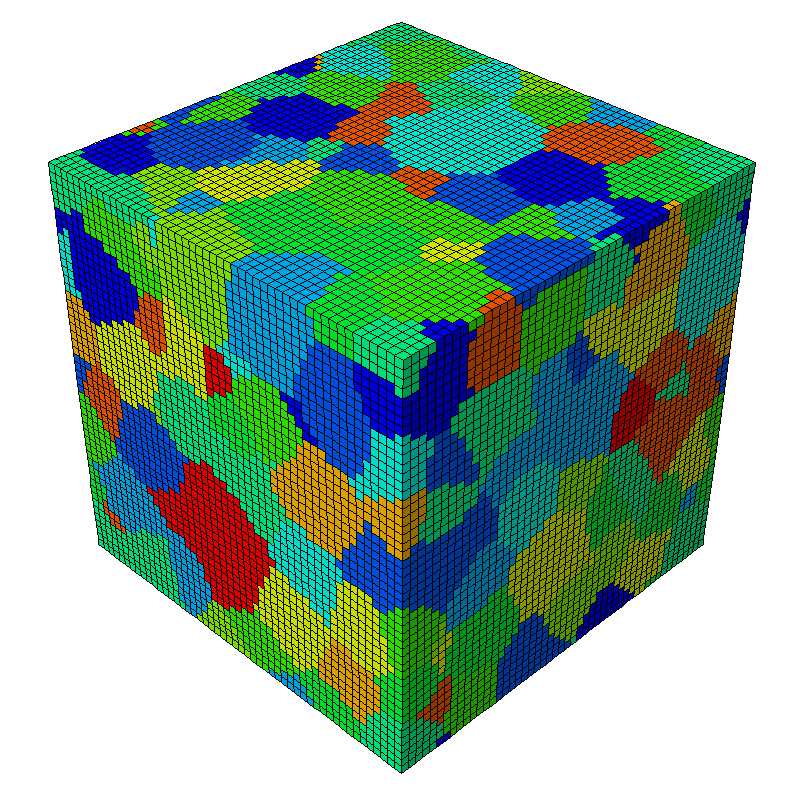}
\includegraphics[scale=0.90]{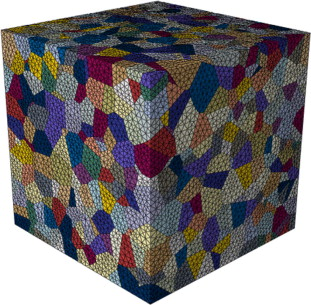}
\caption{Digital representations of an RVE of a polycrystal. (a) Voxel-based representation. (b) Voronoi-based representation.}  
\label{DRM}
\end{figure}

Once all the critical information about the microstructure of the polycrystal has been gathered, it is necessary to build a digital model of the RVE to perform computational homogenization. The RVEs can be constructed as a one-to-one representation of an actual microstructure measured from X-ray computed tomography or serial-sectioning data, or by generating synthetic microstructures from the statistical descriptors representing the microstructure. Two different types of discretizations are usually employed to define the geometry of the RVE. In the first one, the RVE is divided in voxels and grains are formed by groups of contiguous voxels that have the same crystallographic orientation (Fig. \ref{DRM}a). Voxel-based discretizations are the best option for direct representation of microstructures measured by three dimensional techniques because they can be directly extracted from the measured data. In addition, voxel-based discretizations can be directly exported into FFT-based codes as grid points or into FEM as regular hexaedral elements. Such voxel-based finite element meshes have good quality metrics and can be deformed up to very large strains \citep{SL13}. These voxel-based discretizations present, however, two major drawbacks. Firstly, the grain boundaries are stepped surfaces and, thus, this type of representation is not appropriate  to simulate phenomena localized at GBs, like grain boundary sliding. Secondly, the voxel-based FEM discretization often leads to very large number of elements because the voxel size is controlled by the dimensions of the smallest features that have to be resolved within the RVE.

These limitations of the voxel-based digital models can be overcome with representations of the grain structure based on tessellation. A tessellation is a subdivision of the 3-D space into convex polyhedra that intersect only at their boundaries, which are flat surfaces. The most popular one is the Voronoi tessellation, which is generated by a set $S$ of points in the three dimensional space by assigning a volume $V_{{\bf{x}}_i}$ to the point  $P_i({\bf{x}}_i)   \in S$ formed by all points $P({\bf{y}})$  in the space  which have $P_i$ as their nearest neighbour \citep{R12}. Cell boundaries in the Voronoi construction are, however, always equidistant from the generators of their cells and the range of cell patterns which can be generated is limited. Thus, Voronoi tessellations are not always suitable to reproduce the actual grain size distribution and weighted generalizations of the Voronoi model are frequently used. One possible generalization is the Laguerre tessellation in which the volume $V_{{\bf{x}_i}}$ in the space associated to the point $P_i \in S$  is formed by the points $P(\bf{y})$ that fullfill the condition 

\begin{equation}
P({\bf{y}})  \in V_{{\bf{x}}_i} \quad {\rm if} \quad d_L({\bf{y}}, {\bf{x}}_i) < d_L({\bf{y}}, {\bf{x}}_j), \quad j\ne i \quad {\rm and} \quad  {\bf{x}}_j \in S
\end{equation}

\noindent where $d_L({\bf{y}}, {\bf{x}}_i)$ is the "Laguerre" distance between points ${\bf{y}}$ and ${\bf{x}}_i$, which is given by 

\begin{equation}
d_L({\bf{y}}, {\bf{x}}_i) = \| {\bf{x}}_i - {\bf{y}} \|^2 - r_i^2
\end{equation}

\noindent where $r_i$ ($>$ 0) is the weight associated to point $P_i$. This definition leads to a partition of the space formed by convex, space-filling polyhedrons. It can be shown \citep{L2007, Xin1997Philo} that if $S$ is chosen as a system of nonoverlapping spheres characterized by the coordinates of their centers,  ${\bf{x}}_i$ and the corresponding radius, $r_i$, each cell of Laguerre tesselation completely contains its generating sphere and the volume distribution of Laguerre cells is almost equal to volume distribution of  their generating spheres. 

Thus, the strategy to generate grain structures within the RVE begins with the experimental grain size distribution, which can be often approximated by a log-normal function. A polydisperse sphere distribution, following the experimental grain size distribution, is then introduced in the RVE and densely packed using collective rearrangement algorithms \citep{T01} such as the force biased algorithm \citep{bargiel1991c}.  This algorithm starts with an initial distribution of spheres $S($\textit{\textbf{x}}$_i, r_i)$  characterized by the position of the center \textit{\textbf{x}}$_i$ and the radius $r_i$  distributed in the RVE. In this stage, overlapping of spheres is possible and allowed. Then, the algorithm attempts to reduce the overlaps between spheres by pushing apart overlapped spheres while small spheres are pushed to fill the empty spaces between large ones. After certain number of iterations, repositioning of overlapped spheres is stopped and the spheres gradually shrank to reduce the total amount of overlaps below a certain threshold. Finally, the coordinates of the centers of the spheres and their diameter are provided as output for the Laguerre tesselation. In other cases, Monte Carlo algorithms were used to obtain the spatial distribution of the set points for the tessellation, so the final cell size distribution coincides with the experimental grain size distribution \citep{CBJ15, MLD18}.

Once the grain size distribution has been reproduced in the RVE, the crystallographic texture can be introduced by assigning different orientations to the grains to reproduce the statistical distribution given by the ODF. It should be noted that the minimum number of grains in the RVE should be large enough to reproduce accurately both the grain size distribution and the texture of the polycrystal. Moreover, more sophisticated grain orientation algorithms can be used to account for the presence of a given fraction of low- or high-angle grain boundaries, which may lead to important differences in the mechanical behavior at both microscopic and macroscopic levels.

The digital representation of the microstructure is an important and time consuming task and microstructure builders have been developed, such as Neper \citep{Neper} and Dream3D \citep{dream3d}. They also provide tools to clean up the voxelized microstructures obtained from tomography or serial-sectioning or from tessellation and to discretize the microstructure for full-field simulations.

\section{Crystal plasticity models}

The first model that described the plastic deformation of metallic single crystals as a result of crystallographic slip was proposed by \cite{Taylor1923,Taylor1928}. In this seminal work, the deformation of Al single crystals was analysed and explained as the result of the shear deformation along twelve slip systems and the driving force for the shear deformation was the resolved shear stress on each slip system. A few years later, this model was used as to analyze the deformation of a polycrystal as an aggregate of grains \citep{T38}. The ideas of Taylor for the deformation of single crystals were adapted into the framework of continuum mechanics by \cite{Hill196695}, in the case of small strains. The theory, based on the general internal variable thermodynamic formalism, was extended to finite deformations in the 70s by \cite{Rice1971433} and \cite{Hill1972401}. They used the concept of the multiplicative decomposition of the deformation gradient into elastic and plastic parts, introduced by \cite{Lee196719}. Different constitutive models based on this framework were developed for single crystals in the 80's using either rate-independent formulations \citep{PEIRCE19821087,PEIRCE1983133} or viscoplasticity \citep{Asaro1985923}. In parallel, attention was also paid to the development of rigorous numerical implementations of the models, including efficient and well-posed integration methods for the highly nonlinear viscoplastic laws \citep{Cuitino1992437} and rigorous integration for the finite deformation framework \citep{Miehe19963367}. The result of all these studies -- and many more not reviewed here --  is a well-stablished theory of CP which will be summarized below. 

The starting point of most CP models is the multiplicative decomposition of the deformation gradient $\mathbf{F}$ into its elastic ($\mathbf{F}^e$)  and plastic ($\mathbf{F}^p$) parts (Fig. \ref{Fig:decomp})

\begin{equation}
\label{eq:FeFp}
\mathbf{F}=\mathbf{F}^e\mathbf{F}^p
\end{equation}

\noindent where the configuration defined by $\mathbf{F}^p$ is called the relaxed or intermediate configuration. In the context of CP, it is assumed that $\mathbf{F}^p$ leaves the crystal lattice undistorted and unrotated \citep{Rice1971433,Hill1972401} and the rotation of the lattice is determined only by $\mathbf{F}^e$. Although this decomposition is generally accepted  for CP, several issues as the existence and uniqueness of the decomposition and its connection with the microscopic distortion generated by the dislocations are still under debate \citep{REINA201440,REINA2016231}.  

\begin{figure}
\includegraphics[width=400pt,keepaspectratio]{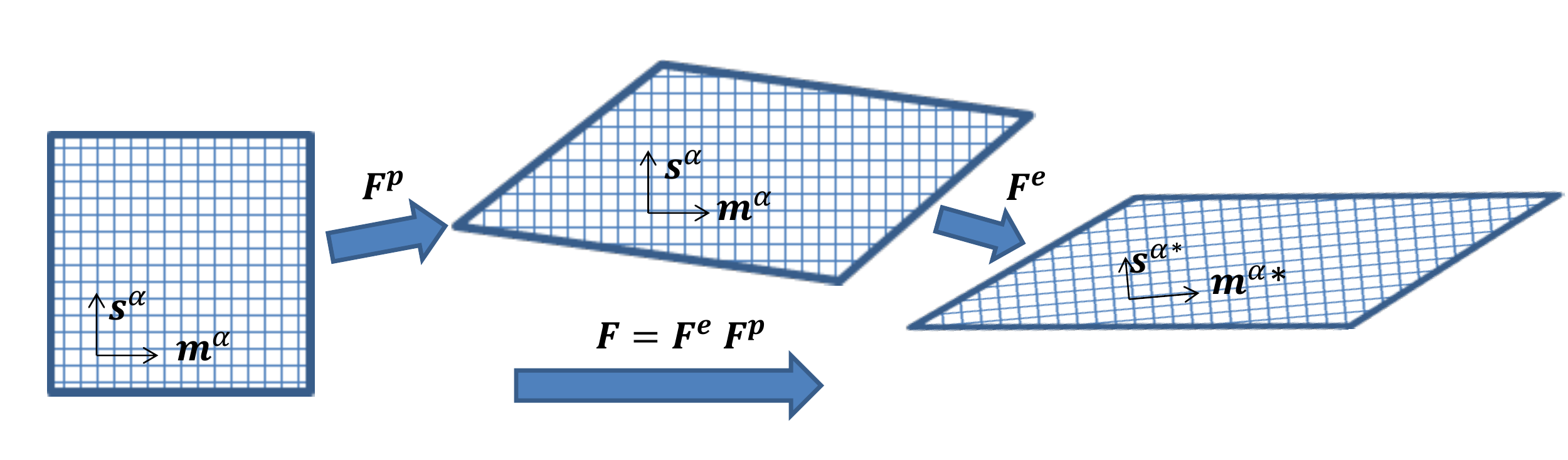}
\caption{Schematic of the multiplicative decomposition of the deformation gradient}
\label{Fig:decomp}
\end{figure}

From the definition of the velocity gradient, $\mathbf{L}$, expression (\ref{eq:FeFp}) leads to

\begin{equation}
\label{eq:Lp}
\mathbf{L}=\dot{\mathbf{F}}\mathbf{F}^{-1}=\dot{\mathbf{F}}^e\mathbf{F}^{e^{-1}}+\mathbf{F}^e\dot{\mathbf{F}}^p\mathbf{F}^{p^{-1}}\mathbf{F}^{e^{-1}}
\end{equation}

\noindent where $\mathbf{L}^p =\dot{\mathbf{F}}^p\mathbf{F}^{p^{-1}}$ stands for the plastic deformation rate in the intermediate configuration.

The constitutive equations can be obtained from the energy density per unit volume expressed in the intermediate configuration \citep{Cuitino1992437,HAN20051188}. Following the internal variable formalism, the free energy density, $\psi$, can be written as

\begin{equation}
\label{eq:free_energy}
\psi=\psi(\mathbf{F};\mathbf{F^p},\mathbf{q})
\end{equation}

\noindent where $\mathbf{q}$ is a set of internal variables. The free energy density can be split into the elastic and plastic energy densities \citep{ Cuitino1992437}, 

\begin{equation}
\label{eq:free_energy2}
\psi=\psi^e( \mathbf{F}\mathbf{F}^{p^{-1}})+\psi^p( \mathbf{F^p},\mathbf{q}).
\end{equation}

\noindent and the second Piola-Kirchoff stress tensor in the intermediate configuration $\mathbf{S}$ can then be obtained as

\begin{equation}
\mathbf{S}=\frac{\partial \psi^e }{\partial \mathbf{E}^e}
\end{equation}

\noindent where $\mathbf{E}^e$ stands for the Green-Lagrange elastic strain in the intermediate configuration and is given by

\begin{equation}
\label{eq:green}
\mathbf{E}^e=\frac{1}{2}(\mathbf{F}^{e^T}\mathbf{F^e}-\mathbf{I}).
\end{equation}  

The second Piola-Kirchoff stress in the intermediate configuration $\mathbf{S}$ is related with the Cauchy stress, $\boldsymbol{\sigma}$, according to

\begin{equation}
\mathbf{S}=\frac{1}{J}\mathbf{F}^{e^{-1}}\boldsymbol{\sigma}{\mathbf{F}^{e^{-T}}}
\end{equation}  

\noindent where $J$ is the determinant of $\mathbf{F}$.  Assuming a quadratic potential for the elastic energy $\psi^e$  in eq. \eqref{eq:free_energy2}, $\mathbf{S}$ can be expressed as a linear function of the Green-Lagrange elastic strain according to
 
\begin{equation}
\label{eq:hook}
\mathbf{S}=\mathbb{L}\mathbf{E}^e
\end{equation}

\noindent where $\mathbb{L}$ stands for the fourth order elastic stiffness tensor of the single crystal.

The crystallographic nature of the plastic deformation of single crystals is described by two orthogonal unit vectors which define the slip system $k$, $\mathbf{s}^k$ and  $\mathbf{m}^k$, that stand for the slip direction and slip plane normal, respectively and that remain invariant in the intermediate configuration (Fig. \ref{Fig:decomp}). The available slip systems of a given crystalline material, $k=1,2,..., n$ are determined by its lattice. For instance, FCC materials presented 12 slip systems characterized by the $\{111\}$ planes and the $<$110$>$ directions.

\cite{Rice1971433} proposed the conventional flow rule of a single crystal based on this geometrical description according to

\begin{equation}
\mathbf{L}^p =\sum_k \dot{\gamma}^k \mathbf{s}^k \otimes \mathbf{m}^k = \sum_k \dot{\gamma}^k \mathbf{Z}^k
\label{eq:Lp_def}
\end{equation}

\noindent where $\gamma^k$ are internal variables of the model ($\gamma^k \in \mathbf{q}$) that account for the accumulated plastic slip in each slip system $k$. The dyadic product $\mathbf{s}^k \otimes \mathbf{m}^k$ is the non-symmetric Schmid tensor of the system $k$, $\mathbf{Z}^k$.
 
The evolution of $\gamma^k$ is normally assumed to be dependent on the internal variables chosen to describe the state $\mathbf{q}$ and on the stress through the resolved shear stress $\tau^k$ 

\begin{equation}
\dot{\gamma}^k=\dot{\gamma}^k(\tau^k;\mathbf{q}).
\label{eq:shearingrate}
\end{equation}

The resolved shear stress $\tau^k$ is defined in the framework of finite strains as \citep{Cuitino1992437}

\begin{equation} 
\label{eq:resolved}
\tau^k = \mathbf{F}{^e{^T}}\mathbf{F}^e\mathbf{S}:\mathbf{s}^k \otimes \mathbf{m}^k
\end{equation} 

\noindent which can be simplified in the usual case of small elastic strains  ($\mathbf{F}^{e^T} \mathbf{F}^e\approx\mathbf{I} $)  as

\begin{equation} 
\label{eq:resolved2}
\tau^k \approx \mathbf{S} :\mathbf{s}^k \otimes \mathbf{m}^k  = \mathbf{S} :\mathbf{Z}^k
\end{equation} 

The last two ingredients of a CP model are the particular functions that dictate the shear rate $\dot{\gamma}^k$ of each slip system (eq. \eqref{eq:shearingrate}) as well as the evolution of the internal variables $\mathbf{q}$ during deformation. Many different flavours of CP have been developed in the last decades depending on these functions and the most relevant ones will be reviewed below grouped in three categories: phenomenological, physically-based and strain gradient plasticity models.

\subsection{Phenomenological crystal plasticity models}
\label{subsection_phen}

Crystal plasticity, in contrast with classical macroscopic plasticity, has a clear physical basis and always includes explicit microscopic information of the material, such as the geometrical definition of the active slip systems. However, CP models can use phenomenological expressions to define both the slip rates and the evolution of the internal variables. This phenomenological approach is based on the classical constitutive equation theory, and the internal variables that determine the state of the crystal, as well as their evolution laws, are not directly related to the microscopic physical magnitudes or processes. The first approaches  \citep{Rice1971433,Hill1972401} were rate-independent but they 
were relatively complex and prone to convergence problems because multiple combinations of shear increments can lead to the same plastic strain \citep{PEIRCE19821087}. Although most of the problems associated with rate-independent formulations have been overcome \citep{ANAND1996525}, viscoplastic formulations became a very popular alternative since they were introduced by \cite{PEIRCE19831951}. 

The set of internal variables, $\mathbf{q}$, contains the information about the accumulated plastic slip, $\gamma^k$, and the critical resolved shear stress (CRSS), $g^k$, in each slip system. For a given state, $g^k$ corresponds to the minimum value of the resolved shear stress, eq. \eqref{eq:resolved}, to activate the plastic flow in that system. The slip rate, eq. \eqref{eq:shearingrate}, is given by a power-law function according to

\begin{equation} 
\dot{\gamma}^k=\dot{\gamma}_0 \left ( \frac{|\tau^k|}{g^k} \right )^{1/m} \mathrm{sign}(\tau^k)
\label{eq:power_law}
\end{equation} 

\noindent where $\dot{\gamma}_0$ and $m$ stand for the reference strain rate and the strain rate sensitivity parameter, respectively. The strain rate sensitivity parameter is equivalent to $1/n$, where $n$ stands for the strain rate sensitivity exponent. Note that only positive values of $\dot{\gamma}^k$ were considered in the original work of \cite{PEIRCE19831951} by including positive and negative slip directions $\pm \mathbf{s}^k$ while in eq. \eqref{eq:power_law}, as in most of the actual formulations, $\dot{\gamma}^k$ can take both signs and only one direction of $\mathbf{s}^k$ is considered.

Regarding strain hardening, the model by \cite{PEIRCE19831951} was developed for monotonic loading and therefore only considered isotropic hardening. The CRSS on each system included the explicit contributions of the slip in the same system (self hardening) and of the slip on all the other slip systems (latent hardening) and the evolution of the CRSS can be written as

\begin{equation} 
\label{eq:hard1}
\dot{g^k}=\sum_j h_{kj}| \dot{\gamma}^j|
\end{equation} 

\noindent where $h_{kj}$ are the latent hardening moduli, with $h_{kk}$ the self hardening modulus. The model proposed by \cite{PEIRCE19831951} defined the hardening modulus, eq. \eqref{eq:hard1}, as

\begin{equation} 
 h_{kj} =h(\Gamma)[q+(1-q)\delta_{kj}]
\label{eq:hard2}
\end{equation} 

\noindent where $q$ is a  parameter defining the latent hardening, $\Gamma$ is the accumulated shear on all the slip systems

\begin{equation}\label{eq:Gamma}
\Gamma=\sum_k\int{|\dot\gamma^k| dt}
\end{equation}

\noindent and the function $h(\Gamma)$ proposed for FCC single crystals takes the form 

\begin{equation}
h(\Gamma)=h_0\; {\rm sech}^2{\bigg(\frac{h_0\Gamma}{\tau_s-\tau_0}\bigg)}
\label{eq:asaro_needleman}
\end{equation}

\noindent where $h_0$ is the initial hardening modulus and $\tau_0$ and $\tau_s$ stand for the initial and the saturation values of CRSS, respectively. 

Alternative expressions of $h(\Gamma)$ have been proposed to account for non-vanishing hardening rates at large plastic strains. This is the case of the Voce hardening law proposed in \cite{Tome1984} that defines $h$ as

\begin{equation}
\label{Voce}
h(\Gamma)=h_s\left[1-\exp\bigg(\frac{-h_0\Gamma}{\tau_s-\tau_0}\bigg)\right]+\left(\tau_s-\tau_0+h_s\Gamma\right)\frac{h_0}{\tau_s-\tau_0}\exp\bigg(\frac{-h_0\Gamma}{\tau_s^-\tau_0}\bigg)
\end{equation}

\noindent where a new parameter $h_s$ is introduced to define the hardening slope at large plastic strains. An alternative hardening model, developed by \cite{Bassani199121}, proposed a similar non-vanishing hardening rate through a parameter $h_s$ using eq. \eqref{eq:asaro_needleman} as starting point. The corresponding hardening moduli can be expressed as

\begin{equation}
\begin{array}{ll}
  h_{kk} = &\left[ (h_0-h_s) \mathrm{sech}^2\bigg(\frac{h_0-h_s}{\tau_s-\tau_0}\gamma_k\bigg) + h_s \right ] \left[1+\sum_{j \neq k} f_{kj}\tanh\bigg(\frac{\gamma_j}{\gamma_k}\bigg) \right] \\
 h_{kj}  = &q h_{kk}
\end{array}
\end{equation}

\noindent where $h_{kk}$ and and $h_{kj}$ stand for the self hardening and the latent hardening coefficients in eq. \eqref{eq:hard1}. In contrast to \cite{PEIRCE19831951},  the dependence of the self hardening coefficient on the slip accumulated in the different slip systems is not uniform in this formulation, but given by some interaction coefficients $f_{kj}$ that depend on the nature of the dislocation junctions between the slip systems. Most of the phenomenological CP models developed since then for monotonic loading are based on either of the three hardening models summarized above. 

The extension of this phenomenological framework to account for cyclic deformation is done by the introduction in the constitutive model of the effect of a backstress to formulate kinematic hardening laws at the crystal level  \citep{CAILLETAUD1991,CAILLETAUD199255,HU1992839} and similar approaches have been used to model creep \citep{HGM03, VGM07}. The plastic shear rate depends in this case on both the CRSS and the backstress according to \citep{CAILLETAUD199255}

\begin{equation} 
\dot{\gamma}^k=
\left\{ 
\begin{array}{cc}
\dot{\gamma}_0 \left ( \frac{|\tau^k-\chi^k|-g^k}{K} \right )^{1/m} \mathrm{sign}(\tau^k-\chi^k) & \mathrm{if} \quad |\tau^k-\chi^k|\geq g^k \\
0 & \mathrm{if} \quad  |\tau^k-\chi^k|< g^k
\end{array}
\right.
\label{eq:power_law_cailletaud}
\end{equation}

\noindent where $\chi^k$ stands for the backstress of system $k$ and $K$ is a numerical parameter. 

The kinematic hardening is defined by the evolution of the backstress and \cite{CAILLETAUD199255} proposed the following expression  for FCC metals

 \begin{equation} 
 \dot{\chi}^k=c \dot{\gamma}^k - d| \dot{\gamma}^k|
\end{equation}

\noindent where $c$ and $d$ are material parameters that define the hardening rate. 

Other expressions of non-linear phenomenological kinematic hardening laws at the crystal level have been proposed based in most cases in well-stablished relations at the macroscale. One of the most common expressions is the Frederick-Armstrong law, adapted from macroscopic plasticity to the crystal level in \cite{CAILLETAUD1991} according to

 \begin{equation} 
 \dot{\chi}^k=c \dot{\gamma}^k- d| \dot{\gamma}^k |{\chi}^k.
 \end{equation} 

\noindent to account for the Bauschinger effect in single crystal Ni-based superalloys. More recently, \cite{CRUZADO2017148} presented a phenomenological CP model for cyclic loading that includes the effect of cyclic softening and an alternative evolution of the backstress to account for the mean stress relaxation. The plastic slip rate in this model is expressed as

\begin{equation}
\label{eq:power_law_cruzado}
\dot{\gamma}^k=\dot{\gamma}_0 \left(
  \frac{| \tau^k-\chi^k |}{g^k}\right)^{1/m}\mathrm{sign} (\tau^k-\chi^k). 
\end{equation} 

\noindent The evolution of the backstress is obtained as a simplified version of the Ohno-Wang macroscopic model \citep{Ohno1993} limited to the first two terms 

\begin{equation}
\label{eq_(9):kinematic_OWM}
\dot{\chi}^k=c\dot{\gamma}^k-d\chi^k|\dot{\gamma}^k|\left( \frac{|\chi^k|}{c/d}\right)^{r}   
\end{equation}

\noindent  where $c$ and $d$ are the parameters of the Frederick-Armstrong model while ${r}$ is an extra parameter that controls the mean stress relaxation velocity. Finally, the cyclic softening was accounted for through a new internal variable, the cyclic plastic slip, $\gamma_{cyc}$, is given by,

\begin{equation}
\label{eq_(15):cyclic_plast}
\gamma_{cyc}=\sum_k \int_0^t |\dot{\gamma}^k | \mathrm{d}t - \sum_k \left|\int_0^t \dot{\gamma}^k \mathrm{d}t\right|.
\end{equation}

\noindent which was taken into account in the evolution of the CRSS, $\dot{g}^k=\dot{g}^k(\gamma_{cyc})$, using a Voce type law with negative slope.

\subsection{Physically-based crystal plasticity models}

Physically-based CP models (as opposed to phenomenological ones) contain a stronger physical connection with the microscopic mechanisms of plastic deformation. Thus, some microscopic physical quantities are included as internal variables (i.e. dislocation densities) and/or rate equations are based on the active microscopic deformation mechanisms. In addition, these models usually include as input some additional microstructure information such as the initial dislocation density, the volume fraction of precipitates or second phases and, very often, include the effect of temperature. 

The first ingredient of physically-based models is the relationship between the plastic slip rate and the dislocation movement. This is introduced through the Orowan equation \citep{Orowan1934} that connects the plastic slip rate on a given slip system, $\dot{\gamma}^k$, with the mobile dislocation density, $\rho_m^k$, the Burgers vector, $b^k$, and the average dislocation velocity, $\bar{v}^k$,  according to

\begin{equation}
\dot{\gamma}^k = \rho_m^k b^k \bar{v}^k.
\label{eq:orowan}
\end{equation}

This equation replaces the power law rate equations in the phenomenological rate-dependent models, i.e. eqs. \eqref{eq:power_law}, \eqref{eq:power_law_cailletaud}, \eqref{eq:power_law_cruzado}. The driving force for the dislocation movement in eq. \eqref{eq:orowan} is the resolved shear stress, eq. \eqref{eq:resolved}, that is introduced through the average velocity.  In metals with compact lattices (FCC and for some slip systems in HCP), lattice friction is negligible and slip occurs at very low stresses. In this case, a linear viscous relation can be postulated between the resolved stress and the velocity of a single dislocation \citep{hirth1982theory} leading to

\begin{equation}
v^k=\frac{\tau^k b^k}{B}
\label{eq:Dragg}
\end{equation}

\noindent where $B$ is the drag coefficient, a material parameter that depends on temperature. If the density of the dislocations is very low, the average dislocation velocity in eq. \eqref{eq:orowan} can be obtained directly from the CRSS using eq. \eqref{eq:Dragg}, and a linear viscous relationship is established between the CRSS and the plastic slip rate. However, this relationship is not realistic in most cases because dislocations have to overcome different barriers during slip, leading to an average dislocation velocity different from the one given by eq. \eqref{eq:Dragg}. 

The barriers to dislocation movement can be classified as temperature-dependent (thermal) or independent (athermal), depending on whether thermal activation can help to overcome the obstacle. For instance,  long-range elastic interactions among dislocations introduce an athermal threshold CRSS for the dislocation movement $\tau_a$. \cite{T38} determined this threshold stress for pure metals, which is given by

\begin{equation}
\label{eq:taylor0}
\tau_a \propto  \mu b \sqrt{\rho} 
\end{equation}

\noindent where $\mu$ is the shear modulus and $1/\sqrt{\rho}$ stands for the average distance between dislocations.

The thermal barriers are due to the short range interactions of dislocations with other dislocations (jogs created by the intersection  of forest dislocations and their movement by vacancy generation) and point defects. The strength of the barrier at 0 K is given by $\tau_t$ and the CRSS necessary to overcome the barrier is given by $\tau_a+\tau_t$. At finite temperatures, thermal energy helps the dislocation to jump over the barrier, and the average dislocation velocity (eq. \ref{eq:orowan}) becomes dependent on the temperature. 

The influence of temperature on dislocation slip under short range interactions was studied in detail in \cite{K75} with the framework of the transition state theory and this work is the basis of most temperature-dependent physically-based CP models in the literature \citep{KOTHARI199851,MA20043603,Cheong20045665,RODRIGUEZGALAN2015191, SG16}. In summary,  the  average dislocation velocity  to be inserted in eq. \eqref{eq:orowan} can be written as

\begin{equation}
\label{thermal_kocks}
\bar{v}^{k} = \left \{
\begin{array}{lll}
0 & \mathrm{if} & 0 \leq  |\tau^{k}| \leq  {\tau}_{a} \\ 
\bar{l}^{k} \nu_{0} \exp{\left(-\frac{\Delta{G}}{kT} \right)} & \mathrm{if} & {\tau}_{a} <  |\tau^{k}| < \tau_a+\tau_t
\end{array}\right. 
\end{equation}

\noindent where $\bar{l}^{k}$ is the average distance between the obstacles in the slip system $k$, $\nu_{0}$ the attempt frequency, $k$ the Boltzmann constant and $T$ the absolute temperature.  $\Delta{G}(\tau^k)$ stands for the Gibbs free energy that has to be supplied by thermal fluctuations to overcome the obstacle, which depends on the applied shear stress $\tau^k$, and the exponential term expresses the probability of the occurrence of a jump over a short-range barrier. 

The evolution of the Gibbs free energy with the applied shear stress in the presence of a general array of obstacles in the slip plane can be expressed as \citep{K75},

\begin{equation}
\label{G_Kocks}
\Delta G(\tau^{k}) = \Delta{F} \left[1-\left< \frac{ |\tau^{k}| - {\tau}_{a}}{\tau_{t}}\right>^{p}\right]^{q}
\end{equation}

\noindent where $\Delta{F}$ is the activation free energy necessary to overcome the obstacles without the aid of an applied shear stress, and $<x>$ stand for the Macaulay brackets, which return $x$ if $x > 0$ and 0 otherwise. Finally, $p$ and $q$ (in the range $0 \leq p \leq 1$ and $ 1 \leq q \leq 2$) are two parameters that define the strength of the obstacle as a function of the distance propagated by the dislocation. A simplification of eq. \eqref{G_Kocks} often found in CP models \citep{MA20043603,DUNNE20071061} assumes that the obstacle strength is constant ($p=q=1$) and eq. \eqref{G_Kocks} can be written as

\begin{equation}
\label{G_Kocks2}
\Delta G(\tau^k) = \Delta{F} -\tau^\alpha V
\end{equation}

\noindent where $V$ is the activation volume, a material constant that determines the actual volume of the material affected by the short range dislocation-obstacle interaction.

It should be finally noted that the most important contribution to the thermal strength, $\tau_t$, in metals with non-compact lattices (such as BCC or low density planes in HCP crystals) is the lattice friction. In this case, $\tau_t$  at 0 K is the Peierls stress, that can be obtained from the simple model by Peierls and Nabarro \citep{Peierls1940, Nabarro1947} or from atomistic or {\it ab initio} simulations (see, for instance, \cite{Stukowski2015108} for BCC W and  \cite{YasiCurtin2009} for pyramidal slip in HCP Mg).

In addition to a physically-based model of $\dot{\gamma}^k$, many physically-based CP models incorporate micromechanical internal variables, such as dislocation densities, to account directly for the strain hardening from physical considerations. The relationship between dislocation densities and the athermal strength is based on the Taylor  model, eq. \eqref{eq:taylor0}. This expression is usually enriched by including several terms that account for the interactions between dislocations to obtain the flow stress in the different slip systems (here denoted as $g^k$ to emphasize the relation with the CRSS introduced in the phenomenological models) . In the case of FCC crystals, this can be expressed as \citep{FRANCIOSI19821627}
 
 \begin{equation}
\label{eq:taylor1}
g^k = \mu^k b \sqrt{\sum_j a_{kj} \rho^j} 
\end{equation}

\noindent where $\rho^j$ is dislocation density in the slip system $j$ and $a_{kj}$ are a set of non-dimensional coefficients that determine the self-hardening and the latent hardening due to the interactions among dislocations (similar to the phenomenological expression, eq. \eqref{eq:hard1}). The values $a_{kj}$ for different crystal lattices can be  obtained from dislocation dynamics simulations \citep{Devincre1745,BERTIN201472}. 

The dislocation densities in eq. \eqref{eq:taylor1},  $\rho^k$, are introduced as local internal variables and correspond to the average of dislocation length per unit volume at each point of the crystal. The evolution of these internal variables is normally based in the  Kocks-Mecking model \citep{MECKING19811865,ESTRIN198457}. This model considers that hardening is controlled by the competition between storage and annihilation of dislocations and that both processes are additive. Thus, the corresponding evolution law for one slip system $k$ is given by

\begin{equation}
\label{eq:kocksmecking}
\dot{\rho^k}=\bigg[\frac{1}{\ell(\rho^k)}-2y_c\rho^k\bigg]|\dot{\gamma^k}|.
\end{equation}

\noindent where the first term, $1/\ell(\rho^k)$, is athermal and controls the storage of dislocations and $\ell(\rho^k)$ is the dislocation mean free path,  which corresponds to the distance travelled by a dislocation segment before it is stopped by an obstacle. In the absence of precipitates or other obstacles to the dislocation motion, $\ell(\rho^k) = k_1/\sqrt{\rho^k}$, where $k_1$ is a material constant. The second term, $2y_c\rho^k$, is associated with the dislocation annihilation due to dynamic recovery. It depends on the  temperature, and is characterised by $y_c$,  a constant that depends on the critical annihilation distance between dislocations. It  should be noted that eq. \eqref{eq:kocksmecking}  can also be extended to alloys with a distribution of impenetrable obstacles. In this case, the dislocation mean free path will be determined by the obstacle spacing \citep{ESTRIN198457}. 

The generalization of the Kocks-Mecking model for multiple slip systems reads

\begin{equation}
\label{eq:kocksmecking2}
\dot{\rho^k}=\bigg[\frac{1}{\ell^k(\rho^1,\cdots,\rho^n)}-2y_c\rho^k\bigg]|\dot{\gamma}^k|
\end{equation}

\noindent and

\begin{equation}
\ell^k \approx \frac{k_1}{\sqrt{\sum_{j \neq k} \rho^k}}.
\end{equation} 

Several modifications of the Kocks-Mecking model can be found in the literature based the results obtained  from simplified dislocation mechanics models \citep{Cheong20045665} or dislocation dynamics simulations \citep{Devincre1745,sansal2010}.  Moreover, the introduction of the distance to grain boundary as an upper bound to the dislocation mean free path has been successfully used to simulate the effect of grain size on the strength of Cu polycrystals \citep{Hauoala2018}. 

\subsection{Strain gradient crystal plasticity models}

The CP framework presented above is local, i.e. the material response at a given point depends only on the local values of both state and internal variables at that point. Therefore, the constitutive equations are size-independent because there are not length scales involved. However, the experimental evidence as well as dislocation dynamics simulations show that the strength of single crystals is size-dependent when they are subjected to homogeneous  \citep{E15} and inhomogeneous plastic deformation, such as nanoindentation
\citep{STELMASHENKO19932855,SanchezMartin2014283} or bending tests of single crystal cantilever beams \citep{MOTZ20054269,KIENER2008580,GONG20115970}. The development of plastic strain heterogeneities in the crystal deformation is also the reason behind the well-known dependency of the plastic strength of polycrystals on grain size \citep{Hall1951-1,Petch1953-1}. 

The effect of the plastic heterogeneities in the mechanical response of a crystal was first rationalized by \cite{Nye1953-1} and \cite{Ashby1970-1} as a result of the interaction between statistically stored dislocations (SSDs), which evolve from random trapping processes during plastic deformation, and geometrically necessary dislocations (GNDs) induced by the presence of plastic strain gradients. From the modelling viewpoint, the most common approach to account for the effect of plastic strain heterogeneities in the material response is by introducing  the influence of  some plastic strain gradient measure in the constitutive equation, leading to the so-called strain gradient plasticity. This idea was introduced by \cite{AIFANTIS1987211} for macroscopic plasticity and further developed to account for size effects in polycrystals \citep{FMAH94,NIX1998411}. This modelling framework was extended to CP by \cite{ACHARYA1995} and \cite{SHU1999297}, who in these seminal papers introduced the two different strategies followed in the strain gradient crystal plasticity (SGCP) models developed since then. 

In the first modelling approach, \emph{lower-order} SGCP, strain gradients enter only in the instantaneous hardening moduli, while the thermodynamic consistency is preserved \citep{Acharya2003-1}. This approximation allows the use of the classical mathematical framework of boundary value problems in standard plasticity \citep{ACHARYA1995,Dai1997-1,Busso2000-1,HAN20051188,HAN20051204,Ma20062169,CHEONG20051797,DUNNE20071061}. The second approach stands for the \emph{higher order} SGCP models in which some internal variables are chosen as kinematic variables. This implies  the introduction of stresses conjugated to these kinematic variables as well as the corresponding "higher order" boundary conditions \citep{GURTIN20025,GURTIN20051,GURTIN2008702, BARDELLA2006128, Bardella_Segurado_etal_2013, yefimov2004, EVERS20042379, BAYLEY20067268, BORG2008688, NIORDSON201431}.
 
The obvious benefit of lower-order formulations is their simple structure, so they can be easily implemented in existing general-purpose finite element codes. In addition, they avoid the additional higher-order boundary conditions, which are not always easy to interpret physically. Due to these facts, lower-order SGCP models are the basis of most physically-based CP models that account for size effects. In addition, they have been coupled to vacancy diffusion models to account for the effect of dislocation climb, a critical process to simulate creep deformation \citep{GEERS2014136}. On the other hand, the main limitation of lower-order formulations is the impossibility of accounting for boundary layers in constrained plastic flow, as the development of these layers requires the use of higher order boundary conditions on the plastic slip fields. In this sense, higher order models are capable of accounting for the gradient development due to the presence of a passivation layer or a grain boundary within a polycrystal.

\subsubsection{Lower-order SGCP}

The starting point in lower-order SGCP models is the definition of a measure for the plastic strain gradients.  This is usually based on the  relationship between the plastic slip gradients and the density of GNDs. For single slip, the plastic slip gradient gives rise to the development of a density of GNDs, $\rho_{GND}$, to maintain continuity in the crystal  according to \citep{Nye1953-1}
\begin{equation}
\rho_{GND} =-\frac{1}{b}\frac{\partial \gamma}{\partial x_n}
\end{equation}

\noindent where $\gamma$ is the plastic slip, $b$ the Burgers vector and $x_n$ stands for the coordinate normal to the slip direction. The generalization of the previous equation to multiple slip systems is expressed through the Nye's tensor, $\boldsymbol{\alpha}$, which was introduced by \cite{Nye1953-1} and generalized by \cite{ARSENLIS19991597} as

\begin{equation}
\label{defA}
\alpha_{ij} = \sum_a \rho_{GND}^a b_i^a  t_j^a
\end{equation}

\noindent where $a$ stands for a straight dislocation segment of length $l^a$ parallel to $\mathbf{t}^a$ with Burgers vector $\mathbf{b}^a$,  and $\rho_{GND}^a$ stands for the length of segment $a$ per unit volume.  Nye's tensor accounts for the lattice curvature and, therefore,  can be expressed as function of the plastic slip gradients in the case of small strains according to \citep{ARSENLIS19991597} 

\begin{equation}
\label{Nye_small_strain}
\boldsymbol{\alpha}=\nabla \times \sum_\alpha \gamma^k \mathbf{s}^k \otimes \mathbf{m}^k
\end{equation}

\noindent where $\nabla \times$ represents the curl operator. In the case of finite strains, the Nye tensor is usually defined as \citep{Busso2000-1,Ma20062169}

\begin{equation}
\label{Nye_finite_strain}
\boldsymbol{\alpha}=\nabla \times  \mathbf{F}^\mathrm{p}
\end{equation} 

\noindent although other tensorial measures of the plastic incompatibility are defined in the literature, as the finite strain geometric dislocation tensor $\mathbf{G}$ proposed in \cite{CERMELLI20011539} and given by
\begin{equation}
\label{Nye_finite_strain}
\mathbf{G}= \mathbf{F}^\mathrm{p} (\nabla \times  \mathbf{F}^\mathrm{p})
\end{equation} 

In lower-order gradient models, the plastic strain gradient is included in the hardening expression. This is normally done adding to the Taylor hardening model (eq. \eqref{eq:taylor1}) that relates the CRSS with the dislocation density, a term that accounts for the GND density and depends on some plastic gradient measure. Along these lines, a first model inspired on physical considerations, was presented by \cite{HAN20051188,HAN20051204}, who defined a GND density for each slip system, $\eta^k$, based in a projection of the dislocation tensor (eq. \ref{Nye_finite_strain}) in the system. Then, the effective gradient-dependent CRSS, $g_{eff}$, was given by

\begin{equation}
\label{eq:mech_SGSCP}
g^k_{eff}= g_0 \sqrt{(g^k / g_0)+ l \eta^k}
\end{equation}

\noindent where $g^k$ is the CRSS of the system $k$ in the absence of gradients and $l$ stands for an intrinsic length scale. It should be noted that  eq. \eqref{eq:mech_SGSCP} resembles the phenomenological approaches of macroscopic gradient plasticity developed by \cite{NIX1998411} but the physical origin of $l$ is clearer and can be obtained from physical considerations.

If the density of GNDs in each system is explicitly resolved, the CRSS can be defined as \citep{CHEONG20051797}

\begin{equation}
g^k \propto \mu b \sqrt{\sum_j \bigg( a^{SSD}_{kj} \rho^j_{SSD}+a^{GND}_{kj} \rho^j_{GND}} \bigg)
\end{equation}

\noindent where  $\rho^j_{SSD}$ stands for the SSDs density and $\rho^j_{GND}$ corresponds to the GNDs density in system $j$. The coefficients $a^{SSD}_{kj}$ and $a^{GND}_{kj}$ define the latent hardening interactions among the different slip systems. The distribution of GND densities on the different systems $\rho^k_{SSD}$ is not uniquely determined by the dislocation tensor and additional constraints have to be imposed. \cite{ARSENLIS19991597} proposed to minimize either the dislocation density or the dislocation length to obtain the actual GND distribution. Alternatively, the GND evolution can be determined by integrating some evolution laws obtained by expressing the Nye tensor in terms of the spatial gradient of the slip rate \citep{Busso2000-1,Ma20062169,CHEONG20051797,DUNNE20071061}.

In order to implement these models in a finite element framework, the slip gradients or Nye tensor at the integration point level have to be determined. This task has been traditionally done using a very efficient local element approach \citep{Busso2000-1,DUNNE20071061} in which the internal variables included in the gradient term are extrapolated  from the integration points to the nodes in each element. The gradients are obtained by deriving the element shape functions. However, it has been recently shown by \cite{Rodriguez-Galan2017} that this simple approach might present convergence problems, and alternative methods to compute the gradients based on recovery techniques are more reliable \citep{Han2007-1, Rodriguez-Galan2017}.

\subsubsection{Higher-order SGCP}

There are several alternative formulations of higher-order SGCP models. One of the most extended ones was proposed by \cite{GURTIN20025,GURTIN20051,GURTIN2008702}. The principal concepts and equations of this theory will be reviewed here because they are common to many other gradient models \citep{BARDELLA2006128, Bardella_Segurado_etal_2013, BORG2008688, NIORDSON201431}. In this theory, the plastic slip in each system, $\gamma^k$, is included in the constitutive equation as a kinematic variable (independent state variable). A central point to the model is that the work associated with each independent kinematic process should be accounted for in the energy balance. Therefore, microforces conjugated with the slip and slip gradients appear in the formulation. In the absence of body forces, the virtual power principle for a domain $\Omega$ with boundary $\partial \Omega$ under external macroscopic surface traction $\mathbf{t}$ and a microscopic surface traction $ \Xi^k$ for each slip system reads 

\begin{equation}
\label{eq:gurtin_virtual}
\int_{\partial \Omega} \mathbf{t} \cdot \mathbf{\tilde{v}} \mathrm{d}A +\sum_\alpha \int_{\partial \Omega} \Xi^k \tilde{\dot{\gamma}}^k  \mathrm{d}A=
\int_{\Omega} \boldsymbol{\sigma}: \mathrm{grad} (\mathbf{\tilde{v}}) \mathrm{d}V +\sum_k \int_{\partial \Omega} (\pi^k  \tilde{\dot{\gamma}}^k+ \boldsymbol{\kappa}^k \cdot \mathrm{grad}( \tilde{\dot{\gamma}}^k))  \mathrm{d}V
\end{equation}

\noindent where the left- and right-hand sides of this equation correspond to the external and the internal power, respectively, and the fields $\tilde{\mathbf{v}}$ and $\tilde{\dot{\gamma}}^k$ are the virtual fields of the velocity and slip rate, respectively.  $\pi^k$ and $\boldsymbol{\kappa}^k$ are the higher order stresses conjugated with the slip rate (scalar microforce) and the gradient of the slip rate (vector microforce). From the virtual power expression, the resulting balance equations are the classical macroscopic force and momentum balances and the microforce balance,

\begin{equation}
\label{eq:gurtin_microforce}
\mathrm{div} \boldsymbol{\kappa}^k+\tau^k -\pi^k=0 \ \mathrm{for} \ k=1,..,n.
\end{equation}

Note that the macroscopic stress tensor enters in the microforce balance through the resolved shear stress $\tau^k$, eq. \eqref{eq:resolved}. In this framework, $\pi^k$ can be viewed as an internal resistance force to the slip caused by the other dislocations and the microforce vector $\boldsymbol{\kappa}^k$ represents the interaction of dislocations through surfaces.

The second ingredient of the theory is the constitutive equation for the microstresses. In \cite{GURTIN20025}, the elastic free energy, eq. \eqref{eq:free_energy2}, is augmented by a defect energy $\Psi$ that depends on the dislocation tensor $\mathbf{G}$ eq. (\ref{Nye_finite_strain}), defined in \cite{CERMELLI20011539}. If a quadratic expression is chosen, the defect energy $\Psi$ is given by

\begin{equation}
\label{eq:defect_energy}
\Psi=\frac{1}{2}\lambda |\mathbf{G}|^2
\end{equation}

\noindent where $\lambda$ is an scalar material parameter, with force dimensions, that can be split in a reference modulus of strength $\pi_0$ with dimensions of stress and the square of a characteristic length scale $l$ \citep{BARDELLA2006128}, 

\begin{equation}
\lambda=\pi_0 l^2.
\end{equation}

From the expression of the defect energy, eq. \eqref{eq:defect_energy}, a linear relation is found between the microforce vector and the dislocation tensor,

\begin{equation}
\boldsymbol{\kappa}^k=\lambda J^{-1} \mathbf{F}^e (\mathbf{m}^k \otimes \mathbf{G}\mathbf{s}^k).
\end{equation}

Finally, higher order boundary conditions should be applied in the external boundaries. Two type of microscopic boundary conditions are usually considered in higher order SGCP, \emph{microfree} and \emph{microhard} boundary conditions. A surface $S$ has microfree boundary conditions if 

\begin{equation}
\boldsymbol{\kappa}^k  \cdot \mathbf{n}=0 \ \mathrm{on} \ S, \ k = 1, 2,...,n
\end{equation}

\noindent and this implies that slip through that surface $S$ is not restricted. Thus, the microforce vector is directly linked to the applied macroscopic traction on that boundary. The microhard boundary condition establishes that the plastic slip on a surface $S$ is restricted,

\begin{equation}
\dot{\gamma}^k=0 \ \mathrm{on} \ S, \ k = 1, 2,...,n
\end{equation}

\noindent This condition emulates, e.g. a passivation layer that does not allow the dislocation flux. Other special boundary conditions can be applied at the grain boundaries based on the flow of Burgers vector at and across the boundary surface, as discussed by \cite{GURTIN20051}.

The finite element implementation of higher-order SGCP models is a complicated task. Firstly, it requires higher order continuity associated with higher order strain gradient terms, and this condition is normally addressed by means of mixed finite elements formulations \citep{SHU1999297}. Secondly, the application of standard implicit integration algorithms to higher-order theories presents difficulties in both efficiency and accuracy for some relevant boundary value problems, as shown in the implementation of Gurtin macroscopic theory of 2005 proposed by \cite{Lele2008}. Different numerical alternatives have been used to improve the efficiency of finite element implementations, using discontinuous Galerkin  \citep{Ostien2008}, explicit integration \citep{BORG2008688}, or the recent implicit viscoplastic approach proposed by \cite{Panteghini2016} using a special viscoplastic potential. In addition, the implementation of higher-order SGCP models implies  the additional computational cost of introducing a very large number of kinematic variables. For instance,  the number of kinematic variables per node will increase from 3 to 15 if a theory similar to the one presented by \cite{GURTIN20025} is implemented in a three dimensional model of a FCC material. Thus, higher-order SGCP models  have not been applied to simulate the behavior of complex three dimensional RVEs of polycrystals and most of the computational homogenization studies in the literature using SGCP are based in lower-order theories. An interesting and promising alternative to overcome some of the limitations of the higher-order SGCP models is the use of an FFT-based framework to solve the boundary value problem of higher-order models, as shown by \cite{LN16}, who used the small strain version of \cite{GURTIN20025} theory to analyze grain size effects in polycrystals.

\subsection{Deformation by twinning}
In addition to dislocation slip, plastic deformation in some metals with low symmetry crystal structures (such as HCP Ti, Mg and Zr) can occur by twinning \citep{MW73}. This deformation mechanism is present when there are not enough  slip systems for an arbitrary shape change of the crystal and twinning provides an additional mechanism to accommodate deformation. A mechanical twin formally corresponds to a sheared volume for which the lattice orientation is transformed into its mirror image across a so-called twin or habit plane (oblique dividing plane defined by the twinning direction). The sheared region of the crystal undergoes an irreversible shear deformation whose value is determined by the lattice geometry and twin plane.

Mechanical twinning is a process that involves two steps. The first one is the nucleation and propagation of a thin twin band across the grain, starting normally from a grain boundary. Afterwards, the twinned region propagates in the direction perpendicular to the twin plane and eventually occupies most of the parent grain \citep{MW73}. 

The geometrical description of deformation twinning was stablished in 1965 \citep{Bilby240}. Since then, many formulations have been published, in which the geometrical description and the driving forces for twinning have been studied for different materials \citep{CHRISTIAN19951}. The introduction of deformation by twinning in CP models was taken into account since the early developments of CP \citep {HOUTTE1978591,TOME19912667}, specially due to its relevance on the deformation of  HCP metallic alloys. The mechanical process of twinning is very complex and strong simplifications are needed for the introduction in a CP framework. The first models were applied in the context on mean-field approximations \citep{HOUTTE1978591,TOME19912667} and one of the main challenges of introducing twinning was accounting for the large number of crystal orientations that appear due the formation of twins within each grain. \cite{HOUTTE1978591} tracked the evolution of the volume fraction of the twinned regions in each grain, but reorientation was performed only on selected full grains following a statistical criterion based on the evolution of volume fraction of the twinned regions in the grain and in the entire polycrystal. Thus, the number of grains was kept constant in this approach. An improvement with respect to this approach was proposed by \cite{TOME19912667} leading to the predominant twinning reorientation (PTR) model that has been extensively used within the framework of the VPSC formulation to  account for deformation twinning \citep{PTK07}. In addition, an alternative model was also proposed in \cite{TOME19912667} based on a discretization of the orientation space $SO3$ and the representation of the orientations as volume fractions in the discrete space. With this framework, the accommodation of deformation by twinning does not increase the number of orientations in the model and non-predominant orientations can also be represented. Nevertheless, the main limitaiton of these models is that they were developed within the context of mean-field approximations for polycrystal plasticity and, therefore, cannot be easily included in single crystal plasticity formulations. 

The first model that was able to describe twinning deformation of the individual single crystals was developed by \cite{K98}, and other similar models have been developed  afterwards \citep{SA03, KOWALCZYKGAJEWSKA201028,AD13a, AD13b, CK15, MD16}. Kalidindi's model has been extensively used \citep{AZ31_Joshi,HLD14, HHP14, HHS15,KHAN2016772,Jung2017}, and  will be briefly described here.

The twinning model of \cite{K98} is a two-scale model. Each material point has a substructure and is divided into two phases, a parent region and a twinned region (Fig. \ref{Fig:decomp}), which might be potentially formed by $N_{tw}$ subregions. Each subregion belongs to a given twinning system $\alpha$ and its volume fraction is $f^\alpha$. Thus, the parent region volume fraction is given by $1-\sum_{\alpha =1}^{N_{tw}} f^\alpha$. The material point is considered a \emph{composite} material in which the iso-strain hypothesis holds ($\mathbf{F}$ and $\mathbf{F}^e$ are the same in all regions). The plastic deformation is the result of three mechanisms, and the plastic velocity gradient in the intermediate configuration contains three terms, the standard one for dislocation slip, eq. \eqref{eq:Lp_def}, and two new contributions. The first extra contribution, $\mathbf{L}_{tw}^p$, stands for the rate of deformation due to the twin transformation of a  volume fraction  $\mathrm{d}f^\alpha$ of the parent phase

\begin{figure}
\includegraphics[width=400pt,keepaspectratio]{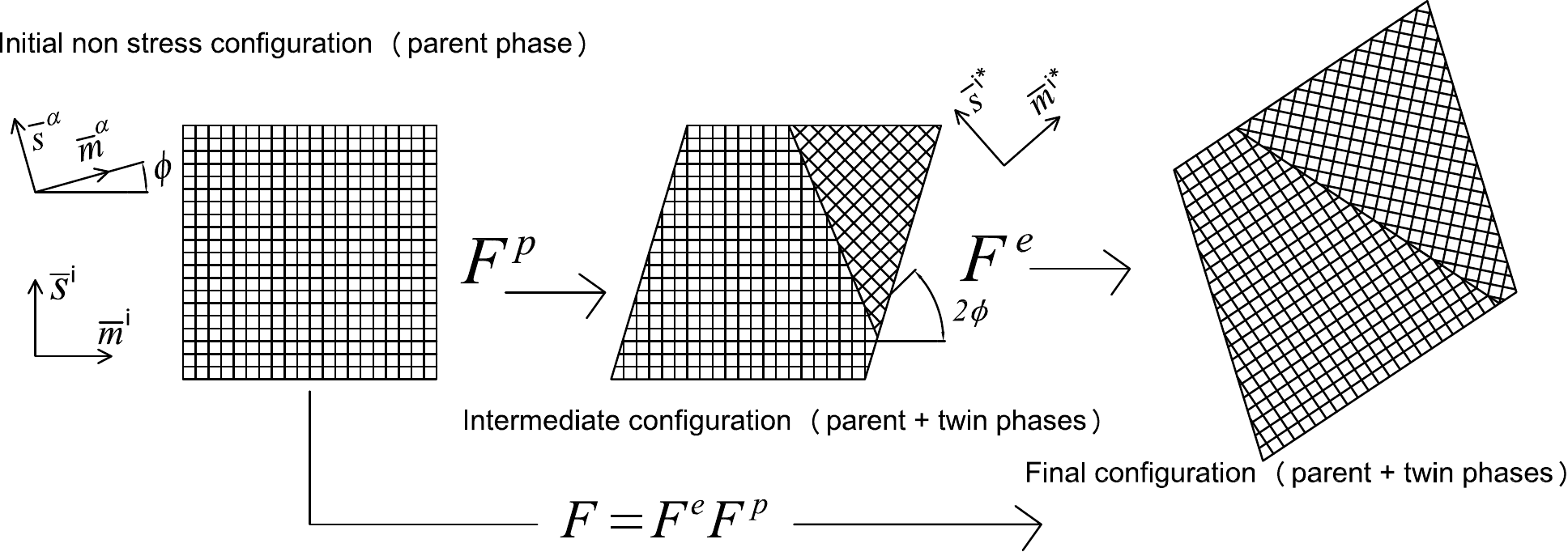}
\caption{Multiplicative decomposition indicating material point subdivision in parent and twin phases. Reprinted  from \cite{HLD14}}
\label{Fig:decomp}
\end{figure}

\begin{equation}
\mathbf{L}_{tw}^p=\sum_{\alpha=1}^{N_{tw}} \dot{f}^\alpha \gamma_{tw}\ 
\mathbf{s}_{tw}^\alpha\otimes \mathbf{m}_{tw}^\alpha
\label{eq:Ltwin}
\end{equation}

\noindent where $\dot{f}^\alpha=\mathrm{d}f^\alpha/\mathrm{d}t$ is the transformation rate  in the twin system $\alpha$, $\mathbf{m}_{tw}^\alpha$ and $\mathbf{s}_{tw}^\alpha$ are unit vectors along the twin plane normal and the twining direction, respectively, and $\gamma_{tw}$ stands for the characteristic shear strain of the twinning mode. It is interesting to note that, contrary to dislocation slip, the accumulated plastic deformation by twinning is limited, and the maximum plastic shear that can be by accommodated is $\gamma_{tw}$ that corresponds to the full transformation of a material point into a given twin variant. 

The second contribution to plastic slip corresponds to the slip in the $N_{sl-tw}$ slip systems of the transformed regions, $\mathbf{L}_{sl-tr}^p$, which can be expressed as,

\begin{equation}
 \mathbf{L}_{sl-tr}^p= \sum_{\alpha=1}^{N_{tw}} f^\alpha\left( \sum_{k^*=1}^{N_{sl-tw}}  \dot{\gamma}^{k^*} \mathbf{s}_{sl}^{k^*}\otimes\mathbf{m}_{sl}^{k^*}\right) 
\end{equation}

\noindent where $\mathbf{s}_{sl}^{k^*}$ and $\mathbf{m}_{sl}^{k*}$ stand for the unit vectors in the slip and normal directions to the slip plane. They can be computed from the orientation of the slip planes  $k$ in the parent grain by means of  the  matrix, $\mathbf{Q}^\alpha$, 

\begin{equation}
\mathbf{Q}^\alpha=2 \mathbf{m}_{tw}^\alpha\otimes \mathbf{m}_{tw}^\alpha-\mathbf{I}.
\end{equation}

\noindent that takes into account the rotation of the crystal due to twinning in the $\alpha$ plane.

An evolution equation defining $\dot{f}^\alpha$, eq. (\ref{eq:Ltwin}), has to be specified to complete the flow rule for twinning. It is well accepted that the driving force for twin growth is the resolved shear stress on the twinning system. Accordingly, a viscous law depending on the resolved shear stress was proposed in \cite{K98}, which is equivalent to the expression used for slip in \cite{Asaro1985923}. Thus,

\begin{equation}\label{eq:twin}
\dot{f}^\alpha=\dot{f}_0
\left(\frac{\langle\tau^\alpha\rangle}{g^\alpha}\right)^\frac{1}{m} \qquad
\mathrm{with} \qquad \langle \tau \rangle=\left\{\begin{array}{c} \tau \
    \mathrm{if} \ \tau\geq 0 \\ 0 \ \mathrm{if} \ \tau<0 \end{array}\right.
\end{equation}

\noindent where $\dot{f}_0$ is a reference transformation rate, $m$ the strain rate sensitivity parameter and $<x>$ stand for the Macaulay brackets that are introduced to account for the polar nature of twinning deformation. $\tau^\alpha$ is the resolved stress in the twinning system $\alpha$ and its value is given by 
\begin{equation}
 \tau^\alpha=\mathbf{S}^{parent}:\mathbf{s}_{tw}^\alpha \otimes\mathbf{m}_{tw}^\alpha
\end{equation}
where $\mathbf{S}^{parent}$ denotes the value of the second Piola-Kirchoff stress in the parent region. Note that  the stresses in the parent and twinned region are different due to the isostrain approach. In particular, $\mathbf{S}^{parent}$ is obtained directly from the linear relation between the elastic strain in the intermediate configuration, $\mathbf{E}^e$ (common for both parent and twin phases)
\begin{eqnarray}
\label{eq:hook}
\mathbf{S}^{parent}=\mathbb{L}: \mathbf{E}^e
\end{eqnarray}

\noindent where $\mathbb{L}$ stands for the fourth-rank elastic stiffness tensor of the crystal in its original orientation. The Piola-Kirchhoff stress tensor for the full integration point (containing the parent and all the twinned phases) in the intermediate configuration, $\mathbf{S}$,  is obtained  in this case from the volume-averaged stress tensors in the different phases
\begin{equation}
\mathbf{S}=\bigg(1-\sum_{\alpha=1}^{N_{tw}}f^\alpha \bigg)\mathbf{S}^{parent}+\sum_{\alpha=1}^{N_{tw}} f^\alpha \mathbf{S}^{\alpha}
\end{equation}
\noindent where $\mathbf{S}^{\alpha}$ is the stress in each of the twinned regions, and is obtained from an expression equivalent to eq. (\ref{eq:hook}).

\section{Viscoplastic self-consistent homogenization of polycrystals}

The computation of the effective mechanical response and texture evolution of polycrystalline materials using homogenization theory has a long tradition \citep{S28, T38} and self-consistent approximations have been extensively used to deal with this problem. The 1-site VPSC theory of polycrystal deformation can be traced back to the work of \cite{MCA87}, who established a homogenization procedure based on an iterative method involving the computation of integrals in ellipsoidal domains of the infinite-medium Green's function, customarily used in the solution of the PDEs governing the micromechanical response of heterogeneous materials. This formulation was implemented numerically by \cite{LT93}  taking into account the polycrystal anisotropy, leading to the first version of the VPSC code. Since its inception, the VPSC code has experienced several improvements and extensions, e.g. recrystallization \citep{WCB97}; 2-site approximation for 2-phase polycrystals \citep{LC97}; VPSC modelling of lamellar structures \citep{LUH98}; relative directional compliance interaction \citep{T99}; second-order linearization \citep{LTP07};  improved VPSC modelling of twinning using the PTR approach \citep{PTK07}; dislocation density-based hardening models \citep{BT08};  climb and glide VPSC model \citep{LHT10}; dilatational VPSC for porous polycrystals \citep{LIP11}; lattice rotation rate fluctuation calculation \citep{LZK16};  improved VPSC for arbitrarily low rate sensitivities \citep{KZB16}; improved hardening laws for strain-path changes \citep{WBT16}; VPSC prediction of intragranular misorientation evolution \citep{ZPL17}, etc. The VPSC homogenization strategy is nowadays extensively used to simulate plastic deformation of polycrystalline aggregates and for interpretation of experimental results in metals, minerals and polymers.

The self-consistent theory is one of the most commonly used homogenization methods to estimate the mechanical response behavior of polycrystals and was originally proposed by \cite{H54} for linear elastic materials. For nonlinear aggregates (as those formed by grains deforming in the viscoplastic regime), several self-consistent (SC) approximations have been proposed. They  differ in the procedure used to linearize the non-linear mechanical response at the grain level, but all of them eventually make use of the original linear SC theory. They include the secant \citep{H65, Hutchinson1976101}, the tangent \citep{Molinari19872983, LT93} and the affine \citep{MBS00}  first-order approximations, which are based on linearization schemes that use the information on field averages at the grain level, and disregard higher-order statistical information inside the grains. However, the above assumption may be questionable in materials which present strong directionality and/or large variations in local properties. This is the case of low rate-sensitivity materials, aggregates made of highly anisotropic grains, voided and/or multiphase polycrystals. In all these cases, strong deformation gradients are likely to develop inside grains because of differences in properties with neighbour crystals or phases, including voids.

More accurate nonlinear homogenization methods were developed, mainly due to the work of Ponte Casta\~{n}eda and collaborators, to overcome the above limitations. These methods use linearization schemes at grain level that also incorporate accessible information on the second moments of the stress field distributions in the grains. These more elaborate SC formulations are based on the concept of a linear comparison material, which expresses the effective potential of the nonlinear viscoplastic polycrystal in terms of that of a linearly viscous aggregate whose properties are determined from suitably-designed variational principles. Ponte Casta\~{n}eda's first variational method was originally proposed for nonlinear composites \citep{P91} and then extended to viscoplastic polycrystals \citep {BP95}. 
It makes use of the SC approximation for linearly viscous polycrystals to obtain bounds and estimates for nonlinear viscoplastic polycrystals. The more recent second-order method, proposed for nonlinear composites \citep{P02}, and later extended to visoplastic polycrystals \citep{LP04}, uses the SC approximation for a more general class of linearly viscous polycrystals, having a non-vanishing strain-rate at zero stress, to generate even more accurate SC estimates for viscoplastic polycrystals. The implementation of a fully anisotropic second-order approach inside the VPSC code \citep{LTP07} has been a necessary step towards improving its predictive capability for polycrystalline materials that exhibit high contrast in local properties. Unavoidably, this improved capability came at the expense of more complex and numerically demanding algorithms. In what follows, the VPSC formulation is first presented using the affine linearization scheme \citep{MBS00}, and the second-order linearization procedure \citep{LP04} is described next.

The self-consistent formulation represents the polycrystal by means of weighted, ellipsoidal, statistically-representative (SR) grains. Each of these SR grains represent the average behavior of all the grains with a particular crystallographic orientation and morphology but different local environments. These SR grains should be regarded as representing the behavior of mechanical phases, i.e. all the single crystals with a given orientation ($r$) belong to mechanical phase ($r$) and are represented by the SR grain ($r$). Note the difference between \textit{mechanical phases}, which differ from each other only in terms of crystallographic orientation and/or morphology, and actual \textit{phases}, which differ from each other in crystallographic structure and/or composition. In what follows, \textit{SR grain ($r$)} and \textit{mechanical phase ($r$)} will be used interchangeably. The weights represent volume fractions and are chosen to reproduce the initial texture of the material. Each representative grain is treated as an ellipsoidal viscoplastic inclusion embedded in an effective viscoplastic medium. Both inclusion and medium are anisotropic. Plastic deformation in the inclusion is accommodated by dislocation slip activated by a resolved shear stress. As a consequence of all the above assumptions, the representation of the polycrystalline aggregate under the SC model is non-space-resolved, and corresponds to an entire class of polycrystals with microstructures consistent with a given statistical distribution. 

In general, homogenization models for viscoplastic deformation assumes that the elastic strains are much smaller than the plastic ones and, thus, are neglected. The plastic deformation rates are constitutively related to stress in the current configuration using small strain kinematics, resulting in relations between Cauchy stress and velocity gradient (instead of the second Piola-Kirchoff stress and the deformation gradient). Once the velocity gradient is obtained, the evolution of microstructure and micromechanical variables can be calculated by integrating the velocity gradient field, or local (grain) averages of the latter, in small time increments to update the current configuration of the material. 

\subsection{Local constitutive behavior and homogenization}

Asumming small-strain kinematics for the deformation rates in the current configuration, the macroscopic velocity gradient $V_{i,j} $ applied to a polycrystalline aggregate can be decomposed into an average symmetric strain-rate $\dot{E}_{ij} =\frac{1}{2} \left(V_{i,j} +V_{j,i} \right)$ and an average antisymmetric rotation-rate $\dot{\Omega }_{ij} =\frac{1}{2} \left(V_{i,j} -V_{j,i} \right)$. The plastic component of the deformation is assumed to be much larger than the elastic part and, therefore, the flow is incompressible.  The viscoplastic constitutive behavior at each material point $\mathbf{x}$ is described by means of a non-linear rate-sensitive equation as

\begin{equation} \label{GrindEQ__1_} 
\dot{\boldsymbol{\epsilon} }\left(\mathbf{x}\right)=\sum _{k=1}^{N_{k} }\mathbf{Z}_s^{k} \left(\mathbf{x}\right) \dot{\gamma }^{k} \left(\mathbf{x}\right)  \end{equation} 
\noindent where $\dot{\boldsymbol\epsilon}$ is the local strain rate and
 $\mathbf{Z}_s^{k} \left(\mathbf{x}\right)=\frac{1}{2} \left(\mathbf{s}^{k} \left(\mathbf{x}\right)\otimes \mathbf{m}^{k} \left(\mathbf{x}\right)+\mathbf{m}^{k} \left(\mathbf{x}\right)\otimes \mathbf{s}^{k} \left(\mathbf{x}\right)\right)$  is the symmetric Schmid tensor associated with slip system $k$ and $\dot{\gamma }^{k} $ stands for the local shear rate on slip system $k$. The shear rate on each system follows a power-law relation \citep{PEIRCE1983133}
\begin{multline} \label{GrindEQ__2_} 
\dot{\gamma }^{k} \left(\mathbf{x}\right)=\dot{\gamma }_{0}  \left(\frac{|\tau ^{k} \left(\mathbf{x}\right)|}{g^{k} \left(\mathbf{x}\right)} \right)^{n} \mathrm{sign}\left(\tau ^{k} \left(\mathbf{x}\right)\right)= \\ = \dot{\gamma }_{0} \left(\frac{|\mathbf{Z}_s^{k} \left(\mathbf{x}\right):\, \boldsymbol\sigma'\left(\mathbf{x}\right)|}{g^{k} \left(\mathbf{x}\right)} \right)^{n} {\rm sign}\left(\mathbf{Z}_s^{k} \left(\mathbf{x}\right): \boldsymbol\sigma'\left(\mathbf{x}\right)\right) 
\end{multline} where $g^{k} $ is the CRSS of slip system $k$ , $\dot{\gamma }_{0} $ is the reference strain rate,  $n$ the rate-sensitivity exponent and $\dot{\boldsymbol\epsilon}$ and $\boldsymbol\sigma'$ stand for the strain rate and the deviatoric part of the Cauchy stress. If the shear rates are known, the lattice rotation rate, $\dot{\boldsymbol\omega}^{p}$ (or plastic spin) associated with slip activity at a single crystal material point $\mathbf{x}$ is given by:

\begin{equation} \label{GrindEQ__3_} 
\dot{\boldsymbol\omega }^{p} \left(\mathbf{x}\right)=\sum _{k}\mathbf{Z}_{a}^{k} \left(\mathbf{x}\right)   \dot{\gamma }^{k} \left(\mathbf{x}\right) 
\end{equation} 

\noindent where $\mathbf{Z}_{a}^{k}\left(\mathbf{x}\right)=\frac{1}{2} \left(\mathbf{s}^{k} \left(\mathbf{x}\right)\otimes \mathbf{m}^{k} \left(\mathbf{x}\right)-\mathbf{m}^{k} \left(\mathbf{x}\right)\otimes \mathbf{s}^{k} \left(\mathbf{x}\right)\right)$  is the antisymmetric Schmid tensor. Note that although eqs. \eqref{GrindEQ__1_} -  \eqref{GrindEQ__3_} can be used to deal with crystal deformation by slip and twinning, only slip will be considered in the examples that follow in the context of both mean-field and full-field approaches, to avoid the additional complication of twinning reorientation. Moreover, the constitutive behavior described by Eqs.  \eqref{GrindEQ__1_} -  \eqref{GrindEQ__3_} does not consider other possible crystal deformation mechanisms and microstructural evolution processes, such as climb, grain-boundary sliding or recrystallization.

Assuming that the following linear relations, i.e. approximations of the actual non-linear relations, eqs.  \eqref{GrindEQ__1_} -  \eqref{GrindEQ__2_}, hold for the SR grain ($r$), it can be written

\begin{equation} \label{GrindEQ__4_} 
\dot{\boldsymbol\epsilon }\left(\mathbf{x}\right)=\mathbb{M}^{(r)} \boldsymbol\sigma'\left(\mathbf{x}\right)+\dot{\boldsymbol\epsilon }^{0(r)}  
\end{equation} 

\noindent where $\mathbb{M}^{(r)} $and $\dot{\boldsymbol\epsilon }^{0(r)} $ are the viscous compliance tensor and back-extrapolated strain rate (strain rate under zero stress) of grain ($r$), respectively, which depend on the linearization assumption. For example, under the affine linearization, they are given by

\begin{equation} \label{GrindEQ__5_} 
\mathbb{M}^{(r)} =n\dot{\gamma}_{0}  \sum _{k=1}^{N_{k} }\frac{\mathbf{Z}_s^{k(r)} \otimes  \mathbf{Z}_s^{k(r)}}{g^{k(r)}}  \left(\frac{|\mathbf{Z}_s^{k(r)} :\boldsymbol{\sigma}'^{(r)}|}{g^{k(r)} } \right)^{n-1}  
\end{equation} 

\begin{equation} \label{GrindEQ__6_} 
\dot{\boldsymbol\epsilon }^{0(r)} =\left(1-n\right)\; \dot{\gamma }_{0} \, \sum _{k=1}^{N_{k} }\left(\frac{|\mathbf{Z}_s^{k(r)} : \boldsymbol\sigma'^{(r)}|}{g^{k(r)} } \right)^{n} {\rm sign} \left(\mathbf{Z}_s^{k(r)} : \boldsymbol{\sigma}'^{(r)} \right) 
\end{equation} 

\noindent where the index ($r$) on the magnitudes on the right-hand side of these equations indicates average over the SR ($r$), e.g. $\boldsymbol{\sigma}'^{(r)} =\left\langle \boldsymbol\sigma'\left(\mathbf{x}\right)\right\rangle ^{(r)} =\left\langle \boldsymbol\sigma'\right\rangle ^{(r)} $. 

Following \cite{H65} and \cite{Hutchinson1976101}, the homogenized behavior of a linear heterogeneous medium whose local behavior is described by eq. \eqref{GrindEQ__4_} can also be described by an analogous linear relation at the effective (macroscopic) level

\begin{equation} \label{GrindEQ__7_} 
\dot{\mathbf E}=\overline{\mathbb M} \boldsymbol\Sigma'+\dot{\mathbf E}^{0}  
\end{equation} 

\noindent where $\dot{\mathbf{E}}$ and $\boldsymbol\Sigma'$ are the overall (macroscopic) deviatoric strain rate and stress tensor, respectively, and $\overline{\mathbb M}$ and $\dot{\mathbf E}^{0} $ stand for the viscous compliance tensor and the back-extrapolated strain rate, respectively, of an {\it a priori} unknown Homogeneous Equivalent Medium (HEM). The response of this HEM is obtained by the linear SC method. The problem underlying the SC method is that of an inhomogeneous domain (r) of moduli $\mathbb{M}^{(r)}$ and $\dot{\boldsymbol\epsilon }^{0 (r)} $, embedded in an infinite medium of moduli $\overline{\mathbb M}$ and $\dot{\mathbf E}^{0} $. Invoking the concept of the equivalent inclusion \citep{M87}, the local constitutive behavior in grain ($r$) can be rewritten as

\begin{equation} \label{GrindEQ__8_} 
\dot{\boldsymbol\epsilon }\left(\mathbf{x}\right)=\overline{\mathbb M} \boldsymbol\sigma'\left(\mathbf{x}\right)+\dot{\mathbf E}^{0} +\dot{\boldsymbol\epsilon }^{*} \left(\mathbf{x}\right) 
\end{equation} 

\noindent where $\dot{\boldsymbol\epsilon }^{*} \left(\mathbf{x}\right)$ is an eigenstrain-rate field, which follows from replacing the inhomogeneity by an equivalent inclusion. Rearranging and subtracting eq. \eqref{GrindEQ__7_} from Eq. \eqref{GrindEQ__8_} gives

\begin{equation} \label{GrindEQ__9_} 
\tilde{\boldsymbol\sigma}'\left(\mathbf{x}\right)=\overline{\mathbb{L}} \left[\tilde{\dot{\boldsymbol\epsilon }}\left(\mathbf{x}\right)-\dot{\boldsymbol\epsilon }^{*} \left(\mathbf{x}\right)\right].
\end{equation}

\noindent where the symbol "$\sim$" denotes local deviations from macroscopic values of the corresponding magnitudes, and $\overline{\mathbb{L}}=\overline{\mathbb M}^{-1} $. Combining eq. \eqref{GrindEQ__9_} with the equilibrium condition gives:

\begin{equation} \label{GrindEQ__10_} 
\sigma _{ij,j}^{} \left(\mathbf{x}\right)=\tilde{\sigma }_{ij,j}^{} \left(\mathbf{x}\right)=\tilde{\sigma}'_{ij,j} \left(\mathbf{x}\right)+\tilde{\sigma }_{,i}^{m} \left(\mathbf{x}\right) 
\end{equation} 

\noindent where $\sigma _{ij}^{} $ and $\sigma _{}^{m} $ are the Cauchy stress tensor and Cauchy the mean stress, respectively. From the relation between the strain rate and the velocity gradient deviations, $\tilde{\dot{\epsilon }}_{ij} \left(\mathbf{x}\right)=\frac{1}{2} \left(\tilde{v}_{i,j} \left(\mathbf{x}\right)+\tilde{v}_{j,i} \left(\mathbf{x}\right)\right)$, and taking into account the incompressibility condition associated with plastic deformation, 

\begin{equation} \label{GrindEQ__11_} 
\left|\begin{array}{l} {\; \overline{L}_{ijkl} \; \tilde{v}_{k,lj} \left(\mathbf{x}\right)+\tilde{\sigma }_{,i}^{m} \left(\mathbf{x}\right)+\varphi _{ij,j}^{} \left(\mathbf{x}\right)=0} \\ {\; \tilde{v}_{k,k} \left(\mathbf{x}\right)=0} \end{array}\right.  
\end{equation} 

\noindent where

\begin{equation} \label{GrindEQ__12_} 
\varphi _{ij}^{} \left(\mathbf{x}\right)=-\overline{L}_{ijkl} \dot{\epsilon }_{kl}^{*} \left(\mathbf{x}\right) 
\end{equation} 

\noindent is a heterogeneity or \textit{polarization field}, and its divergence $f_{i} \left(\mathbf{x}\right)=\varphi _{ij,j} \left(\mathbf{x}\right)$ is an artificial external force field applied to the material. 

The set of equations \eqref{GrindEQ__11_} consists of four differential equations with four unknowns: three are the components of the deviation from the average value of the velocity vector $\tilde{v}_{i} \left(\mathbf{x}\right)$, and one is the deviation of the mean stress $\tilde{\sigma}^{m}\left(\mathbf{x}\right)$. A set of N linear differential equations with N unknown functions and a polarization term can be solved using the Green function method. $G_{km} \left(\mathbf{x}\right)$ and  $H_{m} \left(\mathbf{x}\right)$ are the Green functions associated with $\tilde{v}_{i} \left(\mathbf{x}\right)$ and $\tilde{\boldsymbol\sigma}^{m} \left(\mathbf{x}\right)$, respectively, which solve the auxiliary problem of a unitary volumetric force with a single non-vanishing {\it m}-component,

\begin{equation} \label{GrindEQ__13_} 
\left|\begin{array}{l} {\; \overline{L}_{ijkl}^{} \; G_{km,lj} \, \left(\mathbf{x}-\mathbf{x}'\right)+H_{m,i} \, \left(\mathbf{x}-\mathbf{x}'\right)+\delta _{im} \, \delta \, \left(\mathbf{x}-\mathbf{x}'\right)=0} \\ {\; G_{km,k} \left(\mathbf{x}-\mathbf{x}'\right)=0} \end{array}\right.  
\end{equation}

\noindent  where the Green's function $G_{km} \left(\mathbf{x}-\mathbf{x}'\right)$ is the velocity component in the $x_{k}$-direction at point $\mathbf{x}$ when a unit body force in the $x_{m}$-direction is applied at point $\mathbf{x}'$ in an infinitely extended material and $H_{m} \left(\mathbf{x}-\mathbf{x}'\right)$ is the corresponding response in mean stress.

Once the solution of eq. \eqref{GrindEQ__13_} is obtained, the velocity field is given by the convolution integral:

\begin{equation} \label{GrindEQ__14_} 
\tilde{v}_{k} \left(\mathbf{x}\right)=\int _{\mathbb{R}^3 }G_{ki} \left(\mathbf{x}-\mathbf{x}'\right) f_{i} \left(\mathbf{x}'\right) d\mathbf{x}'  
\end{equation} 

The set of equations \eqref{GrindEQ__13_} can be solved using Fourier transforms. Expressing the Green functions in terms of their inverse Fourier transforms, the differential equations \eqref{GrindEQ__13_} can be transformed into an algebraic set 

\begin{equation} \label{GrindEQ__15_} 
\left|\begin{array}{l} {\; \alpha _{j} \, \alpha _{l} \, \overline{L}_{ijkl} \; k^{2} \hat{G}_{km} \, \left(\boldsymbol\xi \right)+\alpha _{i} \, ik\hat{H}_{m} \, \left(\boldsymbol\xi \right)=\delta _{im} } \\ {\; \alpha _{k} \, k^{2} \hat{G}_{km} \left(\boldsymbol\xi \right)=0} \end{array}\right.  
\end{equation} 

\noindent where $k$ and $\boldsymbol\alpha $ are the modulus and the unit vector associated with a point of Fourier space $\boldsymbol\xi =k\boldsymbol\alpha $, respectively. Calling $A_{ik}^{'} \left(\alpha \right)=\alpha _{j} \, \alpha _{l} \, \overline{L}_{ijkl}^{} $, the set \eqref{GrindEQ__15_} can be expressed as a matrix product $\mathbf{A} \mathbf{B}= \mathbf{C}$ where $\mathbf{A}$, $\mathbf{B}$ and $\mathbf{C}$ are given by

\begin{multline} \label{GrindEQ__16_} \quad \quad \quad  \quad \quad
\mathbf{A}\left(\boldsymbol\alpha \right) = \left[
\begin{array}{cccc}  A_{11}^{'}  & A_{12}^{'}  & A_{13}^{'} & \alpha_{1} \\ A_{21}^{'}  & A_{22}^{'} & A_{23}^{'}  & \alpha_{2}  \\ A_{31}^{'}  & A_{32}^{'} & A_{33}^{'}  & \alpha_{3}  \\ \alpha _{1}  & \alpha _{2}  & \alpha _{3} & {0} 
\end{array} \right] \\
\mathbf{B} = \left[
\begin{array}{ccc}  {k^{2} \hat{G}_{11} } & {k^{2} \hat{G}_{12} } & {k^{2} \hat{G}_{13} }  \\ {k^{2} \hat{G}_{21} } & {k^{2} \hat{G}_{22} } & {k^{2} \hat{G}_{23} }  \\ {k^{2} \hat{G}_{31} } & {k^{2} \hat{G}_{32} } & {k^{2} \hat{G}_{33} }  \\ {ik\hat{H}_{1} } & {ik\hat{H}_{2} } & {ik\hat{H}_{3} } 
\end{array} \right] \quad \quad
\mathbf{C} = \left[\begin{array}{ccc}
{ 1} & { 0} & {0} \\ { 0} & {1} & { 0} \\ { 0} & {0} & {1} \\ {0} & {0} & {0}
\end{array} \right]
\end{multline} 

\noindent Using the explicit form of matrix $\mathbf{C}$, 

\begin{equation} \label{GrindEQ__17_} 
\mathbf{B} =\mathbf{A}^{-1}  \mathbf{C}=\left[\begin{array}{ccc} {A_{11}^{-1} } & {A_{12}^{-1} } & {A_{13}^{-1} } \\ {A_{21}^{-1} } & {A_{22}^{-1} } & {A_{23}^{-1} } \\ {A_{31}^{-1} } & {A_{32}^{-1} } & {A_{33}^{-1} } \\ {A_{41}^{-1} } & {A_{42}^{-1} } & {A_{43}^{-1} } \end{array}\right].
\end{equation} 

\noindent gives: 

\begin{equation} \label{GrindEQ__18_} 
k^{2} \hat{G}_{ij} \left(\boldsymbol\xi \right)=A_{ij}^{-1} \left(\boldsymbol\alpha \right)\; \; \; \; \; (i,j=1,3) 
\end{equation} 

\noindent Comparing eqs. \eqref{GrindEQ__16_} and \eqref{GrindEQ__17_}, since the components of $\mathbf{A}$ are real functions of $\boldsymbol\alpha$, so are those of $k^{2} \hat{G}_{ij} \left(\boldsymbol\xi \right)$. This property leads to real integrals in the derivation that follows.

Knowing the Green's function expression in Fourier space, the solution of the eigenstrain-rate problem can be obtained using convolution integrals. Taking partial derivatives to eq. \eqref{GrindEQ__14_} leads to

\begin{equation} \label{GrindEQ__19_} 
\tilde{v}_{k,l}^{} \left(\mathbf{x}\right)=\int _{\mathbb{R}^3 }G_{ki,l} \left(\mathbf{x}-\mathbf{x}'\right)\; f_{i} \left(\mathbf{x}'\right)\; d\mathbf{x}'  .
\end{equation} 

Replacing the expression of the artificial volumetric force field in eq. \eqref{GrindEQ__19_}, recalling that $\partial G_{ij} \left(\mathbf{x}-\mathbf{x}'\right)/\partial \mathbf{x}=-\partial G_{ij} \left(\mathbf{x}-\mathbf{x}'\right)/\partial \mathbf{x}'$, integrating by parts, and using the divergence theorem \citep{M87}, 
 
\begin{equation} \label{GrindEQ__20_} 
\tilde{v}_{k,l}^{} \left(\mathbf{x}\right)=\int _{\mathbb{R}^3 }G_{ki,jl} \left(\mathbf{x}-\mathbf{x}'\right)\; \varphi _{ij}^{} \left(\mathbf{x}'\right)\; d\mathbf{x}'.  
\end{equation} 

The integral eq. \eqref{GrindEQ__20_} provides an exact implicit solution to the problem. Furthermore, it is known from Eshelby's elastic inclusion formalism that the stress and strain are constant over the domain of the inclusion ($r$), $\Omega_r$, if the eigenstrain is uniform over an ellipsoidal domain where the stiffness tensor is uniform. This suggests the use of an {\it a priori} unknown constant polarization within the volume  of the ellipsoidal inclusion and allows  to average the local field of eq. \eqref{GrindEQ__20_} over the domain  and obtain an average strain rate inside the inclusion of the form

\begin{equation} \label{GrindEQ__21_} 
\tilde{v}_{k,l}^{(r)} =\left(-\frac{1}{\Omega_r } \int _{\Omega_r }\int _{\Omega_r }G_{ki,jl} \left(\mathbf{x}-\mathbf{x}'\right)\; d\mathbf{x}\, d\mathbf{x}'  \right)\; \overline{L}_{ijmn}^{} \dot{\epsilon }_{mn}^{*(r)}  
\end{equation} 

\noindent where $\tilde{v}_{k,l}^{(r)}$ and $\dot{\epsilon }_{mn}^{*(r)} $ have to be interpreted as average quantities inside the inclusion. Expressing the Green's function in terms of the inverse Fourier transform and taking derivatives,

\begin{multline} \label{GrindEQ__22_} 
\tilde{v}_{k,l}^{(r)} =\left(\frac{1}{8\pi ^{3} \Omega_r } \int _{\Omega_r }\int _{\Omega_r }\int _{\mathbb{R}^3 }\alpha _{j} \alpha _{l} \left(\, k^{2} \hat{G}_{ki} \left(\boldsymbol\xi \right)\right)\; \exp \left[-i\boldsymbol\xi \left(\mathbf{x}-\mathbf{x}'\right)\right]\, d\boldsymbol\xi d\mathbf{x}\, d\mathbf{x}'   \right)  \\ \overline{L}_{ijmn}^{} \dot{\epsilon }_{mn}^{*(r)} = T_{klij}^{} \, \overline{L}_{ijmn}\dot{\epsilon }_{mn}^{*(r)} .
\end{multline}

Writing $d\boldsymbol\xi $ in spherical coordinates ($d\boldsymbol\xi =k^{2}\sin \theta dk d\theta d\varphi $) and using the relation \eqref{GrindEQ__18_}, the Green's interaction tensor $T_{klij}$ can be expressed as
\begin{equation} \label{GrindEQ__23_} 
T_{klij} =\frac{1}{8\pi ^{3} \Omega_r } \int _{0}^{2\pi }\; \int _{0}^{\pi }\; \alpha _{j} \, \alpha _{l} \, A_{ki}^{-1} \left(\boldsymbol\alpha \right)  \; \Lambda \left(\boldsymbol\alpha \right)\; \sin \theta \, d\theta \, d\varphi  
\end{equation} 

\noindent where

\begin{equation} \label{GrindEQ__24_} 
\Lambda \left(\boldsymbol\alpha \right)=\int _{0}^{\infty }\; \left(\int _{\Omega_r }\int _{\Omega_r }\exp \left[-i\boldsymbol\xi \left(\mathbf{x}-\mathbf{x}'\right)\right]\; d\mathbf{x}\, d\mathbf{x}'  \right)\; k^{2} dk  
\end{equation} 

Integrating eq. \eqref{GrindEQ__24_} inside an ellipsoidal grain of radii $\left(a,b,c\right)$ (Berveiller et al 1987) and replacing in eq. \eqref{GrindEQ__23_} leads to

\begin{equation} \label{GrindEQ__25_} 
T_{klij} =\frac{abc}{4\pi } \int _{0}^{2\pi }\int _{0}^{\pi }\frac{\alpha _{j} \, \alpha _{l} \, A_{ki}^{-1} \left(\boldsymbol\alpha \right)}{\left[\rho \left(\boldsymbol\alpha \right)\right]^{3} } \sin \, \theta \, d\theta \, d\varphi    
\end{equation} 

\noindent where $\rho \left(\boldsymbol\alpha \right)=\left[(a\alpha _{1} )^{2} +(b\alpha _{2} )^{2} +(c\alpha _{3} )^{2} \right]^{1/2} $. The symmetric and antisymmetric Eshelby tensors (functions of  and the shape of the ellipsoidal inclusion, representing the morphology of the SR grains) are defined as

\begin{equation} \label{GrindEQ__26_} 
S_{ijkl}^{} =\frac{1}{4} \left(T_{ijmn}^{} +T_{jimn}^{} +T_{ijnm}^{} +T_{jinm}^{} \right)\; \overline{L}_{mnkl}^{}  
\end{equation} 
\begin{equation} \label{GrindEQ__27_} 
P _{ijkl}^{} =\frac{1}{4} \left(T_{ijmn}^{} -T_{jimn}^{} +T_{ijnm}^{} -T_{jinm}^{} \right)\; \overline{L}_{mnkl}^{}  
\end{equation} 

\noindent and the average strain rate and rotation rate deviations, $\tilde{\dot{\boldsymbol\epsilon }}^{(r)}$ and $\tilde{\dot{\boldsymbol\omega }}_{}^{(r)}$, in the ellipsoidal domain are obtained by taking symmetric and antisymmetric components to eq. \eqref{GrindEQ__22_} and using eqs. \eqref{GrindEQ__26_} - \eqref{GrindEQ__27_}

\begin{equation} \label{GrindEQ__28_} 
\tilde{\dot{\boldsymbol\epsilon }}_{}^{(r)} =\mathbb{S}\dot{\boldsymbol\epsilon }_{}^{*(r)}  
\end{equation} 

\begin{equation} \label{GrindEQ__29_} 
\tilde{\dot{\boldsymbol\omega }}_{}^{(r)} =\mathbb{P} \dot{\boldsymbol\epsilon }^{*(r)} =\mathbb{P} \mathbb{S}^{-1} \tilde{\dot{\boldsymbol\epsilon }}^{(r)}  
\end{equation} 

\noindent where $\tilde{\dot{\boldsymbol\epsilon }}^{(r)} =\dot{\mathbf E}-\dot{\boldsymbol\epsilon }^{(r)} $ and $\tilde{\dot{\boldsymbol\omega }}^{(r)} =\dot{\boldsymbol\Omega }-\dot{\boldsymbol\omega }^{(r)} $ are deviations of the average strain rate and rotation rate inside the inclusion, respectively, with respect to the corresponding overall magnitudes, and $\dot{\boldsymbol\epsilon}^{*(r)} $ is the average eigenstrain-rate in the inclusion. Therefore, the lattice rotation rate field is given by

\begin{equation} \label{GrindEQ__30_} 
\dot{\boldsymbol\omega }\, \left(\mathbf{x}\right)=\dot{\boldsymbol\Omega }+\tilde{\dot{\boldsymbol\omega }}^{\left(r\right)} -\dot{\boldsymbol\omega }^{p} \left(\mathbf{x}\right) 
\end{equation}

\subsection{Interaction and localization equations}

Taking volume averages over the domain of the inclusion on both sides of eq. \eqref{GrindEQ__9_} gives

\begin{equation} \label{GrindEQ__31_} 
\tilde{\boldsymbol{\sigma}}'^{(r)} =\overline{\mathbb{L}} \left(\tilde{\dot{\boldsymbol\epsilon }}^{(r)} -\dot{\boldsymbol\epsilon }^{*(r)} \right) 
\end{equation} 

\noindent and replacing the eigenstrain-rate given by eq. \eqref{GrindEQ__28_} into eq. \eqref{GrindEQ__31_}, leads to the \textit{interaction equation}:

\begin{equation} \label{GrindEQ__32_} 
\tilde{\dot{\boldsymbol\epsilon}}^{(r)} =-\tilde{\mathbb{M}} \tilde{ \boldsymbol\Sigma}'^{(r)} 
\end{equation} 

\noindent where the interaction tensor is given by:

\begin{equation} \label{GrindEQ__33_} 
\tilde{{\mathbb M}}=\left({\mathbb I}-{\mathbb S}\right)^{-1} {\mathbb S} \overline{{\mathbb M}}.
\end{equation} 

Replacing the constitutive relations of the inclusion and the effective medium in the interaction equation and, after some manipulation, leads to the \textit{localization equation}:

\begin{equation} \label{GrindEQ__34_} 
\boldsymbol{\sigma}'^{(r)} ={\mathbb B}^{(r)} \boldsymbol{\Sigma}'+\mathbf{b}^{(r)}  
\end{equation} 

\noindent where the localization tensors are defined as:

\begin{equation} \label{GrindEQ__35_} 
{\mathbb B}^{(r)} =\left({\mathbb M}^{{(r)}} +\tilde{{\mathbb M}}\right)^{-1} \left(\overline{{\mathbb M}}+\tilde{{\mathbb M}}\right) 
\end{equation} 
\begin{equation} \label{GrindEQ__36_} 
{\mathbf b}^{{(r)}} =\left({\mathbb M}^{{(r)}} +\tilde{{\mathbb M}}\right)^{-1} \left(\dot{{\mathbf E}}^{{0}} - \dot{{\boldsymbol\epsilon }}^{{0(r)}} \right) 
\end{equation} 

\subsection{Self-consistent equations}

The derivation presented in the previous sections solves the problem of an equivalent inclusion embedded in an effective medium. This result is used in this section to construct a polycrystal model, in which each SR grain ($r$) stands for an inclusion embedded in an effective medium that represents the polycrystal. The properties of such medium are not known {\it a priori} and have to be found through an iterative procedure. If the stress localization equation, eq. \eqref{GrindEQ__34_} is introduced in the local constitutive equation, eq. \eqref{GrindEQ__4_},  averaged over the SR grain ($r$),

\begin{equation} \label{GrindEQ__37_} 
\boldsymbol\epsilon^{(r)} =\mathbb{M}^{(r)} {\mathbb  B}^{(r)}  \boldsymbol\Sigma +\mathbb{M}^{(r)} { \mathbf b}^{(r)} +\boldsymbol\epsilon^{0(r)}. 
\end{equation} 

Taking volumetric average to eq. \eqref{GrindEQ__37_} and enforcing the condition that the average of the strain-rates over the aggregate has to coincide with the macroscopic quantities, leads to,
\begin{equation} \label{GrindEQ__38_} 
\dot{\mathbf E}=\left\langle \dot{\boldsymbol\epsilon }^{(r)} \right\rangle  
\end{equation} 

\noindent where the brackets denote average over the SR grains weighted by the associated volume fraction. From the macroscopic constitutive relation, eq. \eqref{GrindEQ__7_}, the following \textit{self-consistent equations}  are obtained for the HEM's compliance and back-extrapolated term:

\begin{equation} \label{GrindEQ__39_} 
\overline{\mathbb M} =\left\langle \mathbb{M}^{(r)} {\mathbb  B}^{(r)} \right\rangle   
\end{equation} 
\begin{equation} \label{GrindEQ__40_} 
\dot{\mathbf E}^{0} =\left\langle \mathbb{M}^{(r)} {\mathbf  b}^{(r)} +\dot{\boldsymbol\epsilon }^{0(r)} \right\rangle  
\end{equation} 

\subsection {Linearization assumptions}

As indicated above, different choices are possible for the linearized behavior at the grain level, and the results of the homogenization scheme depend on this choice. Several first-order linearization schemes are presented below, defined in terms of the stress first-order moment (average) inside SR grain ($r$).

The \textit{secant approximation} \citep{H65, Hutchinson1976101} assumes the following linearized secant moduli:

\begin{equation} \label{GrindEQ__41_} 
{\mathbb M}_{\sec }^{(r)} =\dot{\gamma }_{0} \sum_{k}\frac{\mathbf{Z}_s^{k(r)} \otimes \mathbf{Z}_s^{k(r)} }{g^{k(r)} } \left(\frac{|\mathbf{Z}_s^{k(r)} : \boldsymbol{\sigma}'^{(r)}| }{g^{k(r)} } \right)^{n-1}  
\end{equation} 
\begin{equation} \label{GrindEQ__42_} 
\dot{\boldsymbol\epsilon }_{\sec }^{0(r)} =0 
\end{equation}

\noindent where the index ($r$) in $\mathbf{Z}_s^{k(r)} $ and $g^{k (r)} $ indicates uniform (average) values of these magnitudes, corresponding to a given orientation and hardening state associated with the SR grain ($r$).  

Under the \textit{affine approximation} \citep{MBS00}, the affine moduli are given by

\begin{equation} \label{GrindEQ__43_} 
{\mathbb M}_{aff}^{(r)} =n\dot{\gamma }_{0}  \sum _{k}\frac{\mathbf{Z}_s^{k(r)} \otimes  \mathbf{Z}_s^{k(r)} }{g^{k(r)} } \left(\frac{|\mathbf{Z}_s^{k(r)} :\boldsymbol{\sigma}'^{(r)} |}{g^{k(r)} } \right)^{n-1}  
\end{equation} 

\begin{equation} \label{GrindEQ__44_} 
\dot{\boldsymbol\epsilon }_{aff}^{0(r)} =\left(1-n\right) \dot{\gamma }_{0}  \sum _{k} \left(\frac{|\mathbf{Z}_s^{k(r)} : \boldsymbol{\sigma}'^{(r)} |}{g^{k(r)} } \right) ^{n} {\rm sign} \left(\mathbf{Z}_s^{k(r)} : \boldsymbol{\sigma}'^{(r)} \right) 
\end{equation} 

In the case of the \textit{ tangent approximation} \citep{Molinari19872983, LT93},  the tangent moduli are formally the same as in the affine case: ${\mathbb M}_{tg}^{(r)} ={\mathbb M}_{aff}^{(r)} $ and $\dot{\boldsymbol\epsilon }_{tg}^{0(r)} =\dot{\boldsymbol\epsilon }_{aff}^{0(r)} $. However, \cite{Molinari19872983} used the secant compliance, eq. \eqref{GrindEQ__41_}, instead of the affine one  in order to avoid the iterative adjustment of the macroscopic back-extrapolated term to adjust $\overline{\mathbb M}$ (to be denoted $\overline{\mathbb M}_{\sec } $), in combination with the tangent-secant relation $\overline{\mathbb M}_{tg} =n\overline{\mathbb M}_{\sec } $ \citep{Hutchinson1976101}. Then, the expression of the interaction tensor is given by

\begin{equation} \label{GrindEQ__45_} 
\tilde{\mathbb M}=\left({\mathbb I}-{\mathbb S}\right)^{-1}{\mathbb S}\overline{\mathbb M}_{tg} = n\left({\mathbb I}-{\mathbb S}\right)^{-1}{\mathbb S}\overline{\mathbb M}_{\sec }  
\end{equation} 

Qualitatively, the interaction  eq. \eqref{GrindEQ__32_},  indicates that the larger the interaction tensor, the smaller the deviation of grain stresses with respect to the average stress. As a consequence,  the tangent approximation tends to a uniform stress state or \textit{ lower-bound approximation} for $n\to \infty $ \citep{S28}. This rate-insensitive limit of the tangent formulation is an artefact created by the use of the above tangent-secant relation of the non-linear polycrystal in the self-consistent solution of the linear comparison polycrystal. On the contrary, the secant interaction has been proven to tend to a uniform strain-rate state or \textit{upper-bound approximation} \citep{T38} in the rate insensitive limit.

\subsection {Second-order formulation}

The more accurate \textit{second-order approximation} to linearize the behavior of the mechanical phase provides improved micromechanical predictions, which are obtained from the calculation of average fluctuations of the stress distribution inside the linearized SR grains. The methodology to obtain these fluctuations was derived by \cite {BD87},Ê\cite{ K90} and \cite{PB90}  for composites and extended by \cite{LTP07} for polycrystals. The effective stress potential $\overline{U}_{T} $ of a linearly viscous polycrystal described by eq. \eqref{GrindEQ__10_} may be written in the form \citep{W81, L73},

\begin{equation} \label{GrindEQ__46_} 
\overline{U}_{T} =\frac{1}{2} \overline{\mathbb M}:\left(\boldsymbol\Sigma'\otimes \boldsymbol\Sigma'\right)+\dot{\mathbf E}^{0} :\boldsymbol\Sigma'+\frac{1}{2} \overline{G} 
\end{equation} 

\noindent where $\overline{G}$ is the power under zero applied stress. The self-consistent expression for $\overline{\mathbb M}$ and $\dot{\mathbf E}^{0} $,  eqs. \eqref{GrindEQ__39_} and \eqref{GrindEQ__40_}, can be re-written as

\begin{equation} \label{GrindEQ__47_} 
\overline{\mathbb M}=\left\langle \mathbb{M}^{(r)} \mathbb{B}^{(r)} \right\rangle =\sum _{r}c^{(r)}  \mathbb{M}^{(r)} \mathbb{B}^{(r)}   
\end{equation} 
\begin{equation} \label{GrindEQ__48_} 
\dot{\mathbf E}^{0} =\left\langle \mathbb{M}^{(r)} \mathbf{b}^{(r)} +\dot{\boldsymbol\epsilon}^{0(r)} \right\rangle =\sum _{r}c^{(r)}  \left(\mathbb{M}^{(r)} \mathbf{b}^{(r)} +\dot{\boldsymbol\epsilon }^{0(r)} \right) =\sum _{r}c^{(r)} \dot{\boldsymbol\epsilon }^{0(r)} \mathbb{B}^{(r)}   
\end{equation} 

\noindent where $c^{(r)} $ is the volume fraction associated with SR grain ($r$). The corresponding expression for $\overline{G}$ is

\begin{equation} \label{GrindEQ__49_} 
\overline{G}=\sum _{r}c^{(r)} \dot{\boldsymbol\epsilon }^{0(r)} : \mathbf{b}^{(r)}.  
\end{equation} 

The average second-order moment of the stress field over a SR grain ($r$) of the polycrystal is a fourth-rank tensor given by:

\begin{equation} \label{GrindEQ__50_} 
\left\langle \boldsymbol\sigma'\otimes \boldsymbol\sigma'\right\rangle ^{(r)} =\frac{2}{c^{(r)} } \frac{\partial \overline{U}_{T} }{\partial \mathbb{M}^{(r)} }  
\end{equation}

Replacing eqs. \eqref{GrindEQ__47_} - \eqref{GrindEQ__49_} in \eqref{GrindEQ__50_}, leads to 

\begin{equation} \label{GrindEQ__51_} 
\left\langle \boldsymbol\sigma'\otimes \boldsymbol\sigma'\right\rangle ^{(r)} =\frac{1}{c^{(r)} } \frac{\partial \overline{\mathbb M}}{\partial \mathbb{M}^{(r)} } :\left(\boldsymbol\Sigma'\otimes \boldsymbol\Sigma'\right)+\frac{1}{c^{(r)} } \frac{\partial \dot{\mathbf E}^{0} }{\partial \mathbb{M}^{(r)} } :\boldsymbol\Sigma'+\frac{1}{c^{(r)} } \frac{\partial \overline{G}}{\partial \mathbb{M}^{(r)} }  
\end{equation} 

The algorithmic expressions to calculate the partial derivatives on the right-hand side can be found in \cite{LTP07}.

Once the average second-order moments of the stress field over each SR grain ($r$) are obtained, the implementation of the second-order procedure follows the work of \cite{LP04}. The covariance tensor of stress fluctuations in the SR grains of the linear comparison polycrystal is given by:

\begin{equation} \label{GrindEQ__52_} 
{\mathbb C}_{\sigma'}^{(r)} =\left\langle \boldsymbol\sigma'\otimes \boldsymbol\sigma'\right\rangle ^{(r)} -\boldsymbol{\sigma}'^{(r)} \otimes \boldsymbol{\sigma}'^{(r)}  
\end{equation} 

\noindent and the average and the average fluctuation of resolved shear stress on slip system $k$ of SR grain ($r$) is given by

\begin{equation} \label{GrindEQ__53_} 
\overline{\tau }^{k(r)} =\mathbf{Z}_s^{k(r)} : \boldsymbol{\sigma}'^{(r)}  
\end{equation} 
\begin{equation} \label{GrindEQ__54_} 
\hat{\tau}_{}^{k(r)} =\overline{\tau }^{k(r)} \pm \left(\mathbf{Z}_s^{k(r)} : {\mathbb C}_{\sigma'}^{(r)} :\mathbf{Z}_s^{k(r)} \right)^{{1/2} }  
\end{equation} 

\noindent where the positive (negative) branch should be selected if $\bar{\tau }^{k(r)} $ is positive (negative). The slip potential associated with slip system $k$ of the nonlinear polycrystal is defined as:

\begin{equation} \label{GrindEQ__55_} 
\phi ^{k} \left(\tau \right)=\frac{\tau _{0}^{k} }{n+1} \left(\frac{\left|\tau \right|}{g^{k} } \right)^{n+1}.  
\end{equation} 

\noindent The linearized local behavior associated with SR grain ($r$) is then given by:

\begin{equation} \label{GrindEQ__58_} 
\dot{\boldsymbol\epsilon }^{(r)} =\mathbb{M}_{SO}^{(r)} \boldsymbol{\sigma}'^{(r)} +\dot{\boldsymbol\epsilon }_{SO}^{0(r)}  
\end{equation} 

\noindent with:

\begin{equation} \label{GrindEQ__59_} 
\mathbb{M}_{SO}^{(r)} =\sum _{k}\alpha^{k(r)} \left(\mathbf{Z}_s^{k(r)} \otimes \mathbf{Z}_s^{k(r)} \right)  
\end{equation} 
\begin{equation} \label{GrindEQ__60_} 
\dot{\boldsymbol\epsilon }_{SO}^{0\, (r)} =\sum _{k} e^{k(r)}  \mathbf{Z}_s^{k(r)} 
\end{equation} 

\noindent where $\phi'^{k} \left(\tau \right)=\mathrm{d}\phi^{k} /\mathrm{d}\tau$ stand for
two scalar magnitudes associated with each slip system $k$ of each SR grain ($r$) are defined as
\begin{equation} \label{GrindEQ__56_} 
\alpha^{k\left(r\right)} =\frac{\phi'^{k\left(r\right)} \left(\hat{\tau }_{}^{k(r)} \right)-\phi'^{k\left(r\right)} \left(\bar{\tau }_{}^{k(r)} \right)}{\hat{\tau }^{k(r)} -\bar{\tau }^{k(r)} }  
\end{equation} 
\begin{equation} \label{GrindEQ__57_} 
e^{k\left(r\right)} =\phi'^{k\left(r\right)} \left(\bar{\tau }^{k(r)} \right)-\alpha^{k(r)} \bar{\tau }^{k(r)}  
\end{equation} 

Once the linear comparison polycrystal is defined by eqs. \eqref{GrindEQ__59_}-\eqref{GrindEQ__60_}, different second-order estimates of the effective behavior of the nonlinear aggregate can be obtained. Approximating the potential of the nonlinear polycrystal in terms of the potential of the linear comparison polycrystal and a suitable measure of the error, \cite{LP04} generated the following expression (corresponding to the so-called \textit{energy} version of the second-order theory) for the effective potential of the nonlinear polycrystal

\begin{equation} \label{GrindEQ__61_} 
\overline{U}\left(\boldsymbol\Sigma'\right)=\sum_{r}c^{(r)} \sum _{k}\left\{ \phi^{k\left(r\right)} \left(\hat{\tau }^{k(r)} \right)+\phi'^{k\left(r\right)} \left(\overline{\tau }^{k(r)} \right) \left(\overline{\tau }_{}^{k(r)} -\hat{\tau }^{k(r)} \right)\right\}   
\end{equation} 

\noindent from where the effective response of the homogenized polycrystal can be obtained as ${\dot{\mathbf E}=\partial \overline{U}\left(\boldsymbol\Sigma'\right) \mathord{\left/{\vphantom{\dot{\mathbf E}=\partial \bar{U}\left(\boldsymbol\Sigma'\right) \partial \boldsymbol\Sigma'}}\right.\kern-\nulldelimiterspace} \partial \boldsymbol\Sigma'} $. 

The alternative \textit{constitutive equation} version of the second-order theory simply consists in making use of the effective stress-strain-rate relations for the linear comparison polycrystal. The effective strain is obtained as

\begin{equation} \label{GrindEQ__62_} 
\dot{\mathbf E}=\sum _{r}c^{(r)} \sum _{k}\mathbf{Z}_s^{k(r)} \phi'^{k\left(r\right)} \overline{\tau}^{k(r)}
\end{equation} 

Both versions of the second order theory give slightly different results, depending on non-linearity and local anisotropic contrast. Such gap is relatively small compared with the larger differences obtained with the different SC approaches. The \textit{constitutive equation} version is -- in principle -- less rigorous since it does not derive from a potential function, but has the advantage that can be obtained by simply following the affine algorithm described in the previous sections, using the linearized moduli defined by eqs. \eqref{GrindEQ__59_}-\eqref{GrindEQ__60_}.

\subsection {Numerical implementation}

To illustrate the practical use of the VPSC formulation, the steps required to predict the local and overall viscoplastic response of a polycrystal are detailed below. Starting for convenience with an initial Taylor guess, i.e. $\dot{\boldsymbol\epsilon }^{(r)} =\dot{\mathbf E}$ for all grains,  the following non-linear equation is solved to get $\boldsymbol{\sigma}'^{(r)} $,

\begin{equation} \label{GrindEQ__63_} 
\dot{\mathbf E}=\dot{\gamma }_{0} \sum _{k}\mathbf{Z}_s^{k(r)} \left(\frac{\mathbf{Z}_s^{k(r)} : \boldsymbol{\sigma}'^{(r)} }{g^{k(r)} } \right) ^{n} {\rm sign} \left(\mathbf{Z}_s^{k(r)} : \boldsymbol{\sigma}'^{(r)} \right) 
\end{equation}

\noindent and an appropriate first-order linearization scheme is used to obtain initial values of $\mathbb{M}^{(r)}$ and  $\dot{\boldsymbol\epsilon }^{0(r)} $, for each SR grain ($r$). Next, initial guesses for the macroscopic moduli $\overline{\mathbb M}$ and $\dot{\mathbf E}^{0} $ are obtained (usually as simple averages of the local moduli). The initial guess for the macroscopic stress $\boldsymbol\Sigma'$ can be obtained from them and the applied strain rate, eq. \eqref{GrindEQ__7_}, while the Eshelby tensors ${\mathbb S}$ and $\mathbb{P} $ can be calculated using the macroscopic moduli and the ellipsoidal shape of the SR grains by means of the procedure described in section 4.1. Subsequently, the interaction tensor $\tilde{{\mathbb M}}$, eq. \eqref{GrindEQ__33_}, and the localization tensors ${\mathbb B}^{{\rm (r)}} $ and ${\mathbf b}^{{\rm (r)}} $, eqs. \eqref{GrindEQ__36_} and \eqref{GrindEQ__37_}, can be calculated as well. With these tensors, new estimates of  $\overline{\mathbb M}$ and $\dot{\mathbf E}^{0} $ are obtained by solving iteratively the self-consistent eqs. \eqref{GrindEQ__39_} and \eqref{GrindEQ__40_}. After achieving convergence on the macroscopic moduli (and, consequently, also on the macroscopic stress and the interaction and localization tensors), a new estimation of the average grain stresses can be obtained using the localization relation, eq. \eqref{GrindEQ__34_}. If the recalculated average grain stresses are different  from the input values within certain tolerance, a new iteration should be started, until convergence is reached. If the chosen linearization scheme is the second-order formulation, an additional loop on the linearized moduli is needed, using the improved estimates of the second-order moments of the stress in the grains, obtained by means of the methodology described in section 4.4. This additional loop roughly increases the calculation time by one order of magnitude with respect to first-order linearizations. When the iterative procedure is completed, the average shear rates on the slip system ($k$) in each grain ($r$) are calculated as:

\begin{equation} \label{GrindEQ__64_} 
\dot{\gamma }^{k(r)} =\dot{\gamma }_{0} \left(\frac{\mathbf{Z}_s^{k(r)} : \boldsymbol{\sigma}'^{(r)} }{g^{k(r)} } \right)^{n} {\rm sign} \left(\mathbf{Z}_s^{k(r)} : \boldsymbol{\sigma}'^{(r)} \right) 
\end{equation} 

These average shear rates are in turn used to calculate the lattice rotation rates associated with each SR grain:

\begin{equation} \label{GrindEQ__65_} 
\dot{\boldsymbol\omega }^{(r)} =\dot{\boldsymbol{\Omega}} +\tilde{\dot{\boldsymbol\omega }}^{(r)} -\dot{\boldsymbol\omega }^{p(r)}  
\end{equation} 

\noindent where, following eq. \eqref{GrindEQ__3_},

\begin{equation} \label{GrindEQ__66_} 
\dot{\boldsymbol\omega }^{p(r)} =\sum _{k}\mathbf{Z}_{a}^{k(r)}\dot{\gamma }^{k(r)}   
\end{equation} 

\noindent where $\mathbf{Z}_{a}^{k(r)}$ is the (uniform) antisymmetric Schmid tensor of slip system $k$ in SR grain ($r$). 

The above numerical scheme can be used to predict the texture development by applying the viscoplastic deformation to the polycrystal in incremental steps. The latter is done by assuming constant rates during a time interval $\Delta t$ (such that $\dot{\mathbf E} \Delta t$ corresponds to a macroscopic strain increment of the order of a few percents) and using: a) the strain rates and rotation rates (times $\Delta t$) to update the shape and orientation of the SR grains, and b) the shear rates (times $\Delta t$) to update the critical stress of the deformation systems due to strain hardening using one the models described in section 3, after each deformation increment. Note that the above explicit update schemes relies on the fact that the orientation and hardening variables evolve slowly within the adopted time interval. Otherwise, $\Delta t$ should be chosen smaller.

\section{Full-field homogenization of polycrystals}

Full-field homogenization aims to predict the macroscopic response and microscopic field distribution of heterogenous materials based on the numerical simulation of the mechanical response of an RVE of the material microstructure \citep{Bohm}. The method is computationally expensive because it involves the solution of a boundary value problem that might contain a large number of degrees of freedom. Nevertheless, it can provide more accurate predictions of the macroscopic response than mean-field homogenization models (such as VPSC) because it includes more accurate information of the details of the microstructure. In addition,  full-field homogenization of polycrystals  can account for the effect of microstructural details that cannot be easily included in mean-field models such as relative grain sizes and grain shape distribution, as well as grain boundary misorientation distributions. Finally, it provides very accurate information for the local values of the stress and strain fields as well as of the state variables throughout the microstructure. This information is critical for predicting damage localization and failure of heterogeneous materials. 

Several methods have been used to simulate the mechanical response of an RVE in within the framework of full-field homogenization. Among them, the two most common approaches are the FFT-based method  \citep{MS98}, extended to viscoplastic polycrystals in \cite{L01}, and the FEM.

\subsection {Homogenization using the finite element method}

The first attempts to use the FEM to obtain the mechanical response of a polycrystal were carried out in the 90s by \cite{Kalidindi1992537}, \cite{BKA92} and \cite{MSS99}. These first works were limited to 2-D and the RVEs used for the simulations were just a regular arrangement of finite elements in which each element represented a whole grain. These studies were able to capture the macroscopic response and texture evolution under large strains but thedistribution of local fields was missing due to the rudimentary representation of the microstructure. Indeed, these simulations were closer to mean-field polycrystalline models (such as Taylor or VPSC) than to full-field simulations because the fields in each grain were constant (or just vary linearly) and did not take into account the strong gradients found in actual polycrystals. The use of complex RVEs with more realistic representations of the polycrystalline microstructure started at the early 2000s (many years after the development of CP theory and of the first numerical implementations) due to the  high computational cost of these type of models. The first 3-D simulations used regular arrangements of grains (for example, rhombic dodecahedra in \cite{MIKA19991355}) but each grain was represented using several finite elements and, thus, the strong deformation gradients due to the accommodation of the plastic strain incompatibility between adjacent grains were accounted for. Finally, the use of several elements per grain was combined with more complex representation of the microstructure in 3-D, leading to the first realistic full-field simulations \citep{BARBE2001513,BARBE2001537}. Since then, full-field simulation using the FEM has been a very active area of research.  The finite element implementation of CP models has been reviewed by \cite{Roters20101152, REB10} and therefore will not presented here. Only some notes on the boundary conditions used in the boundary value problem will be reviewed due to their implication in obtaining the macroscopic fields as result of the full field simulations. 

The boundary conditions used in the simulation of the RVE deformation are a key issue in full-field homogenization. These conditions are energetically consistent if the stress and strain microscopic fields and the corresponding macroscopic fields fulfill the so-called Hill-Mandel principle of macro-homogeneity \citep{H65}. This principle states that the macroscopic stress power should be equal to the volume average of the microscopic stress power. Four different types of boundary conditions can be imposed to the RVE resulting in microfields compatible with this condition \citep{Bohm,Peric2011,OSTOJASTARZEWSKI2006112}: (1) statically uniform boundary conditions where homogeneous surface tractions are applied to the RVE faces; (2) kinematically uniform boundary conditions where constant displacements are applied to the RVE boundary; (3) mixed conditions combining uniform tractions and constant displacements on the different RVE boundaries and (4) periodic boundary conditions. For any given RVE, the results obtained under periodic boundary conditions  (even in the case of non-periodic microstructures) lie between the corresponding lower and upper estimates obtained with homogeneous boundary conditions (1) and (2) and show a faster convergence towards the actual effective response with increasing the RVE size \citep{HAZANOV}. Moreover, periodic boundary conditions reproduce exactly the deformation of an infinite media constructed by a periodic arrangement of the RVE. For these reasons, periodic boundary conditions are the common choice for full-field homogenization and will be described here.

\begin{figure}[h]
\includegraphics[width=1.1\textwidth]{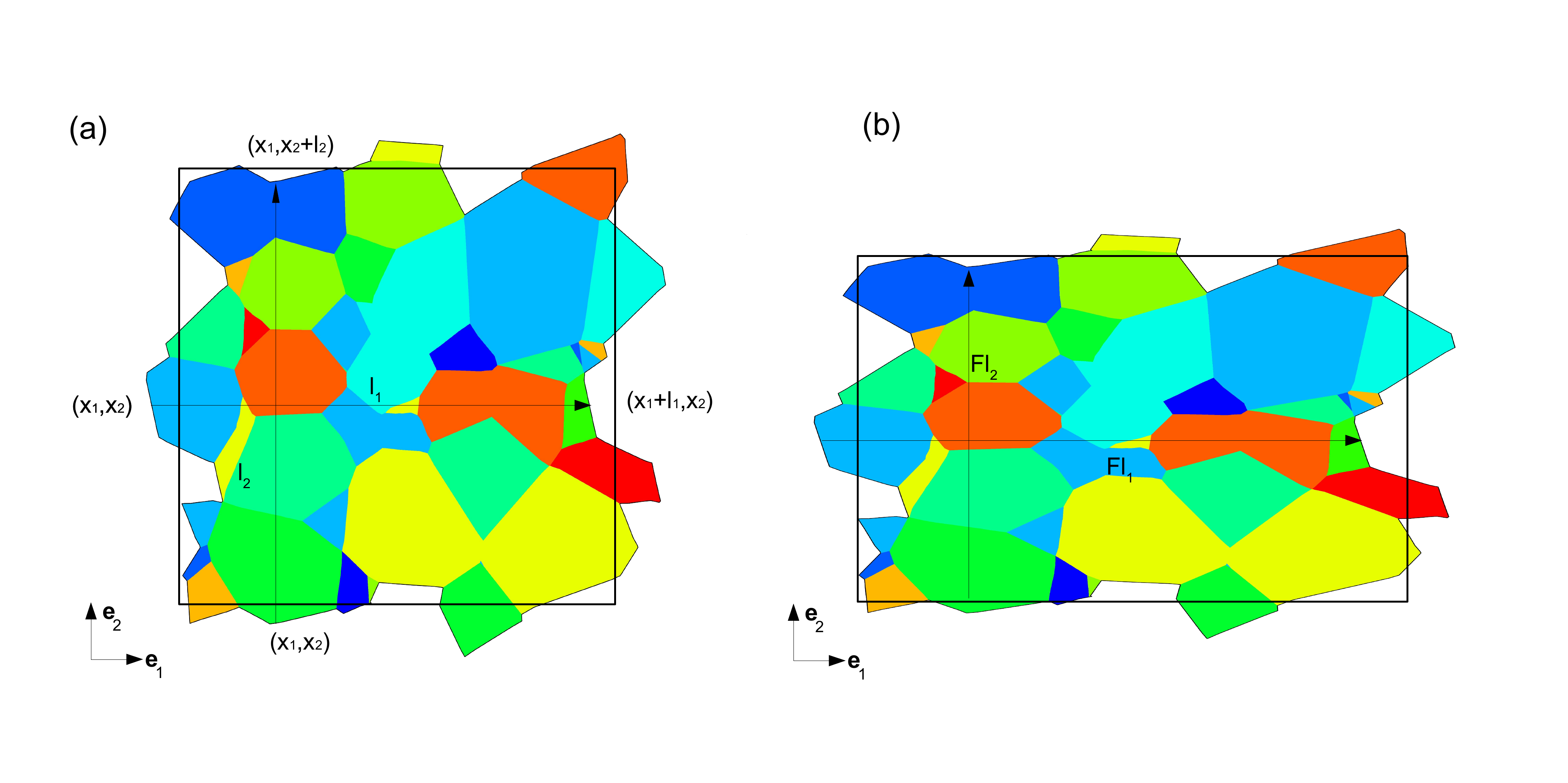}
\caption{Deformation of a periodic RVE of a polycrystal under periodic boundary conditions in 2-D. (a) Undeformed configuration. (b) Deformed shape under biaxial deformation.}  
\label{figure:PBC}
\end{figure}

Periodic boundary conditions assume that the whole RVE deforms as a jigsaw puzzle. If a RVE is selected such that the periodicity of the microstructure is cubic, then the periodic translation of the cell in the three directions will fill the whole space (Figure \ref{figure:PBC}(a)). Let  $\mathbf{l}_i=l_i\mathbf{e}_i$ be the three orthogonal vectors defining the cubic periodicity and let $\mathbf{e}_i$ be the corresponding unit vectors defining the basis. If $x_1, x_2, x_3$ are the coordinates of a point in the RVE in the system defined by $\mathbf{e}_i$, the periodic boundary conditions link the local displacement vector $\mathbf{u}$ of the nodes on opposite faces of the RVE with the far-field macroscopic deformation gradient $\bar{\mathbf{F}}$ according to,

\begin{equation}\label{pbc}
\begin{split}
\mathbf{u}(x_1,x_2,0)-\mathbf{u} (x_1,x_2,L_3)=(\bar{\mathbf{F}}-\mathbf{1})\mathbf{l}_3\\
\mathbf{u}(x_1,0,x_3)-\mathbf{u} (x_1,L_2,x_3)=(\bar{\mathbf{F}}-\mathbf{1})\mathbf{l}_2\\
\mathbf{u}(0,x_2,x_3)-\mathbf{u} (L_1,x_2,x_3)=(\bar{\mathbf{F}}-\mathbf{1})\mathbf{l}_1.\\
\end{split}
\end{equation}

The resulting deformed cell preserves the cubic periodicity, but the new vectors defining the periodicity are given by $\mathbf{\bar{F}}\mathbf{l}_i$, as it can be observed in Figure \ref{figure:PBC}(b).

These boundary conditions are implemented in the FEM by coupling the displacement degrees of freedom of the nodes in the RVE surface. The most direct implementation of periodic boundary conditions requires a periodic mesh, i.e. the 2D discretization of two opposite faces of the RVE boundary should be identical, although some interpolation methods were proposed in the literature to overcome this requirement \citep{NGUYEN2012390}. The application of the eq. \eqref{pbc} to the nodes of a periodic mesh can be done by constraint elimination \citep{Peric2011,MMS99} or by means of the Lagrange multiplier method \citep{Bohm_Han2001,SL02}. Standard finite element codes do not allow to apply direct constraint elimination and the common approach is then to use Lagrange multipliers by imposing a relation in the displacement of some nodes in the model \citep{SL02,SL13}. In particular, a master node $M_i$ ($i=1, 2 ,3$) can be defined for each pair of opposite cube faces and the value of the far-field macroscopic deformation gradient is imposed to the RVE through the displacement of those master nodes according to 
 
\begin{equation}
\mathbf{u}(M_i)=(\bar{\mathbf{F}}-\mathbf{1})\mathbf{l}_i.
\end{equation}

If some components of the far-field deformation gradient are not known {\it a priori} (e.g. under uniaxial tension), the corresponding  effective stresses $\bar{\boldsymbol{\sigma}}$ are set instead. This task is carried out by applying nodal forces $P$ to the corresponding master node $i$ and degree of freedom $j$ according to
\begin{equation}
P_j(M_i)=(\bar{\boldsymbol{\sigma}}\mathbf{e}_i)_j A_i
\label{pbc_stress}
\end{equation} 

\noindent where $A_i$ is the \emph{area} of the cell perpendicular to direction $\mathbf{e}_i$. Under small strains, the area $A_i$ corresponds to the undeformed geometry and eq. \eqref{pbc_stress} is used to obtain the value of the forces to be applied as function of the target stress. If finite deformations are applied, $A_i$ corresponds to the actual area and the relation between the force and the corresponding Cauchy stress component is not known {\it a priori} because the actual value of the area changes with the deformation. In this case, to impose a given history of any stress component, an iterative procedure should be applied to correct the applied forces with the cell deformation.

\subsection {Homogenization using the fast Fourier transform}

The large number of degrees of freedom required by full-field homogenization limits the size of the microstructures that can be simulated using the FEM. Conceived as a very efficient alternative to the FEM, a numerical formulation based on the use of the FFT algorithm was originally proposed by \cite {MS94} to compute convolution integrals over the entire periodic RVE involving the periodic Green's function of a reference medium. In the original method, known as the basic scheme, the solution is obtained using a fixed-point iterative scheme. The contrast in phase properties has a strong impact in the convergence rate (the higher the contrast, the more numerically demanding the convergence). Moreover, the properties of the reference medium, although do not influence the final results, also strongly affect the convergence rate. Several modifications to the Moulinec-Suquet scheme have been proposed afterwards \citep{EM99,MMS00,zeman2010accelerating,MONCHIET2013276,Kabel2014} based on modifying the iterative algorithm in order to improve the convergence rate. All these approaches rely on the use of a reference medium and, therefore, their performance is linked to an appropriate election of its properties.

An alternative approach to FFT-based homogenization in which the problem is derived using a weak formulation, analogous to finite elements, was recently proposed \citep{VONDREJC2014156,Geers2016,Geers2017}. This approach does not require a reference medium and the Green's function of the classical schemes are replaced here by projection operators that ensure the compatibility of the strain fields. The similarities of this approach with Galerkin finite elements make very easy the introduction of generic non-linear constitutive equations, including their corresponding tangents to accelerate the convergence.

FFT-based methods were originally developed \cite {MS94} to carry out computational homogenization in periodic composites and have been profusely used for this class of materials \citep{MS98, EM99, MMS99, IMP06, BD10, MS14, W15}. In this type of problems, the source of heterogeneity is related to the spatial distribution of phases with different mechanical properties and the phase behavior is given in general by simple constitutive equations (in many cases, by isotropic linear elastic relations). The extension of FFT-based homogenization to polycrystals was proposed first in \cite{L01} to exploit the superior numerical performance of CP-FFT with respect to CP-FE, allowing a large reduction of the computational time and/or the use of very complex and detailed polycrystalline RVEs. Since then, different numerical implementations of the FFT-based method for polycrystals have been developed, depending on the constitutive equation adopted to represent the behavior of each single crystal material point. Classic schemes have been used to solve the linearized problem In almost all the cases, requiring the definition of a reference medium. An exception is the recent work by \cite{LS18} who adapted the Galerkin FFT-based approach to study the cyclic behavior and fatigue performance of polycrystals.

A summary of the type of problems and constitutive regimes in which FFT-based homogenization has been applied to polycrystals includes thermo-elasticity \citep{BLC09}, rigid-viscoplasticity \citep{L01, LCB05, LBC08, LMM09}, elasto-viscoplasticity for small strains \citep{LKE12, GMC12} and large strains \citep{EDL13,LS18}. More complex behaviors include dilatational plasticity \citep{LEC13}, strain-gradient plasticity  \citep{LN16}, curvature-driven plasticity \citep{UCT16}, and transformation plasticity \citep{RLB13, OBB18}. Recent implementations of field dislocation mechanics theories \citep{BBS14, BTD14} and discrete dislocation mechanics formulations \citep{GRL16, BUP15} are also examples of the great potential of FFT-based formulations to provide the numerical efficiency needed for implementation of these very powerful, but also very complex and numerically-demanding, emerging micromechanical formulations.

In what follows, the rigid-viscoplastic FFT-based formulation (VPFFT) for fully-dense, incompressible polycrystals, and the dilatational extension of the latter (DVPFFT) to predict microstructural effects on porosity evolution in polycrystals with intregranular voids are presented in order to illustrate the methodology and provide the theory and algorithms behind one of the examples of section 7. For details of other numerical implementations of the FFT-based formulation, the reader is referred to the specific papers listed above.

Briefly, the viscoplastic FFT-based formulation consists in iteratively adjusting a compatible strain rate field, related with an equilibrated stress field through a constitutive potential, such that the average of local work rates is minimized. The method is based on the fact that the local mechanical response of a heterogeneous medium can be calculated as a convolution integral between Green functions associated with appropriate fields of a linear reference homogeneous medium and the actual heterogeneity field. For a periodic medium, use can be made of the discrete Fourier transform to reduce convolution integrals in real space to simple products in Fourier space. Thus, the FFT algorithm can be utilized to transform the heterogeneity field into Fourier space and, in turn, to get the mechanical fields by transforming that product back to real space. However, the actual heterogeneity field depends precisely on the {\it a priori} unknown mechanical fields. Therefore, an iterative scheme has to be implemented to obtain, upon convergence, a compatible strain rate field and a stress field in equilibrium.

A periodic RVE of the polycrystal is discretized into $N_{1} \times N_{2} \times N_{3} $ Fourier points. This discretization determines a regular grid in the Cartesian space $\left\{\mathbf{x}^{d} \right\}$ and a corresponding grid in Fourier space $\left\{\boldsymbol\xi^{d} \right\}$. Velocities and tractions along the boundary of the unit cell are left undetermined. A velocity-gradient $V_{i,j} $ (which can be decomposed into a symmetric strain rate and a antisymmetric rotation rate: $V_{i,j} =\dot{E}_{ij}^{} +\dot{\Omega }_{ij} $) is imposed to the unit cell. The local strain rate field is a function of the local velocity field, i.e. $\dot{\epsilon }_{ij} \left(v_{k} \left(\mathbf{x}\right)\right)$, and can be split into its average and a fluctuation term: $\dot{\boldsymbol\epsilon }_{ij} \left(v_{k} \left(\mathbf{x}\right)\right)=\dot{E}_{ij} +\tilde{\dot{\epsilon }}_{ij} \left(\tilde{v}_{k} \left(\mathbf{x}\right)\right)$, where $v_{i} \left(\mathbf{x}\right)=\dot{E}_{ij} x_{j} +\tilde{v}_{i} \left(\mathbf{x}\right)$. By imposing periodic boundary conditions, the velocity fluctuation field $\tilde{v}_{k} \left(\mathbf{x}\right)$ is assumed to be periodic across the boundary of the unit cell, while the traction field is antiperiodic, to meet equilibrium on the boundary between contiguous unit cells.

The local constitutive relation between the strain-rate $\dot{\boldsymbol\epsilon} (\mathbf{x})$ and the deviatoric stress $\boldsymbol\sigma' (\mathbf{x})$ is given by the same rate-sensitivity relation used within the VPSC framework, eq. \eqref{GrindEQ__1_}. If a fourth rank tensor $\mathbb{L}^{0} $ is chosen as the stiffness of a linear reference medium (the choice of $\mathbb{L}^{0} $ can be quite arbitrary, but the speed of convergence of the method will depend on this choice). The polarization field $\varphi_{ij} (\mathbf{x})$, eq. \eqref{GrindEQ__12_}, is defined as

\begin{equation} \label{GrindEQ__67_} 
\varphi _{ij} (\mathbf{x})=\tilde{\sigma }'_{ij} \left(\mathbf{x}\right)-L_{ijkl}^{0}  \tilde{\dot{\epsilon }}_{kl} (\mathbf{x}) 
\end{equation} 

Then, the Cauchy stress deviation can be written as

\begin{equation} \label{GrindEQ__68_} 
\tilde{\sigma }_{ij} \left(\mathbf{x}\right)=L_{ijkl}^{0} \tilde{\dot{\epsilon }}_{kl} (\mathbf{x})+\varphi _{ij} (\mathbf{x})+\tilde{\sigma }^{m} \left(\mathbf{x}\right) \delta _{ij}  
\end{equation} 

\noindent and combining Eq. \eqref{GrindEQ__68_} with the equilibrium ($\sigma _{ij,j} \left(\mathbf{x}\right)=\tilde{\sigma }_{ij,j} \left(\mathbf{x}\right)=0$), the incompressibility condition, and the relation $\tilde{\dot{\epsilon }}_{ij} \left(\mathbf{x}\right)=\frac{1}{2} \left(\tilde{v}_{i,j} \left(\mathbf{x}\right)+\tilde{v}_{j,i} \left(\mathbf{x}\right)\right)$ leads to,

\begin{equation} \label{GrindEQ__69_} 
\left|\begin{array}{l} {\; L_{ijkl}^{0} \tilde{v}_{k,lj} \left(\mathbf{x}\right)+\tilde{\sigma }_{,i}^{m} \left(\mathbf{x}\right)+\varphi _{ij,j} \left(\mathbf{x}\right)=0} \\ {\; \tilde{v}_{k,k} \left(\mathbf{x}\right)=0} \end{array}\right.  
\end{equation} 

This system of differential equations is formally equivalent to system \eqref{GrindEQ__11_}. However, both systems actually differ in that: a) the HEM's stiffness modulus $\overline{\mathbb L}$ of eq. \eqref{GrindEQ__11_} is replaced in eq. \eqref{GrindEQ__69_} by the stiffness of a linear reference medium $\mathbb{L}^{0} $, and b) the polarization field in Eq. \eqref{GrindEQ__69_} has in general non-vanishing values throughout the unit cell and is periodic (owing to the RVE periodicity), while the polarization field in Eq. \eqref{GrindEQ__11_} vanishes outside the domain of the inclusion. The auxiliary system involving periodic Green functions is, then, given by eq. \eqref{GrindEQ__15_},

\begin{equation} \label{GrindEQ__70_} 
\left|\begin{array}{l} {\; L_{ijkl}^{0} \; G_{km,lj} \left(\mathbf{x}-\mathbf{x}'\right)+H_{m,i} \left(\mathbf{x}-\mathbf{x}'\right)+\delta _{im} \delta \left(\mathbf{x}-\mathbf{x}'\right)=0} \\ {\; G_{km,k} \left(\mathbf{x}-\mathbf{x}'\right)=0} \end{array}\right.  
\end{equation} 

After some manipulation,  the velocity and the velocity gradient deviation fields are given by the convolution integrals

\begin{equation} \label{GrindEQ__71_} 
\tilde{v}_{k} \left(\mathbf{x}\right)=\int _{\mathbb{R}^{\rm 3}}{\rm G}_{{\rm ki,j}} \left(\mathbf{x}-\mathbf{x}'\right)\; \varphi _{{\rm ij}} \left(\mathbf{x}'\right)\; {\rm d}\mathbf{x}'  
\end{equation} 
\begin{equation} \label{GrindEQ__72_} 
\tilde{{v}}_{{i,j}} \left(\mathbf{x}\right)=\int _{{\mathbb{R}}^{{\rm 3}} }{\rm G}_{{\rm ik,jl}} \left(\mathbf{x}-\mathbf{x}'\right)\; \varphi _{{\rm kl}} \left(\mathbf{x}'\right)\; {\rm d}\mathbf{x}' . 
\end{equation} 

Convolution integrals in direct space are simply products in Fourier space 

\begin{equation} \label{GrindEQ__73_} 
\hat{\tilde{v}}_{k} \left(\xi \right)=\left(-i\xi _{j} \right)\; \hat{G}_{ki} \left(\xi \right)\; \hat{\varphi }_{ij} \left(\xi \right) 
\end{equation} 

\begin{equation} \label{GrindEQ__74_} 
\hat{\tilde{v}}_{i,j} \left(\xi \right)=\hat{\Gamma }_{ijkl}^{} \left(\xi \right)\; \hat{\varphi }_{kl} \left(\xi \right) 
\end{equation} 

\noindent where the symbol "$\wedge$" indicated a Fourier transform. The Green operator in eq. \eqref{GrindEQ__74_} is defined as  ${\Gamma }_{ijkl} = G_{ik,jl}$.

The tensors $\hat{G}_{ij} \left(\xi \right)$ and $\hat{\Gamma }_{ijkl} \left(\xi \right)$ can be calculated by taking Fourier transform to the set of eqs. \eqref{GrindEQ__70_}

\begin{equation} \label{GrindEQ__75_} 
\left|\begin{array}{l} {\; \xi _{l} \xi _{j} \; L_{ijkl}^{0} \hat{G}_{km} \left(\xi \right)+i\xi _{i} \hat{H}_{m} \left(\xi \right)=\delta _{im} } \\ {\; \xi _{k} \hat{G}_{km} \left(\xi \right)=0} \end{array}\right.  
\end{equation} 

Defining the 3x3 matrix $A'_{ik} \left(\xi \right)=\xi _{l} \xi{j} L_{ijkl}^{0} $ and embedding it in the 4x4 matrix $A\left(\xi \right)$ as

\begin{equation} \label{GrindEQ__76_} 
A\left(\xi \right)=\left|\begin{array}{cccc} {A'_{11} } & {A'_{12} } & {A'_{13} } & {\xi _{1} } \\ {A'_{21} } & {A'_{22} } & {A'_{23} } & {\xi _{2} } \\ {A'_{31} } & {A'_{32} } & {A'_{33} } & {\xi _{3} } \\ {\xi _{1} } & {\xi _{2} } & {\xi _{3} } & {0} \end{array}\right|
\end{equation} 

\noindent it leads to (see eqs. \eqref{GrindEQ__16_}-\eqref{GrindEQ__18_})

\begin{equation} \label{GrindEQ__77_} 
\hat{G}_{ij} \left(\xi \right)=A_{ij}^{-1} \; \left(i,j=1,3\right) 
\end{equation} 
\begin{equation} \label{GrindEQ__78_} 
\hat{\Gamma }_{ijkl}^{} \left(\xi \right)=-\xi _{j} \xi _{l} \hat{G}_{ik} \left(\xi \right) 
\end{equation} 

\subsection*{Numerical implementation} 

Assigning an initial guess values to the strain-rate field in the regular grid $\left\{\mathbf{x}^{d} \right\}$ (e.g. $\tilde{\dot{\epsilon }}_{ij}^{0} \left(\mathbf{x}^{d} \right)=0\Rightarrow \dot{\epsilon }_{ij}^{0} \left(\mathbf{x}^{d} \right)=\dot{E}_{ij} $), and computing the corresponding stress field $\sigma'^{0}_{ij} (\mathbf{x}^{d})$ from the local constitutive relation, eq. \eqref{GrindEQ__1_} (which requires to know the initial values of the critical stresses $g^{k} \left(\mathbf{x}^{d} \right)$, and the Schmid tensors $\mathbf{Z}_{a} \left(\mathbf{x}^{d} \right)$), an initial guess for the polarization field in direct space $\varphi _{ij}^{0} \left(\mathbf{x}^{d} \right)$, eq. \eqref{GrindEQ__1_}, can be obtained which in turn can be Fourier-transformed to obtain $\hat{\varphi }_{ij}^{0} \left(\boldsymbol\xi ^{d} \right)$. Furthermore, assuming that $\lambda _{ij}^{0} \left(\mathbf{x}^{d} \right)=\sigma _{ij}^{0} \left(\mathbf{x}^{d} \right)$ is the initial guess for a field of Lagrange multipliers associated with the compatibility constraints, the iterative procedure based on Augmented Lagrangians proposed by \cite{MMS00} can be used. If the polarization field after iteration \textit{n } is known, the \textit{n}+1-th iteration starts by computing the new guess for the kinematically-admissible strain-rate deviation field

\begin{equation} \label{GrindEQ__79_} 
\hat{\tilde{d}}_{ij}^{n+1} \left(\boldsymbol\xi ^{d} \right)=-\hat{\Gamma }_{ijkl}^{sym} \left(\boldsymbol\xi ^{d} \right)\; \hat{\varphi }_{kl}^{n} \left(\boldsymbol\xi ^{d} \right),\quad \forall \boldsymbol\xi ^{d} \ne \mathbf{0};\quad {\rm and} \quad \hat{\tilde{d}}_{ij}^{n+1} \left(\mathbf{0}\right)= \mathbf{0} 
\end{equation} 

\noindent where $\hat{\Gamma }_{ijkl}^{sym} $ is the Green operator, appropriately symmetrized. The corresponding field in real space is, thus, obtained by application of the inverse FFT, i.e.

\begin{equation} \label{GrindEQ__80_} 
\tilde{d}_{ij}^{n+1} \left(\mathbf{x}^{d} \right)={\rm fft}^{-1} \left\{\hat{\tilde{d}}_{ij}^{n+1} \left(\boldsymbol\xi ^{d} \right)\right\} 
\end{equation} 

\noindent and the new guess for the deviatoric stress field is calculated from (omitting subindices):

\begin{equation} \label{GrindEQ__81_} 
\begin{array}{l} {\boldsymbol\sigma'^{n+1} \left(\mathbf{x}^{d} \right)+ \mathbb{L}^{0} :\dot{\gamma }_{0} \sum _{k} \mathbf{Z}_s^{k} \left(\mathbf{x}^{d} \right) \left(\frac{\mathbf{Z}_s^{k} \left(\mathbf{x}^{d} \right):\boldsymbol\sigma'^{n+1} \left(\mathbf{x}^{d} \right)}{g^{k} \left(\mathbf{x}^{d} \right)} \right)^{n}  {\rm sign} \left(\mathbf{Z}_s^{k} \left(\mathbf{x}^{d} \right):\boldsymbol\sigma'^{n+1} \left(\mathbf{x}^{d} \right)\right)=} \\ {\quad \quad \quad \quad \quad \quad \quad =\lambda ^{n} \left(\mathbf{x}^{d} \right)+\mathbb{L}^{0} :\left(\dot{\mathbf E}+\tilde{d}^{n+1} \left(\mathbf{x}^{d} \right)\right)} \end{array} 
\end{equation} 

The iteration is completed with the calculation of the new guess for the Lagrange multiplier field according to

\begin{equation} \label{GrindEQ__82_} 
\lambda ^{n+1} \left(\mathbf{x}^{d} \right)=\lambda ^{n} \left(\mathbf{x}^{d} \right)+\mathbb{L}^{0} :\left(\tilde{\dot{\boldsymbol\epsilon }}^{n+1} \left(\mathbf{x}^{d} \right)-\tilde{d}^{n+1} \left(\mathbf{x}^{d} \right)\right) 
\end{equation} 

Equations \eqref{GrindEQ__81_}-\eqref{GrindEQ__82_} guarantee the convergence of $\dot{\boldsymbol\epsilon }\left(\mathbf{x}^{d} \right)$ (i.e. the strain-rate field related with the stress through the constitutive equation) towards $d\left(\mathbf{x}^{d} \right)$ (i.e. the kinematically-admissible strain-rate field) to fulfil compatibility, and of the Lagrange multiplier field $\lambda \left(\mathbf{x}^{d} \right)$ towards the stress field $\boldsymbol\sigma'\left(\mathbf{x}^{d} \right)$ to fulfil equilibrium.

\subsection*{Time integration} 

Upon convergence, the microstructure can be updated using an explicit scheme. The resulting shear strain-rate field

\begin{equation} \label{GrindEQ__83_} 
\dot{\gamma }^{k} \left(\mathbf{x}^{d} \right)=\dot{\gamma }_{0} \left(\frac{\mathbf{Z}_s^{k} \left(\mathbf{x}^{d} \right):\boldsymbol\sigma'\left(\mathbf{x}^{d} \right)}{g^{k} \left(\mathbf{x}^{d} \right)} \right)^{n}  {\rm sign} \left(\mathbf{Z}_s^{k} \left(\mathbf{x}^{d} \right):\boldsymbol\sigma'\left(\mathbf{x}^{d} \right)\right) 
\end{equation} 

\noindent can be assumed to be constant during a time interval $\left[{\rm t,t}+{\rm \Delta t}\right]$. The macroscopic and local strain increments are then calculated as $\Delta E_{ij} =\dot{E}_{ij}  \Delta t$ and $\Delta\epsilon _{ij} \left(\mathbf{x}^{d} \right)=\dot{\epsilon }_{ij} \left(\mathbf{x}^{d} \right) \Delta t$ and the local crystallographic orientations are updated according to the following local lattice rotation

\begin{equation} \label{GrindEQ__84_} 
\omega _{ij} \left(\mathbf{x}^{d} \right)=\left(\dot{\Omega }_{ij} \left(\mathbf{x}^{d} \right)+\tilde{\dot{\omega }}_{ij} \left(\mathbf{x}^{d} \right)-\dot{\omega }_{ij}^{p} \left(\mathbf{x}^{d} \right)\right)\Delta t 
\end{equation} 

\noindent where $\dot{\omega }_{ij}^{p} \left(\mathbf{x}^{d} \right)$ can be obtained from eqs. \eqref{GrindEQ__3_} and \eqref{GrindEQ__108_}, and $\tilde{\dot{\boldsymbol\omega }}\left(\mathbf\mathbf{x}^{d} \right)$ is obtained back-transforming the converged antisymmetric field

\begin{equation} \label{GrindEQ__85_} 
\hat{\tilde{\dot{\omega }}}_{ij}^{} \left(\xi ^{d} \right)=-\hat{\Gamma }_{ijkl}^{antisym} \left(\xi ^{d} \right)\; \hat{\varphi }_{kl}^{} \left(\xi ^{d} \right),\quad \forall \xi ^{d} \ne 0;\quad {\rm and} \quad \hat{\tilde{\dot{\omega }}}_{ij}^{} \left(\mathbf{0}\right)=\mathbf{0}.
\end{equation} 

The CRSSs of the deformation systems associated with each material point should also be updated after each deformation increment due to strain hardening using one the models described in section 3.

After each time increment, the new position of the Fourier points can be determined calculating the velocity fluctuation term $\tilde{v}_{k} \left(\mathbf{x}^{d} \right)$ back-transforming eq. \eqref{GrindEQ__71_}, and

\begin{equation} \label{GrindEQ__86_} 
X_{i} \left(\mathbf{x}^{d} \right)=x_{i}^{d} +\left(\dot{E}_{ij} x_{j}^{d} +\tilde{v}_{i} \left(\mathbf{x}^{d} \right)\right) \Delta t .
\end{equation} 

Evidently,  the set of advected Fourier points no longer forms a regular grid, after the very first deformation increment due to the heterogeneity of the medium. A rigorous way of dealing with this situation was proposed by \cite{LMM01}  based on the particle-in-cell method \citep{SZS95}.

\subsection*{Extension to dilatational viscoplasticity}

The previous FFT-based algorithm can be adapted to the case of dilatational viscoplasticity  of voided materials with polycrystalline or homogeneous isotropic matrix. The constitutive behavior of Fourier points belonging to a void is simply  $\boldsymbol\sigma$ = 0 . In the cases of porous materials with crystalline or isotropic matrix, the constitutive equation for Fourier points with material properties is given, respectively, by eq. \eqref{GrindEQ__1_}, or by

\begin{equation}\label{iso}
\dot {\boldsymbol{\epsilon}} \left( {{x}^d} \right) = \frac{{3{{\dot \gamma }_0}}}{{2{\sigma _0}}}{\left( {\frac{{{\sigma _{eq}}\left( {\mathbf{x}^d} \right)}}{{{\sigma _0}}}} \right)^{n - 1}}\boldsymbol{\sigma} \left( {\mathbf{x}^d} \right)
\end{equation}

\noindent where $\sigma _0$ is the flow stress.

Given the effective compressibility of the material due to the presence of voids, the differential equation whose solution is the Green's function  ${G_{km}}\left( {\mathbf{x}^d} \right)$ associated with the velocity field is given by  eq. \eqref{GrindEQ__70_}

\begin{equation}
L_{ijkl}^o{G_{km,lj}}\,\left( {{\mathbf{x}} - {\mathbf{x'}}} \right) + {\delta _{im}}\,\delta \left( {{\mathbf{x}} - {\mathbf{x'}}} \right) = {\bf{0}}
\end{equation}

\noindent hence (see eqs. \eqref{GrindEQ__77_}-\eqref{GrindEQ__78_})

\begin{equation}
{\hat G_{ij}}\left( \xi  \right) = A_{ij}^{ - 1}\left( \xi  \right) \quad {\rm and} \quad {A_{ik}}\left( \boldsymbol\xi  \right) = {\xi _l}{\xi _j}L_{ijkl}^o
\end{equation}

\noindent and 

\begin{equation}\label{Gam}
\hat \Gamma _{ijkl}^{}\left( \xi  \right) =  - {\xi _j}{\xi _l}{\hat G_{ik}}\left( \xi  \right).
\end{equation}

The algorithm for a full stress tensor $\boldsymbol{\Sigma}$ imposed to a unit cell representing a porous material requires an initial guess for the average strain rate

\begin{equation}
\dot{E}_{ij}^0 = \dot{E'}_{ij}^0 + \frac{{\dot{E}_{kk}^0}}{3}{\delta _{ij}}
\end{equation}

\noindent where $\dot E'_{ij}$  and $\dot E_{kk}$ are the deviatoric and hydrostatic parts, which will be adjusted iteratively. Initial guess values also need to be assigned to the local strain rate field. The deviatoric part is taken to be uniform

\begin{equation}
\tilde{\dot{\epsilon}}_{ij}^{'0}\left( {\mathbf{x}} \right) = {\bf{0}} \Rightarrow \dot{\epsilon}_{ij}^{'0}\left( {\mathbf{x}} \right) = \dot{E}_{ij}^{'0}.
\end{equation}

An initial guess of uniform dilatation is assumed for the Fourier points representing voids,

\begin{equation}
\tilde {\dot\epsilon}_{kk}^o\left( {\bf{x}} \right) = \left( {\frac{1}{f} - 1} \right)\dot E_{kk}^o
\end{equation}

\noindent where $f$ is the current porosity in the RVE. For these initial values, the corresponding stress field in the material points  ${\boldsymbol{\sigma}^0}\left( {\bf{x}} \right)$ is obtained by inverting the local constitutive relation given by eq. \eqref{GrindEQ__1_}  (for polycrystals) or eq. \eqref{iso} (for isotropic matrix), whereas for points belonging to voids, the stress simply vanishes. The initial specification of these fields and reference stiffness allows us to calculate the initial guess for the polarization field in direct space, eq. \eqref{GrindEQ__67_}, which in turn is Fourier-transformed, used in combination with eq. \eqref{Gam} to solve the convolution integral in Fourier space, and anti-transformed to obtain a new guess for the strain-rate field, etc. Furthermore, assuming   $\lambda _{ij}^o\left( {\bf{x}} \right) = \sigma _{ij}^o\left( {\bf{x}} \right)$ as the initial guess for an auxiliary stress field associated with the compatibility constraint, the Augmented Lagrangian procedure (eqs. \eqref{GrindEQ__81_}-\eqref{GrindEQ__82_}) can be readily applied to iteratively determine the local fields. Microstructural changes, including porosity evolution, can then be tracked using the updating algorithms described in \cite{LEC13}.

\section{Multiscale modelling of polycrystals}

Simulation of polycrystals by means of either mean-field homogenization schemes or full-field homogenization of an RVE of the microstructure can only consider homogeneous stress or strain boundary conditions. Thus, it cannot be used to determine the mechanical response of polycrystals with arbitrary geometry or subjected to complex mechanical loads. Multiscale frameworks in which the single crystal (microscopic) behavior is connected with the specimen (macroscopic) response through an intermediate (mesoscopic) scale are the most direct solution to link the non-homogeneous deformation of an structure with its microstructure. The behavior at mesoscopic scale is obtained from the homogenized response of a polycrystal, that represents the behavior of each (polycrystalline) material point in the macroscale. These models allow considering microstructural effects on the response of a polycrystalline specimen and they are only applicable to problems with a clear separation of scales, in which the length-scale associated with the  gradients of the mechanical fields at the macroscale has to be large compared with the length-scale of the polycrystalline microstructure, i.e. the grain or sub-grain size.

The models and numerical methods at both ends of this multiscale problem are well-established and readily available. The mechanical behavior of single crystals is modelled using any of the CP models presented in section 3, while the macroscopic behavior at the macroscopic scale, possibly of complicated geometry and undergoing complex boundary conditions, is modeled with the FEM using large-strain kinematics. Different flavours of the multiscale simulation strategy are determined by the model used at the mesoscopic scale. For example, they can be based on full-field homogenization, in which the mechanical response at each integration point is given by numerical simulation of a polycrystalline RVE associated to that integration point. Within this group, finite elements have been used to perform this mesoscale computational homogenization for different polycrystalline microstructures \citep{MSS99, WC00, KG08}. In the general case of heterogeneous materials (of which a polycrystal is a particular case), these approaches are sometimes referred to as FE$^{2}$ models \citep{F03}. 

However, FEM-based homogenization at the mesoscale requires a huge computational effort for the whole multiscale calculation. An obvious direction for reducing this cost is the use of more efficient full-field models based on FFT to obtain the average behavior of the RVE in each integration point (section 5.2). A more complex alternative is the use of model order reduction for the mecanical behavior at the microscale. Model order reduction has been mainly used for linear heterogeneous materials or more simple constitutive models at the microscale, but it has been recently applied to polycrystals \citep{Michel2016}. Finally, the computational cost of concurrent multiscale models is strongly reduced if, instead of a numerical model, a mean-field homogenization approach (section 4) is used at the microscale. In this latter case, the polycrystal response at each integration point is obtained from an estimate of its effective behavior, based a non-space resolved statistical representation of the microstructure and homogenization theory to connect the single crystal and polycrystal levels.

Another important aspect of the implementation at mesoscopic level within the multiscale analysis, as discussed by \cite{VKV06}, concerns whether the polycrystal model is embedded in the FE computation (i.e. interrogated on the fly each time the behavior of the corresponding material point is needed by the FE analysis) or used in a hierarchical fashion (i.e. the FE model utilizes a number of parameters which are identified in advance (pre-computed), using the mesoscale model) \citep{VKV06, VYV09}. Embedded models are obviously more accurate to follow the microstructural evolution and how the later affects the macroscopic behavior, while hierarchical approximations are more efficient. In connection with these two different strategies, it is also worth mentioning a promising intermediate approach, known as the adaptative sampling method \citep{BKA08}, which consists in embedding a lower length-scale model (e.g. at polycrystal level) and storing the response obtained for a given microstructure each time this model is interrogated. In this way, a microstructure-response database is populated, which can be used instead of a new call to the polycrystal model when the current microstructure is not far from an already computed point of this microstructure-response space.

Currently, one of the most efficient multiscale approaches is based in the combination of FEM at the macroscopic level with a VPSC approximation at the mesoscale. Along these lines, \cite{SLL12} developed a semi-implicit elasto-visco-plastic multiscale framework to simulate the plastic deformation of polycrystalline structures. This modelling framework have been later extended for different type crystals and applied to simulate thermomechanical processes on full components \citep{KCB15, ZBK17, ZB18}. Due to the relevance of this work, the main equations of the framework developed in  \cite{SLL12} will be reviewed here, including some implementation aspects in a standard finite element code.

\subsection {Kinematics}
The mesoscopic material model based on VPSC \citep{SLL12} was developed using both small-strain and finite-deformation frameworks. For finite deformations, a \emph{hypoelastic} constitutive formulation was used. Hypoelastic models are a common approach to extend small-strain plasticity models to the finite strain range by recasting the original evolution equations in terms of objective stress rates that preserve frame invariance \citep{Peric_plasticity}. This approach to define user materials is a common option in most of commercial codes as ABAQUS, i.e. the one chosen for the numerical implementation of the model, due to its simplicity and usability. The application of the hypoelastic framework in this multiscale model allowed using small-strain kinematics inside the VPSC-based user-defined material subroutine in combination with a co-rotational reference frame that is determined and stored to account for large strains and rotations. In small strains, the total strain increment ($\Delta \boldsymbol\epsilon $) is decomposed in elastic ($\Delta \boldsymbol\epsilon _{el} $) and viscoplastic ($\Delta \boldsymbol\epsilon_{vp} $) parts, such that the constitutive relation at each polycrystalline integration point is given by:

\begin{equation} \label{GrindEQ__87_} 
\Delta \boldsymbol\epsilon =\Delta \boldsymbol\epsilon _{el} +\Delta \boldsymbol\epsilon _{vp} = {\mathbb L}^{-1} \Delta \boldsymbol\sigma +\Delta \boldsymbol\epsilon _{vp}  
\end{equation} 

\noindent where ${\mathbb L}$ is the elastic stiffness tensor of the polycrystalline material point, $\Delta \boldsymbol\sigma $ is the  stress increment and $\Delta \boldsymbol\epsilon _{vp} =\Delta \boldsymbol\epsilon _{vp} \left(\boldsymbol\sigma \right)$ is computed with the VPSC model for each polycrystalline material point. For large deformations, the stress increment is based on the interval integration of the Jaumann rate of the Cauchy stress.
in consistency with ABAQUS's treatment of rotations. 

A \textit{local coordinate system}, defined by the unit vectors $\mathbf{e}_{\alpha }  \left(\alpha =1,3\right)$ in which the polycrystal behavior is computed, undergoes rotations that need to be accounted for by the FEM. This system will be referred to as \textit{co-rotational system}.  In the adopted hypoelastic solution of the finite-deformation problem, the material model is formulated incrementally, using the Jaumann rate of the Cauchy stress as the objective rate to define the stress increments. In this case, the evolution of the unit vectors of the local coordinate system rotating with the polycrystal is given by:

\begin{equation} \label{GrindEQ__88_} 
\dot{\mathbf e}_{\alpha } =\dot{\boldsymbol\omega } {\mathbf e}_{\alpha }  
\end{equation}

\noindent where $\dot{\boldsymbol\omega }$ is the total rotation-rate of the polycrystalline material point. The integration  in a time increment (from time ${t}$ to $t+\Delta t$) gives an incremental rotation matrix ${\rm \Delta }{\mathbf R}$ that rotates the local reference system from ${t}$ to $t+\Delta t$,

\begin{equation} \label{GrindEQ__89_} 
{\mathbf R}^{t+\Delta t} ={\Delta }{\mathbf R} {\mathbf R}^{t}  
\end{equation}

\noindent where ${\mathbf R}^{t+\Delta t} $ is the rotation matrix from the local coordinate system at $t=0$ to its orientation at time $t+\Delta t$, e.g.
 
\begin{equation} \label{GrindEQ__90_} 
{\mathbf e}_{\alpha }^{t+\Delta t} ={\mathbf R}^{t+\Delta t} {\mathbf e}_{\alpha }^{0}  
\end{equation}

Similarly to eq. \eqref{GrindEQ__87_},  the velocity-gradient ${\mathbf L}$ for large strains is decomposed into elastic and viscoplastic components

\begin{equation} \label{GrindEQ__91_} 
{\mathbf L}={\mathbf L}_{el} +{\mathbf L}_{vp}  
\end{equation} 

\noindent where:

\begin{equation} \label{GrindEQ__92_} 
{\mathbf L}_{el} =\dot{\boldsymbol\epsilon}_{el} +\dot{\boldsymbol\omega } 
\end{equation} 

\noindent and

\begin{equation} \label{GrindEQ__93_} 
{\mathbf L}_{vp} =\dot{\boldsymbol\epsilon}_{vp}  
\end{equation}

Elastic and viscoplastic constitutive laws involve only the symmetric component of the deformation, while the antisymmetric component is ambiguous. Consequently,  the full rotation-rate in eq. \eqref{GrindEQ__92_} was assigned to the elastic part of the velocity-gradient for convenience.

The Jaumann rate of the Kirchoff stress, $ \mathop{\boldsymbol\tau }\limits^{\nabla }$, which is the stress measure work-conjugate with the rate of deformation, is defined by

\begin{equation} \label{GrindEQ__94_} 
\mathop{\boldsymbol\tau }\limits^{\nabla } =\dot{\boldsymbol\sigma }-\dot{\boldsymbol\omega }\boldsymbol\sigma +\boldsymbol\sigma \dot{\boldsymbol\omega }. 
\end{equation} 

If plastic strains are much larger than elastic ones, deformation can be considered isochoric ($J=\det \left(\mathbf{F}\right)=1$ and $\tau = J\sigma \approx \sigma $). Thus, $\mathop{\boldsymbol{\sigma} }\limits^{\nabla } $ is related to the elastic strain-rate by

\begin{equation} \label{GrindEQ__95_} 
\mathop{\boldsymbol\sigma }\limits^{\nabla } ={\mathbb L} \left(\dot{\boldsymbol\epsilon }-\dot{\boldsymbol\epsilon }_{vp} \right).
\end{equation} 

The FE solution requires that the constitutive relation in eq. \eqref{GrindEQ__95_} is expressed and integrated in a fixed coordinate system. If the coordinate system associated to every polycrystalline material point is unique and given by $\mathbf{e}_{\alpha }^{0} $,  this reference frame is used as the required fixed coordinate system. Integrating eq. \eqref{GrindEQ__95_} from  $t$ to $t+\Delta t$ gives

\begin{equation} \label{GrindEQ__96_} 
\mathop{\boldsymbol\sigma }\limits^{\nabla } \Delta t={\mathbb L}\left(\Delta \boldsymbol\epsilon -\Delta \boldsymbol\epsilon_{vp} \right) 
\end{equation} 

In this formulation, the polycrystal elastic stiffness and the viscoplastic strain increment   are calculated by means of the VPSC model in the co-rotational system. Therefore, use is made of ${\mathbf R}^{t+\Delta t}$ to express these magnitudes  in the fixed coordinate system, e.g. if $\Delta \boldsymbol\epsilon_{vp}^{*} $ is the viscoplastic strain increment computed by means of the VPSC model (the symbol ``*'' denotes magnitudes expressed in the co-rotational system), $\Delta \boldsymbol\epsilon_{vp} $ in eq. \eqref{GrindEQ__96_} is obtained as

\begin{equation} \label{GrindEQ__97_} 
\Delta \boldsymbol\epsilon _{vp} =\left({\mathbf R}^{t+\Delta t} \right)^{T} \Delta \boldsymbol\epsilon _{vp}^{*} {\mathbf R}^{t+\Delta t}.  
\end{equation}

In eq. \eqref{GrindEQ__96_}, $\Delta \boldsymbol\epsilon $ is the logarithmic strain increment, obtained from $\Delta \boldsymbol\epsilon ^{*} $ given by the FE analysis as

\noindent 
\begin{equation} \label{GrindEQ__98_} 
\Delta \boldsymbol\epsilon^{*} =\log \left(\Delta {\mathbf V}\right) 
\end{equation} 

\noindent with $\Delta{\mathbf V}$ being the left Cauchy stretch tensor associated with the deformation gradient increment from $t$ to $t+\Delta t$, such that $\Delta {\mathbf F}=\Delta {\mathbf V}\Delta {\mathbf R}$. Hence,

\begin{equation} \label{GrindEQ__99_} 
\Delta \boldsymbol\epsilon =\left({\mathbf R}^{t+\Delta t} \right)^{T} \Delta \boldsymbol\epsilon ^{*} {\mathbf R}^{t+\Delta t}.
\end{equation} 

Finally, for completeness, once the right-hand of eq. \eqref{GrindEQ__96_} is obtained by computing the left-hand magnitudes in the fixed reference system, the stress expressed in co-rotational axes (as needed by VPSC as part of the proposed algorithm, see next section) is obtained as

\begin{equation} \label{GrindEQ__100_} 
\boldsymbol\sigma^{t+\Delta t*} ={\mathbf R}^{t+\Delta t} \left(\boldsymbol\sigma^{t} +\mathop{\boldsymbol\sigma }\limits^{\nabla } \Delta t\right) \left({\mathbf R}^{t+\Delta t} \right)^{T}.  
\end{equation} 

\subsection {VPSC-based UMAT implementation}

The approach at the macroscopic level in the FEM implementation is standard: the applied load is divided in increments, and the equilibrium at each increment is obtained by means of the FEM analysis in an iterative fashion, using a global non-linear solver. The load increment is controlled by the time, and once the problem has being solved at time ${t}$, the solution for the next time increment requires the polycrystal model to provide a tangent stiffness (Jacobian) matrix $\mathbb{L}^{tg} ={\partial \Delta \boldsymbol\sigma  \mathord{\left/{\vphantom{\partial \Delta \boldsymbol\sigma  \partial \Delta \boldsymbol\epsilon }}\right.\kern-\nulldelimiterspace} \partial \Delta \boldsymbol\epsilon } $ for each material point, so the FEM can compute an initial guess for the nodal displacements at $t+\Delta t$. The strain increments obtained from that prediction for each material point, $\Delta \boldsymbol\epsilon^{FE} $, together with the stress $\boldsymbol\sigma^{t} $ and the set of internal state variables $q_{i}^{t} $ corresponding to the previous increment, are used inside the UMAT to calculate a new guess for the stress and the Jacobian at $t+\Delta t$. When convergence in stress equilibrium is achieved by the global non-linear scheme, the new values (at $t+\Delta t$) of the stresses, the internal variables, and the Jacobian matrix are accepted for every node, and the calculation advances to the next increment. For a given $\Delta \boldsymbol\epsilon^{FE} $, the VPSC-UMAT is based on the minimization of an {\it ad hoc} residual, defined below.

\begin{equation} \label{GrindEQ__101_} 
\boldsymbol\sigma^{t+\Delta t} =\boldsymbol\sigma ^{t} +\mathbb{L}\Delta \boldsymbol\epsilon _{el} =\boldsymbol\sigma ^{t} +\mathbb{L}\left(\Delta \boldsymbol\epsilon -\Delta \boldsymbol\epsilon _{vp} \right) 
\end{equation}

\noindent where $\mathbb{L}$ is the elastic stiffness of the polycrystal. In the present context, the natural choice for $\mathbb{L}$ is to use the elastic self-consistent (ELSC) estimate, given by  eq. \eqref{GrindEQ__39_},

\begin{equation} \label{GrindEQ__102_} 
\mathbb{L} =\left\langle {\mathbf{L}^{(r)}}^{-1}  \mathbb{B}^{(r)} \right\rangle^{-1}   
\end{equation} 

\noindent where $\mathbb{L}^{(r)} $ and $\mathbb{B}^{(r)} $ are the elastic stiffness and stress localization tensors, eq. \eqref{GrindEQ__24_}, of SR grain ($r$). Note that an ELSC calculation for the determination of $\mathbb{L}$ is implemented in the VPSC code at the beginning of each deformation increment. In this way, the texture changes are also accounted for in the determination of the stiffness of the polycrystalline material element.

Combining eq. \eqref{GrindEQ__38_} with the viscoplastic constitutive relation coming from VPSC, eq. \eqref{GrindEQ__35_}, leads to,

\begin{equation} \label{GrindEQ__103_} 
\Delta \boldsymbol\epsilon =\mathbb{L}^{-1} \Delta \boldsymbol\sigma +\Delta t \dot{\boldsymbol\epsilon }_{vp}^{(px)} \left(\boldsymbol\sigma ^{t} +\Delta \boldsymbol\sigma ; q_i^t\right).
\end{equation} 

The residual $X\left(\Delta \boldsymbol\sigma \right)$ can be defined  as a (non-linear) function of the stress increment $\Delta \boldsymbol\sigma =\boldsymbol\sigma ^{t+\Delta t} -\boldsymbol\sigma ^{t} $, which the UMAT should return to the FEM, as the constitutive response of the polycrystalline material point to the trial strain increment $\Delta \boldsymbol\epsilon ^{FE} $

\begin{equation} \label{GrindEQ__104_} 
{\mathbf X}\left(\Delta \boldsymbol\sigma \right)=\Delta \boldsymbol\epsilon -\Delta \boldsymbol\epsilon^{FE} ={\mathbb L}^{-1}\Delta \boldsymbol\sigma +\Delta t \dot{\boldsymbol\epsilon }_{vp}^{(px)} \left(\boldsymbol\sigma ^{t} +\Delta \boldsymbol\sigma ; \beta_{i}^{t} \right)-\Delta \boldsymbol\epsilon ^{FE}  
\end{equation} 

The condition ${\mathbf X}\left(\Delta \boldsymbol\sigma \right)=0$ (i.e. $\Delta \boldsymbol\epsilon =\Delta \boldsymbol\epsilon^{FE} $), is obtained using a Newton-Raphson (NR) scheme to solve this non-linear algebraic equation. The corresponding Jacobian, $\mathbb{J}_{NR}, $ is given by:

\begin{multline} \label{GrindEQ__105_} 
\frac{\partial {\mathbf X}\left(\Delta \boldsymbol\sigma \right)}{\partial \left(\Delta \boldsymbol\sigma \right)} ={\mathbb J}_{NR} \left(\Delta \boldsymbol\sigma \right)={\mathbb L}^{-1} +\Delta t \frac{\partial \dot{\boldsymbol\epsilon }_{vp}^{(px)} }{\partial \left(\Delta \boldsymbol\sigma \right)} \left(\boldsymbol\sigma ^{t} +\Delta \boldsymbol\sigma ; q_{i}^{t} \right) = \\ ={\mathbb L}^{-1} +\Delta t {\mathbb M}^{(px)} \left(\boldsymbol\sigma ^{t} +\Delta \boldsymbol\sigma ; q_{i}^{t} \right) 
\end{multline} 

Hence, given a guess $\Delta \boldsymbol\sigma^{k-1} $ for the stress increment, the new guess is obtained as:

\begin{equation} \label{GrindEQ__106_} 
\Delta \boldsymbol\sigma^{k} =\Delta \boldsymbol\sigma ^{k-1} -{\mathbb J}_{NR}^{-1} \left(\Delta \boldsymbol\sigma^{k-1} \right){\mathbf X}\left(\Delta \boldsymbol\sigma ^{k-1} \right). 
\end{equation} 

Moreover, using the first equalities of eqs. \eqref{GrindEQ__104_} and \eqref{GrindEQ__105_},

\begin{equation} \label{GrindEQ__107_} 
{\mathbb J}_{NR} =\frac{\partial \mathbf{X}\left(\Delta \boldsymbol\sigma \right)}{\partial \left(\Delta \boldsymbol\sigma \right)} =\frac{\partial \left(\Delta \boldsymbol\epsilon -\Delta \boldsymbol\epsilon ^{FE} \right)}{\partial \left(\Delta \boldsymbol\sigma \right)} =\frac{\partial \left(\Delta \boldsymbol\epsilon \right)}{\partial \left(\Delta \boldsymbol\sigma \right)}  
\end{equation}

\noindent the Jacobian matrix that the VPSC-based UMAT should pass to the FEM is obtained,

\begin{equation} \label{GrindEQ__108_} 
{\mathbb L}^{tg} =\frac{\partial \left(\Delta \boldsymbol\sigma \right)}{\partial \left(\Delta \boldsymbol\epsilon \right)} ={\mathbb J}_{NR}^{-1} =\left[{\mathbb C}^{-1} +\Delta t\; {\mathbb M}^{(px)} \right]^{-1}.  
\end{equation}

Eq. \eqref{GrindEQ__108_} provides a closed expression for the FEM Jacobian, as a function of the viscoplastic tangent moduli (which is calculated as part of the VPSC algorithm), the elastic stiffness of the aggregate, and the FE time increment. The use of this expression greatly reduces the overall computational cost because the polycrystal stress and the elasto-viscoplastic tangent stiffness tensor are obtained from the same calculation loop. Moreover, the use of eq.  \eqref{GrindEQ__108_} for the FEM Jacobian usually provides quadratic convergence to the macroscopic non-linear equations.

Finally, it must be remarked that the above procedure is semi-implicit: implicit in the stress value, because internal equilibrium is checked at the end of the increment, but explicit in the internal variables $q_{i} $ (such as grain orientations, morphology and hardening variables). The reason for the proposed algorithm to remain explicit in those internal variables is related to the difficulty and computational cost of deriving a residual and a Jacobian matrix that would contain literally thousands of internal variables. In any case, the explicit updating of the internal variables does not appear to change significantly the convergence of the macroscopic model in most cases,  as shown by \cite{SLL12}.

\section{Application to virtual testing of polycrystals}

In order to demonstrate the potential of CHP, applications of the tools and strategies presented above are illustrated in this section, with examples of simulations of the mechanical behavior of polycrystals under different conditions. 

\subsection{Strength}

The mechanical behavior of materials is usually characterized through a uniaxial stress-strain curves that contain the information about the most important mechanical parameters: elastic modulus, yield strength, strain hardening rate and ductility. Thus, the prediction of uniaxial stress-strain curves is obviously the first step to validate the capabilities of CHP and there are many examples in the literature for different polycrystalline materials (see, among many others, \cite{Roters20101152, REB10, ZLE15, CBJ15, GST16, SG16}). From a Materials Science perspective, the information required to carry out an accurate prediction of the stress-strain curve includes a detailed description of the microstructure, and of the hardening parameters that dictate the plastic deformation of the constituent single  crystals, as well. While microstrutural information can be {\it easily} obtained using modern characterization techniques, the quantification of hardening behavior has to be achieved using different strategies based on: a) micromechanical tests, b) inverse optimization, or c) multiscale models. While multiscale approaches to predict the CRSSs of the different slip systems in engineering materials are not yet mature, one example of each of the other two strategies is detailed below.

\subsubsection{Strength of IN718 Ni-based superalloy}

\begin{figure}[h]
\includegraphics[scale=0.55]{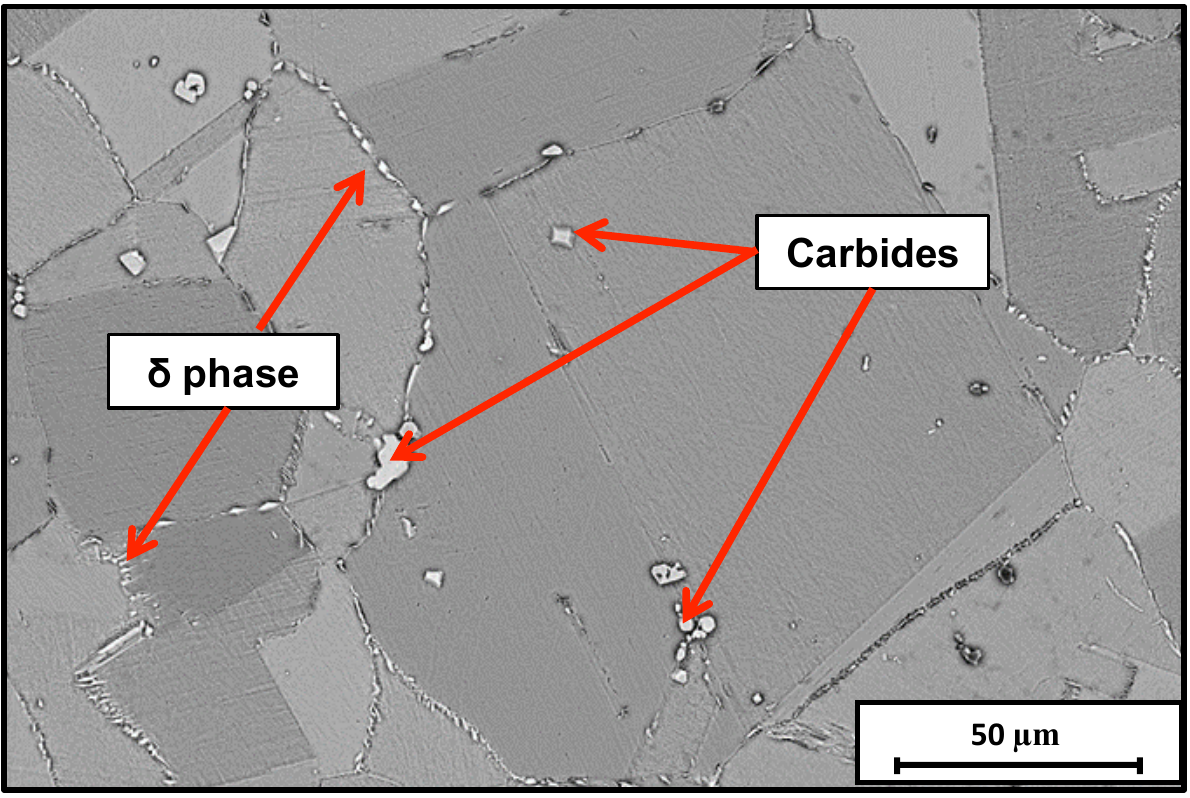}
\includegraphics[scale=0.55]{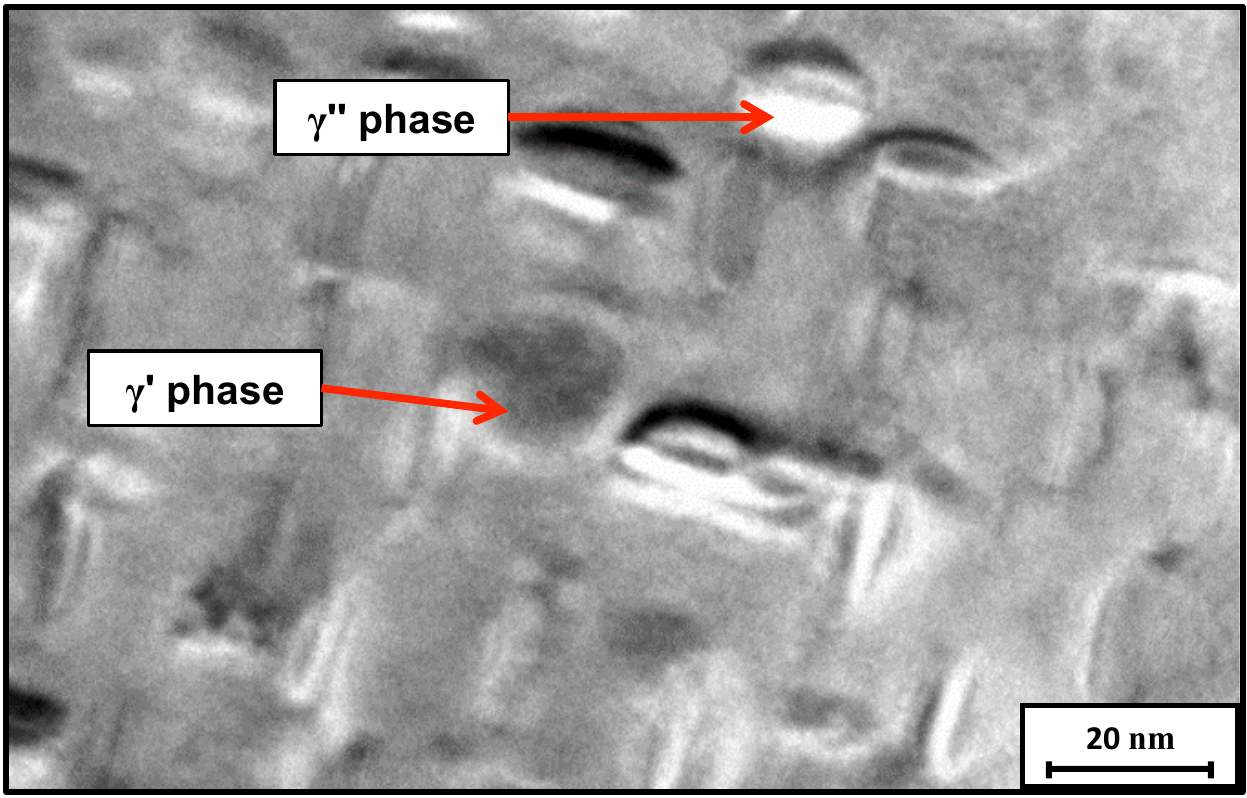}
\caption{Microstructure of IN718 Ni-based superalloy. (a) Polycrystal grain structure showing the distribution of metal carbides and $\delta$ phase at the grain boundaries, (b) Distribution of $\gamma'$ and $\gamma''$ precipitates within the Ni FCC solid solution. Reprinted from \cite{CBJ15}.}  
\label{IN718}
\end{figure}

Polycrystalline IN718 is a Ni-Fe based superalloy widely used for structural applications up to 650-700$^\circ$C because of its good castability and weldability, high strength and corrosion resistance. The microstructure of IN718 is made up by a Ni FCC solid solution which contains a dispersion of nm-sized $\gamma'$ (Ni$_3$(Al,Ti)) and $\gamma''$ (Ni$_3$Nb) coherent precipitates within the grains together with $\mu$m-sized metal carbides and $\delta$ phase (Ni$_3$Nb) particles at grain boundaries (Fig. \ref{IN718}) \citep{R89, SLK05}. The volume fractions of $\gamma'$ and $\gamma''$ phases are in the range 3-5\% and 10-20\%, respectively, depending on the bulk alloy composition, the heat-treatment and the degree of element segregation \citep{F87}. The strength of the alloy in this case (and in the case of many other Ni-based superalloys) is provided by the interaction of the dislocations with the fine distribution of $\gamma'$ and $\gamma''$. The precipitate size and spacing is of the order of 10 - 20 nm in wrought IN718 (Fig. \ref{IN718}b), which stands for the critical length scale that controls the mechanical behavior and, under these circumstances, micromechanical testing techniques that probe the mechanical response of the material in volumes of several $\mu$m$^3$ (much larger than the spacing between precipitates) can be used to determine the CRSS. Examples of these strategies in metallic alloys include micropillar compression \citep{CBJ15} and nanoindentation \citep{HHS15, SPS14, ZZR15, VK15, PK17}.

\begin{figure}[h]
\includegraphics[scale=0.55]{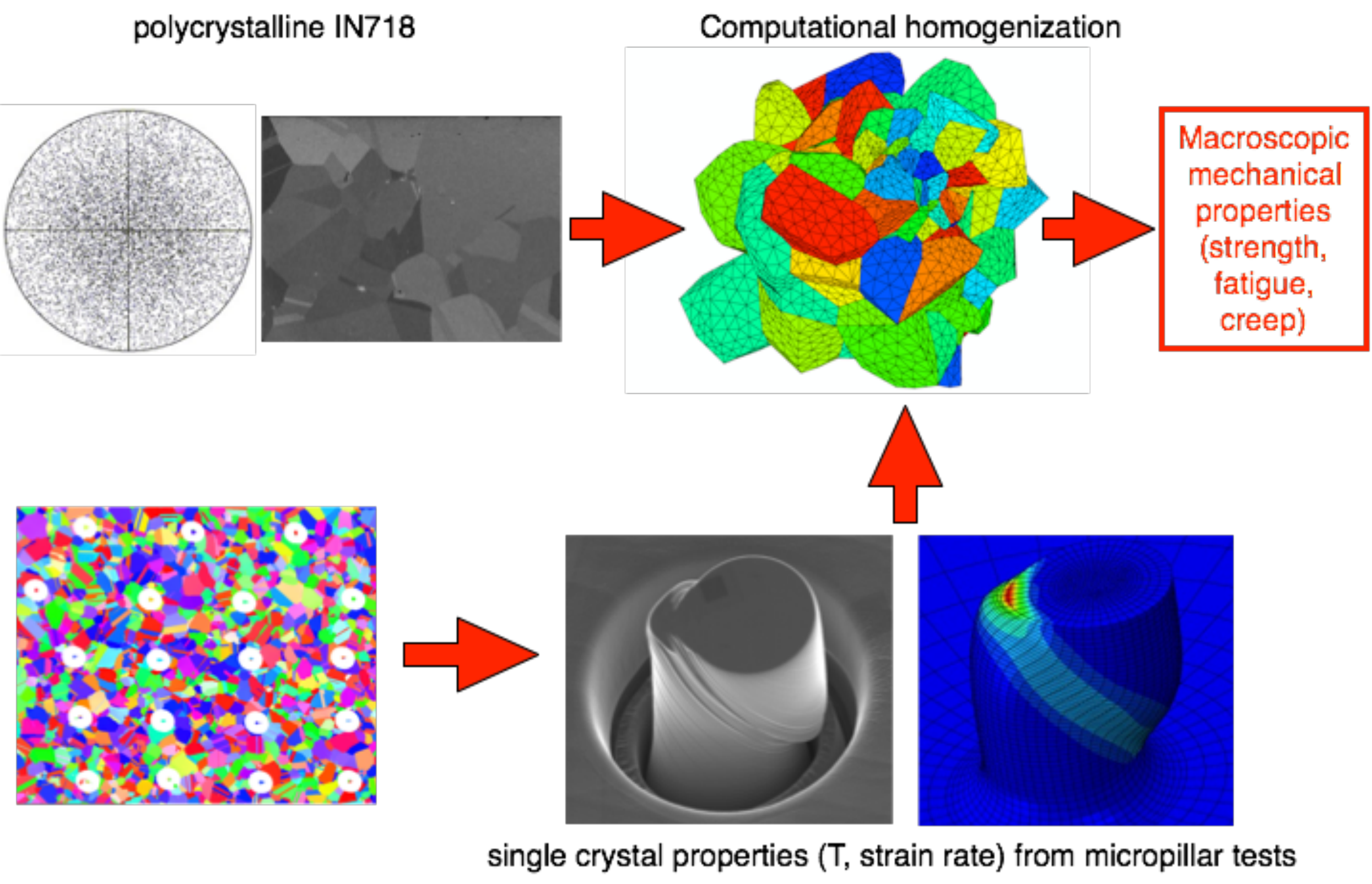}
\caption{Coupled experimental-simulation strategy to predict the mechanical properties of IN718 Ni-based superalloy using computational homogenization of polycrystals. \citep{CBJ15}.}  
\label{VTIN718}
\end{figure}

The strategy to predict the mechanical behavior of polycrystalline Ni-based superalloys is depicted in Fig. \ref{VTIN718} \citep{CBJ15}. The microstructural information included the grain size distribution obtained from optical microscopy images, that can be easily translated into the actual 3-D distribution in the case of equiaxed grains \citep{HB98}, while the grain orientation was provided by X-ray diffraction  measurements. The cubic RVE of the microstructure was generated from the Voronoi tessellation of a set of points, which was obtained from a Monte Carlo algorithm, so the grain size distribution in the RVE followed the experimental log-normal grain size distribution, following the strategy presented in section 2.1.
The grain orientations within the RVE were assigned to match the measured ODF.

Deformation of the Ni solid solution occurred by dislocation slip along the 12 \{111\}$<$110$>$ slip systems of the FCC lattice. The crystal was assumed to behave as an elasto-viscoplastic solid in which the plastic slip rate in the slip system $k$, $\dot \gamma^k$,  follows a phenomenological power-law dependency according to (eq. \eqref{eq:power_law})
 
\begin{equation}\label{eq:IN718i}
\dot \gamma^k  = {\dot \gamma _0}\;{\left| {\frac{{{\tau ^k }}}{{g^k }}} \right|^{1/m}}{\mathop{\rm sgn}} ({\tau ^k })
\end{equation}

\noindent where  $\dot \gamma _0$ stands for the reference strain rate, $g^k $ is CRSS of  the slip system $k$ at the reference strain rate and $m$ the strain rate sensitivity parameter. The evolution of the CRSS of a given slip system, $g^k $, is expressed as, 

\begin{equation}
\dot{g}^k  = \sum\limits_k  {{h\,q_{kj }}} \left| {{{\dot \gamma }^{j}}} \right|
\end{equation}

\noindent where $h$ stands for the self hardening modulus and $q_{kj}$ are the interaction parameters that stand for the influence of hardening between different slip systems. The evolution of the self hardening was described according to the Voce  model, 

\begin{equation}
h\left( {{\Gamma}} \right) = {h_s} + \bigg({h_0} - {h_s} + \frac{{{h_0}{h_s}{\Gamma}}}{{{\tau _s} - {\tau _0}}}\bigg)\exp\bigg\{\frac{-\Gamma h_0}{\tau_s-\tau _0}\bigg\} \label{eq:IN718f}
\end{equation}

\noindent where $h_0$ is the initial hardening modulus, $\tau_0$ the initial yield shear stress, $\tau_s$ the saturation yields shear stress, $h_s$ the saturation hardening modulus at large strains and $\Gamma$ stands for the accumulated shear strain in all slip systems, which is given  by eq. \eqref{eq:Gamma}.

\begin{figure}[h]
\includegraphics[scale=0.56]{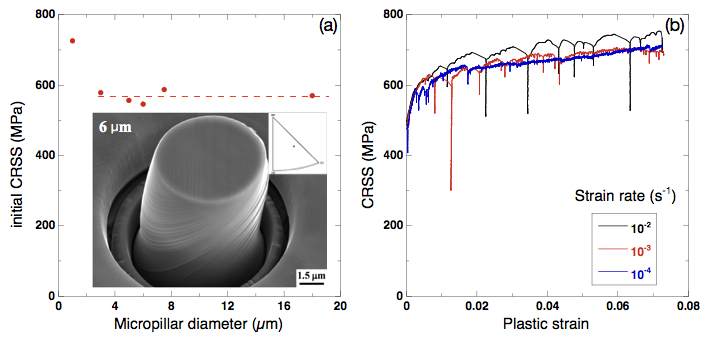}
\caption{(a) Initial CRSS as a  function of the micropillar diameter for micropillars oriented for single slip along the $<235>$ or $<245>$ directions. One deformed micropillar showing the slip traces is depicted within the plot. (b) Effect of strain rate on the stress {\it vs.} plastic strain curve in micropillars of 5 $\mu$m in diameter oriented for single slip. From \cite{CBJ15}.}  
\label{MP1}
\end{figure}

The parameters of this phenomenological constitutive equation were obtained from compression tests in single crystals  carried out on micropillars with circular section extracted by means of focus ion beam milling from crystals oriented in different orientations. Initially, micropillars of different diameters oriented for single slip along the $<235>$ or $<245>$ directions were tested and the initial CRSS is plotted as a function of the micropillar diameter in Fig. \ref{MP1}a. It was found the the initial CRSS (and the strain hardening rate, see \cite{CBJ15}) were independent of the micropillar size (and, thus, representative of the bulk properties of the alloy) when the micropillar diameter was $>$ 3 $\mu$m. Thus, micropillars of 5 $\mu$m in diameter were tested at different strain rates to determine the strain rate sensitivity parameter $m$ (Fig. \ref{MP1}b). In addition, micropillars oriented for double slip (coplanar and non coplanar) as well as for multiple slip were tested. These experimental curves were used to determine the parameters of the constitutive model for IN718 (which are shown in Table \ref{t:IN718}) by comparison with finite element simulations of the micropillar compression tests using the constitutive equation presented in eqs. (\ref{eq:IN718i}) -- (\ref{eq:IN718f}). 

\begin{table}
\centering
\begin{tabular}{|c|c|c|c|c|c|c|}
\hline
 $\dot \gamma_0$ & $m$  & $\tau_0$  & $\tau_s$ & $h_0$ & $h_s$ & q \\ 
  s$^{-1}$ &  & (MPa) &     (MPa) & (GPa) & (GPa) &   \\ \hline
 10$^{-3}$ & 0.017 &  466 & 599 & 6 & 0.3 & 1\\ \hline
\end{tabular}
\caption{Parameters of the crystal plasticity model for IN718 Ni-based superalloy at ambient temperature obtained from micropillar compression tests. From \cite{CBJ15}.}\label{t:IN718}
\end{table}

\begin{figure}[h]
\includegraphics[scale=0.6]{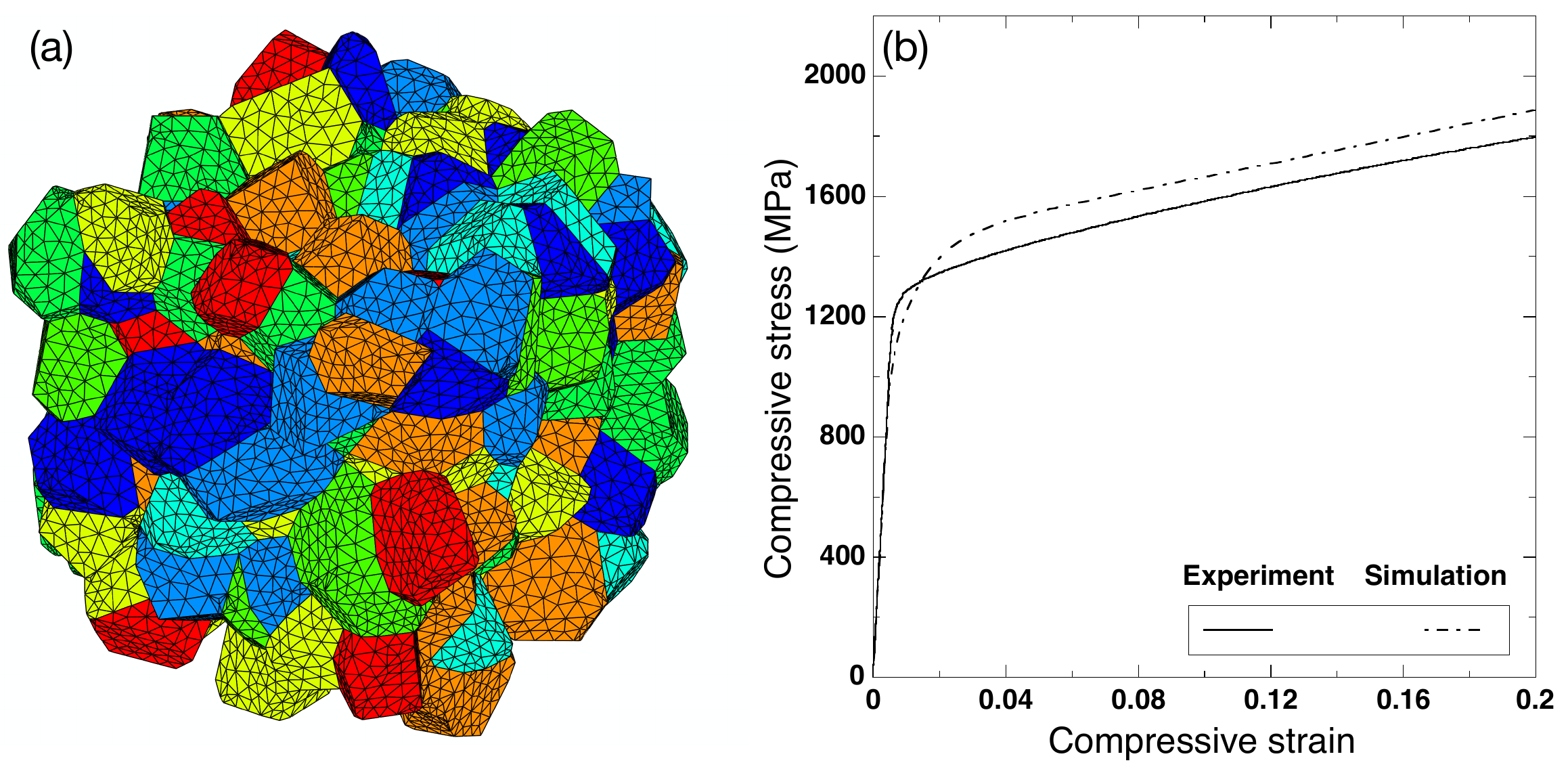}
\caption{(a) RVE of  polycrsytalline IN718 showing the finite element discretization (b) Experimental  and computational homogenization results of the true stress {\it vs.} logarithmic strain curve in uniaxial compression of IN718 at ambient temperature. From \cite{CBJ15}.}  
\label{CH}
\end{figure}

The RVE used to determine the mechanical behavior of polycrystalline IN718 by means of computational homogenization is depicted in Fig. \ref{CH}a. It contained 210 grains and each grain discretized with at least 610 10-node quadratic tetrahedral elements. The grain size distribution and grain orientation followed the experimental data and simulations under periodic boundary conditions were carried out with four  different realizations  of the grain distribution. The differences among them were below 1.3\% and the average true stress {\it vs.} logarithmic strain curve under uniaxial compression obtained from computational homogenization is compared with the experimental data in Fig. \ref{CH}b. The agreement
between them was fairly good: the maximum difference in the compressive flow stress was below 4\% and the prediction of the strain hardening rate was very accurate, validating the approach to predict the mechanical behavior of the alloy. 

\subsubsection{Strength of AZ31 Mg alloys}

Computational homogenization of HCP polycrystals to determine the strength presents several additional difficulties, as compared with cubic materials. Firstly, HCP have different slip systems (basal, prismatic, pyramidal, etc.) with different slip resistances that have to be determined independently. Secondly, they often exhibit deformation twinning, which leads to very different hardening rates. Finally, most HCP metallic alloys present a very strong size effect during nanomechanical characterization (either using nanoindentation  \citep{SPS14, ERR14, NPS17} or micropillar compression \citep{BR13, LLA17}). Thus, the actual values of the CRSS for slip and twinning in the bulk alloy cannot be obtained using the strategy presented in the previous section.

\begin{figure}[h]
\includegraphics[scale=0.55]{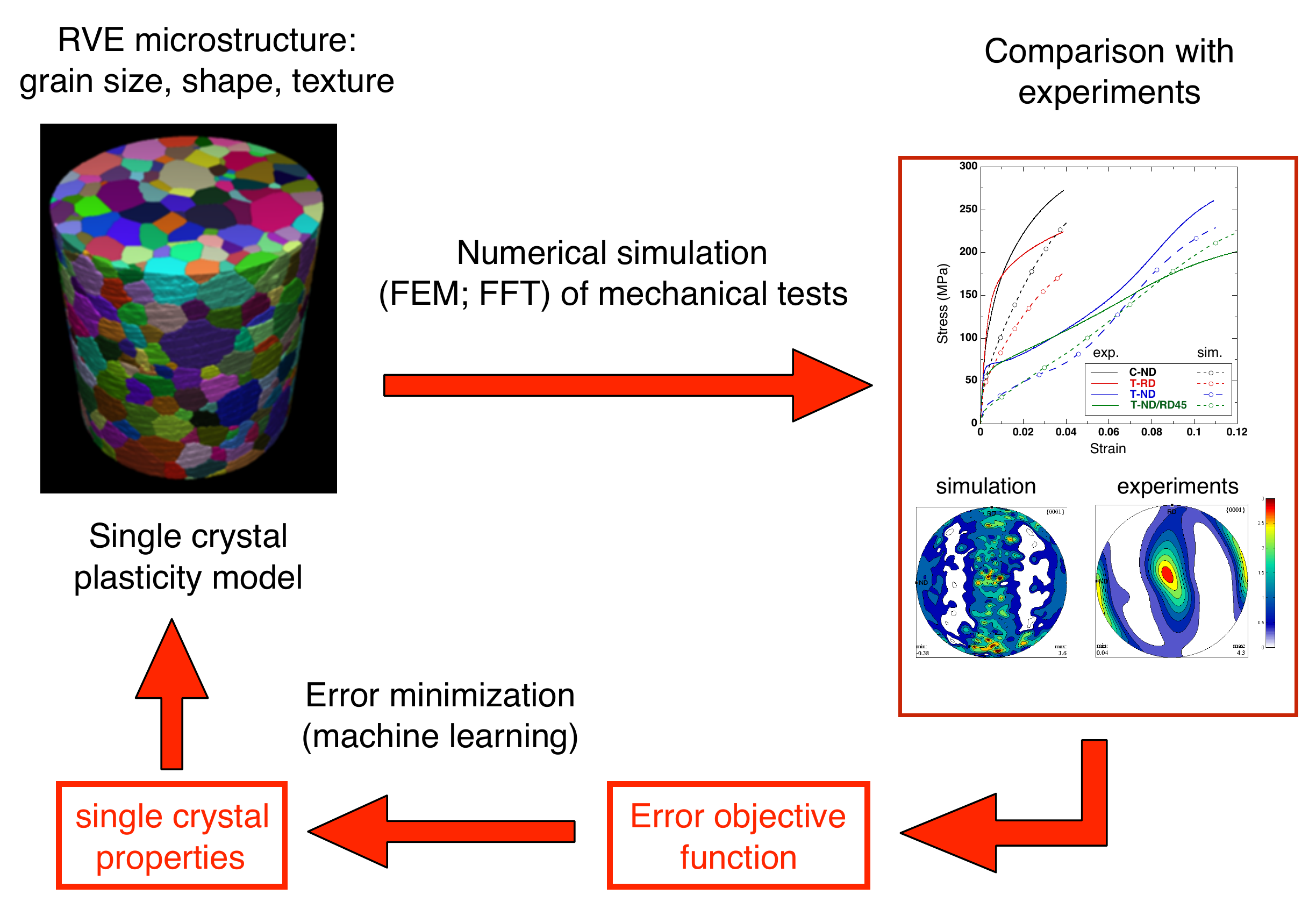}
\caption{Inverse optimization strategy to predict the mechanical properties of AZ31 Mg alloy using computational homogenization of polycrystals. \citep{HLD14}.}  
\label{IO}
\end{figure}

Under these circumstances, the best option to obtain the actual values of the CRSS for slip and twining is to use inverse optimization strategies, as the one depicted in Fig. \ref{IO} for AZ31 Mg alloy \citep{HLD14}. 
The first step of the computational homogenization is the construction of an RVE which includes all the relevant microstructural details (grain size, shape, orientation) and a CP model that takes into account all the actual deformation mechanisms, namely basal, prismatic and pyramidal slip, together with tensile twinning. The plastic slip rate for a given slip system  $k$ follows a power-law  according to eq. \eqref{eq:power_law}. Similarly, the twinning rate, $\dot{f}$, also follows a viscous law given by eq. \eqref{eq:twin}.

The activation of each deformation mode depends on the CRSS ($g^k$ or $g^\alpha$) for each slip or twinning system. The initial values (in absence of previous plastic deformation) of the CRSSs are given by $\tau^k_0$ or $\tau^\alpha_0$ for the slip system $k$ or the twin system $\alpha$, respectively. A phenomenological hardening model was considered for the evolution of the CRSSs, which is able to
reproduce the different stages of single crystal deformation \citep{Hardening_Anand}. The evolution of the CRSSs, $g^k$  for slip and $g^\alpha$ for twinning, are expressed by equations (\ref{eq:slip_hardening}) and  (\ref{eq:twin_hardening}), respectively, 

\begin{equation}
\label{eq:slip_hardening}
\dot{g}^k=q_{sl-sl}\sum_{j=1}^{3} h_0^{j}\left(  1-\frac{\tau^j}{\tau^j_{sat}}\right)^{a_{sl}}|\dot{\gamma}^j|
+q_{tw-sl}h_0^{tw}\left( 1-\frac{\tau^\beta}{\tau^{tw}_{sat}}
\right)^{a_{tw}}|\dot{\gamma}^\beta|
\end{equation}

\begin{equation}
\label{eq:twin_hardening}
\dot{g}^{\alpha}=q_{tw-tw} h_0^{tw}\left( 1-\frac{\tau^\alpha}{\tau^{tw}_{sat}}
\right)^{a_{tw}}\dot{f}^\alpha \gamma_{tw}
\end{equation}

\noindent where $\tau^j$ and $\tau^\alpha$ stand for the resolved shear stress on the corresponding slip (either basal, prismatic or pyramidal) and twinning system, $\dot{\gamma}^j$ is the plastic shear strain rate along system $j$, $\dot{f}^\alpha$ is the rate of the volume fraction transformation in the twin system $\alpha$, and $\gamma_{tw}$ is the characteristic shear of the twinning mode ($\gamma_{tw}=$ 0.129 for extension twinning of Mg alloys, \cite{HLD14}). It should be noted that extension twinning is a polar mechanism and it will only take place when the applied deformation leads to extension of the $c$ axis of the HCP lattice.

The different parameters in these equations define the contributions arising from self-hardening and latent hardening. The self-hardening of a given slip ($k$) or twinning ($\alpha$) system is defined by three terms: the saturation stress, $\tau_{sat}$, the initial hardening rate $h_0$ and the hardening exponent $a$. The latent-hardening contribution to slip due to slip in other systems is introduced with the coefficient $q_{sl-sl}$ whereas the contribution induced by twinning is given by $q_{tw-sl}$. The model only takes into account the effect of twinning on slip and it is assumed that slip does not influence twinning ($q_{sl-tw}=0$) following \cite{CBT09b} and \cite{AZ31_Joshi}.

One obvious difficulty is to find the actual values of the many parameters of the model. In most cases, they have been fitted by comparison with experimental results on polycrystals using a trial and error approach \citep{MSS99,Miehe20022123,SLL12,TML01,TLN02,AZ31_Liu_Pei,AZ31_Kalidindi,AZ31_Tome,AZ31_Tome_Agnew, SL13} but finding the optimum parameter set is neither easy nor a unique result is guaranteed. In fact, it is not unusual to find that different authors report different (or even contradictory) values  of the CRSSs of slip and twinning for similar materials. To overcome this limitation, \cite{HLD14, HSL15} proposed an optimization algorithm which is based in the construction of an  objective error function $O(\boldsymbol{\beta})$ defined as 

\begin{equation}\label{error}
 O(\boldsymbol \beta) = \sum_{i=1}^p | y_i - f(x_i, \ \boldsymbol \beta) |=\left\|\mathbf{y}-\mathbf{f}(\boldsymbol\beta)\right\|. 
\end{equation}

\noindent where $x_i,y_i$ are a set of $p$ points defining an experimental data set of the polycrystal (for instance, stress-strain curves in different orientations, pole figures after deformation, evolution of the fraction of twinned material, etc.) and $y_i=f(x_i;\boldsymbol{\beta})$ are the predictions of the corresponding data obtained by means of the numerical simulation of an RVE of the microstructure in which the mechanical behavior of each grain within the polycrystal is defined by a CP model,  which depends on a set of parameters $\boldsymbol{\beta}$. The minimization of the error function can be carried out using different algorithms, such as Levenberg-Marquardt \citep{HLD14, HSL15} or NelderÐMead \citep{ZYB12, CE17}. It should be noted that novel machine-learning strategies are very promising to carry out this optimization process \citep{MKR17} and also that the optimization strategy often requires several iteration loops. Thus, efficient numerical algorithms based on FFT will become critical within this framework.

\begin{figure}[h]
\centering
\includegraphics[scale=0.55]{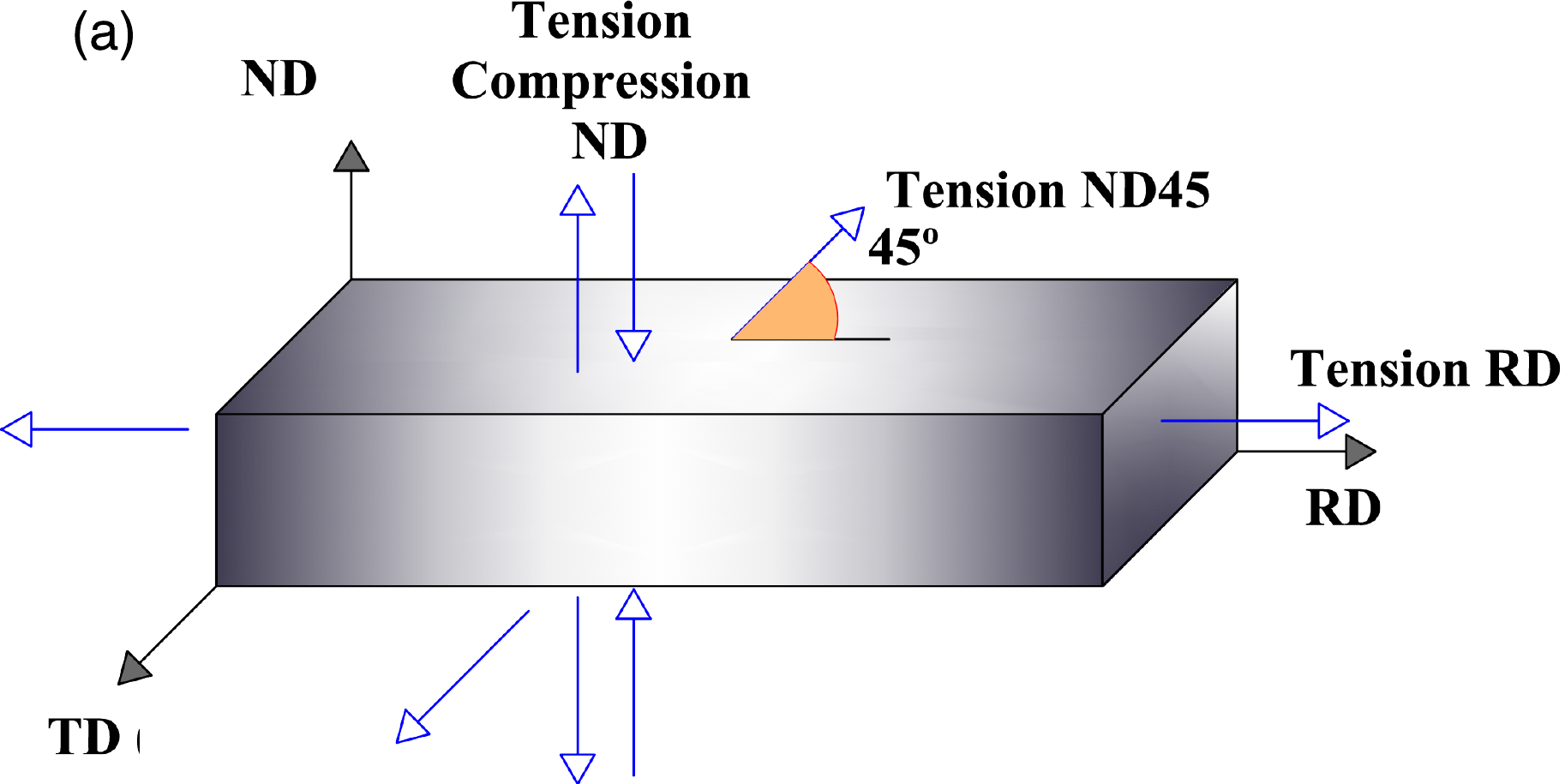} 
\includegraphics[scale=0.55]{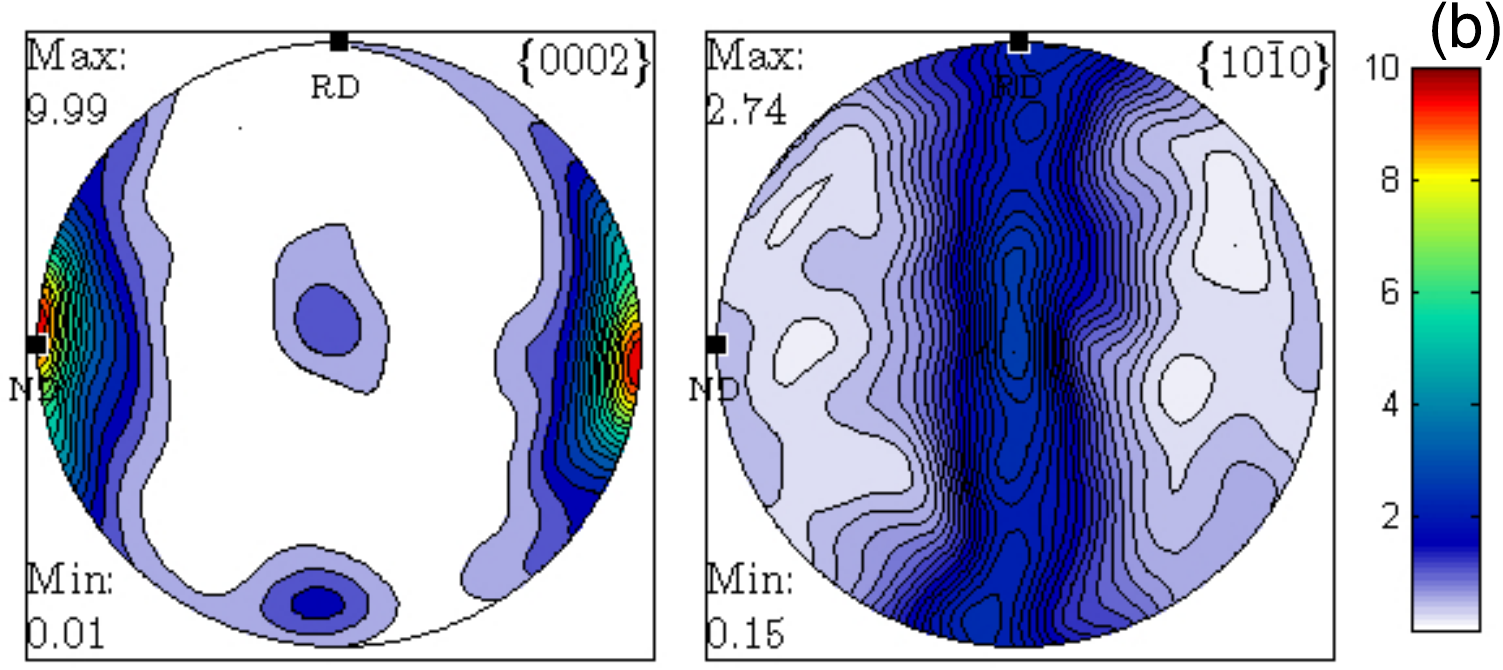}
\caption{(a) Slab of rolled AZ31 Mg alloy showing the RD, ND and TD orientations as well as the orientation of the sample for the mechanical tests. (b) Pole figure of the rolled AZ31 alloy showing the strong basal texture. Reprinted from \cite{HLD14}.}  
\label{AZ31}
\end{figure}

\begin{figure}[h]
\centering
\includegraphics[scale=0.8]{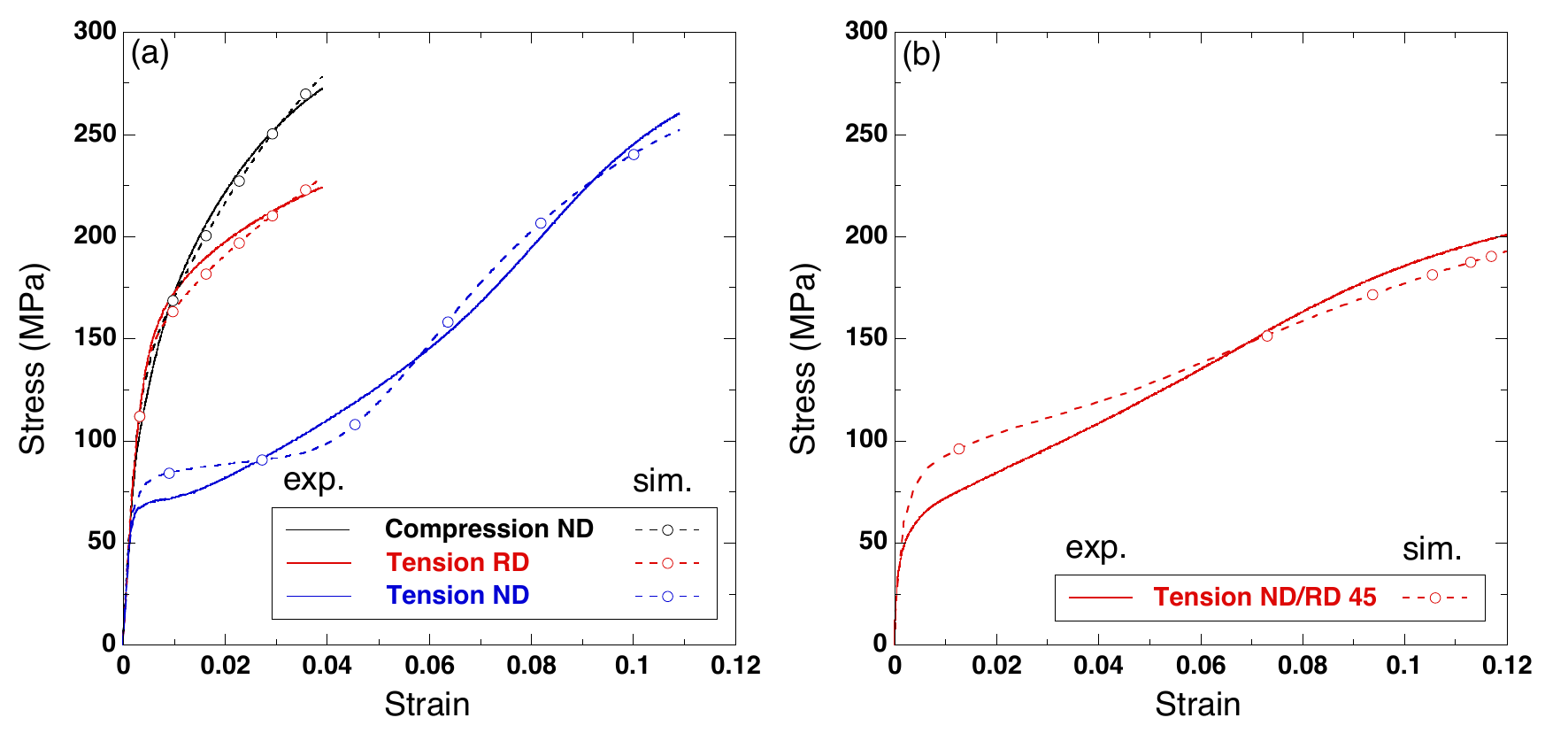} 
\caption{(a) Experimental (solid lines) and numerical (broken lines with symbol) stress-strain curves resulting from the optimization procedure. (b) Model prediction of the tensile test  in the RD-ND plane at 45$^\circ$ from both orientations. Solid lines correspond to experimental results while broken lines with symbols stand for the numerical simulations. Reprinted from \cite{HLD14}.}
\label{TestAZ31}
\end{figure}

The ability of the inverse optimization strategy to provide accurate values of the CRSS in complex HCP materials, like rolled AZ31Mg alloys processed by hot rolling, was demonstrated by \cite{HLD14}, Fig. \ref{AZ31}a. The average grain size of the rolled AZ31 Mg alloy was 25 $\mu$m and the pole figure of the as-rolled material is plotted in Fig. \ref{AZ31}b. It shows the strong basal texture  typical of rolled Mg alloys, with the $c$ axis parallel to the normal direction (ND). Also, the spread prismatic poles  appear along  the rolling (RD) and transverse (TD) directions. Specimens for tension and compression experiments along different orientations indicated in Fig. \ref{AZ31}a were machined from the plate and representative stress-strain curves of the mechanical tests can be found in Fig. \ref{TestAZ31}. These curves show the strong plastic anisotropy of Mg alloys, which is triggered by the limited number of slip systems and by the polar nature of extension twinning, which is only activated when deformation leads to an extension of the $c$ axis. As a result, deformation of wrought Mg alloys is  markedly dependent on the orientation, and different slip systems (and in different order) are activated as a function of the loading direction (either tension or compression). For instance, tensile tests along ND show the typical S-shape of the stress-strain curve characteristic of twinning-dominated deformation, while the compression test along ND was dominated by pyramidal slip.

The stress-strain curves corresponding to the tests along ND in tension and compression and in tension along RD were used to determine the parameters of the CP model using the Levenberg-Marquardt method. Full-field homogenization was carried out on a RVE of the microstructure discretized with cubic finite elements under periodic boundary conditions. The RVE contained 584 grains and was created with the microstructure generator \cite{dream3d}. On average, each grain was discretized with 7 voxels. The grains were equiaxed and the grain size followed a log-normal distribution with an average grain volume equal to the RVE divided by 584. The experimental stress-strain curves in the three different cases used to calibrate the constitutive equation (tension and compression along ND and tension along RD) were in very good agreement with the simulation predictions. as shown in Fig. \ref{TestAZ31}a. Moreover, this set of parameters was used to predict the stress-strain behaviour of the specimens tested in tension at 45$^\circ$ from the ND/RD orientations (Fig. \ref{TestAZ31}b). The agreement was again good and demonstrates the capabilities of  inverse optimization approach to perform virtual test of polycrystals.

Further investigations \citep{HSL15} were carried out to analyze the robustness of the Levenberg-Marquardt optimization strategy to obtain the CRSSs of the different slip modes and extension twinning in an AZ31 Mg alloy. It was found that the results were not influenced by the properties of the single crystals chosen as starting point for the optimization process in so far they were reasonable for the Mg alloy considered. Very similar values of the CRSSs were obtained regardless of whether the initial plastic behavior of the single crystals was isotropic, anisotropic or highly anisotropic. In addition, it was found that there is a minimum critical information that has to be supplied to the optimization strategy in order to achieve meaningful results. In the case of AZ31 Mg alloy, plastic deformation in the basal plane is controlled by basal slip while plastic deformation along the $c$ axis accommodated by either twinning or a combination of prismatic and pyramidal $\langle$c+a$\rangle$ slip. Thus, a minimum of two stress-strain curves (one in which deformation along the $c$ is controlled by twinning and another in which is controlled by  prismatic and pyramidal $\langle$c+a$\rangle$ slip) are necessary as input for the inverse optimization algorithm. Including a third input stress-strain curve in the optimization process improved the accuracy of the results from the quantitative viewpoint.

\subsection{Fatigue life}

Fatigue crack nucleation in metallic alloys is a process that involves several length scales, including the features of dislocation slip and dislocation/dislocation interactions at the nm scale and the collective motion of dislocations and the formation of particular dislocation patterns at the microscale. They give rise to the  localization of plastic deformation in slip bands within the grains that will be the origin of embryonic cracks \citep{S12}. This process is very dependent on the local details of the microstructure (grain size, twins, texture) as well as on the presence of defects (pores or intermetallic inclusions). Thus, the role of the microstructure is much more critical in fatigue that in the case of strength. In the latter,  the yield strength and the strain hardening are mainly controlled by the average value of the microstructural features (as given by the average grain size, pole figure, etc.) and little differences are found in polycrystals with the same average features until localization of damage occurs at the last stages of deformation. On the contrary, fatigue crack nucleation occurs by the localization of damage and mainly depends  on the extreme values of the statistical distribution of the microstructural features, rather than on the average values. This leads to a very large experimental scatter among "nominally" identical polycrystals (from the viewpoint of average features) and makes very difficult to establish a solid link between microstructure and properties.

Obviously, only simulation techniques that take into account the microstructural details are able to provide the link between microstructure and fatigue properties. This is the reason why microstructure-sensitive computational modelling of fatigue\citep{MCDOWELL20101521, PINEAU2016484} is indeed among the applications of CHP that has sparked more interest in the last decade because it provides a unique framework to understand the influence of the microstructure and of the defects on the fatigue behavior of engineering materials. Moreover, metrics can be develop to predict the fatigue life as a function of the loading conditions and of the microstructure of the polycrystal, providing an engineering tool to relate microstructure and fatigue performance. This opens the path to account for design allowables that are controlled by the inherent microstructural variability in the material and to design optimum microstructures for a given loading condition.

\begin{figure}[h]
\includegraphics[scale=0.55]{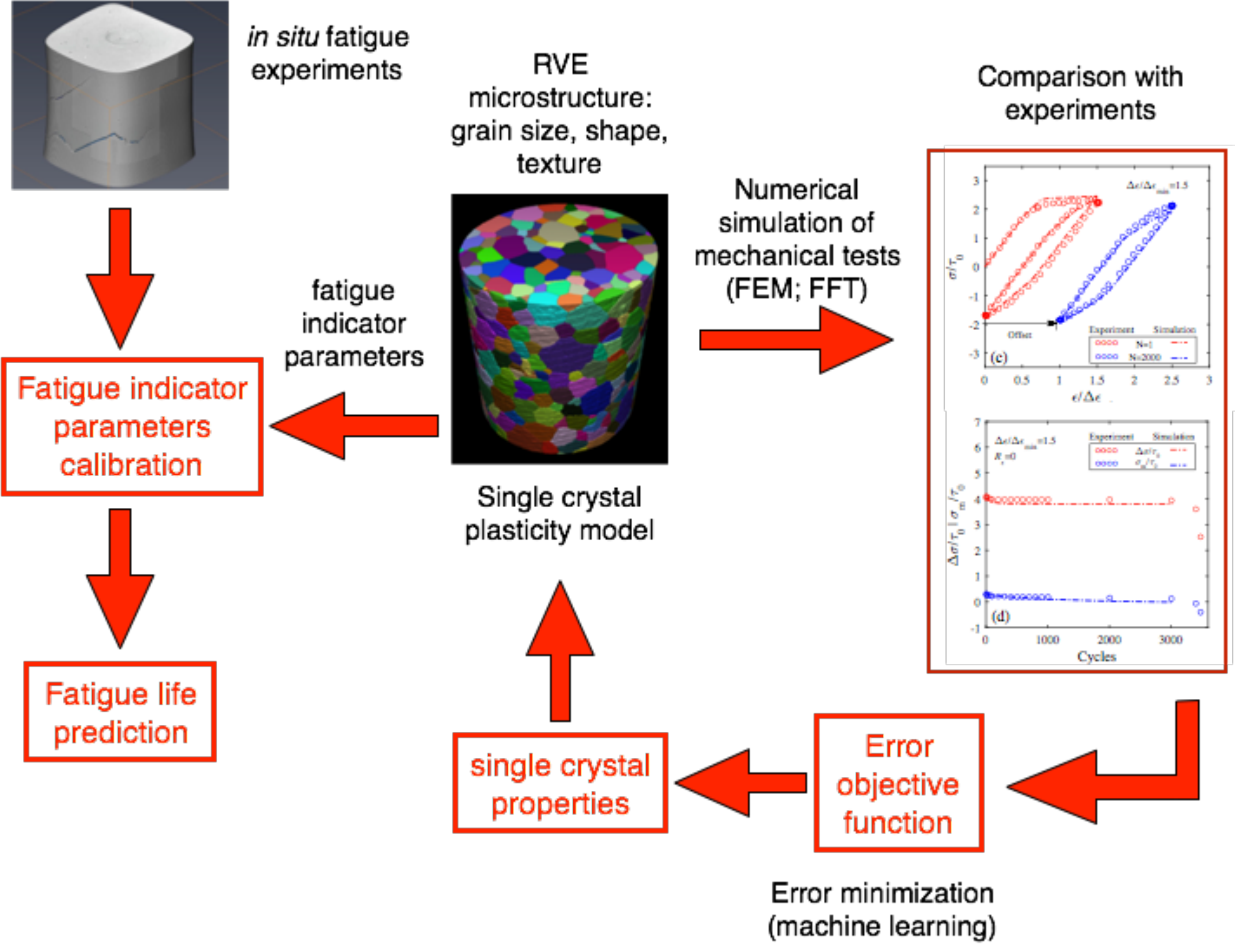}
\caption{Coupled experimental-simulation strategy to predict the fatigue life of polycrystalline metallic alloys using computational homogenization of polycrystals.}  
\label{fig:fatigue}
\end{figure}

The modelling framework to predict the fatigue behavior of polycrystalline metallic alloys by means of CHP is depicted in Fig. \ref{fig:fatigue}. The first step is to obtain the full micromechanical fields during cyclic deformation by means of the numerical simulation of the mechanical behavior of an RVE. The constitutive model for the single crystals should be able to account for the the main features found during cyclic deformation of metallic alloys, namely Bauschinger effect, mean stress relaxation, ratcheting and cyclic softening/hardening. This is normally achieved by means of phenomenological models (section \ref{subsection_phen}), whose parameters have to be calibrated for each alloy by comparison with the experimental cyclic stress-strain curves. This task can be done by means of trial-and-error approaches or optimization algorithms based in the construction of an  objective error function \citep{CRUZADO2017148}.

One critical factor to obtain the parameters of the phenomenological CP model as well as to predict the fatigue life is that the mechanical response of the polycrystal changes with the number of cycles until a stable stress-strain hysteresis loop is attained. Stabilization of the mechanical behavior may take from few of cycles up to a few hundreds or thousands of cycles, depending of the alloy and of the loading conditions  \citep{CRUZADO2017148}. In the latter scenario, the computational cost makes impossible the simulation of the mechanical response of the RVE  and different cyclic jump strategies have been proposed to reach such large number of cycles without  the explicit simulation of each of them \citep{Ghosh2011}. Among them,  \cite{Joseph2010} proposed  a novel wavelet transformation-based multi-time scaling (WATMUS) algorithm  which allows to reduce the number of simulated fatigue cycles by a factor of up to 100. Other simpler approaches can also be found on the literature. For instance,  \cite{CRUZADO2017148} proposed a methodology, following similar approaches in the context of fretting fatigue \citep{MCCOLL20041114},  based  on the linear extrapolation of the evolution of the internal variables at the constitutive equation level, followed by one or several stabilization cycles.

Once the stabilized cyclic stress-strain hysteresis loop has be determined by mean of computational homogenization of the behavior of the RVE,  the prediction of the fatigue life is based  in some Fatigue Indicator Parameter (FIP) obtained from the evolution of mechanical fields and internal variables at the local level within the RVE. Different FIPs have been
reported in the literature to describe the main driving force that controls crack formation and they can be linked, using phenomenological relations, to some stage of the fatigue life of the alloy under study (crack nucleation, small crack propagation or full fatigue life). 

The most common FIPs found in the literature are related to the accumulation of some variables during one fatigue cycle, such as the accumulated plastic strain per cycle,  $P(\mathbf{x})$ \citep{MCDOWELL200349, Manonukul2004, Sweeney2012, Sweeney2014, Sweeney2015}

\begin{equation}
\label{FIP_P}
P(\mathbf{x}) = \int_{cyc} \sqrt{\frac{2}{3}\mathbf{L_p}(\mathbf{x})\mathbf{L_p}(\mathbf{x})} \;\mathrm{d}t
\end{equation}

\noindent or the maximum plastic shear strain $\Delta\gamma(\mathbf{x})$

\begin{equation}
\label{FIP_gamma}
\Delta\gamma(\mathbf{x})= \max_{k} \biggl\{ \int_{cyc} |\dot{\gamma}^k(\mathbf{x}) | \;\mathrm{d}t \biggr\}\end{equation}

\noindent which are based exclusively on the local plastic strain fields. The Fatemi-Socie FIP, $FS$, is given by
 
 \begin{equation}
\label{FIP_FS}
FS(\mathbf{x})= \max_{k} \biggl\{\frac{\Delta\gamma^k(\mathbf{x})}{2}\left[1+k\frac{\sigma^k_{n}(\mathbf{x})}{\sigma_y}\right]\biggr\}
\end{equation}

\noindent where $\Delta\gamma^k$ and $\sigma^k_{n}$ stand for the maximum plastic shear strain and the maximum normal stress perpendicular to the slip plane of the $k$ slip system and $\sigma_y$ is the yield stress. This parameter was originally develop to rationalize the fatigue behavior of laboratory specimens \citep{FFE:FFE149,FS1989} but it was used as the driving force to nucleate a crack in both Low Cycle Fatigue (LCF) and High Cycle Fatigue (HCF) simulations based on CHP \citep{BENNETT200327, Shenoy2007, Musinski2012,Przybyla2013,Castelluccio2013, Castelluccio2014}.

Other popular FIP is the energy dissipated per cycle, $W(\mathbf{x})$, introduced  by  
\cite{FFE:FFE0612,KORSUNSKY20071990}
 \begin{equation}
\label{FIP_W}
W(\mathbf{x})= \max_{k} \biggl\{\int_{cyc} \tau^k(\mathbf{x})\dot{\gamma}^k(\mathbf{x}) \mathrm{d}t \biggr\}\end{equation}

\noindent and the energy density stored in each cycle, $G(\mathbf{x})$ \citep{Sweeney2012, Sweeney2014, Sweeney2015, Wan2016}, which is expressed as 

 \begin{equation}
\label{FIP_G}
G(\mathbf{x})= \int_{cyc} \xi \frac{ \boldsymbol{\sigma}(\mathbf{x}):\dot{\boldsymbol{\epsilon}_p}(\mathbf{x})} {\sqrt{\rho_{SSD}+\rho_{GND}}}\mathrm{d}t\\\end{equation}

\noindent where $\xi$ is a factor that stands for the fraction of the external work in a cycle that has been accumulated though the creation of dislocation structures. Its value is around 0.05 taking into account the typical fraction of the energy dissipated with respect to the external work \citep{Wan2016}. 

Many other FIPs can be found in the literature \citep{ANAHID20112157, SANGID2011595, OSG16} but they are not included here for the sake of brevity. There is no clear rule for selecting one of the different FIPs presented to predict the fatigue life and, indeed, different authors rely on one or another FIP for similar materials. The FIP that better correlates a fatigue process depends on the length scale considered, the characteristics of the alloy and also on the fatigue stage studied (incubation or small crack growth).  Persistent slip bands form driven by the accumulation of plastic strain at the single crystal level and under uniaxial loading and both $P$ and  $\Delta\gamma$ are able to reproduce this observation \citep{Manonukul2004}. The role of stress in the development of persistent slip bands and small cracks is much more important at the polycrystal level and under multiaxial loading and this is better accounted for using  other FIPs including both plastic slip and stress. A critical assessment of the ability of the different FIPs to correlate crack initiation implies comparison with {\it in situ} microscopic experiments \citep{CHEN2018213} or with lower scale models \citep{SANGID201358}.

The FIPs defined in eqs. \eqref{FIP_P}-\eqref{FIP_G} are local values computed at each point $\mathbf{x}$ of the RVE. This local value depends, however, on the details of the discretization and it is not representative of a fatigue damaged region  \citep{Sweeney2013}. From a physical viewpoint, the FIP should be averaged over a region representative of the crack incubation zone, while  the volume averaging is indeed necessary to avoid spurious stress concentrations and minimize mesh size effects from the computational perspective, as noted in \cite{Castelluccio2015}. Three different averaging volumes are normally used, namely finite element,  grain  \citep{Shenoy2007} or slip band \citep{Castelluccio2015}. In the latter, the FIP is averaged on a slip band of constant thickness parallel to each slip plane $k$ at $\mathbf{x}$. After averaging, the maximum value of the averaged FIP in the whole RVE (either the maximum averaged FIP in all the finite elements or in all the grains or in all the slip bands) is used to determine the fatigue life as indicated below  \citep{Shenoy2007, Castelluccio2013, Cruzado2018a}. 

The last ingredient to predict the fatigue life in the microstructure-sensitive computational fatigue models is to establish the link between the FIPs and the different stages of the fatigue life. In this respect,  micromechanics-based fatigue models can be roughly classified in two groups. In the first one,  different relationships are established between the FIPs and the fatigue crack nucleation and propagation stages. In the second one, it is assumed that most of the fatigue life is spent in the nucleation of a crack and a direct relationship is established between the FIPs and the fatigue life. 

The first approach  was introduced by \cite{MCDOWELL200349} and considers at least three stages in the fatigue life: crack incubation, microstructurally small crack growth (MSC) and long crack growth (LC). The fatigue life $N$ is given by 

\begin{equation}
N=N_{N}+N_{MSC}+N_{LC}.
\label{eq:fatiguestages}
\end{equation} 

\noindent where $N_{N}$ is the number of fatigue cycles to nucleate a crack at a favourable site in the microstructure whose length is of the order of some microstructural length scale such as the grain size. $N_{MSC}$ stands for the number of cycles spent from the nucleation of the crack until it reaches a length of typically 3-10 times the grain size. Crack nucleation and propagation in these two stages are strongly dependent on the microstructure (grain size, texture, size, shape and spatial disitribution of second phases, etc) \citep{S12}. Finally, $N_{LC}$ is the number of the cycles associated with the propagation of a crack which is much longer than the longest critical microstructural length scale (typically the grain size in polycrystals). This stage is not very dependent on the microstructure and can be accounted for with the classical crack propagation theories based on fracture mechanics. Thus, CHP is used in this approach to determine $N_{N}$ and $N_{MSC}$. Different FIPs and different relationships between each FIP and either $N_{N}$ or $N_{MSC}$ can be established to take into account that the driving forces for both process do not have to be equivalent. 

The physical process of crack incubation is not yet fully understood. Thus,  the determination of the principal driving forces and the development of  robust initiation criteria based on these forces is based in more or less phenomenological approximations. Plastic deformation is accumulated progressively during cyclic deformation because of irreversible slip leading to the formation of low-energy dislocation arrangements to accommodate the irreversible slip. This results in the strain localization in a small region and in the formation of the so-called persistent slip bands, that are the precursors to crack initiation. The first relation to link a FIP to the cycles for crack nucleation based on dislocation mechanics was proposed by \cite{TanakaMura1981}, and it has the form

\begin{equation}
N_{N} (FIP)^2=\frac{K_{TM}}{d_g}
\label{nucleation}
\end{equation}

\noindent where the FIP in the original work  \citep{TanakaMura1981} was the accumulated plastic strain per cycle, eq. \eqref{FIP_P}, $d_g$ stands for the grain size and $K_{TM}$ is a material constant. Nevertheless, many other expressions and models can be found in the literature \citep{SANGID201358} and there is no a clear consensus on the form and of variables that should be included.  In most of the models in which the FIPs are explicitly related to $N_N$ 
\citep{Shenoy2006,Shenoy2007,Przybyla2010,PRZYBYLA2012293,Przybyla2013,Castelluccio2013,Castelluccio2014,Castelluccio2015}, this process is linked with the FIPs related to the plastic strain ($P$ and $\Delta\gamma$) following expressions based on the Tanaka and Mura model.

Microstructurally small crack growth is treated  within the framework of CHP following traditional expressions of macroscopic fatigue crack growth in which the cyclic crack tip displacement range ($\Delta CTOD$) is used as driving force instead of the stress intensity factor range \citep{MCDOWELL200349}. However, $\Delta CTOD$ is difficult to obtain in CHP  because it would require to define a crack and a fine mesh around it \citep{TCI15} but FIPs computed in standard computational homogenization simulations (without the crack) can be used as an effective surrogate measure of $\Delta CTOD$. Therefore, the microstructural crack growth rate can be obtained as \citep{Shenoy2007}

\begin{equation}
\frac{\mathrm{d}a}{\mathrm{d}N}=K_{CTOD} \tau_y \ FS\ a
\label{MSC}
\end{equation}

\noindent where $a$ is the crack length, $\tau_y$  an averaged value of the critical resolved shear stresses in all the slip systems, $g^k$, $FS$  the Fatemi-Socie FIP, eq. \eqref{FIP_FS}, and $K_{CTOD}$ a constant. Similar expressions but including a threshold have been used in \cite{Przybyla2013}. Eq. \eqref{MSC} can be integrated from an initial crack length of one grain up to a length in which the crack becomes long with respect to the microstructure. This modelling approach has been widely used for modelling HCF in Ni based superalloys,\citep{Shenoy2007,Musinski2012, Przybyla2013,Castelluccio2013,Castelluccio2014}, Al alloys \citep{MCDOWELL200349} and Ti alloys \citep{PRZYBYLA2012293}.

The constant that appears in the models for crack nucleation or short crack propagation presented above has always to be obtained by comparison with some experimental data. This calibration of the model is, however, problematic because it is very difficult to determine experimentally the actual number of fatigue cycles to nucleate a crack or to propagate the crack in the small crack regime. While this may be feasible by means on {\it in situ} fatigue tests coupled with diffraction and phase-contrast X-ray microtomography \citep{BBG09, HKR11}, they are extremely expensive and time-consuming and the available experimental data are limited to macroscopic fatigue tests in which the crack nucleation event cannot be detected. 

A simple approach to overcome this problem was introduced by \cite{Manonukul2004} assuming that the fatigue life is mainly dominated by crack nucleation.  Thus, the number of cycles necessary for crack nucleation is identified with the total fatigue life and the relationship is established between the FIP and the fatigue life. \cite{Manonukul2004} also showed that this relationship was characteristic of the  polycrystalline C263 Ni-based superalloy and independent of the loading conditions, etc. This direct relationship between the FIP and the fatigue life has been profusely used since then using different FIPs and also considering different expressions to relate the FIP with the  fatigue life. For instance,  FIPs based on accumulated plastic strain have been used to assess the effect of elastic anisotropy and other factors in the fatigue life of a ferritic steel \citep{Sweeney2013}. The cyclic strain energy dissipation FIP, $W$ (eq. \eqref{FIP_W}) was used to study the behaviour under LCF in Inconel 718 superalloy \citep{Cruzado2018a,Cruzado2018b}, showing that a microstructure-based fatigue model was able to predict the bilnear Coffin-Manson regime observed experimentally in this alloy. The same FIP was used to predict the fatigue life of   stainless steel medical devices \citep{Sweeney2012}  and of a CoCr alloy \citep{Sweeney2014}. The stored energy FIP $G$ (eq. \eqref{FIP_G}) was used within a physically-based SGCP framework  to predict the fatigue life in a ferritic steel in \citep{Wan2016} and the same modelling approach was also employed in \cite{SWEENEY2014341} to study fatigue in a CoCr alloy. The concept of stored energy as driving force for crack initiation has also been proposed by \cite{SANGID2011595} for a Ni-based superalloy where the energy stored was computed using a model that includes continuum and atomistic effects.

As a final remark, it is important to note the statical nature of fatigue predictions based in CHP. The number of grains in the RVE is always much smaller than that  in the actual microstructure and therefore the values of the FIPs obtained in the simulation of different RVEs corresponding to the same microstructure can be quite different. Thus, in general, the results obtained with a single RVE are not fully representative of the fatigue life for a given microstructure and do not provide reliable information about the scatter resulting from the particular orientation arrangement. This limitation is often overcome with the concept of SRVEs \citep{JKF04} presented in section 2. Fatigue estimations are usually obtained aggregating the results of many different RVEs which are statistically representative of the same microstructure \citep{Shenoy2006, Castelluccio2013}.

\subsection*{Prediction of the fatigue life in IN718 Ni-based superalloy}
\begin{figure}[h]
\includegraphics[scale=0.90]{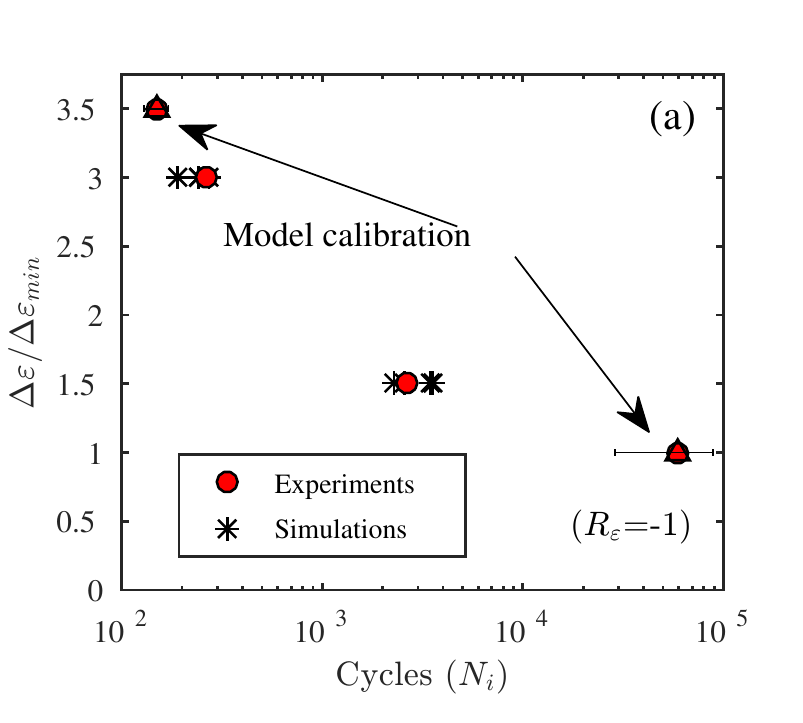}
\includegraphics[scale=0.90]{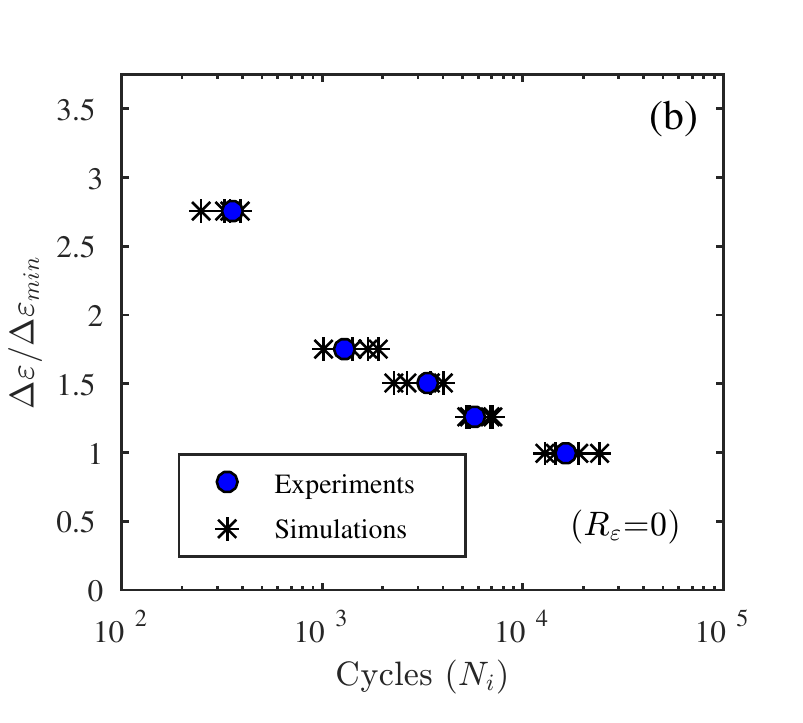}
\caption{Experimental  results and model predictions for the fatigue life of IN718 Ni-based superalloy at 400$^{\circ}$C.
 (a) $R_{\varepsilon}$ = -1 and  (b) $R_{\varepsilon}$ = 0.  The cyclic strain amplitude, $\Delta \epsilon$, is normalized by $\Delta \epsilon_{min}$, the minimum cyclic strain amplitude used in the tests. Reprinted from \cite{Cruzado2018a}.}
\label{Fatigue Life predictions}
\end{figure}

To illustrate the fatigue life prediction process using CHP, an example  of prediction of the fatigue life in IN718 Ni-based superalloy in the LCF regime at 400$^\circ$C is presented next \citep{CRUZADO2017148,Cruzado2018a,Cruzado2018b}. The CP model used to represent the cyclic plastic behavior \citep{CRUZADO2017148} included the mechanisms leading to Bauschinger effect, mean-stress relaxation and cyclic softening and a summary of the constitutive equations was already presented in eqs. \eqref{eq:power_law_cruzado}-\eqref{eq_(15):cyclic_plast}. The fatigue life estimation was based in the plastic energy dissipated by cycle FIP, eq. \eqref{FIP_W}, averaged in slip bands \citep{Castelluccio2015}. The resulting FIP of a particular RVE, was given by
 
\begin{eqnarray}
W_{cyc}=\max_{i=1,nb} \left \{\max_{k_i} \frac{1}{V_i}\int_{V_i} W_{cyc}^{k_i}(\mathbf{x})\mathrm{d}V_i \right \}
\label{eq_(5):FIPfinal}
\end{eqnarray}

\noindent  where $k_i$ (= 1, 2, 3) corresponds to the three different slips systems contained in the slip plane parallel to the band $i$, $V_i$ is the volume of that band and $nb$ is the total number of bands in the microstructure, which is 4 times the number of elements in the RVE. The fatigue  life estimation for this particular RVE (the number of cycles $N_i$) is obtained as function of $W_{cyc}$ using a power-law relation, introduced by \cite{Cruzado2018a}, to account for the change in deformation mechanism that controls the nucleation of fatigue cracks in this material: from localized deformation in a few grains at small cyclic strain ranges to homogeneous plastic deformation at large cyclic strain ranges. Thus,

\begin{equation}
N_i= \frac{W_{crit}} {(W_{cyc})^m}
\label{eq:energy}
\end{equation}

\noindent where the parameters $W_{crit}$ and $m$ were obtained from two independent fatigue experiments corresponding to two different cyclic strain ranges under the same strain ratio $R_\varepsilon$=-1 (Fig. \ref{Fatigue Life predictions}a). To this end, the FIPs ($W_{cyc}$) obtained by averaging the results of 20 different RVEs for each of the two loading conditions selected were used. 

Finally, the model was used to estimate the fatigue life under different loading conditions. Predictions for each loading condition were obtained by averaging the FIPs computed using different RVEs. The ability of the proposed model to predict the fatigue life of IN718 alloy at 400$^\circ$C is depicted in Figs. \ref{Fatigue Life predictions}(a) and (b), in which the experimental and the model results  for the fatigue life are plotted as a function of the cyclic strain range in tests carried out with $R_\varepsilon$ = -1 and 0, respectively. The model was able to predict accurately the fatigue life under small and large cyclic strain ranges for all the loading cases considered. Moreover, the numerical predictions with different RVEs led to estimations of the scatter in fatigue life which followed the same trends that the exerpimental observations:  the scatter decreased as the applied cyclic strain range increased.

\subsection{Void growth in polycrystals}

Ductile failure of polycrystalline metals takes place by the nucleation, growth and coalescence of voids \citep{T90}. Void nucleation is normally triggered by either  fracture or interface decohesion of second phase particles while void growth up to coalescence -- that controls the final ductility of the material -- is driven by the plastic flow and mainly depends on the stress state and on microstructural details. While the effect of stress state on void growth and coalescence has been analyzed in detail within the framework of isotropic plasticity models \citep{BL10}, the influence of the microstructure (grain size, orientation) is far less understood. Nevertheless, it has been experimentally established that grain anisotropy and grain boundaries play an important role on void growth, particularly for low symmetry crystals \citep{LMK14, NW16, PAM16}. 

CHP has been applied to analyze the effect of the microstructural factors on void growth using the dilatational viscoplastic FFT-based model described at the end of section 5.2. This approach was used  to simulate  void growth in damaged polycrystalline materials under high stress triaxiality  loading that resembles that of typical Cu spall experiments \citep{ECD11, ECD13}. The simulation results in two cases in which the matrix behavior was different (FCC polycrystal {\it vs.} isotropic matrix) were compared with post-mortem orientation images from one of such experiments, showing grain orientations and voids in the region of incipient damage. This provides information on the effect of the matrix's polycrystallinity on porosity evolution, and identify microstructural effects on void growth, such as the influence of the Taylor factor of the crystalline ligaments that link interacting voids.

The mechanical behavior of the single crystals in the polycrystal assumed an FCC material with 12 \{111\} $<$110$>$ slip systems, a strain rate sensitivity exponent  $n$ =5 and a reference strain rate $\dot\gamma_0$ = 1, eq. \eqref{eq:power_law}.  The initial critical resolved shear stress  was 100 MPa  with a simple  linear hardening  $h_{kj}$ = 30 MPa, eq. \eqref{eq:hard2}. These values are consistent with the ones measured in shock-compressed Cu with 5\% pre-strain \citep{SMG05} . Meanwhile, for consistency, the constitutive parameters that describe the behavior of an analogous isotropic matrix material, eq. \eqref{iso},  were $n$ = 5, $\dot\gamma_0$ = 1, $\sigma_0$ = 250 MPa and $H$ = 75 MPa.

The polycrystalline RVE included 200 grains with an initial porosity of 1\% distributed in 50 spherical voids randomly seeded at grain boundaries. It represents the spall region of a polycrystalline high-purity Cu target following compressive  shock hardening and subsequent random nucleation of intergranular voids. The  isotropic matrix RVE was built keeping the same void distribution and replacing the grains by a homogenous isotropic matrix with the analogous constitutive behavior given above. 
 
The initial porosity represented spallation voids that nucleate when interacting rarefaction waves produce the required tensile state for damage initiation. From that point on, it was assumed that the material deforms under an axisymmetric stress state, with far-field stresses $ {\Sigma _{zz}} > {\Sigma _{xx}} = {\Sigma _{yy}}$ (where $z$ is the shock direction, normal to the spall plane), and a constant stress-triaxiality  $ X_\sigma  = \Sigma _m / \Sigma _{eq}$, where $ {\Sigma _m} = \left(\Sigma _{zz} + 2\Sigma _{xx} \right)/3$, and  ${\Sigma _{eq}} = {\Sigma _{zz}} - {\Sigma _{xx}}$.

\begin{figure}[!]
\centering
\includegraphics[scale=0.7]{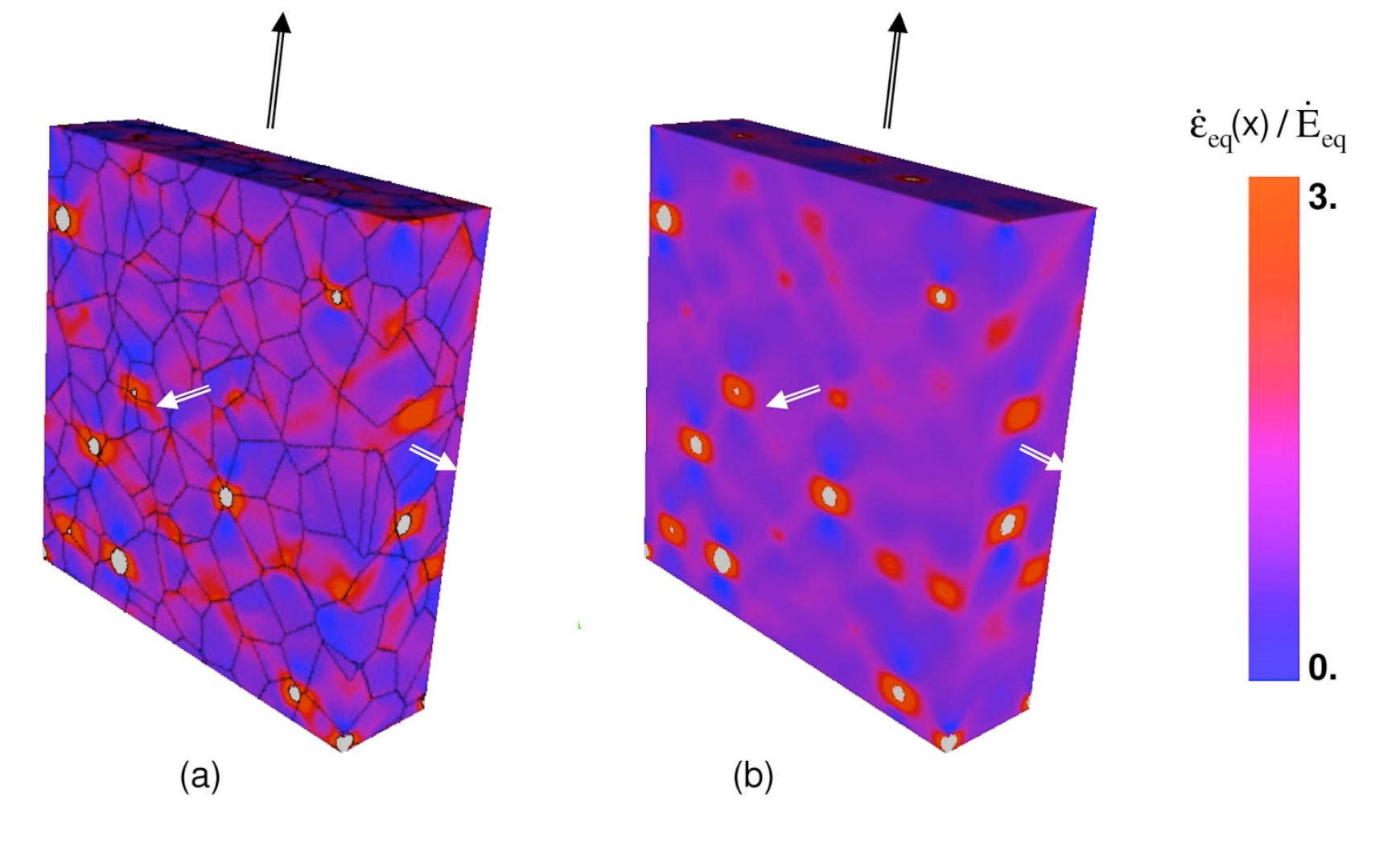}
\includegraphics[scale=0.7]{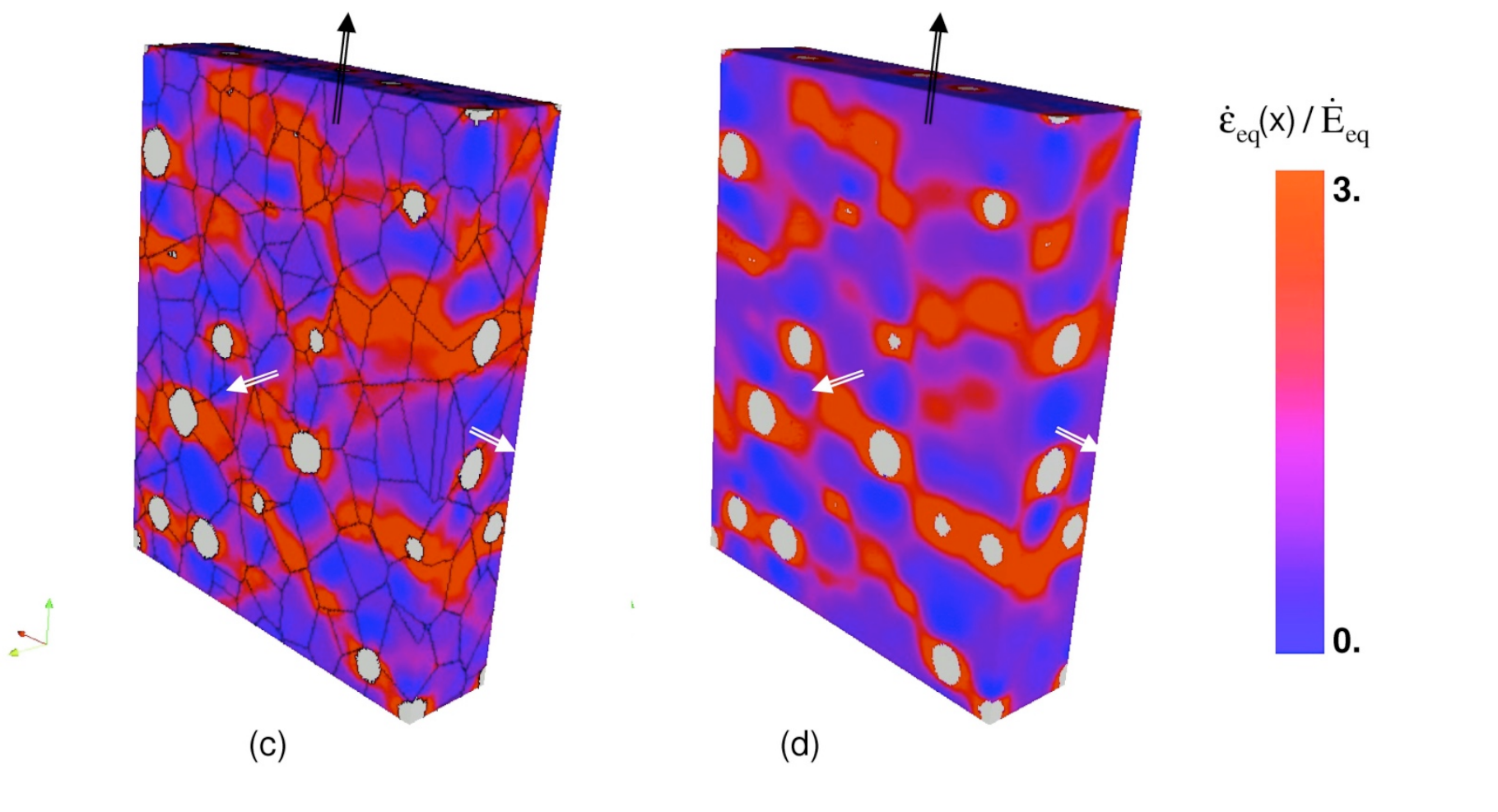}
\caption{Contour plots of the equivalent strain rate fields during spalling of a porous metal when the matrix is modelled as an FCC polycrystal with intergranular cavities (left), or a homogenous isotropic matrix (right), for $X_{\Sigma}$ = 2.5. (a-b) Initial, (c-d) final configurations. The black arrow indicates the direction of the largest principal stress, normal to the spall plane. Simulations were carried out using full-field homogenization with FFT. Reprinted from \cite{LEC13}.}  
\label{fig:porosity}
\end{figure}

The local fields and the porosity evolution for both the polycrystal and the isotropic matrix cases are plotted in Figure  \ref{fig:porosity} for $X_{\Sigma}$ = 2.5 . Figs. \ref{fig:porosity}a and c stand  for the relative equivalent strain rate fields calculated at the initial and final deformation steps, respectively, in the case of the polycrystalline FCC matrix. The corresponding results for the isotropic matrix are plotted in Figs. \ref{fig:porosity}b and d. The black arrows indicate the direction of the largest principal stress, normal to the spall plane. The volumetric deformation is fully accommodated by void growth, resulting in a strong localization of (incompressible) plastic deformation in the material surrounding the cavities. This is revealed by the high values (relative to the macroscopic equivalent strain rate) of the plastic strain rate fields, especially between cavities that are close to each other and whose ligaments are relatively well aligned with the direction of largest principal stress.

\begin{figure}[!]
\centering
\includegraphics[scale=0.85]{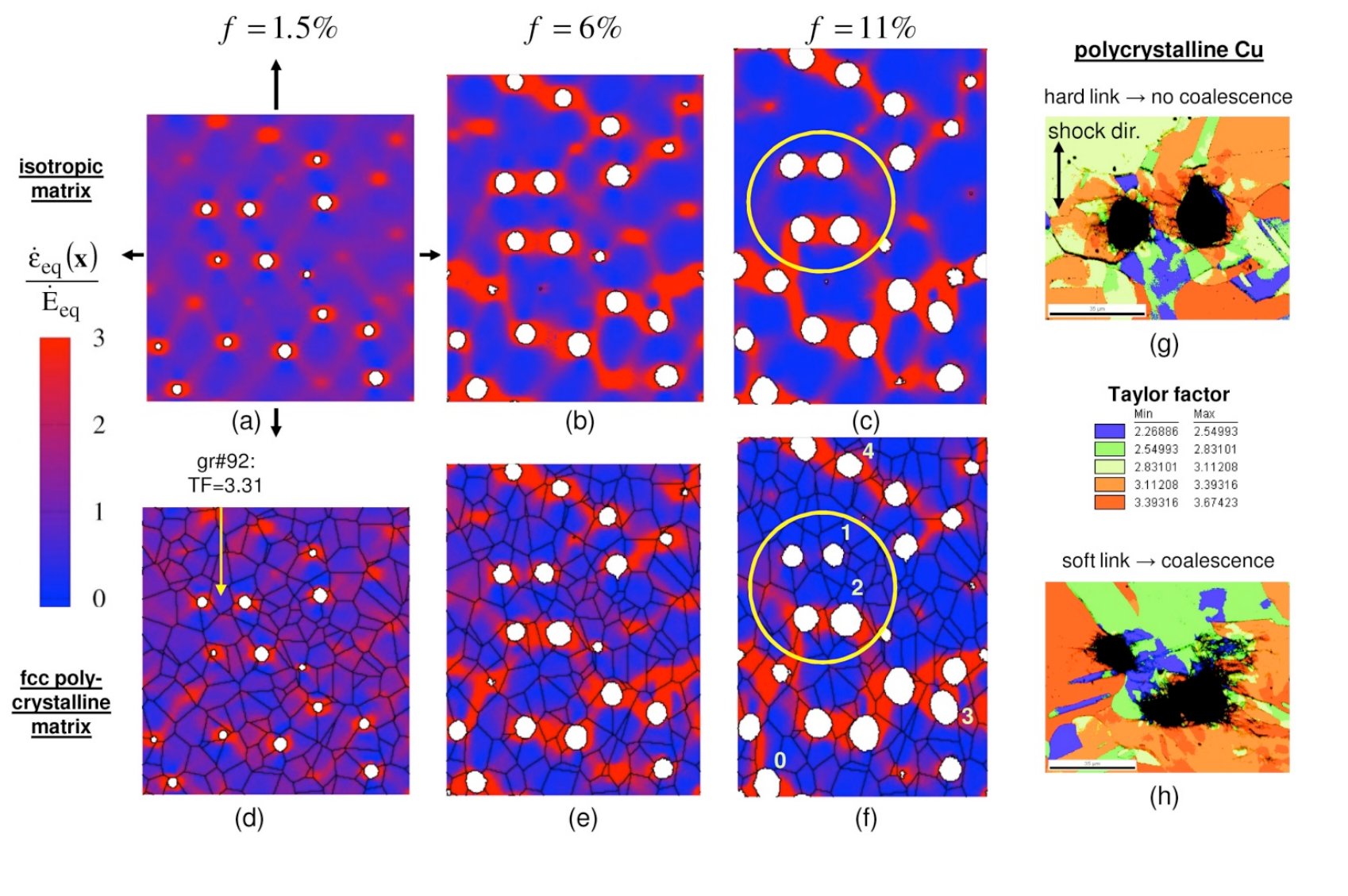}
\caption{2-D sections corresponding to the 3-D simulations in Fig. \ref{fig:porosity}, at different levels of porosity, for porous materials with (a-c) isotropic matrix, and (d-f) FCC polycrystalline matrix. (g-h) Typical post-mortem EBSD-generated Taylor factor maps of dynamically-damaged polycrystalline Cu, recovered from the spall region. Reprinted from \cite{LEC13}.}  
\label{fig:porosity2D}
\end{figure}

In order to better appreciate the differences between both cases,  2-D sections from the 3-D simulations of Fig. \ref{fig:porosity} are plotted in Figs. \ref{fig:porosity2D}a to c for the porous material with isotropic matrix and in Figs. \ref{fig:porosity2D}d to f the case of the FCC polycrystalline matrix at different levels of porosity. The experimental results of the typical post-mortem EBSD-generated Taylor factor maps of a damaged polycrystalline Cu, from the spall region of the target  are also shown in Figs. \ref{fig:porosity2D}g and h \citep{ECD11, ECD13}.  Void growth  in the case of the isotropic matrix is only affected by the distance and relative position of the interacting voids. For example, the two pairs of voids within the yellow circle in Fig. \ref{fig:porosity2D}c interact strongly and grow in a similar way. On the contrary,  the voids in the lower pair (marked Ò2Ó) do interact and grow profusely while the upper pair (marked Ò1Ó) stop growing, likely due to the presence of a hard crystalline ligament, in the polycrystalline matrix case (Fig. \ref{fig:porosity2D}f). Indeed, this ligament is mostly formed by grain 92 in this particular 2-D section, whose initial Taylor factor  with respect to tension along the $z$ direction  is 3.31, i.e. a hard crystallographic orientation for deforming in tension along that axis. Interestingly, the EBSD Taylor factor maps show similar behaviour. Fig. \ref{fig:porosity2D}g shows two voids that are ideally located to interact, but they did not coalesce, likely because of the hard ligament linking them (note that the orange and red colours represent Taylor factors $>$ 3.1). On the other hand, Fig. \ref{fig:porosity2D}h shows three voids (the two on the right have already merged) in similar spatial configuration as those of Fig. \ref{fig:porosity2D}g, but linked by soft orientations (Taylor factor $< $ 2.5), which have coalesced or are in an advanced state of pre-coalescence. It should be emphasized  the qualitative character of the above comparison, which is based on two strong assumptions, namely: 1) that the effect of the 3-D neighbourhood can be neglected, and 2) that the stress state remained axisymmetric throughout the entire deformation process, to justify the use of Taylor factors along the shock (axial) direction.

\subsection{Multiscale modelling of rolling}

Multiscale models with homogenization at the mesoscale (described in section 6) are useful to simulate the mechanical behavior of polycrystals subjected to non-homogeneous stress states. An example of application of this strategy is the analysis of the spatial variation of the texture evolution during rolling of an FCC polycrystal \citep{SLL12}. These variations, in turn, change the local behavior and, as a consequence, determine the redistribution of stresses during deformation. This case constitutes a challenge for the proposed approach, since it includes heterogeneous deformation, large strains and rotations, contact and complex local strain histories.

\begin{figure}[!]
\centering
\includegraphics[scale=0.85]{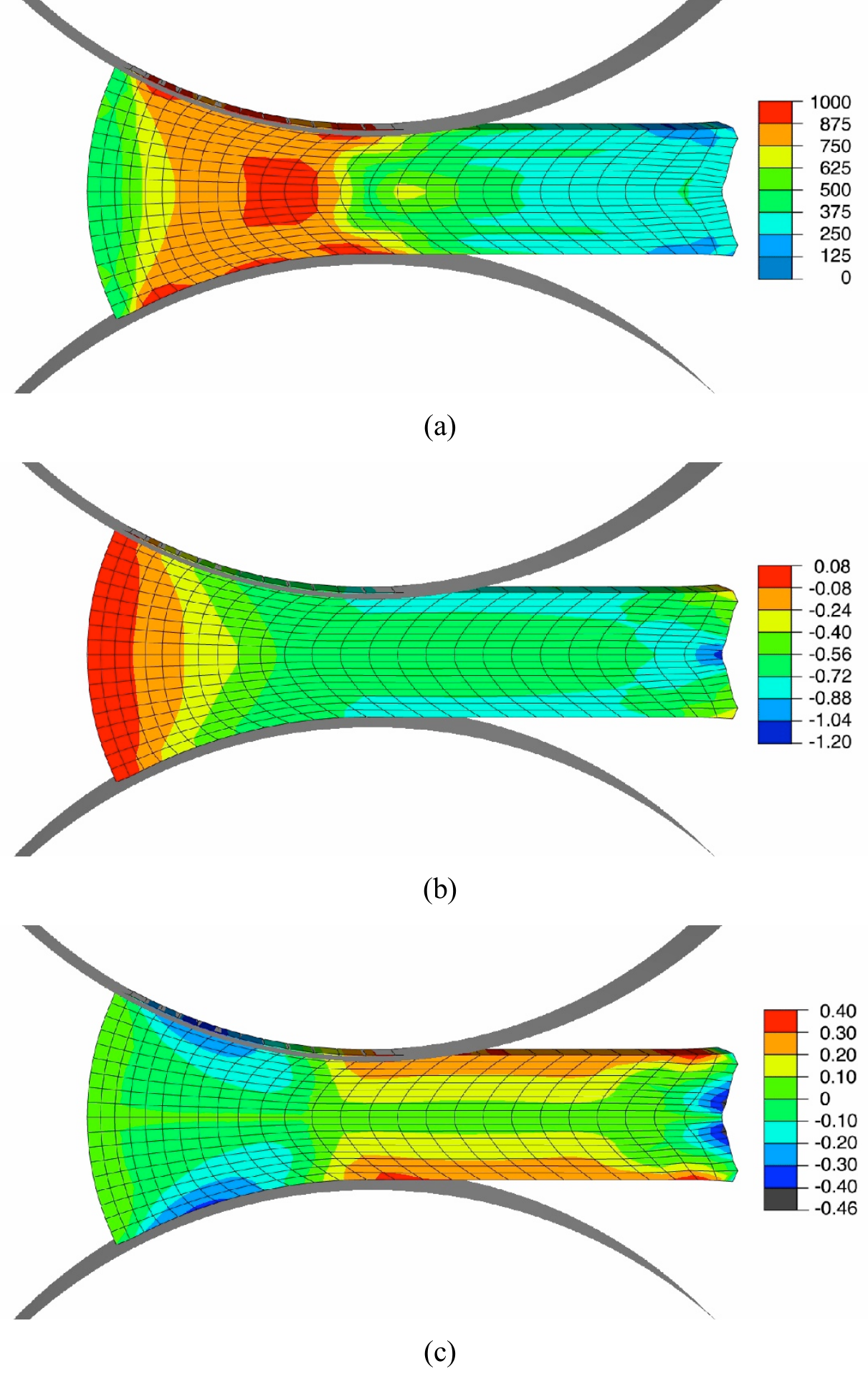}
\caption{Multiscale finite element simulation of rolling of an FCC plate, up to 50\% reduction. Contour plots  of: a) Von Mises stress (in MPa), b) diagonal strain component in ND, $\epsilon_{yy}$. c) shear strain component along RD normal to ND, $\epsilon_{xy}$. Reprinted from \cite{SLL12}.}  
\label{fig:Rolling}
\end{figure}

The simulation of rolling at the macroscopic level is carried out using the FEM. A parallelepipedic slab of material is rolled between two cylinders leading to 50\% reduction in thickness (Fig. \ref{fig:Rolling}). The initial aspect ratio of the slab was 4x1 in the rolling-direction/normal-direction plane (RD-ND plane). Plane-strain was imposed in the transverse direction, TD, and symmetry was used to reduce the size of the FEM model. The RD and TD coincided with the $x$ and $z$ axes of the macrocopic FEM. The slab was  discretrized with a structured mesh of linear cubes with full integration (C3D8 in Abaqus) using 26 x 9 elements in the RD-ND plane and 1 element in TD. Load was applied in two steps. The initial step consisted in a small vertical displacement of the cylinders in order to establish contact conditions between rolls and plate at the entry and the cylinders rotate at 2 rad/s in the second step until the plate has been completely rolled. The movement of the cylinders was transferred to the plate by friction, and the imposed angular velocity corresponds to an average compression strain-rate component of  along the normal direction (ND, $y$ axis).

The mechanical behavior of the material at each Gauss point of the FEM was determined using the VPSC approach. The FCC polycrystalline material was initially represented by 500 randomly-oriented grains which deform plastically in 12 \{111\} $<$110$>$ slip systems. The elastic stiffness tensor of the Al FCC crystal was given by $L_{1111}$ = 108 GPa, $L_{1212}$ = 62 GPa and $L_{4444}$ = 28 GPa. The VPSC constitutive parameters, adjusted to reproduce the experimental behavior of a 6116 Al alloy deformed in uniaxial compression \citep{TLN02} followed a Voce law, eq. \eqref{eq:IN718f}, characterized by $\tau_o$ 116 MPa, $\tau_s$ = 119 MPa, $h_o$ = 793 MPa and $h_s$ = 31 MPa, isotropic hardening ($q_{kj}$ =1, $\forall k,j$), and a strain rate sensitivity exponent $n$ =10.

The geometry of the rolled plate in the final stage is shown in Fig. \ref{fig:Rolling}, together with the corresponding contour plots of the Von Mises stress, the diagonal strain component in ND, $\epsilon_{yy}$, and the shear strain, $\epsilon_{xy}$. Interestingly, the stress does not vanish in the region already rolled and a heterogeneous distribution of residual stresses is left in the material. It can be concluded  from both  Fig. \ref{fig:Rolling}c and the final shape of the elements that elements near the surface underwent significant shear deformation along RD normal to ND, while elements towards the center were subjected to smaller shear distortion. This shear deformation is a consequence of the friction between the metal surface and the cylinders.

\begin{figure}[!]
\centering
\includegraphics[scale=0.85]{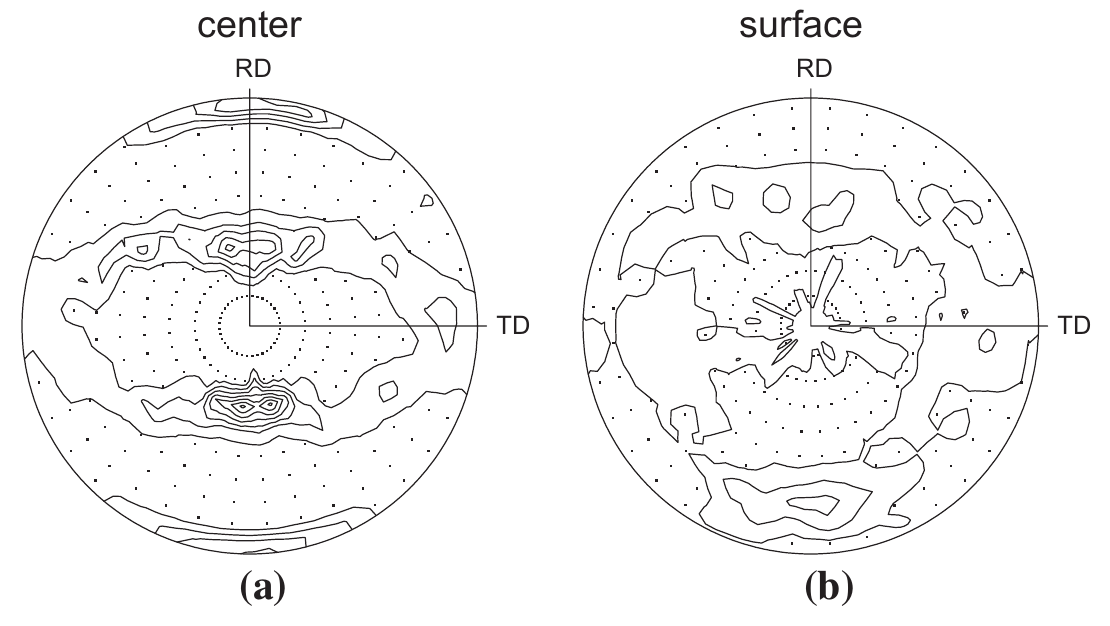}
\caption{$\{111\}$ pole figures predicted by multiscale modeling strategy for two polycrystalline material points located: a) near the center, and b) near the surface of the slab. Lines represent multiple of random distributions (mrd) and dots represent regions with less than 1 mrd. Reprinted from \cite{SLL12}.}  
\label{fig:Texture}
\end{figure}

The different deformation histories of the polycrystalline material points through the section determine different evolution of the texture. This can be appreciated in Figs. \ref{fig:Texture}a and b, which show the $<$111$>$ pole figures of two material points located near the center and near the surface of the plate, respectively. Both points were taken far from the leading tip in the rolling direction in order to minimize edge effects and obtain results that resemble those of a continuum rolling process. The textures near the center resemble typical plane-strain compression textures of high stacking fault energy FCC polycrystals, while the texture associated with a point near the surface shows significant rotation with respect to TD of the rolling texture components, consistent with a plane-strain + shear strain history applied to the material point. 

\section{Concluding Remarks and Future Developments}  

CHP by means of mean-field approximations or full-field simulations of a RVE of the microstructure has become a mature tool in the last twenty years, enabling the simulation of the mechanical behavior of polycrystalline aggregates taking into account the effect of microstructure. This "micromechanical" approach is significantly different from the classical phenomenological plasticity models for polycrystals (for example, J$_2$ plasticity and associated damage and failure models). These models are known to provide an accurate response for the homogenized behavior at the macroscopic level  and the parameters are calibrated by means of mechanical tests of polycrystalline specimens. Nevertheless, this classical methodology presents several limitations:

$\bullet$	Each set of parameters is only valid for a given set of testing conditions (temperature, strain rate, loading mode) and microstructural features (grain size distribution and morphology, texture). Any change in these factors requires the re-calibration of the model.

$\bullet$	These phenomenological plasticity models are accurate to predict the homogenized properties (average behavior of many crystals), but fail to determine the macroscopic properties when they are controlled by microstructural features (fatigue and creep crack initiation, crack propagation, effect of defects, etc.).

$\bullet$ Phenomenological plasticity models assume that the material is homogeneous and do not include information about the microstructural details, such as grain size and orientation, intragranular distribution of metallurgical phases. Hence, they cannot be coupled with models dedicated to predict the formation and/or evolution of microstructure during thermo-mechanical processing.

Obviously, CHP has the potential to overcome these limitations by establishing a clear link between microstructure and mechanical properties that is unfeasible using classical plasticity models. Consequently, CHP is nowadays the standard research tool to analyze the mechanical behavior of polycrystalline aggregates and is starting to be adopted by industry to solve engineering problems \citep{REB10}. This progress in industrial applications has been supported by the availability of different open-source software packages that can be used to represent the microstructure of polycrystals \citep{dream3d, Neper} and to simulate the mechanical behavior using mean-field \citep{LT93, LTP07} and/or full-field methods based on the FEM or the FFT \citep{Damask, Capsul}. The application of "virtual testing" and " virtual design" methodologies based on CHP is expected to have a significant industrial impact, as it has already happened in the field of composite materials for aerospace \citep{LGM11, GVM17}. In particular, it will greatly reduce the number of experimental tests necessary to characterize the mechanical behavior of new materials and will provide guidelines to modify the microstructure of the latter in order to improve their mechanical performance under different loading conditions. In addition, the techniques used in the engineering design will not be based in huge and expensive experimental campaigns but on micromechanical models that can account for the variability in the microstructure.

Nevertheless, as CHP is used to tackle more and more complex situations, several important limitations of the current strategies to simulate the mechanical performance of polycrystals appear, which should be addressed in the future. The first important problem to be analyzed is the slip transfer across grain boundaries during deformation. There is ample experimental evidence showing that grain boundaries may act as strong barriers for dislocations, which are stopped at the boundary leading to the formation of pile-ups, or are easily transferred to the neighbour grain leading to the propagation of deformation on a different slip system \citep{BEZ14, BMR16, MMD17, HNV18}. Similar effects have been found for twinning \citep{HGL16, KBM16}. The nature of the grain boundary from the viewpoint of slip transmission is assumed to depend on three angles that define the misorientation between the slip vectors, the slip plane normals and the slip plane intersections with the grain boundary plane. Different criteria have been proposed but the experimental evidence is still far from being conclusive \citep{BEZ14, BMR16}. Nevertheless, the behavior of grain boundaries may be a determinant factor in the mechanical behavior of strongly textured polycrystals (where one type of grain boundary may be dominant) and in the nucleation of damage by shear band formation \citep{WSD14} or grain boundary cracking \citep{BCG12, CMP15}. This is particularly important in the case of low symmetry crystals, in which the limited availability of slip systems may led to large stress concentrations at the grain boundaries if slip transfer is hindered. Moreover, cracks  are often nucleated at grain boundaries during fatigue and crack propagation across neighbour grains is critical to determine whether a microstructurally small crack  will be arrested or propagate until failure \citep{MM16, WMD16}. 

Grain boundaries are assumed to be transparent for slip transfer in standard (phenomenological) CP models and, thus, the effect of the grain boundary on the slip transmission is not accounted for. Physically-based  CP models can introduce the effect of grain boundaries through different parameters that take into account the increase in dislocation density near the grain boundary \citep{MBB15, SZM16, Hauoala2018}. Finally, lower-order SGCP account for the effect grain boundaries through the density of GNDs generated near the boundaries due to the deformation incompatibility between grains with different orientations, while higher-order SGCP can control the dislocation flux across the boundary by means of the higher-order boundary conditions. Thus, either physically-based or SGCP models can be used to  simulate the effect of grain boundaries on the plastic deformation of polycrystals and the development and experimental validation of these strategies will be an important area of research in the near future.

Another important development in CHP is the introduction of the "physical" twins in the simulations. So far, simulation of twinning during mechanical deformation has been taken into account through models that treat twinning as a pseudo-slip system \citep{K98, AD13a, AD13b, CK15}. Twinning is triggered by a CRSS on the twin plane and leads to a plastic deformation that depends on the volume fraction of twinned material and on the characteristic shear strain of the twin system considered. Moreover, plastic slip within the twinned regions can also be included. While this approach is accurate from the viewpoint of the effective properties of the polycrystal, it neglects the physical formation of a twinned region which is connected to the parent grain through two coherent grain boundaries. The importance of the presence of the twinned region in the deformation mechanisms have become very clear in the analysis of fatigue crack initiation in FCC metals, which show that fatigue cracks tend to nucleate in slip bands on \{111\} planes parallel and, slightly offset from, annealing twins because of the stress concentration associated with the presence of the twin \citep{HN90, SVB15, SLM16}. Obviously, this effect can only be accounted for if the "physical" twin is included within the grain in the numerical simulation. Annealing twins can be introduced into the initial RVE (although they require very fine meshing of the grains, leading to very large models) \citep{CSP15, YGH16}  but twins are often nucleated during deformation and should be introduced {\it on the fly} during the numerical simulation. This is necessary to account for the hardening effect of twinning on dislocation slip in a physical way and not through phenomenological models that cannot include the many different types of interactions. Strategies to achieve this goal have been presented recently \citep{AMB15, CG17, LSD18, CSM18} and they should be further developed.

Finally, the crystals in polycrystalline materials subjected to large plastic deformation do not only undergo a change in shape. Grain fragmentation as well as recrystallization processes lead to dramatic microstructural changes during deformation (particularly at high temperatures) that are used during thermo-mechanical processes to manufacture materials with new microstructures and properties. Coupling microstructure evolution with mechanical deformation will allow the "virtual processing" and "virtual design" of novel microstructures with optimized properties for specific applications and will dramatically increase the add value of CHP. Strategies to achieve these goals have been developed by coupling CHP with phase-field simulations \citep{AEF12a, AEF12b, GBH14, VBS15, LAO16, ADMAL2018} or cellular automata \citep{CHW13, JSY16}. They should be developed further to overcome the computational problems associated with the different time scales for mechanical deformation and microstructural evolution as well as the experimental and theoretical determination of the driving forces that control the microstructural evolution (grain boundary mobility, nucleation, etc.) under these complex conditions.

\section*{Acknowledgements}
This investigation was supported by the European Research Council under the European Unions Horizon 2020 research and innovation programme (Advanced Grant VIRMETAL, grant agreement No. 669141), by the Spanish Ministry of Economy and Competitiveness through the project DPI2015-67667 and by Los Alamos National Laboratory's Laboratory-Directed Research and Development (LDRD) program.


\begin{thebibliography}{351}
\expandafter\ifx\csname natexlab\endcsname\relax\def\natexlab#1{#1}\fi
\expandafter\ifx\csname url\endcsname\relax
  \def\url#1{\texttt{#1}}\fi
\expandafter\ifx\csname urlprefix\endcsname\relax\def\urlprefix{URL }\fi

\bibitem[{Abdolvand and Daymond(2013{\natexlab{a}})}]{AD13a}
Abdolvand, H., Daymond, M.~R., 2013{\natexlab{a}}. Multi-scale modeling and
  experimental study of twin inception and propagation in hexagonal
  close-packed materials using a crystal plasticity finite element approach.
  {Part I: Average behavior}. Journal of the Mechanics and Physics of Solids
  61, 783 -- 802.

\bibitem[{Abdolvand and Daymond(2013{\natexlab{b}})}]{AD13b}
Abdolvand, H., Daymond, M.~R., 2013{\natexlab{b}}. Multi-scale modeling and
  experimental study of twin inception and propagation in hexagonal
  close-packed materials using a crystal plasticity finite element approach.
  {Part II: Local }behavior. Journal of the Mechanics and Physics of Solids 61,
  803 -- 818.

\bibitem[{Abrivard et~al.(2012{\natexlab{a}})Abrivard, Busso, Forest, and
  Appolaire}]{AEF12a}
Abrivard, G., Busso, E., Forest, S., Appolaire, B., 2012{\natexlab{a}}. Phase
  field modelling of grain boundary motion driven by curvature and stored
  energy gradients. {Part I: Theory} and numerical implementation.
  Philosophical Magazine 92, 3618--3642.

\bibitem[{Abrivard et~al.(2012{\natexlab{b}})Abrivard, Busso, Forest, and
  Appolaire}]{AEF12b}
Abrivard, G., Busso, E., Forest, S., Appolaire, B., 2012{\natexlab{b}}. Phase
  field modelling of grain boundary motion driven by curvature and stored
  energy gradients. {Part II: A}pplication to recrystallisation. Philosophical
  Magazine 92, 3643--3664.

\bibitem[{Acharya and Bassani(1995)}]{ACHARYA1995}
Acharya, A., Bassani, J., 1995. Incompatible lattice deformations and crystal
  plasticity. Plastic and Fracture Instabilities in Materials, ASME proceedings
  57, 75--80.

\bibitem[{Acharya et~al.(2003)Acharya, Bassani, and Beaudin}]{Acharya2003-1}
Acharya, A., Bassani, J.~L., Beaudin, A., 2003. Geometrically necessary
  dislocations, hardening, and a simple gradient theory of crystal plasticity.
  Scripta Materialia 48, 167--172.

\bibitem[{Admal et~al.(2018)Admal, Po, and Marian}]{ADMAL2018}
Admal, N.~C., Po, G., Marian, J., 2018. A unified framework for polycrystal
  plasticity with grain boundary evolution. International Journal of
  Plasticity, In press.

\bibitem[{Aifantis(1987)}]{AIFANTIS1987211}
Aifantis, E.~C., 1987. The physics of plastic deformation. International
  Journal of Plasticity 3, 211 -- 247.

\bibitem[{Alkemper and Voorhees(2001)}]{AV01}
Alkemper, J., Voorhees, P.~W., 2001. Quantitative serial sectioning analysis.
  Journal of Microscopy 201, 388--394.

\bibitem[{Allison et~al.(2013)Allison, Cowles, DeLoach, Pollock, and
  Spanos}]{ICME13}
Allison, J., Cowles, B., DeLoach, J., Pollock, T., Spanos, G., 2013. Integrated
  Computational Materials Engineering (ICME): Implementing ICME in the
  Aerospace, Automotive, and Maritime Industries. The Minerals, Metals \&
  Materials Society, Warrendale, PA 15086.

\bibitem[{Allison et~al.(2006)Allison, Li, and Wolverton}]{ALW06}
Allison, J., Li, M., Wolverton, C., 2006. Virtual aluminum castings: An
  industrial application of {ICME}. {JOM} 58, 28--35.

\bibitem[{Anahid et~al.(2011)Anahid, Samal, and Ghosh}]{ANAHID20112157}
Anahid, M., Samal, M.~K., Ghosh, S., 2011. Dwell fatigue crack nucleation model
  based on crystal plasticity finite element simulations of polycrystalline
  titanium alloys. Journal of the Mechanics and Physics of Solids 59, 2157 --
  2176.

\bibitem[{Anand and Kothari(1996)}]{ANAND1996525}
Anand, L., Kothari, M., 1996. A computational procedure for rate-independent
  crystal plasticity. Journal of the Mechanics and Physics of Solids 44, 525 --
  558.

\bibitem[{Ardeljan et~al.(2015)Ardeljan, McCabe, Beyerlein, and
  Knezevic}]{AMB15}
Ardeljan, M., McCabe, R.~J., Beyerlein, I.~J., Knezevic, M., 2015. Explicit
  incorporation of deformation twins into crystal plasticity finite element
  models. Computer Methods in Applied Mechanics and Engineering 295, 396 --
  413.

\bibitem[{Arsenlis and Parks(1999)}]{ARSENLIS19991597}
Arsenlis, A., Parks, D., 1999. Crystallographic aspects of
  geometrically-necessary and statistically-stored dislocation density. Acta
  Materialia 47, 1597 -- 1611.

\bibitem[{Asaro and Needleman(1985)}]{Asaro1985923}
Asaro, R., Needleman, A., 1985. Texture development and strain hardening in
  rate dependent polycrystals. Acta Metallurgica 33, 923--953.

\bibitem[{Ashby(1970)}]{Ashby1970-1}
Ashby, M.~F., 1970. The deformation of plastically non-homogeneous materials.
  Philosophical Magazine 21, 399--424.

\bibitem[{Barbe et~al.(2001{\natexlab{a}})Barbe, Decker, Jeulin, and
  Cailletaud}]{BARBE2001513}
Barbe, F., Decker, L., Jeulin, D., Cailletaud, G., 2001{\natexlab{a}}.
  Intergranular and intragranular behavior of polycrystalline aggregates. {Part
  1: F.E.} model. International Journal of Plasticity 17, 513 -- 536.

\bibitem[{Barbe et~al.(2001{\natexlab{b}})Barbe, Forest, and
  Cailletaud}]{BARBE2001537}
Barbe, F., Forest, S., Cailletaud, G., 2001{\natexlab{b}}. Intergranular and
  intragranular behavior of polycrystalline aggregates. {Part 2: R}esults.
  International Journal of Plasticity 17~(4), 537 -- 563.

\bibitem[{Bardella(2006)}]{BARDELLA2006128}
Bardella, L., 2006. A deformation theory of strain gradient crystal plasticity
  that accounts for geometrically necessary dislocations. Journal of the
  Mechanics and Physics of Solids 54, 128 -- 160.

\bibitem[{Bardella et~al.(2013)Bardella, Segurado, Panteghini, and
  Llorca}]{Bardella_Segurado_etal_2013}
Bardella, L., Segurado, J., Panteghini, A., Llorca, J., 2013. Latent hardening
  size effect in small-scale plasticity. Modelling and Simulation in Materials
  Science and Engineering 21, 055009.

\bibitem[{Bargie{\l} and Mo{\'s}ci{\'n}ski(1991)}]{bargiel1991c}
Bargie{\l}, M., Mo{\'s}ci{\'n}ski, J., 1991. {C}-language program for the
  irregular close packing of hard spheres. Computer Physics Communications 64,
  183--192.

\bibitem[{Barton et~al.(2008)Barton, Knap, Arsenlis, Becker, Hornung, and
  Jefferson}]{BKA08}
Barton, N.~R., Knap, J., Arsenlis, A., Becker, R., Hornung, P.~D., Jefferson,
  D.~R., 2008. mbedded polycrystal plasticity and adaptive sampling.
  International Journal of Plasticity 24, 242--266.

\bibitem[{Bassani and Wu(1991)}]{Bassani199121}
Bassani, J., Wu, T., 1991. Latent hardening in single crystals ii. analytical
  characterization and predictions. Proceedings of the Royal Society of London
  A 435, 21–41.

\bibitem[{Bayerschen et~al.(2016)Bayerschen, McBride, Reddy, and
  B{\"o}hlke}]{BMR16}
Bayerschen, E., McBride, A.~T., Reddy, B.~D., B{\"o}hlke, T., 2016. Review on
  slip transmission criteria in experiments and crystal plasticity models.
  Journal of Materials Science 51, 2243--2258.

\bibitem[{Bayley et~al.(2006)Bayley, Brekelmans, and Geers}]{BAYLEY20067268}
Bayley, C.~J., Brekelmans, W. A.~M., Geers, M. G.~D., 2006. A comparison of
  dislocation induced back stress formulations in strain gradient crystal
  plasticity. International Journal of Solids and Structures 43, 7268 -- 7286.

\bibitem[{Becker(1991)}]{B91}
Becker, R., 1991. Analysis of texture evolution in channel die compression—i.
  effects of grain interaction. Acta Metallurgica et Materialia 39, 1211 --
  1230.

\bibitem[{Bennett and McDowell(2003)}]{BENNETT200327}
Bennett, V.~P., McDowell, D.~L., 2003. Polycrystal orientation distribution
  effects on microslip in high cycle fatigue. International Journal of Fatigue
  25, 27 -- 39.

\bibitem[{Benzerga and Leblond(2010)}]{BL10}
Benzerga, A.~A., Leblond, J.-B., 2010. Ductile fracture by void growth to
  coalescence. Advances in Applied Mechanics 44, 169 -- 305.

\bibitem[{Berbenni et~al.(2014)Berbenni, Taupin, Djaka, and
  Fressengeas}]{BTD14}
Berbenni, S., Taupin, V., Djaka, K.~S., Fressengeas, C., 2014. A numerical
  spectral approach for solving elasto-static field dislocation and
  g-disclination mechanics. International Journal of Solids and Structures 51,
  4157 -- 4175.

\bibitem[{Bertin et~al.(2014)Bertin, Tomé, Beyerlein, Barnett, and
  Capolungo}]{BERTIN201472}
Bertin, N., Tomé, C.~N., Beyerlein, I.~J., Barnett, M.~R., Capolungo, L.,
  2014. On the strength of dislocation interactions and their effect on latent
  hardening in pure magnesium. International Journal of Plasticity 62, 72 --
  92.

\bibitem[{Bertin et~al.(2015)Bertin, Upadhyay, Pradalier, and
  Capolungo}]{BUP15}
Bertin, N., Upadhyay, M.~V., Pradalier, C., Capolungo, L., 2015. A fft-based
  formulation for efficient mechanical fields computation in isotropic and
  anisotropic periodic discrete dislocation dynamics. Modelling and Simulation
  in Materials Science and Engineering 23~(6), 065009.

\bibitem[{Berveiller and Zaoui(1979)}]{BZ79}
Berveiller, M., Zaoui, A., 1979. An extension to the self-consistent scheme to
  plastically-flowing polycristals. Journal of the Mechanics and Physics of
  Solids 26, 325--344.

\bibitem[{Beyerlein and Tom\'e(2008)}]{BT08}
Beyerlein, I.~J., Tom\'e, C., 2008. A dislocation-based constitutive law for
  pure {Zr} including temperature effects. International Journal of Plasticity
  24, 867--895.

\bibitem[{Bhattacharyya et~al.(2001)Bhattacharyya, El-Danaf, Kalidindi, and
  Doherty}]{BEK01}
Bhattacharyya, A., El-Danaf, E., Kalidindi, S.~R., Doherty, R.~D., 2001.
  Evolution of grain-scale microstructure during large strain simple
  compression of polycrystalline aluminum with quasi-columnar grains: Oim
  measurements and numerical simulations. International Journal of Plasticity
  17, 861 -- 883.

\bibitem[{Bieler et~al.(2014)Bieler, Eisenlohr, Zhang, Phukan, and
  Crimp}]{BEZ14}
Bieler, T.~R., Eisenlohr, P., Zhang, C., Phukan, H.~J., Crimp, M.~A., 2014.
  Grain boundaries and interfaces in slip transfer. Current Opinion in Solid
  State and Materials Science 18, 212 -- 226.

\bibitem[{Bilby and Crocker(1965)}]{Bilby240}
Bilby, B., Crocker, A., 1965. The theory of the crystallography of deformation
  twinning. Proceedings of the Royal Society of London A: Mathematical,
  Physical and Engineering Sciences 288, 240--255.

\bibitem[{Birosca et~al.(2009)Birosca, Buffiere, Garcia-Pastor, Karadge,
  Babout, and Preuss}]{BBG09}
Birosca, S., Buffiere, J.-Y., Garcia-Pastor, F., Karadge, M., Babout, L.,
  Preuss, M., 2009. Three-dimensional characterization of fatigue cracks in
  ti-6246 using x-ray tomography and electron backscatter diffraction. Acta
  Materialia 57, 5834 -- 5847.

\bibitem[{Bobeth and Diener(1987)}]{BD87}
Bobeth, M., Diener, G., 1987. Static elastic and thermoelastic field
  fluctuations in multiphase composites. Journal of the Mechanics and Physics
  of Solids 35, 137--149.

\bibitem[{Boehlert et~al.(2012)Boehlert, Chen, Gutiérrez-Urrutia, Llorca, and
  Pérez-Prado}]{BCG12}
Boehlert, C., Chen, Z., Gutiérrez-Urrutia, I., Llorca, J., Pérez-Prado, M.,
  2012. In situ analysis of the tensile and tensile-creep deformation
  mechanisms in rolled {AZ31}. Acta Materialia 60, 1889 -- 1904.

\bibitem[{B\"ohm(2004)}]{Bohm}
B\"ohm, H.~J., 2004. A Short Introduction to Continuum Micromechanics.
  Springer-Verlag Wien.

\bibitem[{B\"ohm and Han(2001)}]{Bohm_Han2001}
B\"ohm, H.~J., Han, W., 2001. Comparisons between three-dimensional and
  two-dimensional multi-particle unit cell models for particle reinforced metal
  matrix composites. Modelling and Simulation in Materials Science and
  Engineering 9, 47--66.

\bibitem[{Borg et~al.(2008)Borg, Niordson, and Kysar}]{BORG2008688}
Borg, U., Niordson, C.~F., Kysar, J.~W., 2008. Size effects on void growth in
  single crystals with distributed voids. International Journal of Plasticity
  24, 688 -- 701.

\bibitem[{Brenner et~al.(2014)Brenner, Beaudoin, Suquet, and Acharya}]{BBS14}
Brenner, R., Beaudoin, A.~J., Suquet, P., Acharya, A., 2014. Numerical
  implementation of static {Field Dislocation Mechanics} theory for periodic
  media. Philosophical Magazine 94, 1764 -- 1787.

\bibitem[{Brenner et~al.(2009)Brenner, Lebensohn, and Castelnau}]{BLC09}
Brenner, R., Lebensohn, R.~A., Castelnau, O., 2009. Elastic anisotropy and
  yield surface estimates of polycrystals. International Journal of Solids and
  Structures 46, 3018--3026.

\bibitem[{Brisard and Dormieux(2010)}]{BD10}
Brisard, S., Dormieux, L., 2010. {FFT}-based methods for the mechanics of
  composites: a general variational framework. Computational Materials Science
  49, 663--671.

\bibitem[{Bronkhorst et~al.(1992)Bronkhorst, Kalidindi, and Anand}]{BKA92}
Bronkhorst, C., Kalidindi, S., Anand, L., 1992. Polycrystalline plasticity and
  the evolution of crystallographic texture in fcc metals. Philosophical
  Transactions: Physical Sciences and Engineering 341, 443--477.

\bibitem[{Busso et~al.(2000)Busso, Meissonnier, and O'Dowd}]{Busso2000-1}
Busso, E.~P., Meissonnier, F.~T., O'Dowd, N.~P., 2000. Gradient-dependent
  deformation of two-phase single crystals. Journal of the Mechanics and
  Physics of Solids 48, 2333--2361.

\bibitem[{Byer and Ramesh(2013)}]{BR13}
Byer, C.~M., Ramesh, K.~T., 2013. Effects of the initial dislocation density on
  size effects in single-crystal magnesium. Acta Materialia 61, 3808 -- 3818.

\bibitem[{Cailletaud(1992)}]{CAILLETAUD199255}
Cailletaud, G., 1992. A micromechanical approach to inelastic behaviour of
  metals. International Journal of Plasticity 8, 55 -- 73.

\bibitem[{Capolungo et~al.(2009)Capolungo, Beyerlein, and Tom\'e}]{CBT09b}
Capolungo, L., Beyerlein, I., Tom\'e, C., 2009. Slip-assisted twin growth in
  hexagonal close-packed metals. Scripta Materialia 60, 32 -- 35.

\bibitem[{CAPSUL(2018)}]{Capsul}
CAPSUL, 2018.
  \url{https://materials.imdea.org/research/simulation-tools/capsul/}.

\bibitem[{Castelluccio and McDowell(2013)}]{Castelluccio2013}
Castelluccio, G.~M., McDowell, D.~L., 2013. A mesoscale approach for growth of
  {3D} microstructurally small fatigue cracks in polycrystals. International
  Journal of Damage Mechanics 23, 791 -- 818.

\bibitem[{Castelluccio and McDowell(2014)}]{Castelluccio2014}
Castelluccio, G.~M., McDowell, D.~L., 2014. Mesoscale modeling of
  microstructurally small fatigue cracks in metallic polycrystals. Materials
  Science and Engineering: A 598, 34 -- 55.

\bibitem[{Castelluccio and McDowell(2015)}]{Castelluccio2015}
Castelluccio, G.~M., McDowell, D.~L., 2015. Microstructure and mesh
  sensitivities of mesoscale surrogate driving force measures for transgranular
  fatigue cracks in polycrystals. Materials Science and Engineering: A 639, 626
  -- 639.

\bibitem[{Cepeda-Jim\'enez et~al.(2015)Cepeda-Jim\'enez, Molina-Aldareguia, and
  P\'erez-Prado}]{CMP15}
Cepeda-Jim\'enez, C.~M., Molina-Aldareguia, J.~M., P\'erez-Prado, M.~T., 2015.
  Effect of grain size on slip activity in pure magnesium polycrystals. Acta
  Materialia 84, 443 -- 456.

\bibitem[{Cermelli and Gurtin(2001)}]{CERMELLI20011539}
Cermelli, P., Gurtin, M.~E., 2001. On the characterization of geometrically
  necessary dislocations in finite plasticity. Journal of the Mechanics and
  Physics of Solids 49, 1539 -- 1568.

\bibitem[{Cerrone et~al.(2015)Cerrone, Stein, Pokharel, Hefferan, Lind, Tucker,
  Suter, Rollett, and Ingraffea}]{CSP15}
Cerrone, A., Stein, C., Pokharel, R., Hefferan, C., Lind, J., Tucker, H.,
  Suter, R., Rollett, A., Ingraffea, A., 2015. Implementation and verification
  of a microstructure-based capability for modeling microcrack nucleation in
  lshr at room temperature. Modelling and Simulation in Materials Science and
  Engineering 23, 035006.

\bibitem[{Chakraborty and Eisenlohr(2017)}]{CE17}
Chakraborty, A., Eisenlohr, P., 2017. Evaluation of an inverse methodology for
  estimating constitutive parameters in face-centered cubic materials from
  single crystal indentations. European Journal of Mechanics - A/Solids 66, 114
  -- 124.

\bibitem[{Chang and Kochmann(2015)}]{CK15}
Chang, Y., Kochmann, D.~M., 2015. A variational constitutive model for
  slip-twinning interactions in hcp metals: Application to single- and
  polycrystalline magnesium. International Journal of Plasticity 73, 39 -- 61.

\bibitem[{Charkaluk et~al.(2002)Charkaluk, Bignonnet, Constantinescu, and
  Dang~Van}]{FFE:FFE0612}
Charkaluk, E., Bignonnet, A., Constantinescu, A., Dang~Van, K., 2002. Fatigue
  design of structures under thermomechanical loadings. Fatigue and Fracture of
  Engineering Materials and Structures 25, 1199--1206.

\bibitem[{Chen et~al.(2018)Chen, Jiang, and Dunne}]{CHEN2018213}
Chen, B., Jiang, J., Dunne, F.~P., 2018. Is stored energy density the primary
  meso-scale mechanistic driver for fatigue crack nucleation? International
  Journal of Plasticity 101, 213 -- 229.

\bibitem[{Cheng and Ghosh(2017)}]{CG17}
Cheng, J., Ghosh, S., 2017. Crystal plasticity finite element modeling of
  discrete twin evolution in polycrystalline magnesium. Journal of the
  Mechanics and Physics of Solids 99, 512 -- 538.

\bibitem[{Cheng et~al.(2018)Cheng, Shen, Mishra, and Ghosh}]{CSM18}
Cheng, J., Shen, J., Mishra, R.~K., Ghosh, S., 2018. Discrete twin evolution in
  mg alloys using a novel crystal plasticity finite element model. Acta
  Materialia 149, 142 -- 153.

\bibitem[{Cheong and Busso(2004)}]{Cheong20045665}
Cheong, K.-S., Busso, E.~P., 2004. Discrete dislocation density modelling of
  single phase fcc polycrystal aggregates. Acta Materialia 52, 5665--5675.

\bibitem[{Cheong et~al.(2005)Cheong, Busso, and Arsenlis}]{CHEONG20051797}
Cheong, K.~S., Busso, E.~P., Arsenlis, A., 2005. A study of microstructural
  length scale effects on the behaviour of fcc polycrystals using strain
  gradient concepts. International Journal of Plasticity 21, 1797 -- 1814.

\bibitem[{Christian and Mahajan(1995)}]{CHRISTIAN19951}
Christian, J., Mahajan, S., 1995. Deformation twinning. Progress in Materials
  Science 39, 1 -- 157.

\bibitem[{Chuan et~al.(2013)Chuan, He, and Wei}]{CHW13}
Chuan, W., He, Y., Wei, L.~H., 2013. Modeling of discontinuous dynamic
  recrystallization of a near-$\beta$ titanium alloy {IMI834} during isothermal
  hot compression by combining a cellular automaton model with a crystal
  plasticity finite element method. Computational Materials Science 79, 944 --
  959.

\bibitem[{Cruzado et~al.(2015)Cruzado, Gan, Jiménez, Barba, Ostolaza, Linaza,
  Molina-Aldaregu{\'\i}a, LLorca, and Segurado}]{CBJ15}
Cruzado, A., Gan, B., Jiménez, M., Barba, D., Ostolaza, K., Linaza, A.,
  Molina-Aldaregu{\'\i}a, J.~M., LLorca, J., Segurado, J., 2015. Multiscale
  modeling of the mechanical behavior of {IN718} superalloy based on
  micropillar compression and computational homogenization. Acta Materialia 98,
  242 -- 253.

\bibitem[{Cruzado et~al.(2017)Cruzado, LLorca, and Segurado}]{CRUZADO2017148}
Cruzado, A., LLorca, J., Segurado, J., 2017. Modeling cyclic deformation of
  inconel 718 superalloy by means of crystal plasticity and computational
  homogenization. International Journal of Solids and Structures 122-123, 148
  -- 161.

\bibitem[{Cruzado et~al.(2018{\natexlab{a}})Cruzado, Lucarini, LLorca, and
  Segurado}]{Cruzado2018b}
Cruzado, A., Lucarini, S., LLorca, J., Segurado, J., 2018{\natexlab{a}}.
  Crystal-plasticity simulation of the effect grain size on the fatigue
  behavior of polycrystalline {Inconel 718}. International Journal of Fatigue
  113, 236--245.

\bibitem[{Cruzado et~al.(2018{\natexlab{b}})Cruzado, Lucarini, LLorca, and
  Segurado}]{Cruzado2018a}
Cruzado, A., Lucarini, S., LLorca, J., Segurado, J., 2018{\natexlab{b}}.
  Microstructure-based fatigue life model of metallic alloys with bilinear
  coffin-manson behavior. International Journal of Fatigue 107, 40--48.

\bibitem[{Cuiti\~no and Ortiz(1992)}]{Cuitino1992437}
Cuiti\~no, A., Ortiz, M., 1992. Material-independent method for extending
  stress update algorithms from small-strain plasticity to finite plasticity
  with multiplicative kinematics. Engineering Computations 9, 437--451.

\bibitem[{Dai and Parks(1997)}]{Dai1997-1}
Dai, H., Parks, D.~M., 1997. Geometrically-necessary dislocation density and
  scale-dependent crystal plasticity. In: Khan, A. (Ed.), Proceedings of Sixth
  International Symposium on Plasticity. Gordon \&\ Breach, pp. 17--18.

\bibitem[{Damask(2018)}]{Damask}
Damask, 2018. \url{https://damask.mpie.de}.

\bibitem[{{de Botton} and Ponte-Casta{\~n}eda(1995)}]{BP95}
{de Botton}, G., Ponte-Casta{\~n}eda, P., 1995. Variational estimates for the
  creep behavior of polycrystals. Proceedings of the Royal Society of London A
  448, 121--142.

\bibitem[{de~Geus et~al.(2017)de~Geus, Vondrejc, Zeman, Peerlings, and
  Geers}]{Geers2016}
de~Geus, T., Vondrejc, J., Zeman, J., Peerlings, R., Geers, M., 2017. Finite
  strain fft-based non-linear solvers made simple. Computer Methods in Applied
  Mechanics and Engineering 318, 412 -- 430.

\bibitem[{de~Sansal et~al.(2010)de~Sansal, Devincre, and Kubin}]{sansal2010}
de~Sansal, C., Devincre, B., Kubin, L., 2010. Grain size strengthening in
  microcrystalline copper: A three-dimensional dislocation dynamics simulation.
  Key Engineering Materials 423, 25--32.

\bibitem[{de~Souza~Neto et~al.(2008)de~Souza~Neto, Peric, and
  Owen}]{Peric_plasticity}
de~Souza~Neto, E., Peric, D., Owen, D. R.~J., 2008. Computational Methods for
  Plasticity. John Wiley and Sons, Ltd.

\bibitem[{Delaire et~al.(2000)Delaire, Raphanel, and Rey}]{DRC00}
Delaire, F., Raphanel, J.~L., Rey, C., 2000. Plastic heterogeneities of a
  copper multicrystal deformed in uniaxial tension: experimental study and
  finite element simulations. Acta Materialia 48, 1075 -- 1087.

\bibitem[{Devincre et~al.(2008)Devincre, Hoc, and Kubin}]{Devincre1745}
Devincre, B., Hoc, T., Kubin, L., 2008. Dislocation mean free paths and strain
  hardening of crystals. Science 320, 1745--1748.

\bibitem[{Do{\v{s}}k{\'a}{\v{r}} et~al.(2014)Do{\v{s}}k{\'a}{\v{r}}, Nov{\'a}k,
  and Zeman}]{dovskavr2014aperiodic}
Do{\v{s}}k{\'a}{\v{r}}, M., Nov{\'a}k, J., Zeman, J., 2014. A periodic
  compression and reconstruction of real-world material systems based on wang
  tiles. Physical Review E 90, 062118.

\bibitem[{Dream.3D(2018)}]{dream3d}
Dream.3D, 2018. \url{http://dream3d.bluequartz.net}.

\bibitem[{Dunne et~al.(2007)Dunne, Rugg, and Walker}]{DUNNE20071061}
Dunne, F., Rugg, D., Walker, A., 2007. Length scale-dependent, elastically
  anisotropic, physically-based hcp crystal plasticity: Application to
  cold-dwell fatigue in {Ti} alloys. International Journal of Plasticity 23,
  1061 -- 1083.

\bibitem[{Echlin et~al.(2011)Echlin, Husseini, Nees, and Pollock}]{EHN11}
Echlin, M.~P., Husseini, N.~S., Nees, J.~A., Pollock, T.~M., 2011. A new
  femtosecond laser-based tomography technique for multiphase materials.
  Advanced Materials 23, 2339 -- 2342.

\bibitem[{Eisenlohr et~al.(2013)Eisenlohr, Diehl, Lebensohn, and
  Roters}]{EDL13}
Eisenlohr, P., Diehl, M., Lebensohn, R.~A., Roters, F., 2013. A spectral method
  solution to crystal elasto-viscoplasticity at finite strains. International
  Journal of Plasticity 46, 37--53.

\bibitem[{El-Awady(2015)}]{E15}
El-Awady, J.~A., 2015. Unraveling the physics of size-dependent
  dislocation-mediated plasticity. Nature Communications 6, 5926.

\bibitem[{Escobedo et~al.(2011)Escobedo, Cerreta, Dennis-Koller, Patterson,
  Bronkhorst, Hansen, Tonks, and Lebensohn}]{ECD11}
Escobedo, J.~P., Cerreta, E.~K., Dennis-Koller, D., Patterson, B.~M.,
  Bronkhorst, C., Hansen, B., Tonks, D., Lebensohn, R.~A., 2011. Effects of
  grain size and boundary structure on the dynamic tensile response of copper.
  Journal of Applied Physics 110, 033513.

\bibitem[{Escobedo et~al.(2013)Escobedo, Cerreta, Dennis-Koller, Trujillo, and
  Bronkhorst}]{ECD13}
Escobedo, J.~P., Cerreta, E.~K., Dennis-Koller, D., Trujillo, C.~P.,
  Bronkhorst, C., 2013. Influence of boundary structure and near neighbor
  crystallographic orientation on the dynamic damage evolution during shock
  loading. Philosophical Magazine 93, 833 -- 846.

\bibitem[{Eshelby(1957)}]{E57}
Eshelby, J., 1957. The determination of the elastic field of an ellipsoidal
  inclusion and related problems. Proceedings of the Royal Society of London A
  252, 561--569.

\bibitem[{Estrin and Mecking(1984)}]{ESTRIN198457}
Estrin, Y., Mecking, H., 1984. A unified phenomenological description of work
  hardening and creep based on one-parameter models. Acta Metallurgica 32, 57
  -- 70.

\bibitem[{{Eswar Prasad} et~al.(2014){Eswar Prasad}, Rajesh, and
  Ramamurty}]{ERR14}
{Eswar Prasad}, K., Rajesh, K., Ramamurty, U., 2014. Micropillar and
  macropillar compression responses of magnesium single crystals oriented for
  single slip or extension twinning. Acta Materialia 65, 316 -- 325.

\bibitem[{Evers et~al.(2004)Evers, Brekelmans, and Geers}]{EVERS20042379}
Evers, L.~P., Brekelmans, W. A.~M., Geers, M. G.~D., 2004. Non-local crystal
  plasticity model with intrinsic ssd and gnd effects. Journal of the Mechanics
  and Physics of Solids 52, 2379 -- 2401.

\bibitem[{Eyer and Milton(1999)}]{EM99}
Eyer, D.~J., Milton, G.~W., 1999. A fast numerical scheme for computing the
  response of composites using grid refinement. The European Physical Journal
  Applied Physics 6, 41--47.

\bibitem[{Fatemi and Socie(1988)}]{FFE:FFE149}
Fatemi, A., Socie, D.~F., 1988. A critical plane approach to multiaxial fatigue
  damage including out-of-phase loading. Fatigue and Fracture of Engineering
  Materials and Structures 11, 149--165.

\bibitem[{Fatemi and Socie(1989)}]{FS1989}
Fatemi, A., Socie, D.~F., 1989. Multiaxial Fatigue: Damage Mechanisms and Life
  Predictions. Vol. 159. Springer.

\bibitem[{Fayman(1987)}]{F87}
Fayman, Y.~C., 1987. Microstructural characterization and element partitioning
  in a direct Ðaged superalloy ({DA}718). Materials Science and Engineering A
  92, 159--171.

\bibitem[{Fern\'andez et~al.(2013)Fern\'andez, J\'erusalem,
  Guti\'errez-Urrutia, and P\'erez-Prado}]{FJG13}
Fern\'andez, A., J\'erusalem, A., Guti\'errez-Urrutia, I., P\'erez-Prado, M.,
  2013. Three-dimensional investigation of the grain boundary-twin interactions
  in a {Mg AZ31} alloy by electron backscatter diffraction and continuum
  modeling. Acta Materialia 61, 7679--7692.

\bibitem[{Feyel(2003)}]{F03}
Feyel, F., 2003. A multilevel finite element method {(FE2)} to describe the
  response of highly non-linear structures using generalized continua.
  Compuational Methods in Applied Mechanics and Engineering 192, 3233--3244.

\bibitem[{Fleck et~al.(1994)Fleck, MUller, Ashby, and Hutchinson}]{FMAH94}
Fleck, N., MUller, G., Ashby, M., Hutchinson, J., 1994. Strain gradient
  plasticity: theory and experiment. Acta Metallurgica et Materialia 42,
  475--487.

\bibitem[{Franciosi and Zaoui(1982)}]{FRANCIOSI19821627}
Franciosi, P., Zaoui, A., 1982. Multislip in f.c.c. crystals a theoretical
  approach compared with experimental data. Acta Metallurgica 30, 1627 -- 1637.

\bibitem[{Fullwood et~al.(2008)Fullwood, Niezgoda, and Kalidindi}]{FNK08}
Fullwood, D.~T., Niezgoda, S.~R., Kalidindi, S.~R., 2008. Microstructure
  reconstructions from 2-point statistics using phase-recovery algorithms. Acta
  Materialia 56, 942 -- 948.

\bibitem[{Geers et~al.(2014)Geers, Cottura, Appolaire, Busso, Forest, and
  Villani}]{GEERS2014136}
Geers, M., Cottura, M., Appolaire, B., Busso, E.~P., Forest, S., Villani, A.,
  2014. Coupled glide-climb diffusion-enhanced crystal plasticity. Journal of
  the Mechanics and Physics of Solids 70, 136 -- 153.

\bibitem[{Ghosh and Dimiduk(2011)}]{Ghosh2011}
Ghosh, S., Dimiduk, D., 2011. Computational Methods for Microstructure Property
  Relationship. Springer.

\bibitem[{Ghosh et~al.(2016)Ghosh, Shahba, Tu, Huskins, and Schuster}]{GST16}
Ghosh, S., Shahba, A., Tu, X., Huskins, E.~L., Schuster, B.~E., 2016. Crystal
  plasticity {FE modeling of Ti} alloys for a range of strain-rates. {Part II:
  I}mage-based model with experimental validation. International Journal of
  Plasticity 87, 69 -- 85.

\bibitem[{Gong and Wilkinson(2011)}]{GONG20115970}
Gong, J., Wilkinson, A.~J., 2011. A microcantilever investigation of size
  effect, solid-solution strengthening and second-phase strengthening for
  〈a〉 prism slip in alpha-ti. Acta Materialia 59, 5970 -- 5981.

\bibitem[{Gonz\'alez and LLorca(2007)}]{GL07}
Gonz\'alez, C., LLorca, J., 2007. Mechanical behavior of unidirectional
  fiber-reinforced polymers under transverse compression: microscopic
  mechanisms and modeling. Composites Science and Technology 67, 2795--2806.

\bibitem[{Gonz\'alez et~al.(2004)Gonz\'alez, Segurado, and LLorca}]{GSL04}
Gonz\'alez, C., Segurado, J., LLorca, J., 2004. Numerical simulation of
  elasto-plastic deformation of composites: Evolution of stress microfields and
  implications for homogenization models. Journal of the Mechanics and Physics
  of Solids 52, 1573--1593.

\bibitem[{Gonz\'alez et~al.(2017)Gonz\'alez, Vilatela, Molina-Aldaregu{\'\i}a,
  Lopes, and LLorca}]{GVM17}
Gonz\'alez, C., Vilatela, J.~J., Molina-Aldaregu{\'\i}a, J.~M., Lopes, C.~S.,
  LLorca, J., 2017. Structural composites for multifunctional applications:
  current challenges and future trends. Progress in Materials Science 89,
  194--251.

\bibitem[{Graham et~al.(2016)Graham, Rollett, and Lesar}]{GRL16}
Graham, J.~T., Rollett, A.~D., Lesar, R., 2016. Fast fourier transform discrete
  dislocation dynamics. Modelling and Simulation in Materials Science and
  Engineering 24, 085005.

\bibitem[{Grennerat et~al.(2012)Grennerat, Montagnat, Castelnau, Vacher,
  Moulinec, Suquet, and Duval}]{GMC12}
Grennerat, F., Montagnat, M., Castelnau, O., Vacher, P., Moulinec, H., Suquet,
  P., Duval, P., 2012. Experimental characterization of the intragranular
  strain field in columnar ice during transient creep. Acta Materialia 60,
  3655--3666.

\bibitem[{Gurtin(2002)}]{GURTIN20025}
Gurtin, M.~E., 2002. A gradient theory of single-crystal viscoplasticity that
  accounts for geometrically necessary dislocations. Journal of the Mechanics
  and Physics of Solids 50, 5 -- 32.

\bibitem[{Gurtin(2008)}]{GURTIN2008702}
Gurtin, M.~E., 2008. A finite-deformation, gradient theory of single-crystal
  plasticity with free energy dependent on densities of geometrically necessary
  dislocations. International Journal of Plasticity 24, 702 -- 725.

\bibitem[{Gurtin and Needleman(2005)}]{GURTIN20051}
Gurtin, M.~E., Needleman, A., 2005. Boundary conditions in small-deformation,
  single-crystal plasticity that account for the burgers vector. Journal of the
  Mechanics and Physics of Solids 53, 1 -- 31.

\bibitem[{Güvenc et~al.(2014)Güvenc, Bambach, and Hirt}]{GBH14}
Güvenc, O., Bambach, M., Hirt, G., 2014. Coupling of crystal plasticity finite
  element and phase field methods for the prediction of srx kinetics after hot
  working. Steel Research International 85, 999--1009.

\bibitem[{Hall(1951)}]{Hall1951-1}
Hall, E.~O., 1951. The deformation and ageing of mild steel: {III} {D}iscussion
  of results. Proceedings of the Physical Society Section B 64, 747--753.

\bibitem[{Han et~al.(2007)Han, Ma, Roters, and Raabe}]{Han2007-1}
Han, C., Ma, A., Roters, F., Raabe, D., 2007. A finite element approach with
  patch projection for strain gradient plasticity formulations. International
  Journal of Plasticity 23, 690--710.

\bibitem[{Han et~al.(2005{\natexlab{a}})Han, Gao, Huang, and Nix}]{HAN20051188}
Han, C.-S., Gao, H., Huang, Y., Nix, W.~D., 2005{\natexlab{a}}. Mechanism-based
  strain gradient crystal plasticity €"i. theory. Journal of the Mechanics and
  Physics of Solids 53, 1188 -- 1203.

\bibitem[{Han et~al.(2005{\natexlab{b}})Han, Gao, Huang, and Nix}]{HAN20051204}
Han, C.-S., Gao, H., Huang, Y., Nix, W.~D., 2005{\natexlab{b}}. Mechanism-based
  strain gradient crystal plasticity €"ii. analysis. Journal of the Mechanics
  and Physics of Solids 53, 1204 -- 1222.

\bibitem[{Hashin and Shtrikman(1963)}]{HS63}
Hashin, Z., Shtrikman, S., 1963. A variational approach to the theory of
  elastic behavior of multiphase materials. Journal of the Mechanics and
  Physics of Solids 11, 127--140.

\bibitem[{Hasija et~al.(2003)Hasija, Ghosh, Mills, and Joseph}]{HGM03}
Hasija, V., Ghosh, S., Mills, M.~J., Joseph, D.~S., 2003. Deformation and creep
  modeling in polycrystalline tiÐ6al alloys. Acta Materialia 51, 4533 -- 4549.

\bibitem[{Hauoala et~al.(2018)Hauoala, Segurado, and LLorca}]{Hauoala2018}
Hauoala, S., Segurado, J., LLorca, J., 2018. An analysis of the influence of
  grain size on the strength of fcc polycrystals by means of computational
  homogenization. Acta Materialia 148, 72--85.

\bibitem[{Hazanov and Huet(1994)}]{HAZANOV}
Hazanov, S., Huet, C., 1994. Order relationships for boundary condition effects
  in heterogeneous bodies smaller that the representative volume. Journal of
  the Mechanics and Physics of Solids 42, 1995--2011.

\bibitem[{Heilbronner and Bruhn(1998)}]{HB98}
Heilbronner, R., Bruhn, D., 1998. The influence of three-dimensional grain size
  distributions on the rheology of polyphase rocks. Journal of Structural
  Geology 20, 695--707.

\bibitem[{Heinz and Neumann(1990)}]{HN90}
Heinz, A., Neumann, P., 1990. Crack initiation during high cycle fatigue of an
  austenitic steel. Acta Metallurgica et Materialia 38, 1933 -- 1940.

\bibitem[{H\'emery et~al.(2018)H\'emery, Nizou, and Villechaise}]{HNV18}
H\'emery, S., Nizou, P., Villechaise, P., 2018. In situ sem investigation of
  slip transfer in {Ti-6Al-4V: E}ffect of applied stress. Materials Science and
  Engineering: A 709, 277 -- 284.

\bibitem[{Herbig et~al.(2011)Herbig, King, Reischig, Proudhon, Lauridsen,
  Marrow, Buffière, and Ludwig}]{HKR11}
Herbig, M., King, A., Reischig, P., Proudhon, H., Lauridsen, E.~M., Marrow, J.,
  Buffière, J.-Y., Ludwig, W., 2011. {3-D} growth of a short fatigue crack
  within a polycrystalline microstructure studied using combined diffraction
  and phase-contrast {X-ray} tomography. Acta Materialia 59, 590 -- 601.

\bibitem[{Herrera-Solaz et~al.(2014{\natexlab{a}})Herrera-Solaz,
  Hidalgo-Manrique, P\'erez-Prado, Letzig, LLorca, and Segurado}]{HHP14}
Herrera-Solaz, V., Hidalgo-Manrique, P., P\'erez-Prado, M.~T., Letzig, D.,
  LLorca, J., Segurado, J., 2014{\natexlab{a}}. Effect of rare earth additions
  on the critical resolved shear stresses of magnesium alloys. Materials
  Letters 128, 199 -- 203.

\bibitem[{Herrera-Solaz et~al.(2014{\natexlab{b}})Herrera-Solaz, LLorca, Dogan,
  Karaman, and Segurado}]{HLD14}
Herrera-Solaz, V., LLorca, J., Dogan, E., Karaman, I., Segurado, J.,
  2014{\natexlab{b}}. An inverse optimization strategy to determine single
  crystal mechanical behavior from polycrystal tests: Application to {AZ31 Mg}
  alloy. International Journal of Plasticity 57, 1 -- 15.

\bibitem[{Herrera-Solaz et~al.(2015)Herrera-Solaz, Segurado, and
  LLorca}]{HSL15}
Herrera-Solaz, V., Segurado, J., LLorca, J., 2015. On the robustness of an
  inverse optimization approach based on the {Levenberg-Marquardt} method for
  the mechanical behavior of polycrystals. European Journal of Mechanics/A 53,
  220 -- 228.

\bibitem[{Hershley(1954)}]{H54}
Hershley, A.~V., 1954. The elasticity of an isotropic aggregate of anisotropic
  cubic crystal. Journal of Applied Mechanics 21, 236--240.

\bibitem[{Hidalgo-Manrique et~al.(2015)Hidalgo-Manrique, Herrera-Solaz,
  Segurado, LLorca, G\'alvez, Ruano, Yi, and P\'erez-Prado}]{HHS15}
Hidalgo-Manrique, P., Herrera-Solaz, V., Segurado, J., LLorca, J., G\'alvez,
  F., Ruano, O.~A., Yi, S., P\'erez-Prado, M.~T., 2015. Origin of the reversed
  yield asymmetry in {Mg}-rare earth alloys at high temperature. Acta
  Materialia 92, 265--277.

\bibitem[{Hill(1965{\natexlab{a}})}]{H65b}
Hill, R., 1965{\natexlab{a}}. Continuum micro-mechanics of elastoplastic
  polycrystals. Journal of the Mechanics and Physics of Solids 13, 89--101.

\bibitem[{Hill(1965{\natexlab{b}})}]{H65}
Hill, R., 1965{\natexlab{b}}. A self consistent mechanics of composite
  materials. Journal of the Mechanics and Physics of Solids 13, 213--222.

\bibitem[{Hill(1966)}]{Hill196695}
Hill, R., 1966. Generalized constitutive relations for incremental deformation
  of metal crystals by multislip. Journal of the Mechanics and Physics of
  Solids 14, 95--102.

\bibitem[{Hill and Rice(1972)}]{Hill1972401}
Hill, R., Rice, J., 1972. Constitutive analysis of elastic-plastic crystals at
  arbitrary strain. Journal of the Mechanics and Physics of Solids 20,
  401--413.

\bibitem[{Hirth and Lothe(1982)}]{hirth1982theory}
Hirth, J., Lothe, J., 1982. Theory of Dislocations. Krieger Publishing Company.

\bibitem[{Hong et~al.(2016)Hong, Godfrey, and Liu}]{HGL16}
Hong, X., Godfrey, A., Liu, W., 2016. Challenges in the prediction of twin
  transmission at grain boundaries in a magnesium alloy. Scripta Materialia
  123, 77 -- 80.

\bibitem[{Houtte(1978)}]{HOUTTE1978591}
Houtte, P., 1978. Simulation of the rolling and shear texture of brass by the
  taylor theory adapted for mechanical twinning. Acta Metallurgica 26, 591 --
  604.

\bibitem[{Hu et~al.(1992)Hu, Rauch, and Teodosiu}]{HU1992839}
Hu, Z., Rauch, E.~F., Teodosiu, C., 1992. Work-hardening behavior of mild steel
  under stress reversal at large strains. International Journal of Plasticity
  8, 839 -- 856.

\bibitem[{Hutchinson(1976)}]{Hutchinson1976101}
Hutchinson, J., 1976. Bounds and self-consistent estimates for creep of
  polycrystalline materials. Proceedings of the Royal Society of London A 348,
  101--127.

\bibitem[{Idiart et~al.(2006)Idiart, Moulinec, and {Ponte Casta\~neda}}]{IMP06}
Idiart, M.~I., Moulinec, H., {Ponte Casta\~neda}, P., 2006. Macroscopic
  behavior and field fluctuations in viscoplastic composites: Second-order
  estimates versus full-field simulations. Journal of the Mechanics and Physics
  of Solids 54, 1029--1063.

\bibitem[{Jeulin et~al.(2004)Jeulin, Kanit, and Forest}]{JKF04}
Jeulin, D., Kanit, T., Forest, S., 2004. Representative volume element: A
  statistical point of view. In: Bergman, D.~J., Inan, E. (Eds.), Continuum
  Models and Discrete Systems. NATO Science Book Series. Springer, pp. 21--27.

\bibitem[{J.Jung et~al.(2017)J.Jung, Yoon, Kim, Latypov, Kim, and
  H.S}]{Jung2017}
J.Jung, Yoon, J.~I., Kim, J., Latypov, M., Kim, J., H.S, K., 2017. Continuum
  understanding of twin formation near grain boundaries of fcc metals with low
  stacking fault energy. npj Computational Materials 3, 21.

\bibitem[{Joseph et~al.(2010)Joseph, Chakraborty, and Ghosh}]{Joseph2010}
Joseph, D.~S., Chakraborty, P., Ghosh, S., 2010. Wavelet transformation based
  multi-time scaling method for crystal plasticity {FE} simulations under
  cyclic loading. Computer Methods in Applied Mechanics and Engineering 199,
  2177 -- 2194.

\bibitem[{Kabel et~al.(2014)Kabel, B\"{o}hlke, and Schneider}]{Kabel2014}
Kabel, M., B\"{o}hlke, T., Schneider, M., 2014. Efficient fixed point and
  newton---krylov solvers for fft-based homogenization of elasticity at large
  deformations. Computational Mechanics 54, 1497--1514.

\bibitem[{Kalidindi et~al.(1992)Kalidindi, Bronkhorst, and
  Anand}]{Kalidindi1992537}
Kalidindi, S., Bronkhorst, C., Anand, L., 1992. Crystallographic texture
  evolution in bulk deformation processing of fcc metals. Journal of the
  Mechanics and Physics of Solids 40, 537--569.

\bibitem[{Kalidindi(1998)}]{K98}
Kalidindi, S.~R., 1998. Incorporation of deformation twinning in crystal
  plasticity models. Journal of the Mechanics and Physics of Solids 46,
  267--271 and 273--290.

\bibitem[{Kalidindi et~al.(2010)Kalidindi, Knezavic, Levinson, Harris, Mishra,
  and Doherty}]{AZ31_Kalidindi}
Kalidindi, S.~R., Knezavic, M., Levinson, A., Harris, R., Mishra, R.~J.,
  Doherty, R.~D., 2010. Deformation twinning in {AZ31}: influence on strain
  hardening and texture evolution. Acta Materialia 58, 6230--6242.

\bibitem[{Khan et~al.(2016)Khan, Zahedi, and Siddiqui}]{KHAN2016772}
Khan, R., Zahedi, F.~I., Siddiqui, A.~K., 2016. Numerical modeling of twinning
  induced plasticity in austenite based advanced high strength steels. Procedia
  Manufacturing 5, 772-- -- 786.

\bibitem[{Kiener et~al.(2008)Kiener, Grosinger, Dehm, and
  Pippan}]{KIENER2008580}
Kiener, D., Grosinger, W., Dehm, G., Pippan, R., 2008. A further step towards
  an understanding of size-dependent crystal plasticity: In situ tension
  experiments of miniaturized single-crystal copper samples. Acta Materialia
  56, 580 -- 592.

\bibitem[{Knezevic and Beyerlein(2018)}]{ZB18}
Knezevic, M., Beyerlein, I.~J., 2018. Multiscale modeling of
  microstructure-property relationships of polycrystalline metals during
  thermo-mechanical deformation. Advanced Engineering Materials, 1700956.

\bibitem[{Knezevic et~al.(2016{\natexlab{a}})Knezevic, Crapps, Beyerlein,
  Coughlin, Clarke, and McCabe}]{KCB15}
Knezevic, M., Crapps, J., Beyerlein, I.~J., Coughlin, D.~R., Clarke, K.~D.,
  McCabe, R.~J., 2016{\natexlab{a}}. Anisotropic modeling of structural
  components using embedded crystal plasticity constructive laws within finite
  elements. International Journal of Mechanical Sciences 105, 227 -- 238.

\bibitem[{Knezevic et~al.(2016{\natexlab{b}})Knezevic, Zecevic, Beyerlein, and
  Lebensohn}]{KZB16}
Knezevic, M., Zecevic, M., Beyerlein, I.~J., Lebensohn, R.~A.,
  2016{\natexlab{b}}. A numerical procedure enabling accurate descriptions of
  strain rate-sensitive flow of polycrystals within crystal visco-plasticity
  theory. Computational Methods in Applied Mechanics and Engineering 308,
  468--482.

\bibitem[{Kocks(1975)}]{K75}
Kocks, W., 1975. Thermodynamics and kinetics of slip. Progress in Materials
  Science 19, 1 -- 291.

\bibitem[{Konrad et~al.(2006)Konrad, Zaefferer, and Raabe}]{KZR06}
Konrad, J., Zaefferer, S., Raabe, D., 2006. Investigation of orientation
  gradients around a hard laves particle in a warm-rolled fe3al-based alloy
  using a 3d ebsd-fib technique. Acta Materialia 54, 1369 -- 1380.

\bibitem[{Korsunsky et~al.(2007)Korsunsky, Dini, Dunne, and
  Walsh}]{KORSUNSKY20071990}
Korsunsky, A.~M., Dini, D., Dunne, F.~P., Walsh, M.~J., 2007. Comparative
  assessment of dissipated energy and other fatigue criteria. International
  Journal of Fatigue 29, 1990 -- 1995.

\bibitem[{Kothari and Anand(1998)}]{KOTHARI199851}
Kothari, M., Anand, L., 1998. Elasto-viscoplastic constitutive equations for
  polycrystalline metals: Application to tantalum. Journal of the Mechanics and
  Physics of Solids 46, 51 -- 83.

\bibitem[{Kothary and Anand(1998)}]{Hardening_Anand}
Kothary, M., Anand, L., 1998. Elasto-viscoplastic constitutive equations for
  polycrystalline metals: {A}pplication to tantalum. Journal of the Mechanics
  and Physics of Solids 46, 51-- 67,69 -- 83.

\bibitem[{Kouznetsova and Geers(2008)}]{KG08}
Kouznetsova, V.~G., Geers, M. G.~D., 2008. A multi-scale model of martensitic
  transformation plasticity. Mechanics of Materials 40, 641--657.

\bibitem[{Kowalczyk-Gajewska(2010)}]{KOWALCZYKGAJEWSKA201028}
Kowalczyk-Gajewska, K., 2010. Modelling of texture evolution in metals
  accounting for lattice reorientation due to twinning. European Journal of
  Mechanics - A/Solids 29, 28 -- 41.

\bibitem[{Kreher(1990)}]{K90}
Kreher, W., 1990. Residual-stresses and stored elastic energy of composites and
  polycrystals. Journal of the Mechanics and Physics of Solids 38, 115--128.

\bibitem[{Kr\"oner(1958)}]{K58}
Kr\"oner, E., 1958. {Berechnung der Elastischen Konstanten des Vielkristalls
  aus den Konstanten des Einkristalls}. Zeitung Physik 51, 504--518.

\bibitem[{Kumar et~al.(2016)Kumar, Beyerlein, McCabe, and Tom\'e}]{KBM16}
Kumar, M.~A., Beyerlein, I.~J., McCabe, R.~J., Tom\'e, C.~N., 2016. Grain
  neighbour effects on twin transmission in hexagonal close packed materials.
  Nature Communications 7, 13826.

\bibitem[{Lahellec et~al.(2001)Lahellec, Michel, Moulinec, and Suquet}]{LMM01}
Lahellec, N., Michel, J.~C., Moulinec, H., Suquet, P., 2001. Analysis of
  inhomogeneous materials at large strains using fast {F}ourier transforms. In:
  Miehe, C. (Ed.), IUTAM Symposium on computational mechanics of solids
  materials. Kluwer Academic, pp. 247--268.

\bibitem[{Lautensack(2007)}]{L2007}
Lautensack, C., 2007. Random {L}aguerre tessellations. {PhD} thesis,
  UniversitŠt Karlsruhe.

\bibitem[{Laws(1973)}]{L73}
Laws, N., 1973. On the thermostatics of composite materials. Journal of the
  Mechanics and Physics of Solids 21, 9--17.

\bibitem[{Lebensohn(2001)}]{L01}
Lebensohn, R.~A., 2001. N-site modelling of a 3d viscoplastic polycrystal using
  {Fast Fourier Transform}. Acta Materialia 49, 2723--2737.

\bibitem[{Lebensohn et~al.(2008)Lebensohn, Brenner, Castelnau, and
  Rollet}]{LBC08}
Lebensohn, R.~A., Brenner, R., Castelnau, O., Rollet, A.~D., 2008. Orientation
  image-based micromechanical modelling of subgrain texture evolution in
  polycrystalline copper. Acta Materialia 56, 3914--3926.

\bibitem[{Lebensohn and Canova(1997)}]{LC97}
Lebensohn, R.~A., Canova, G.~R., 1997. Selfconsistent approach for modelling
  texture development of two-phase polycrystals: application to {T}itanium
  alloys. Acta Materialia 45, 3687 -- 3694.

\bibitem[{Lebensohn et~al.(2005)Lebensohn, Castelnau, Brenner, and
  Gilormini}]{LCB05}
Lebensohn, R.~A., Castelnau, O., Brenner, R., Gilormini, P., 2005. Study of the
  antiplane deformation of linear {2-D} polycrystals with different
  microstructures. International Journal of Solids and Structures 42,
  5441--5449.

\bibitem[{Lebensohn et~al.(2013)Lebensohn, Escobedo, Cerreta, Dennis-Koller,
  Bronkhorst, and Bingert}]{LEC13}
Lebensohn, R.~A., Escobedo, J.~P., Cerreta, E.~K., Dennis-Koller, D.,
  Bronkhorst, C.~A., Bingert, J., 2013. Modelling void growth in
  polycrystalline materials. Acta Materialia 61, 6918--6932.

\bibitem[{Lebensohn et~al.(2010)Lebensohn, Hartley, Tom\'e, and
  Castelnau}]{LHT10}
Lebensohn, R.~A., Hartley, C.~S., Tom\'e, C., Castelnau, O., 2010. Modelling
  the mechanical response of polycrystals deforming by climb and glide.
  Philosophical Magazine 90, 567--583.

\bibitem[{Lebensohn et~al.(2011)Lebensohn, Idiart, {Ponte-Casta{\~n}eda}, and
  Vincent}]{LIP11}
Lebensohn, R.~A., Idiart, M.~I., {Ponte-Casta{\~n}eda}, P., Vincent, P.~G.,
  2011. Dilatational viscoplasticity of polycrystalline solids with
  intergranular cavities. Philosophical Magazine 91, 3038--3067.

\bibitem[{Lebensohn et~al.(2012)Lebensohn, Kanjarla, and Eisenlohr}]{LKE12}
Lebensohn, R.~A., Kanjarla, A.~K., Eisenlohr, P., 2012. An elasto-viscoplastic
  formulation based on fast fourier transforms for the prediction of
  micromechanical fields in polycrystalline materials. International Journal of
  Plasticity 32-33, 59--69.

\bibitem[{Lebensohn et~al.(2009)Lebensohn, Montagnat, Mansuy, Duval,
  Meysonnier, and Philip}]{LMM09}
Lebensohn, R.~A., Montagnat, M., Mansuy, P., Duval, P., Meysonnier, J., Philip,
  A., 2009. Modeling viscoplastic behavior and heterogenous intracrystalline
  deformation of columnar ice polycrystals. Acta Materialia 57, 1405--1415.

\bibitem[{Lebensohn and Needleman(2016)}]{LN16}
Lebensohn, R.~A., Needleman, A., 2016. Numerical implementation of non-local
  polycrystal plasticity using {Fast Fourier Transforms}. Journal of the
  Mechanics and Physics of Solids 97, 333--351.

\bibitem[{Lebensohn and Tom\'e(1993)}]{LT93}
Lebensohn, R.~A., Tom\'e, C.~N., 1993. A self-consistent anisotropic approach
  for the simulation of plastic deformation and texture development of
  polycrystals: Application to zirconium alloys. Acta Metallurgica et
  Materialia 41, 2611 -- 2624.

\bibitem[{Lebensohn et~al.(2007)Lebensohn, Tom\'e, and
  {Ponte-Casta{\~n}eda}}]{LTP07}
Lebensohn, R.~A., Tom\'e, C.~N., {Ponte-Casta{\~n}eda}, P., 2007.
  Self-consistent modeling of the mechanical behavior of viscoplastic
  polycrystals incorporating intragranular field fluctuations. Philosophical
  Magazine 87, 4287 -- 4322.

\bibitem[{Lebensohn et~al.(1998)Lebensohn, Uhlenhut, Hartig, and
  Mecking}]{LUH98}
Lebensohn, R.~A., Uhlenhut, H., Hartig, C., Mecking, H., 1998. Mechanical
  behavior gamma-{TiAl}-based polysinthetically twinned crystals:
  micromechanical modelling and experimental validation. Acta Materialia 46,
  4701 -- 4709.

\bibitem[{Lebensohn et~al.(2016)Lebensohn, Zecevic, Knezevic, and
  {McC}abe}]{LZK16}
Lebensohn, R.~A., Zecevic, M., Knezevic, M., {McC}abe, R.~J., 2016. Average
  intragranular misorientation trends in polycrystalline materials predicted by
  a viscoplastic self-consistent approach. Acta Materialia 104, 228--236.

\bibitem[{Lecarme et~al.(2014)Lecarme, Maire, Kumar, {De Vleeschouwer},
  Jacques, Simar, and Pardoen}]{LMK14}
Lecarme, L., Maire, E., Kumar, K. C.~A., {De Vleeschouwer}, C., Jacques, L.,
  Simar, A., Pardoen, T., 2014. Heterogenous void growth revealed by in situ
  {3-D X}-ray microtomography using automatic cavity tracking. Acta Materialia
  63, 130 -- 139.

\bibitem[{Lee and Liu(1967)}]{Lee196719}
Lee, E.~H., Liu, D.~T., 1967. Finite-strain elastic - plastic theory with
  application to plane-wave analysis. Journal of Applied Physics 38, 19--27.

\bibitem[{Lele and Anand(2008)}]{Lele2008}
Lele, S.~P., Anand, L., 2008. A small-deformation strain-gradient theory for
  isotropic viscoplastic materials. Philosophical Magazine 88, 3655--3689.

\bibitem[{Li et~al.(2016)Li, Sun, and Yang}]{JSY16}
Li, H., Sun, X., Yang, H., 2016. A three-dimensional cellular automata-crystal
  plasticity finite element model for predicting the multiscale interaction
  among heterogeneous deformation, drx microstructural evolution and mechanical
  responses in titanium alloys. International Journal of Plasticity 87, 154 --
  180.

\bibitem[{Lim et~al.(2016)Lim, Abdeljawad, Owen, Hanks, Foulk, and
  Battaile}]{LAO16}
Lim, H., Abdeljawad, F., Owen, S.~J., Hanks, B.~W., Foulk, J.~W., Battaile,
  C.~C., 2016. Incorporating physically-based microstructures in materials
  modeling: Bridging phase field and crystal plasticity frameworks. Modelling
  and Simulation in Materials Science and Engineering 24, 045016.

\bibitem[{Liu et~al.(2018)Liu, Shanthraj, Diehl, Roters, Dong, Dong, Ding, and
  Raabe}]{LSD18}
Liu, C., Shanthraj, P., Diehl, M., Roters, F., Dong, S., Dong, J., Ding, W.,
  Raabe, D., 2018. An integrated crystal plasticityÐphase field model for
  spatially resolved twin nucleation, propagation, and growth in hexagonal
  materials. International Journal of Plasticity 106, 203 -- 227.

\bibitem[{Liu et~al.(2017)Liu, Li, {Arul Kumar}, Pathak, Wang, McCabe, Mara,
  and Tom\'e}]{LLA17}
Liu, Y., Li, N., {Arul Kumar}, M., Pathak, S., Wang, J., McCabe, R.~J., Mara,
  N.~A., Tom\'e, C.~N., 2017. Experimentally quantifying critical stresses
  associated with basal slip and twinning in magnesium using micropillars. Acta
  Materialia 135, 411-- 421.

\bibitem[{Liu and Ponte-Casta{\~n}eda(2004)}]{LP04}
Liu, Y., Ponte-Casta{\~n}eda, P., 2004. Second-order theory for the effective
  behavior and field fluctuations of polycrystals. Journal of the Mechanics and
  Physics of Solids 52, 467--495.

\bibitem[{LLorca et~al.(2011)LLorca, Gonz{\'a}lez, Molina-Aldaregu{\'i}a,
  Segurado, Seltzer, Sket, Rodr{\'i}guez, S{\'a}daba, Mu{\~n}oz, and
  Canal}]{LGM11}
LLorca, J., Gonz{\'a}lez, C., Molina-Aldaregu{\'i}a, J.~M., Segurado, J.,
  Seltzer, R., Sket, F., Rodr{\'i}guez, M., S{\'a}daba, S., Mu{\~n}oz, R.,
  Canal, L.~P., 2011. Multiscale modeling of composite materials: a roadmap
  towards virtual testing. Advanced Materials 23, 5130--5147.

\bibitem[{LLorca and Segurado(2004)}]{SL04}
LLorca, J., Segurado, J., 2004. Three-dimensional multiparticle cell
  simulations of deformation and damage in sphere-reinforced composites.
  Materials Science and Engineering A 365, 267--274.

\bibitem[{Lucarini and Segurado(2018)}]{LS18}
Lucarini, S., Segurado, J., 2018. On the accuracy of spectral solvers for
  micromechanics based fatigue modeling. Computational Mechanics.

\bibitem[{Ma and Roters(2004)}]{MA20043603}
Ma, A., Roters, F., 2004. A constitutive model for fcc single crystals based on
  dislocation densities and its application to uniaxial compression of
  aluminium single crystals. Acta Materialia 52, 3603 -- 3612.

\bibitem[{Ma et~al.(2006)Ma, Roters, and Raabe}]{Ma20062169}
Ma, A., Roters, F., Raabe, D., 2006. A dislocation density based constitutive
  model for crystal plasticity fem including geometrically necessary
  dislocations. Acta Materialia 54, 2169--2179.

\bibitem[{Mahajan and Williams(1973)}]{MW73}
Mahajan, S., Williams, D.~F., 1973. Deformation twinning in metals and alloys.
  International Metallurgical Reviews 18, 43--61.

\bibitem[{Malyar et~al.(2017)Malyar, Micha, Dehm, and Kirchlechner}]{MMD17}
Malyar, N., Micha, J., Dehm, G., Kirchlechner, C., 2017. Size effect in
  bi-crystalline micropillars with a penetrable high angle grain boundary. Acta
  Materialia 129, 312 -- 320.

\bibitem[{Mandal et~al.(2018)Mandal, Lao, Donegan, and Rollett}]{MLD18}
Mandal, S., Lao, J., Donegan, S., Rollett, A.~D., 2018. Generation of
  statistically representative synthetic three-dimensional microstructures.
  Scripta Materialia 146, 128 -- 132.

\bibitem[{Manonukul and Dunne(2004)}]{Manonukul2004}
Manonukul, A., Dunne, F. P.~E., 2004. High{\textendash} and
  low{\textendash}cycle fatigue crack initiation using polycrystal plasticity.
  Proceedings of the Royal Society of London A: Mathematical, Physical and
  Engineering Sciences 460, 1881--1903.

\bibitem[{Mareau and Daymond(2016)}]{MD16}
Mareau, C., Daymond, M.~R., 2016. Micromechanical modelling of twinning in
  polycrystalline materials: Application to magnesium. International Journal of
  Plasticity 85, 156 -- 171.

\bibitem[{Masson et~al.(2000)Masson, Bornert, Suquet, and Zaoui}]{MBS00}
Masson, R., Bornert, M., Suquet, P., Zaoui, A., 2000. Affine formulation for
  the prediction of the effective properties of nonlinear composites and
  polycrystals. Journal of the Mechanics and Physics of Solids 48, 1203--1227.

\bibitem[{Mayeur et~al.(2015)Mayeur, Beyerlein, Bronkhorst, and Mourad}]{MBB15}
Mayeur, J., Beyerlein, I., Bronkhorst, C., Mourad, H., 2015. Incorporating
  interface affected zones into crystal plasticity. International Journal of
  Plasticity 65, 206 -- 225.

\bibitem[{McColl et~al.(2004)McColl, Ding, and Leen}]{MCCOLL20041114}
McColl, I.~R., Ding, J., Leen, S.~B., 2004. Finite element simulation and
  experimental validation of fretting wear. Wear 256, 1114 -- 1127.

\bibitem[{McDowell and Dunne(2010)}]{MCDOWELL20101521}
McDowell, D.~L., Dunne, F. P.~E., 2010. Microstructure-sensitive computational
  modeling of fatigue crack formation. International Journal of Fatigue 32,
  1521 -- 1542.

\bibitem[{McDowell et~al.(2003)McDowell, Gall, Horstemeyer, and
  Fan}]{MCDOWELL200349}
McDowell, D.~L., Gall, K., Horstemeyer, M.~F., Fan, J., 2003.
  Microstructure-based fatigue modeling of cast a356-t6 alloy. Engineering
  Fracture Mechanics 70, 49 -- 80.

\bibitem[{Mecking and Kocks(1981)}]{MECKING19811865}
Mecking, H., Kocks, U., 1981. Kinetics of flow and strain-hardening. Acta
  Metallurgica 29, 1865 -- 1875.

\bibitem[{M\'eric et~al.(1991)M\'eric, Poubanne, and
  Cailletaud}]{CAILLETAUD1991}
M\'eric, L., Poubanne, P., Cailletaud, G., 1991. Single crystal modeling for
  structural calculations: Part 1 model presentation. ASME Journal of
  Engineering Materials and Technology 113, 162--170.

\bibitem[{Michel et~al.(1999)Michel, Moulinec, and Suquet}]{MMS99}
Michel, J.~C., Moulinec, H., Suquet, P., 1999. Effective properties of
  composite materials with periodic microstructure: a computational approach.
  Compuational Methods in Applied Mechanics and Engineering 172, 109--143.

\bibitem[{Michel et~al.(2000)Michel, Moulinec, and Suquet}]{MMS00}
Michel, J.~C., Moulinec, H., Suquet, P., 2000. A computational method based on
  augmented lagrangians and fast {F}ourier transforms for composites with high
  contrast. Computer Modeling in Engineering and Sciences 1, 79--88.

\bibitem[{Michel and Suquet(2016)}]{Michel2016}
Michel, J.-C., Suquet, P., Mar 2016. A model-reduction approach to the
  micromechanical analysis of polycrystalline materials. Computational
  Mechanics 57, 483--508.

\bibitem[{Miehe(1996)}]{Miehe19963367}
Miehe, C., 1996. Exponential map algorithm for stress updates in anisotropic
  multiplicative elastoplasticity for single crystals. International Journal
  for Numerical Methods in Engineering 39, 3367--3390.

\bibitem[{Miehe et~al.(2002)Miehe, Schotte, and Lambrecht}]{Miehe20022123}
Miehe, C., Schotte, J., Lambrecht, M., 2002. Homogenization of inelastic solid
  materials at finite strains based on incremental minimization principles.
  application to the texture analysis of polycrystals. Journal of the Mechanics
  and Physics of Solids 50, 2123 -- 2167.

\bibitem[{Miehe et~al.(1999)Miehe, Schro\"der, and Schotte}]{MSS99}
Miehe, C., Schro\"der, J., Schotte, J., 1999. Computational homogenization
  analysis in finite plasticity simulation of texture development in
  polycrystalline materials. Computational Methods in Applied Mechanics and
  Engineering 171, 387--418.

\bibitem[{Mika and Dawson(1999)}]{MIKA19991355}
Mika, D.~P., Dawson, P.~R., 1999. Polycrystal plasticity modeling of
  intracrystalline boundary textures. Acta Materialia 47, 1355 -- 1369.

\bibitem[{Molinari et~al.(1987{\natexlab{a}})Molinari, Canova, and
  Ahzi}]{MCA87}
Molinari, A., Canova, G.~R., Ahzi, S., 1987{\natexlab{a}}. Self-consistent
  approach of the large deformation polycrystal viscoplasticity. Acta
  Metallurgica 35, 2983--2994.

\bibitem[{Molinari et~al.(1987{\natexlab{b}})Molinari, Canova, and
  Ahzi}]{Molinari19872983}
Molinari, A., Canova, G.~R., Ahzi, S., 1987{\natexlab{b}}. A self consistent
  approach of the large deformation polycrystal viscoplasticity. Acta
  Metallurgica 35, 2983--2994.

\bibitem[{Monchiet and Bonnet(2013)}]{MONCHIET2013276}
Monchiet, V., Bonnet, G., 2013. Numerical homogenization of nonlinear
  composites with a polarization-based fft iterative scheme. Computational
  Materials Science 79, 276 -- 283.

\bibitem[{Mori and Tanaka(1973)}]{MT73}
Mori, T., Tanaka, K., 1973. Average stress in the matrix and average elastic
  energy of materials with misfitting inclusions. Acta Metallurgica et
  Materialia 21, 571--574.

\bibitem[{Motz et~al.(2005)Motz, Schoeberl, and Pippan}]{MOTZ20054269}
Motz, C., Schoeberl, T., Pippan, R., 2005. Mechanical properties of micro-sized
  copper bending beams machined by the focused ion beam technique. Acta
  Materialia 53, 4269 -- 4279.

\bibitem[{Moulinec and Silva(2014)}]{MS14}
Moulinec, H., Silva, F., 2014. Comparison of three accelerated {FFT}-based
  schemes for computing the mechanical response of composite materials.
  International Journal of Numerical Methods in Engineering 97, 960--985.

\bibitem[{Moulinec and Suquet(1994)}]{MS94}
Moulinec, H., Suquet, P., 1994. A fast numerical method for computing the
  linear and nonlinear mechanical properties of composites. Comptes Rendus de
  l'Acad\'emie des Sciences 318, 1417--1423.

\bibitem[{Moulinec and Suquet(1998)}]{MS98}
Moulinec, H., Suquet, P., 1998. A numerical method for computing the overall
  response of nonlinear composites with complex microstructure. Compuational
  Methods in Applied Mechanics and Engineering 157, 69--94.

\bibitem[{Mueller et~al.(2017)Mueller, Kusne, and Ramprasad}]{MKR17}
Mueller, T., Kusne, A., Ramprasad, R., 2017. Machine learning in materials
  science: Recent progress and emerging applications. Reviews in Computational
  Chemistry 29, 1866 -- 273.

\bibitem[{Mura(1987)}]{M87}
Mura, T., 1987. Micromechanics of defects in solids. Martinus-Nijhoff
  Publishers.

\bibitem[{Musinski and McDowell(2012)}]{Musinski2012}
Musinski, W.~D., McDowell, D.~L., 2012. Microstructure-sensitive probabilistic
  modeling of {HCF} crack initiation and early crack growth in {Ni}-base
  superalloy {IN100} notched components. International Journal of Fatigue 37,
  41 -- 53.

\bibitem[{Musinski and McDowell(2016)}]{MM16}
Musinski, W.~D., McDowell, D.~L., 2016. Simulating the effect of grain
  boundaries on microstructurally small fatigue crack growth from a focused ion
  beam notch through a three-dimensional array of grains. Acta Materialia 112,
  20 -- 39.

\bibitem[{Nabarro(1947)}]{Nabarro1947}
Nabarro, F. R.~N., 1947. Dislocations in a simple cubic lattice. Proceedings of
  the Physical Society 59, 256.

\bibitem[{Nayyeri et~al.(2017)Nayyeri, Poole, Sinclair, and Zaefferer}]{NPS17}
Nayyeri, G., Poole, W.~J., Sinclair, C.~W., Zaefferer, S., 2017. The role of
  indenter radius on spherical indentation of high purity magnesium loaded
  nearly parallel to the c-axis. Scripta Materialia 137, 119 -- 122.

\bibitem[{Nemat-Nasser and Hori(1993)}]{NH93}
Nemat-Nasser, S., Hori, M., 1993. Micromechanics: Overall Properties of
  Heterogeneous Materials. Elsevier Science Publishers.

\bibitem[{Nemcko and Wilkinson(2016)}]{NW16}
Nemcko, M.~J., Wilkinson, D.~S., 2016. On the damage and fracture of
  commercially pure magnesium using x-ray microtomography. Materials Science
  and Engineering: A 676, 146 -- 155.

\bibitem[{Neper(2018)}]{Neper}
Neper, 2018. \url{http://neper.sourceforge.net}.

\bibitem[{Nguyen et~al.(2012)Nguyen, Béchet, Geuzaine, and
  Noels}]{NGUYEN2012390}
Nguyen, V.-D., Béchet, E., Geuzaine, C., Noels, L., 2012. Imposing periodic
  boundary condition on arbitrary meshes by polynomial interpolation.
  Computational Materials Science 55, 390 -- 406.

\bibitem[{Niezgoda et~al.(2010)Niezgoda, Turner, Fullwood, and
  Kalidindi}]{NTF10}
Niezgoda, S.~R., Turner, D.~M., Fullwood, D.~T., Kalidindi, S.~R., 2010.
  Optimized structure based representative volume element sets reflecting the
  ensemble-averaged 2-point statistics. Acta Materialia 58, 4432 -- 4445.

\bibitem[{Niordson and Kysar(2014)}]{NIORDSON201431}
Niordson, C.~F., Kysar, J.~W., 2014. Computational strain gradient crystal
  plasticity. Journal of the Mechanics and Physics of Solids 62, 31 -- 47.

\bibitem[{Nix and Gao(1998)}]{NIX1998411}
Nix, W.~D., Gao, H., 1998. Indentation size effects in crystalline materials: A
  law for strain gradient plasticity. Journal of the Mechanics and Physics of
  Solids 46, 411 -- 425.

\bibitem[{Nye(1953)}]{Nye1953-1}
Nye, J.~F., 1953. Some geometrical relations in dislocated crystals. Acta
  metallurgica 1, 153--162.

\bibitem[{Ohno and Wang(1993)}]{Ohno1993}
Ohno, N., Wang, J.-D., 1993. Kinematic hardening rules with critical state of
  dynamic recovery, part {II: A}pplication to experiments of ratcheting
  behavior. International Journal of Plasticity 9, 391 -- 403.

\bibitem[{Olson(2013)}]{O13}
Olson, G.~B., 2013. Genomic materials design: The ferrous frontier. Acta
  Materialia 61, 771--781.

\bibitem[{Orowan(1934)}]{Orowan1934}
Orowan, E., 1934. Zur kristallplastizit\"at. Zeitschrift f\"ur Physik 89, 605
  -- 613.

\bibitem[{Ostien and Garikipati(2008)}]{Ostien2008}
Ostien, J., Garikipati, K., 2008. Discontinuous galerkin method for an
  incompatibility-based strain gradient plasticity theory. In: Reddy, B.~D.
  (Ed.), IUTAM Symposium on Theoretical, Computational and Modelling Aspects of
  Inelastic Media. Springer, pp. 217--226.

\bibitem[{Ostoja-Starzewski(2006)}]{OSTOJASTARZEWSKI2006112}
Ostoja-Starzewski, M., 2006. Material spatial randomness: From statistical to
  representative volume element. Probabilistic Engineering Mechanics 21, 112 --
  132.

\bibitem[{Otsuka et~al.(2018)Otsuka, Brenner, and Bacroix}]{OBB18}
Otsuka, T., Brenner, R., Bacroix, B., 2018. Fft-based modelling of
  transformation plasticity in polycrystalline materials during diffusive phase
  transformation. International Journal of Engineering Science 127, 92 -- 113.

\bibitem[{Ozturk et~al.(2016)Ozturk, Shahba, and Ghosh}]{OSG16}
Ozturk, D., Shahba, A., Ghosh, S., 2016. Crystal plasticity fe study of the
  effect of thermo-mechanical loading on fatigue crack nucleation in titanium
  alloys. Fatigue and Fracture of Engineering Materials and Structures 39,
  752--769.

\bibitem[{Panteghihi and Bardella(2016)}]{Panteghini2016}
Panteghihi, A., Bardella, L., 2016. On the finite element implementation of
  higher-order gradient plasticity, with focus on theories based on plastic
  distortion incompatibility. Computer Methods in Applied Mechanics and
  Engineering 310, 810--865.

\bibitem[{Parton and Buryachenko(1990)}]{PB90}
Parton, V.~Z., Buryachenko, V.~A., 1990. Stress fluctuations in elastic
  composites. Soviet Physics Doklady 35, 191--193.

\bibitem[{Patel and Kalidindi(2017)}]{PK17}
Patel, D.~K., Kalidindi, S.~R., 2017. Estimating the slip resistance from
  spherical nanoindentation and orientation measurements in polycrystalline
  samples of cubic metals. International Journal of Plasticity 92, 19--30.

\bibitem[{Pei et~al.(2011)Pei, Yun-Chang, and Qing}]{AZ31_Liu_Pei}
Pei, L., Yun-Chang, X., Qing, L., 2011. Plastic anisotropy and fracture
  behaviour of {AZ31} magnesium alloy. Transactions of Nonferrous Metals
  Society of China 21, 880--884.

\bibitem[{Peierls(1940)}]{Peierls1940}
Peierls, R., 1940. The size of a dislocation. Proceedings of the Physical
  Society 52, 34.

\bibitem[{Peirce(1983)}]{PEIRCE1983133}
Peirce, D., 1983. Shear band bifurcations in ductile single crystals. Journal
  of the Mechanics and Physics of Solids 31, 133 -- 153.

\bibitem[{Peirce et~al.(1982{\natexlab{a}})Peirce, Asaro, and
  Needleman}]{PAN82}
Peirce, D., Asaro, R.~J., Needleman, A., 1982{\natexlab{a}}. An analysis of
  nonuniform and localized deformation in ductile single crystals. Acta
  Metallurgica 30, 1087--1119.

\bibitem[{Peirce et~al.(1982{\natexlab{b}})Peirce, Asaro, and
  Needleman}]{PEIRCE19821087}
Peirce, D., Asaro, R.~J., Needleman, A., 1982{\natexlab{b}}. An analysis of
  nonuniform and localized deformation in ductile single crystals. Acta
  Metallurgica 30, 1087 -- 1119.

\bibitem[{Peirce et~al.(1983)Peirce, Asaro, and Needleman}]{PEIRCE19831951}
Peirce, D., Asaro, R.~J., Needleman, A., 1983. Material rate dependence and
  localized deformation in crystalline solids. Acta Metallurgica 31, 1951 --
  1976.

\bibitem[{Peric et~al.(2011)Peric, de~Souza~Neto, Feijóo, Partovi, and
  Molina}]{Peric2011}
Peric, D., de~Souza~Neto, E.~A., Feijóo, R.~A., Partovi, M., Molina, A. J.~C.,
  2011. On micro-to-macro transitions for multi-scale analysis of non-linear
  heterogeneous materials: unified variational basis and finite element
  implementation. International Journal for Numerical Methods in Engineering
  87, 149--170.

\bibitem[{Petch(1953)}]{Petch1953-1}
Petch, N.~J., 1953. The cleavage strength of polycrystals. Journal of the Iron
  and Steel Institute 174, 25--28.

\bibitem[{Pineau et~al.(2016)Pineau, McDowell, Busso, and
  Antolovich}]{PINEAU2016484}
Pineau, A., McDowell, D.~L., Busso, E.~P., Antolovich, S.~D., 2016. Failure of
  metals {II: F}atigue. Acta Materialia 107, 484 -- 507.

\bibitem[{{Ponte-Casta{\~n}eda}(2002)}]{P02}
{Ponte-Casta{\~n}eda}, P., 2002. Second-order homogenization estimates for
  nonlinear composites incorporating field fluctuations: {I- T}heory. Journal
  of the Mechanics and Physics of Solids 50, 737 -- 757.

\bibitem[{Ponte-Casta{\~n}eda(1991)}]{P91}
Ponte-Casta{\~n}eda, P., 1991. The effective mechanical properties of
  non-linear composites. Journal of the Mechanics and Physics of Solids 37,
  45--71.

\bibitem[{Ponte-Casta{\~n}eda(1996)}]{P96}
Ponte-Casta{\~n}eda, P., 1996. Exact second-order estimates for the effective
  mechanical properties of nonlinear composite materials. Journal of the
  Mechanics and Physics of Solids 44, 1757--1788.

\bibitem[{Proudhon et~al.(2015)Proudhon, Li, Reischig, Gueninchault, Forest,
  and Ludwig}]{PLR15}
Proudhon, H., Li, J., Reischig, P., Gueninchault, N., Forest, S., Ludwig, W.,
  2015. Coupling diffraction contrast tomography with the finite element
  method. Advanced Engineering Materials 18, 903 -- 912.

\bibitem[{Proust et~al.(2007)Proust, Tom\'e, and Kaschner}]{PTK07}
Proust, G., Tom\'e, C., Kaschner, G.~C., 2007. Modeling texture, twinning and
  hardening evolution during deformation of hexagonal materials. Acta
  Materialia 55, 2137--2148.

\bibitem[{Przybyla and McDowell(2012)}]{PRZYBYLA2012293}
Przybyla, C., McDowell, D., 2012. Microstructure-sensitive extreme-value
  probabilities of high-cycle fatigue for surface vs. subsurface crack
  formation in duplex {Ti-6Al-4V}. Acta Materialia 60, 293 -- 305.

\bibitem[{Przybyla and McDowell(2010)}]{Przybyla2010}
Przybyla, C.~P., McDowell, D.~L., 2010. Microstructure-sensitive extreme value
  probabilities for high cycle fatigue of {Ni}-base superalloy {IN100}.
  International Journal of Plasticity 26, 372 -- 394.

\bibitem[{Przybyla et~al.(2013)Przybyla, Musinski, Castelluccio, and
  McDowell}]{Przybyla2013}
Przybyla, C.~P., Musinski, W.~D., Castelluccio, G.~M., McDowell, D.~L., 2013.
  Microstructure-sensitive {HCF} and {VHCF} simulations. International Journal
  of Fatigue 57, 9 -- 27.

\bibitem[{Pushkareva et~al.(2016)Pushkareva, Adrien, Maire, Segurado, Llorca,
  and Weck}]{PAM16}
Pushkareva, M., Adrien, J., Maire, E., Segurado, J., Llorca, J., Weck, A.,
  2016. Three-dimensional investigation of grain orientation effects on void
  growth in commercially pure titanium. Materials Science and Engineering: A
  671, 221 -- 232.

\bibitem[{Raabe et~al.(2001)Raabe, Sachtleber, Zhao, Roters, and
  Zaefferer}]{RSZ01}
Raabe, D., Sachtleber, M., Zhao, Z., Roters, F., Zaefferer, S., 2001.
  Micromechanical and macromechanical effects in grain scale polycrystal
  plasticity experimentation and simulation. Acta Materialia 49, 3433 -- 3441.

\bibitem[{Radavich(1989)}]{R89}
Radavich, J.~F., 1989. The physical metallurgy of cast and wrought alloy 718.
  In: Loria, E.~A. (Ed.), Superalloy 718-Metallurgy and Applications. TMS, pp.
  229--240.

\bibitem[{Redenbach et~al.(2012)Redenbach, Shklyar, and Andr{\"a}}]{R12}
Redenbach, C., Shklyar, I., Andr{\"a}, H., 2012. {L}aguerre tessellations for
  elastic stiffness simulations of closed foams with strongly varying cell
  sizes. International Journal of Engineering Science 50, 70 -- 78.

\bibitem[{Reina and Conti(2014)}]{REINA201440}
Reina, C., Conti, S., 2014. Kinematic description of crystal plasticity in the
  finite kinematic framework: A micromechanical understanding of {F=FeFp}.
  Journal of the Mechanics and Physics of Solids 67, 40 -- 61.

\bibitem[{Reina et~al.(2016)Reina, Schlmerkemper, and Conti}]{REINA2016231}
Reina, C., Schlmerkemper, A., Conti, S., 2016. Derivation of {F=FeFp} as the
  continuum limit of crystalline slip. Journal of the Mechanics and Physics of
  Solids 89, 231 -- 254.

\bibitem[{Rice(1971)}]{Rice1971433}
Rice, J., 1971. Inelastic constitutive relations for solids: An
  internal-variable theory and its application to metal plasticity. Journal of
  the Mechanics and Physics of Solids 19, 433--455.

\bibitem[{Richards et~al.(2013)Richards, Lebensohn, and Bhattacharya}]{RLB13}
Richards, A.~W., Lebensohn, R.~A., Bhattacharya, K., 2013. Interplay of
  martensitic phase transformation and plastic slip in polycrystals. Acta
  Materialia 61, 4384 -- 4397.

\bibitem[{Rodr{\'\i}guez-Gal\'an et~al.(2015)Rodr{\'\i}guez-Gal\'an, Sabirov,
  and Segurado}]{RODRIGUEZGALAN2015191}
Rodr{\'\i}guez-Gal\'an, D., Sabirov, I., Segurado, J., 2015. Temperature and
  stain rate effect on the deformation of nanostructured pure titanium.
  International Journal of Plasticity 70, 191 -- 205.

\bibitem[{Rodr{\'\i}guez-Gal\'an et~al.(2017)Rodr{\'\i}guez-Gal\'an, Segurado,
  and Romero}]{Rodriguez-Galan2017}
Rodr{\'\i}guez-Gal\'an, D., Segurado, J., Romero, I., 2017. working document.

\bibitem[{Roters et~al.(2010{\natexlab{a}})Roters, Eisenlohr, Bieler, and
  Raabe}]{REB10}
Roters, F., Eisenlohr, P., Bieler, T.~R., Raabe, D., 2010{\natexlab{a}}.
  Crystal Plasticity Finite Element Methods: in Materials Science and
  Engineering. Wiley-VCH.

\bibitem[{Roters et~al.(2010{\natexlab{b}})Roters, Eisenlohr, Hantcherli,
  Tjahjanto, Bieler, and Raabe}]{Roters20101152}
Roters, F., Eisenlohr, P., Hantcherli, L., Tjahjanto, D.~D., Bieler, T.~R.,
  Raabe, D., 2010{\natexlab{b}}. Overview of constitutive laws, kinematics,
  homogenization and multiscale methods in crystal plasticity finite-element
  modeling: theory, experiments, applications. Acta Materialia 58, 1152--1211.

\bibitem[{Sachs(1928)}]{S28}
Sachs, G., 1928. On the derivation of a condition of flowing. Verein Deutscher
  Ingenieure 72, 734 -- 736.

\bibitem[{S\'anchez-Mart\'in et~al.(2014{\natexlab{a}})S\'anchez-Mart\'in,
  P\'erez-Prado, Segurado, Bohlen, Guti\'errez-Urrutia, LLorca, and
  Molina-Aldareguia}]{SanchezMartin2014283}
S\'anchez-Mart\'in, R., P\'erez-Prado, M.~T., Segurado, J., Bohlen, J.,
  Guti\'errez-Urrutia, I., LLorca, J., Molina-Aldareguia, J.~M.,
  2014{\natexlab{a}}. Measuring the critical resolved shear stresses in {M}g
  alloys by instrumented nanoindentation. Acta Materialia 71, 283 -- 292.

\bibitem[{S\'anchez-Mart\'in et~al.(2014{\natexlab{b}})S\'anchez-Mart\'in,
  P\'erez-Prado, Segurado, Bohlen, Guti\'errez-Urrutia, LLorca, and
  Molina-Aldareguia}]{SPS14}
S\'anchez-Mart\'in, R., P\'erez-Prado, M.~T., Segurado, J., Bohlen, J.,
  Guti\'errez-Urrutia, I., LLorca, J., Molina-Aldareguia, J.~M.,
  2014{\natexlab{b}}. Measuring the critical resolved shear stresses in {M}g
  alloys by instrumented nanoindentation. Acta Materialia 71, 283 -- 292.

\bibitem[{Sangid(2013)}]{SANGID201358}
Sangid, M.~D., 2013. The physics of fatigue crack initiation. International
  Journal of Fatigue 57, 58 -- 72.

\bibitem[{Sangid et~al.(2011)Sangid, Maier, and Sehitoglu}]{SANGID2011595}
Sangid, M.~D., Maier, H.~J., Sehitoglu, H., 2011. An energy-based
  microstructure model to account for fatigue scatter in polycrystals. Journal
  of the Mechanics and Physics of Solids 59, 595 -- 609.

\bibitem[{Schmidt et~al.(2008)Schmidt, Lauridsen, and Poulsen}]{LSL08}
Schmidt, W. L.~S., Lauridsen, E.~M., Poulsen, H.~F., 2008. X-ray diffraction
  contrast tomography: A novel technique for three-dimensional grain mapping of
  polycrystals. 1. {D}irect beam case. Journal of Applied Crystallography 41,
  302 -- 309.

\bibitem[{Segurado et~al.(2003)Segurado, Gonz\'alez, and LLorca}]{SGL03}
Segurado, J., Gonz\'alez, C., LLorca, J., 2003. A numerical investigation of
  the effect of particle clustering on the mechanical properties of composites.
  Acta Materialia 51, 2355--2369.

\bibitem[{Segurado et~al.(2012)Segurado, Lebensohn, Llorca, and Tom\'e}]{SLL12}
Segurado, J., Lebensohn, R., Llorca, J., Tom\'e, C., 2012. Multiscale modeling
  of plasticity based on embedding the viscoplastic self-consistent formulation
  in implicit finite elements. International Journal of Plasticity 28,
  124--140.

\bibitem[{Segurado and LLorca(2002)}]{SL02}
Segurado, J., LLorca, J., 2002. A numerical approximation to the elastic
  properties of sphere-reinforced composites. Journal of the Mechanics and
  Physics of Solids 50, 2107--2121.

\bibitem[{Segurado and LLorca(2005)}]{SL05}
Segurado, J., LLorca, J., 2005. Computational micromechanics of composites: The
  effect of particle spatial distribution. Mechanics of Materials 38, 873--883.

\bibitem[{Segurado and LLorca(2013)}]{SL13}
Segurado, J., LLorca, J., 2013. Simulation of the deformation of
  polycrystalline nanostructured {Ti} by computational homogenization.
  Computational Materials Science 76, 3--11.

\bibitem[{Sencer et~al.(2005)Sencer, Maloy, and Gray}]{SMG05}
Sencer, B.~H., Maloy, S.~A., Gray, G.~T., 2005. The influence of shock-pulse
  shape on the structure/property behavior of copper and 316l austenitic
  stainless steel. Acta Materialia 53, 3293 -- 3303.

\bibitem[{Shahba and Ghosh(2016)}]{SG16}
Shahba, A., Ghosh, S., 2016. Crystal plasticity {FE modeling of Ti} alloys for
  a range of strain-rates. {Part I: A} unified constitutive model and flow
  rule. International Journal of Plasticity 87, 48 -- 68.

\bibitem[{Shenoy(2006)}]{Shenoy2006}
Shenoy, M., 2006. Constitutive Modeling and Life Prediction in Ni-Base
  Superalloys. Ph D thesis, Georgia Institute of Technology.

\bibitem[{Shenoy et~al.(2007)Shenoy, Zhang, and McDowell}]{Shenoy2007}
Shenoy, M., Zhang, J., McDowell, D., 2007. Estimating fatigue sensitivity to
  polycrystalline {Ni}-base superalloy microstructures using a computational
  approach. Fatigue and Fracture of Engineering Materials and Structures 30,
  889--904.

\bibitem[{Shu and Fleck(1999)}]{SHU1999297}
Shu, J.~Y., Fleck, N.~A., 1999. Strain gradient crystal plasticity:
  size-dependent deformation of bicrystals. Journal of the Mechanics and
  Physics of Solids 47, 297 -- 324.

\bibitem[{Song et~al.(2005)Song, Lee, and Kim}]{SLK05}
Song, Y.~S., Lee, M.~R., Kim, J.~T., 2005. Effect of grain size for the tensile
  strength and the low cycle fatigue at elevated temperature of alloy 718
  cogged by open die forging press. In: Loria, E.~A. (Ed.), Superalloys 718,
  625, 706 and Derivatives. TMS, pp. 539--549.

\bibitem[{Spowart(2006)}]{S06}
Spowart, J.~E., 2006. Automated serial sectioning for 3-d analysis of
  microstructures. Scripta Materialia 55, 5--10.

\bibitem[{Spowart et~al.(2003)Spowart, Mullens, and Puchala}]{SMP03}
Spowart, J.~E., Mullens, H.~M., Puchala, B.~T., 2003. Collecting and analyzing
  microstructures in three dimensions: a fully automated approach. JOM 55,
  35--37.

\bibitem[{Staroselsky and Anand(2003)}]{SA03}
Staroselsky, A., Anand, L., 2003. A constitutive model for hcp materials
  deforming by slip and twinning: application to magnesium alloy az31b.
  International Journal of Plasticity 19, 1843 -- 1864.

\bibitem[{Stelmashenko et~al.(1993)Stelmashenko, Walls, Brown, and
  Milman}]{STELMASHENKO19932855}
Stelmashenko, N.~A., Walls, M.~G., Brown, L.~M., Milman, Y., 1993.
  Microindentations on {W and Mo} oriented single crystals: An {STM} study.
  Acta Metallurgica et Materialia 41, 2855 -- 2865.

\bibitem[{Stinville et~al.(2016)Stinville, Lenthe, Miao, and Pollock}]{SLM16}
Stinville, J., Lenthe, W., Miao, J., Pollock, T., 2016. A combined grain scale
  elastic–plastic criterion for identification of fatigue crack initiation
  sites in a twin containing polycrystalline nickel-base superalloy. Acta
  Materialia 103, 461 -- 473.

\bibitem[{Stinville et~al.(2015)Stinville, Vanderesse, Bridier, Bocher, and
  Pollock}]{SVB15}
Stinville, J.~C., Vanderesse, N., Bridier, F., Bocher, P., Pollock, T.~M.,
  2015. High resolution mapping of strain localization near twin boundaries in
  a nickel-based superalloy. Acta Materialia 98, 29 -- 42.

\bibitem[{Stukowski et~al.(2015)Stukowski, Cereceda, Swinburne, and
  Marian}]{Stukowski2015108}
Stukowski, A., Cereceda, D., Swinburne, T.~D., Marian, J., 2015.
  Thermally-activated non-schmid glide of screw dislocations in w using
  atomistically-informed kinetic monte carlo simulations. International Journal
  of Plasticity 65, 108 -- 130.

\bibitem[{Su et~al.(2016)Su, Zambaldi, Mercier, Eisenlohr, Bieler, and
  Crimp}]{SZM16}
Su, Y., Zambaldi, C., Mercier, D., Eisenlohr, P., Bieler, T., Crimp, M., 2016.
  Quantifying deformation processes near grain boundaries in α titanium using
  nanoindentation and crystal plasticity modeling. International Journal of
  Plasticity 86, 170 -- 186.

\bibitem[{Sulsky et~al.(1995)Sulsky, Zhou, and Schreyer}]{SZS95}
Sulsky, D., Zhou, S.~J., Schreyer, H.~L., 1995. Application of a
  particle-in-cell method to solid mechanics. Computational Physics
  Communications 87, 236--252.

\bibitem[{Suresh(2012)}]{S12}
Suresh, S., 2012. Fatigue of Materials. Cambridge University Press.

\bibitem[{Sweeney et~al.(2012)Sweeney, McHugh, McGarry, and Leen}]{Sweeney2012}
Sweeney, C.~A., McHugh, P.~E., McGarry, J.~P., Leen, S.~B., 2012.
  Micromechanical methodology for fatigue in cardiovascular stents.
  International Journal of Fatigue 44, 202 -- 216.

\bibitem[{Sweeney et~al.(2014{\natexlab{a}})Sweeney, O'Brien, McHugh, and
  Leen}]{Sweeney2014}
Sweeney, C.~A., O'Brien, B., McHugh, P.~E., Leen, S.~B., 2014{\natexlab{a}}.
  Experimental characterisation for micromechanical modelling of {CoCr} stent
  fatigue. Biomaterials 35, 36 -- 48.

\bibitem[{Sweeney et~al.(2015)Sweeney, O׳Brien, Dunne, McHugh, and
  Leen}]{Sweeney2015}
Sweeney, C.~A., O׳Brien, B., Dunne, F. P.~E., McHugh, P.~E., Leen, S.~B.,
  2015. Micro-scale testing and micromechanical modelling for high cycle
  fatigue of {CoCr} stent material. Journal of the Mechanical Behavior of
  Biomedical Materials 46, 244 -- 260.

\bibitem[{Sweeney et~al.(2014{\natexlab{b}})Sweeney, O'Brien, Dunne, McHugh,
  and Leen}]{SWEENEY2014341}
Sweeney, C.~A., O'Brien, B., Dunne, F. P.~E., McHugh, P.~E., Leen, S.~B.,
  2014{\natexlab{b}}. Strain-gradient modelling of grain size effects on
  fatigue of cocr alloy. Acta Materialia 78, 341 -- 353.

\bibitem[{Sweeney et~al.(2013)Sweeney, Vorster, Leen, Sakurada, McHugh, and
  Dunne}]{Sweeney2013}
Sweeney, C.~A., Vorster, W., Leen, S.~B., Sakurada, E., McHugh, P.~E., Dunne,
  F. P.~E., 2013. The role of elastic anisotropy, length scale and
  crystallographic slip in fatigue crack nucleation. Journal of the Mechanics
  and Physics of Solids 61, 1224 -- 1240.

\bibitem[{Tanaka and Mura(1981)}]{TanakaMura1981}
Tanaka, K., Mura, T., 1981. A dislocation model for fatigue crack initiation.
  ASME Journal of Applied Mechanics 48, 97--103.

\bibitem[{Tandon and Weng(1988)}]{TW88}
Tandon, G., Weng, G., 1988. A theory of particle-reinforced plasticity. Journal
  of Applied Mechanics 55, 126--135.

\bibitem[{Taylor(1938)}]{T38}
Taylor, G.~I., 1938. Plastic strain in metals. Journal of the Institute of
  Metals 62, 307 -- 324.

\bibitem[{Taylor and Elam(1923)}]{Taylor1923}
Taylor, G.~I., Elam, C.~F., 1923. The distortion of an aluminium crystal during
  a tensile test. Proceedings of the Royal Society of London A: Mathematical,
  Physical and Engineering Sciences 102, 643--667.

\bibitem[{Taylor and Elam(1925)}]{Taylor1928}
Taylor, G.~I., Elam, C.~F., 1925. The plastic extension and fracture of
  aluminium crystals. Proceedings of the Royal Society of London A:
  Mathematical, Physical and Engineering Sciences 108, 28--51.

\bibitem[{Thomason(1990)}]{T90}
Thomason, P.~F., 1990. Ductile Fracture of Metals. Pergamon Press.

\bibitem[{Tom\'e(1999)}]{T99}
Tom\'e, C., 1999. Self-consistent polycrystal models: a directional compliance
  criterion to describe grain interactions. Modelling and Simulation in
  Material Science and Engineering 7, 723--728.

\bibitem[{Tom\'e et~al.(1984)Tom\'e, Canova, Kocks, Christodoulou, and
  Jonas}]{Tome1984}
Tom\'e, C., Canova, G.~R., Kocks, U.~F., Christodoulou, N., Jonas, J.~J., 1984.
  The relation between macroscopic and microscopic strain hardening in f.c.c.
  polycrystals. Acta Metallurgica 32~(10), 1637 -- 1653.

\bibitem[{Tom\'e et~al.(1991)Tom\'e, Lebensohn, and Kocks}]{TOME19912667}
Tom\'e, C., Lebensohn, R., Kocks, U., 1991. A model for texture development
  dominated by deformation twinning: Application to zirconium alloys. Acta
  Metallurgica et Materialia 39~(11), 2667 -- 2680.

\bibitem[{Tom\'e et~al.(2001{\natexlab{a}})Tom\'e, Agnew, and
  Yoo}]{AZ31_Tome_Agnew}
Tom\'e, C.~N., Agnew, S.~R., Yoo, M.~H., 2001{\natexlab{a}}. Application of
  texture simulation to understanding mechanical behavior of {Mg} and solid
  solution alloy containing {Li} or {Y}. Acta Materialia 49, 4277--4289.

\bibitem[{Tom\'e et~al.(2002)Tom\'e, Lebensohn, , and Necker}]{TLN02}
Tom\'e, C.~N., Lebensohn, R.~A., , Necker, C.~T., 2002. Orientation
  correlations and anisotropy of recrystallized aluminum. Metallurgical and
  Materials Transactions A 33, 2635 -- 2648.

\bibitem[{Tom\'e et~al.(2001{\natexlab{b}})Tom\'e, Maudlin, Lebensohn, and
  Kaschner}]{TML01}
Tom\'e, C.~N., Maudlin, P.~J., Lebensohn, R.~A., Kaschner, G.~C.,
  2001{\natexlab{b}}. Mechanical response of zirconium: {I. D}erivation of a
  polycrystal constitutive law and finite element analysis. Acta Materialia 49,
  3085--3096.

\bibitem[{Torquato(2001)}]{T01}
Torquato, S., 2001. Random heterogeneous materials. Springer.

\bibitem[{Totry et~al.(2008)Totry, Gonz\'alez, and LLorca}]{TGL08}
Totry, E., Gonz\'alez, C., LLorca, J., 2008. Failure locus of fiber-reinforced
  composites under transverse compression and out-of-plane shear. Composites
  Science and Technology 68, 829--839.

\bibitem[{Tucker et~al.(2015)Tucker, {Cerrone III}, Ingraffea, and
  Rollett}]{TCI15}
Tucker, J.~C., {Cerrone III}, A.~R., Ingraffea, A.~R., Rollett, A.~D., 2015.
  Crystal plasticity finite element analysis for rené88dt statistical volume
  element generation. Modelling and Simulation in Materials Science and
  Engineering 23, 035003.

\bibitem[{Uchic et~al.(2004)Uchic, Groeber, Wheeler, Scheltens, and
  Dimiduk}]{UGW04}
Uchic, M.~D., Groeber, M., Wheeler, R., Scheltens, F., Dimiduk, D., 2004.
  Augmenting the 3d characterization capability of the dual beam {FIB-SEM}.
  Microscopy and Microanalysis 10, 1136 -- 1137.

\bibitem[{Uchic et~al.(2006)Uchic, Groeber, Dimiduk, and Simmons}]{UGD06}
Uchic, M.~D., Groeber, M.~A., Dimiduk, D.~M., Simmons, J.~P., 2006. 3d
  microstructural characterization on nickel superalloys via serial-sectioning
  using a dual-beam {FIB-SEM}. Scripta Materialia 55, 23--28.

\bibitem[{Upadhyay et~al.(2016)Upadhyay, Capolungo, Taupin, Fressengeas, and
  Lebensohn}]{UCT16}
Upadhyay, M.~V., Capolungo, L., Taupin, V., Fressengeas, C., Lebensohn, R.~A.,
  2016. A higher order elasto-viscoplastic model using fast fourier transforms:
  effects of lattice curvatures on mechanical response of nanocrystalline
  metals. International Journal of Plasticity 83, 126--152.

\bibitem[{Vachhani and Kalidindi(2015)}]{VK15}
Vachhani, S.~J., Kalidindi, S.~R., 2015. Grain-scale measurement of slip
  resistances in aluminum polycrystals using spherical nanoindentation. Acta
  Materialia 60, 27--36.

\bibitem[{{Van Houtte} et~al.(2006){Van Houtte}, Kanjarla, {Van Bael},
  Seefeldt, and Delannay}]{VKV06}
{Van Houtte}, P., Kanjarla, A.~K., {Van Bael}, A., Seefeldt, M., Delannay, L.,
  2006. Multiscale modelling of the plastic anisotropy and deformation texture
  of polycrystalline materials. European Journal of Mechanics A/Solids 25,
  634--648.

\bibitem[{{Van Houtte} et~al.(2009){Van Houtte}, Yerra, and {Van Bael}}]{VYV09}
{Van Houtte}, P., Yerra, S.~K., {Van Bael}, A., 2009. The {F}acet method: A
  hierarchical multilevel modelling scheme for anisotropic convex plastic
  potentials. International Journal of Plasticity 25, 332--360.

\bibitem[{Venkatramani et~al.(2007)Venkatramani, Ghosh, and Mills}]{VGM07}
Venkatramani, G., Ghosh, S., Mills, M., 2007. A size-dependent crystal
  plasticity finite-element model for creep and load shedding in
  polycrystalline titanium alloys. Acta Materialia 55, 3971 -- 3986.

\bibitem[{Vondrejc et~al.(2014)Vondrejc, Zeman, and Marek}]{VONDREJC2014156}
Vondrejc, J., Zeman, J., Marek, I., 2014. An fft-based galerkin method for
  homogenization of periodic media. Computers and Mathematics with Applications
  68, 156 -- 173.

\bibitem[{Vondrous et~al.(2015)Vondrous, Bienger, Schreij{\"a}g, Selzer,
  Schneider, Nestler, Helm, and M{\"o}nig}]{VBS15}
Vondrous, A., Bienger, P., Schreij{\"a}g, S., Selzer, M., Schneider, D.,
  Nestler, B., Helm, D., M{\"o}nig, R., 2015. Combined crystal plasticity and
  phase-field method for recrystallization in a process chain of sheet metal
  production. Computational Mechanics 55, 439--452.

\bibitem[{Wan et~al.(2016{\natexlab{a}})Wan, MacLachlan, and Dunne}]{WMD16}
Wan, V., MacLachlan, D., Dunne, F., 2016{\natexlab{a}}. Integrated experiment
  and modelling of microstructurally-sensitive crack growth. International
  Journal of Fatigue 91, 110 -- 123.

\bibitem[{Wan et~al.(2016{\natexlab{b}})Wan, Jiang, MacLachlan, and
  Dunne}]{Wan2016}
Wan, V. V.~C., Jiang, J., MacLachlan, D.~W., Dunne, F. P.~E.,
  2016{\natexlab{b}}. Microstructure-sensitive fatigue crack nucleation in a
  polycrystalline ni superalloy. International Journal of Fatigue 90, 181 --
  190.

\bibitem[{Wang et~al.(2014)Wang, Sandlobes, Diehl, Sharma, Roters, and
  Raabe}]{WSD14}
Wang, F., Sandlobes, S., Diehl, M., Sharma, L., Roters, F., Raabe, D., 2014. In
  situ observation of collective grain-scale mechanics in {M}g and {M}g-rare
  earth alloys. Acta Materialia 80, 77 -- 93.

\bibitem[{Wen et~al.(2016)Wen, Borodachenkova, Tom\'{e}, Vincze, Rauch, Barlat,
  and Gracio}]{WBT16}
Wen, W., Borodachenkova, M., Tom\'{e}, C.~N., Vincze, G., Rauch, E.~F., Barlat,
  F., Gracio, J.~J., 2016. Mechanical behavior of {Mg} subjected to strain path
  changes: Experiments and modeling. International Journal of Plasticity 73,
  171--183.

\bibitem[{Wenk et~al.(1997)Wenk, Canova, Brechet, and Flandin}]{WCB97}
Wenk, H.~R., Canova, G.~R., Brechet, Y., Flandin, L., 1997. A deformation-based
  model for recrystallization of anisotropic materials. Acta Materialia 45,
  3283 -- 3296.

\bibitem[{Werwer and Cornec(2000)}]{WC00}
Werwer, M., Cornec, A., 2000. Numerical simulation of plastic deformation and
  fracture in polysynthetically twinned {(PST) crystals of TiAl}. Computational
  Materials Science 19, 97--107.

\bibitem[{Willis(1981)}]{W81}
Willis, J.~R., 1981. Variational and related methods for the overall properties
  of composites. Advanced Applied Mechanics 21, 1--78.

\bibitem[{Willot(2015)}]{W15}
Willot, F., 2015. Fourier-based schemes for computing the mechanical response
  of composites with accurate local fields. Comptes Rendus M\'{e}canique 34,
  232--245.

\bibitem[{Wu et~al.(2010)Wu, Tom\'e, Agnew, Raeisinia, and Wang}]{AZ31_Tome}
Wu, P.~D., Tom\'e, C.~N., Agnew, S.~R., Raeisinia, B., Wang, H., 2010.
  Evaluation of self-consistent polycrystal plasticity models for magnesium
  alloy {AZ31B} sheet. International Journal of Solids and Structures 47,
  2905--2917.

\bibitem[{Xue et~al.(1997)Xue, Righetti, Telley, Liebling, and
  Mocellin}]{Xin1997Philo}
Xue, X., Righetti, F., Telley, H., Liebling, T.~M., Mocellin, A., 1997. The {
  L}aguerre model for grain growth in three dimensions. Philosophical Magazine
  Part B 75, 567--585.

\bibitem[{Yasi et~al.(2009)Yasi, Nogaret, Trinkle, Qi, {Hector Jr}, and
  Curtin}]{YasiCurtin2009}
Yasi, J.~A., Nogaret, T., Trinkle, D.~R., Qi, Y., {Hector Jr}, L.~G., Curtin,
  W.~A., 2009. Basal and prism dislocation cores in magnesium: comparison of
  first-principles and embedded-atom-potential methods predictions. Modelling
  and Simulation in Materials Science and Engineering 17, 055012.

\bibitem[{Yefimov et~al.(2004)Yefimov, der Giessen, and Groma}]{yefimov2004}
Yefimov, S., der Giessen, E.~V., Groma, I., 2004. Bending of a single crystal:
  discrete dislocation and nonlocal crystal plasticity simulations. Modelling
  and Simulation in Materials Science and Engineering 12, 1069--1086.

\bibitem[{Yeratapally et~al.(2016)Yeratapally, Glavicic, Hardy, and
  Sangid}]{YGH16}
Yeratapally, S.~R., Glavicic, M.~G., Hardy, M., Sangid, M.~D., 2016.
  Microstructure based fatigue life prediction framework for polycrystalline
  nickel-base superalloys with emphasis on the role played by twin boundaries
  in crack initiation. Acta Materialia 107, 152 -- 167.

\bibitem[{Zambaldi et~al.(2012)Zambaldi, Yang, Bieler, and Raabe}]{ZYB12}
Zambaldi, C., Yang, Y., Bieler, T.~R., Raabe, D., 2012. Orientation informed
  nanoindentation of ?-titanium: indentation pileup in hexagonal metals
  deforming by prismatic slip. Journal of Materials Research 27, 356 -- 367.

\bibitem[{Zambaldi et~al.(2015)Zambaldi, Zehnder, and Raabe}]{ZZR15}
Zambaldi, C., Zehnder, C., Raabe, D., 2015. Orientation dependent deformation
  by slip and twinning in magnesium during single crystal indentation. Acta
  Materialia 91, 267 -- 288.

\bibitem[{Zecevic et~al.(2017{\natexlab{a}})Zecevic, Beyerlein, and
  Knezevic}]{ZBK17}
Zecevic, M., Beyerlein, I.~J., Knezevic, M., 2017{\natexlab{a}}. Coupling
  elasto-plastic self-consistent crystal plasticity and implicit finite
  elements: Applications to compression, cyclic tension-compression, and
  bending to large strains. International Journal of Plasticity 93, 187 -- 211.

\bibitem[{Zecevic et~al.(2017{\natexlab{b}})Zecevic, Pantleon, Lebensohn,
  {McC}abe, and Knezevic}]{ZPL17}
Zecevic, M., Pantleon, W., Lebensohn, R.~A., {McC}abe, R.~J., Knezevic, M.,
  2017{\natexlab{b}}. Predicting intragranular misorientation distributions in
  polycrystalline metals using the viscoplastic self-consistent formulation.
  Acta Materialia 140, 398--410.

\bibitem[{Zeman et~al.(2017)Zeman, de~Geus, Vondrejc, Peerlings, and
  Geers}]{Geers2017}
Zeman, J., de~Geus, T. W.~J., Vondrejc, J., Peerlings, R. H.~J., Geers, M.
  G.~D., 2017. A finite element perspective on nonlinear {FFT}-based
  micromechanical simulations. International Journal for Numerical Methods in
  Engineering 110, 903--926.

\bibitem[{Zeman et~al.(2010)Zeman, Vond{\v{r}}ejc, Nov{\'a}k, and
  Marek}]{zeman2010accelerating}
Zeman, J., Vond{\v{r}}ejc, J., Nov{\'a}k, J., Marek, I., 2010. Accelerating a
  fft-based solver for numerical homogenization of periodic media by conjugate
  gradients. Journal of Computational Physics 229~(21), 8065--8071.

\bibitem[{Zhang et~al.(2015)Zhang, Li, Eisenlohr, Liu, Boehlert, Crimp, and
  Bieler}]{ZLE15}
Zhang, C., Li, H., Eisenlohr, P., Liu, W., Boehlert, C.~J., Crimp, M.~A.,
  Bieler, T.~R., 2015. Effect of realistic 3d microstructure in crystal
  plasticity finite element analysis of polycrystalline {Ti-5Al-2.5Sn}.
  International Journal of Plasticity 69, 21 -- 35.

\bibitem[{Zhang and Joshi(2012)}]{AZ31_Joshi}
Zhang, J., Joshi, S.~P., 2012. Phenomenological crystal plasticity modeling an
  detailed micromechanical investigations of pure magnesium. Journal of the
  Mechanics and Physics of Solids 60, 945--972.

\end{thebibliography}

\end{document}